\documentclass[10pt,twoside]{article}
\usepackage[latin1]{inputenc}
\usepackage{graphicx,amsmath,amsfonts,amssymb,mathrsfs}
\usepackage{color}
\usepackage[english]{babel}

\def\YEAR{\year}
\newcount\VOL\VOL=\YEAR\advance\VOL by-1995
\def\firstpage{1}
\def\lastpage{1000}
\def\received{}
\def\revised{}
\def\communicated{}
\makeatletter
\def\magnification{\afterassignment\m@g\count@}
\def\m@g{\mag=\count@\hsize6.5truein\vsize8.9truein\dimen\footins8truein}
\makeatother
\font\eightrm=cmr8
\font\caps=cmcsc10

\font\Caps=cmcsc10 scaled \magstep1

\def\bfseries{\normalsize\caps}
\makeatletter
\@specialpagefalse\headheight=8.5pt
\def\DocMath{}
\renewcommand{\@evenhead}{    \ifnum\thepage>\lastpage\rlap{\thepage}\hfill    \else\rlap{\thepage}\slshape\leftmark\hfill{\caps\SAuthor}\hfill\fi}
\renewcommand{\@oddhead}{    \ifnum\thepage=\firstpage{\DocMath\hfill\llap{\thepage}}    \else{\slshape\rightmark}\hfill{\caps\STitle}\hfill\llap{\thepage}\fi}
\makeatother
\def\TSkip{\bigskip}
\newbox\TheTitle{\obeylines\gdef\GetTitle #1
\ShortTitle  #2
\SubTitle    #3
\Author      #4
\ShortAuthor #5
\EndTitle
{\setbox\TheTitle=\vbox{\baselineskip=20pt\let\par=\cr\obeylines\halign{\centerline{\Caps##}\cr\noalign{\medskip}\cr#1\cr}}
\copy\TheTitle\TSkip\TSkip\def\next{#2}\ifx\next\empty\gdef\STitle{#1}\else\gdef\STitle{#2}\fi\def\next{#3}\ifx\next\empty    \else\setbox\TheTitle=\vbox{\baselineskip=20pt\let\par=\cr\obeylines    \halign{\centerline{\caps##} #3\cr}}\copy\TheTitle\TSkip\TSkip\fi\centerline{\caps #4}\TSkip\TSkip\def\next{#5}\ifx\next\empty\gdef\SAuthor{#4}\else\gdef\SAuthor{#5}\fi\ifx\received\empty\relax
    \else\centerline{\eightrm Received: \received}\fi\ifx\revised\empty\TSkip    \else\centerline{\eightrm Revised: \revised}\TSkip\fi\ifx\communicated\empty\relax
    \else\centerline{\eightrm Communicated by \communicated}\fi\TSkip\TSkip\catcode'015=5}}
\def\Title{\obeylines\GetTitle}
\def\Abstract{\begingroup\narrower
    \parskip=\medskipamount\parindent=0pt{\caps Abstract. }}
\def\EndAbstract{\par\endgroup\TSkip}
\long
\def\MSC#1\EndMSC{\def\arg{#1}\ifx\arg\empty\relax\else
     {\par\narrower\noindent     2000 Mathematics Subject Classification: #1\par}\fi}
\long
\def\KEY#1\EndKEY{\def\arg{#1}\ifx\arg\empty\relax\else
{\par\narrower\noindent Keywords and Phrases: #1\par}\fi\TSkip}
\newbox\TheAdd
\def\Addresses{\vfill\copy\TheAdd\vfill
    \ifodd\number\lastpage\vfill\eject\phantom{.}\vfill\eject\fi}
{\obeylines\gdef\GetAddress #1
\Address #2
\Address #3
\Address #4
\EndAddress
{
\def\xs{4.3truecm}
\parindent=0pt
\setbox0=\vtop{{\obeylines\hsize=\xs#1\par}}
\def\next{#2}
\ifx\next\empty
     \setbox\TheAdd=\hbox to\hsize{\hfill\copy0\hfill}
\else\setbox1=\vtop{{\obeylines\hsize=\xs#2\par}}
\def\next{#3}
\ifx\next\empty
     \setbox\TheAdd=\hbox to\hsize{\hfill\copy0\hfill\copy1\hfill}
\else\setbox2=\vtop{{\obeylines\hsize=\xs#3\par}}
\def\next{#4}
\ifx\next\empty\
     \setbox\TheAdd=\vtop{\hbox to\hsize{\hfill\copy0\hfill\copy1\hfill}
                \vskip20pt\hbox to\hsize{\hfill\copy2\hfill}}
\else\setbox3=\vtop{{\obeylines\hsize=\xs#4\par}}
     \setbox\TheAdd=\vtop{\hbox to\hsize{\hfill\copy0\hfill\copy1\hfill}
        \vskip20pt\hbox to\hsize{\hfill\copy2\hfill\copy3\hfill}}
\fi\fi\fi\catcode'015=5}}\gdef\Address{\obeylines\GetAddress}
\renewcommand{\thesection}{\Roman{section}}
\renewcommand{\theequation}{\thesection.\arabic{equation}}

\newtheorem{theorem}{Theorem}
\newtheorem{lemma}[theorem]{Lemma}
\newtheorem{corollary}[theorem]{Corollary}

\newtheorem{remark}[theorem]{Remark}
\newtheorem{proposition}[theorem]{Proposition}
\input{tcilatex}
\begin{document}

\Title
Diagonalizing Quadratic Bosonic Operators by Non-Autonomous Flow Equation
\ShortTitle
Diagonalizing Quadratic Bosonic Operators
\SubTitle
\Author
Volker Bach and Jean-Bernard Bru
\ShortAuthor
V. Bach and J.-B. Bru
\EndTitle
\Abstract
We study a non--autonomous, non-linear evolution equation on the space of operators on a complex Hilbert space. We specify assumptions that ensure the global existence of its solutions
and allow us to derive its asymptotics at temporal infinity. We demonstrate that these assumptions are optimal in a suitable sense and more general than those used before. The evolution equation derives from the Brocket--Wegner flow that was proposed to diagonalize matrices and operators by a strongly continuous unitary flow. In fact, the solution of the non--linear flow equation leads to a diagonalization of Hamiltonian operators in boson quantum field theory which are quadratic in the field.
\EndAbstract
\MSC
47D06, 81Q10 \\
\EndMSC
\KEY
Quadratic operators, Flow equations for operators,
Evolution equations, Brocket--Wegner flow, Double bracket flow.
\EndKEY

\tableofcontents 
%

\section{Introduction}

\setcounter{equation}{0}%
In theoretical physics, the second quantization formalism is crucial to
treat many--particle problems. In the first quantization formalism, i.e., in
the canonical ensemble, the number of particles of the corresponding wave
function stays fixed. Whereas in second quantization, i.e., in the
grand--canonical ensemble, the particle number is not fixed and their boson
or fermion statistics is incorporated in the well--known
creation/annihilation operators acting on Fock space. Within this latter
framework, we are interested in boson systems, the simplest being the
perfect Bose gas, i.e., a system with no interaction defined by a particle
number--conserving\ quadratic Hamiltonian. By quadratic Hamiltonians we
refer to self--adjoint operators which are quadratic in the creation and
annihilation operators. It is known since Bogoliubov and his celebrated
theory of superfluidity \cite{Bogoliubov1} that such quadratic operators are
reduced to a perfect gas by a suitable unitary transformation, see also \cite%
[Appendix B.2]{BruZagrebnov8}. See also \cite{Williamson} with discussions
of \cite{Hormander} in the finite dimensional case.

Diagonalizations of quadratic boson operators are generally not trivial, and
in this paper, we investigate this question under weaker conditions as
before. Indeed, after Bogoliubov with his $\mathbf{u}$--$\mathbf{v}$ unitary
transformation \cite{Bogoliubov1,BruZagrebnov8}, general results on this
problem have been obtained for real quadratic operators with bounded
one--particle spectrums by Friedrichs \cite[Part V]{Friedrichs}, Berezin
\cite[Theorem 8.1]{Berezin}, Kato and Mugibayashi \cite[Theorem 2]%
{Kato-Mugibayashi}. In the present paper we generalize previous analyses of
real quadratic boson Hamiltonians with positive one--particle spectrum
bounded above and below away from zero to complex, unbounded one--particle
operators without a gap above zero. This generalization is obviously
important since, for most physically interesting applications, the
one--particle spectrum is neither bounded above nor bounded away from zero.
Moreover, our proof is completely different. Its mathematical novelty lies
in the use of\textit{\ }\emph{non--autonomous evolution equations} as a key
ingredient.\

More specifically, we employ the Brocket--Wegner flow, originally proposed
by Brockett for symmetric matrices in 1991 \cite{Brockett1} and, in a
different variant, by Wegner in 1994 for self--adjoint operators \cite%
{Wegner1}. This flow leads to unitarily equivalent operators via a
non--autonomous hyperbolic evolution equation. The mathematical foundation
of such flows \cite{Brockett1,Wegner1} has recently been given in \cite%
{bach-bru}. Unfortunately, the results \cite{bach-bru} do not apply to the
models. In this paper we prove the well--posedness of the Brocket--Wegner
flow $\partial _{t}\mathrm{H}_{t}=\big[\mathrm{H}_{t},[\mathrm{H}_{t},%
\mathrm{N}]\big]$, where for $t\geq 0$, $[\mathrm{H}_{t},\mathrm{N}]:=%
\mathrm{H}_{t}\mathrm{N}-\mathrm{NH}_{t}$ is the commutator of a quadratic
Hamiltonian $\mathrm{H}_{t}$ and the particle number operator $\mathrm{N}$
acting on the boson Fock space. Establishing well--posedness is non--trivial
here because the Brocket--Wegner flow is a (quadratically) non--linear
first--order differential equation for unbounded operators. It is solved by
an auxiliary non--autonomous parabolic evolution equation.

Indeed, non--autonomous evolution equations turn out to be crucial at two
different stages of our proof:

\begin{itemize}
\item[(a)] To show for $t\geq 0$ the well--posedness of the Brocket--Wegner
flow $\partial _{t}\mathrm{H}_{t}=\big[\mathrm{H}_{t},[\mathrm{H}_{t},%
\mathrm{N}]\big]$ for a quadratic operator $\mathrm{H}_{0}$ via an auxiliary
system of non--linear first--order differential equations for operators;

\item[(b)] To rigorously define a family of unitarily equivalent quadratic
operators $\mathrm{H}_{t}=\mathrm{U}_{t}\mathrm{H}_{0}\mathrm{U}_{t}^{\ast }$
as a consequence of the Brocket--Wegner flow.
\end{itemize}

\noindent To be more specific, (a) uses the theory of non--autonomous
\textit{parabolic} evolution equations via an auxiliary systems of
non--linear first--order differential equation for operators. Whereas the
second step (b) uses the theory of non--autonomous \textit{hyperbolic}
evolution equations to define the unitary operator $\mathrm{U}_{t}$ by $%
\mathrm{U}_{0}:=\mathbf{1}$ and $\partial _{t}\mathrm{U}_{t}=-i\mathrm{G}_{t}%
\mathrm{U}_{t}$, for all $t>0$, with generator $\mathrm{G}_{t}:=i\left[
\mathrm{N},\mathrm{H}_{t}\right] $. For bounded generators, the existence,
uniqueness and even an explicit form of their solution is given by the Dyson
series, as it is well--known. It is much more delicate for unbounded
generators, which is what we are dealing with here. It has been studied,
after the first result of Kato in 1953 \cite{Kato1953}, for decades by many
authors (Kato again \cite{Kato,Kato1973} but also Yosida, Tanabe, Kisynski,
Hackman, Kobayasi, Ishii, Goldstein, Acquistapace, Terreni, Nickel,
Schnaubelt, Caps, Tanaka, Zagrebnov, Neidhardt), see, e.g., \cite%
{Katobis,Caps,Schnaubelt1,Pazy,Neidhardt-zagrebnov} and the corresponding
references cited therein. Yet, no unified theory of such linear evolution
equations that gives a complete characterization analogously to the
Hille--Yosida generation theorems is known. By using the Yosida
approximation, we simplify Ishii's proof \cite{Ishii1,Ishii2} and obtain the
well--posedness of this Cauchy problem in the hyperbolic case in order to
define the unitary operator $\mathrm{U}_{t}$.

Next, by taking the limit $t\rightarrow \infty $ of unitarily equivalent
quadratic operators $\mathrm{H}_{t}=\mathrm{U}_{t}\mathrm{H}_{0}\mathrm{U}%
_{t}^{\ast }$, under suitable conditions on $\mathrm{H}_{0}$, we demonstrate
that the limit operator $\mathrm{H}_{\infty }$ is also unitarily equivalent
to $\mathrm{H}_{0}$. This is similar to scattering in quantum field theory
since we analyze the strong limit $\mathrm{U}_{\infty }$ of the unitary
operator $\mathrm{U}_{t}$, as $t\rightarrow \infty $. The limit operator $%
\mathrm{H}_{\infty }=\mathrm{U}_{\infty }\mathrm{H}_{0}\mathrm{U}_{\infty
}^{\ast }$ is a quadratic boson Hamiltonian which commutes with the particle
number operator $\mathrm{N}$, i.e., $[\mathrm{H}_{\infty },\mathrm{N}]=0$ --
a fact which we refer to as $\mathrm{H}_{\infty }$ being $\mathrm{N}$%
--diagonal. In particular, it can be diagonalized by a unitary on the
one--particle Hilbert space, only. Consequently, we provide in this paper a
new mathematical application of evolution equations as well as some general
results on quadratic operators, which are also interesting for mathematical
physicists.

The paper is structured as follows. In Section \ref{Section diagonalization}%
, we present our results and discuss them in the context of previously known
facts. Section \ref{Section evolution equation_our results} contains a
guideline to our approach in terms of theorems, whereas Section \ref{section
illustration} illustrates it on an explicit and concrete case, showing, in
particular, that a pathological behavior of the Brocket--Wegner flow is not
merely a possibility, but does occur. Sections \ref{Section technical proofs}%
--\ref{Section technical proofs copy(1)} are the core of our paper, as all
important proofs are given here. Finally, Section\ \ref{Appendix} is an
appendix with a detailed analysis in Section \ref{Section Non-autonomous
evolution} of evolution equations for unbounded operators of hyperbolic type
on Banach spaces, and with, in Sections \ref{section bogoliubov} and \ref%
{section hilbert schmidt}, some comments on Bogoliubov transformations and
Hilbert--Schmidt operators. In particular, we clarify in Section \ref%
{Section Non-autonomous evolution} Ishii's approach \cite{Ishii1,Ishii2} to
non--autonomous hyperbolic evolution equations.

\section{Diagonalization of Quadratic Boson Hamiltonians\label{Section
diagonalization}}

\setcounter{equation}{0}%
In this section we describe our main results on quadratic boson
Hamiltonians. First, we define quadratic operators in Section \ref{Section
bosonic quadratic operators} and present our findings in Section \ref%
{Section main result} without proofs. The latter is sketched in Section \ref%
{Section evolution equation_our results} and given in full detail in
Sections \ref{Section technical proofs}--\ref{Section technical proofs
copy(1)}. Section \ref{Section historical overview} is devoted to a
historical overview on the diagonalization of quadratic operators.

\subsection{Quadratic Boson Operators\label{Section bosonic quadratic
operators}}

To fix notation, let $\mathfrak{h}:=L^{2}(\mathcal{M})$ be a separable
complex Hilbert space which we assume to be realized as a space of
square--integrable functions on a measure space $(\mathcal{M},\mathfrak{a})$%
. The scalar product on $\mathfrak{h}$ is given by
\begin{equation}
\langle f|g\rangle :=\int_{\mathcal{M}}\overline{f\left( x\right) }g\left(
x\right) \mathrm{d}\mathfrak{a}\left( x\right) \ .  \label{scalar product}
\end{equation}%
For $f\in \mathfrak{h}$, we define its complex conjugate $\bar{f}\in
\mathfrak{h}$ by $\bar{f}\left( x\right) :=\overline{f\left( x\right) }$,
for all $x\in \mathcal{M}$. For any bounded (linear) operator $X$ on $%
\mathfrak{h}$, we define its transpose $X^{\mathrm{t}}$\ and its complex
conjugate $\bar{X}$ by $\langle f|X^{\mathrm{t}}g\rangle :=\langle \bar{g}|X%
\bar{f}\rangle $ and $\langle f|\bar{X}g\rangle :=\overline{\langle \bar{f}|X%
\bar{g}\rangle }$ for $f,g\in \mathfrak{h}$, respectively. Note that the
adjoint of the operator $X$ equals $X^{\ast }=\overline{X^{\mathrm{t}}}=%
\overline{X}^{\mathrm{t}}$, where it exists. The Banach space of bounded
operators acting on $\mathfrak{h}$ is denoted by $\mathcal{B}(\mathfrak{h})$%
, whereas $\mathcal{L}^{1}(\mathfrak{h})$\ and $\mathcal{L}^{2}(\mathfrak{h}%
) $ are the spaces of trace--class and Hilbert--Schmidt operators,
respectively. Norms in $\mathcal{L}^{1}(\mathfrak{h})$ and $\mathcal{L}^{2}(%
\mathfrak{h})$ are respectively denoted by
\begin{equation}
\Vert X\Vert _{1}:=\mathrm{tr}(\left\vert X\right\vert )\ ,\mathrm{\quad
for\ }X\in \mathcal{L}^{1}(\mathfrak{h})\ ,
\end{equation}%
and
\begin{equation}
\Vert X\Vert _{2}:=\sqrt{\mathrm{tr}(X^{\ast }X)}\ ,\mathrm{\quad for\ }X\in
\mathcal{L}^{2}\left( \mathfrak{h}\right) \ .
\end{equation}%
Note that, if there exists a constant $\mathrm{K}<\infty $ such that
\begin{equation}
\left\vert \underset{k=1}{\overset{\infty }{\sum }}\langle \eta _{k}|X\psi
_{k}\rangle \right\vert \leq \mathrm{K}\ ,
\end{equation}%
for all orthonormal bases $\{\eta _{k}\}_{k=1}^{\infty },\{\psi
_{k}\}_{k=1}^{\infty }\subseteq \mathfrak{h}$, then
\begin{equation}
\mathrm{tr}(X)=\underset{k=1}{\overset{\infty }{\sum }}\langle \varphi
_{k}|X\varphi _{k}\rangle
\end{equation}%
for any orthonormal basis $\{\varphi _{k}\}_{k=1}^{\infty }\subseteq
\mathfrak{h}$. Finally, we denote by $\mathbf{1}$ the identity operator on
various spaces. Assume now the following conditions:

\begin{itemize}
\item[A1:] $\Omega _{0}=\Omega _{0}^{\ast }\geq 0$ is a positive operator on
$\mathfrak{h}$.

\item[A2:] $B_{0}=B_{0}^{\mathrm{t}}\in \mathcal{L}^{2}(\mathfrak{h})$ is a
(non--zero) Hilbert--Schmidt operator.

\item[A3:] The operator $\Omega _{0}$ is invertible on $\mathrm{Ran}B_{0}$
and satisfies:
\begin{equation}
\Omega _{0}\geq 4B_{0}(\Omega _{0}^{\mathrm{t}})^{-1}\bar{B}_{0}=4\overline{%
B_{0}^{\ast }\Omega _{0}^{-1}B_{0}}\ .
\end{equation}
\end{itemize}

\noindent Note that neither the operator $\Omega _{0}$ in A1 nor its inverse
$\Omega _{0}^{-1}$ are necessarily bounded. Observe also that $\Omega
_{0}^{2}\geq 4B_{0}\bar{B}_{0}$ implies A3: Use A1--A2, two times $\Omega
_{0}^{2}\geq 4B_{0}\bar{B}_{0}$ in
\begin{equation}
\left( 4B_{0}(\Omega _{0}^{\mathrm{t}})^{-1}\bar{B}_{0}\right) \left(
4B_{0}(\Omega _{0}^{\mathrm{t}})^{-1}\bar{B}_{0}\right) \leq 4B_{0}\bar{B}%
_{0}\leq \Omega _{0}^{2}\ ,
\end{equation}%
and the fact that the map $X\mapsto X^{1/2}$ is operator monotone. However,
the converse does \emph{not} hold, see for instance Remark \ref{remark
conditions copy(3)} in Section \ref{section illustration} where a trivial
example is given.

Next, take some \emph{real} orthonormal basis $\left\{ \varphi _{k}\right\}
_{k=1}^{\infty }$ in the dense domain $\mathcal{D}\left( \Omega _{0}\right)
\subseteq \mathfrak{h}$ of $\Omega _{0}$ and, for any $k\in \mathbb{N}$, let
$a_{k}:=a\left( \varphi _{k}\right) $ be the corresponding boson
annihilation operator acting on the boson Fock space
\begin{equation}
\mathcal{F}_{b}:=\bigoplus_{n=0}^{\infty }\mathcal{S}_{n}\left( \mathfrak{h}%
^{\otimes n}\right) \ ,  \label{Fock}
\end{equation}%
whose scalar product is again denoted by $\left\langle \cdot |\cdot
\right\rangle $. Here, $\mathcal{S}_{n}$ is the orthogonal projection onto
the subspace of totally symmetric $n$--particle wave functions in $\mathfrak{%
h}^{\otimes n}$, the $n$--fold tensor product of $\mathfrak{h}$. Then, for
any fixed $C_{0}\in \mathbb{R}$, the quadratic boson operator is defined
through the operators $\Omega _{0}$ and $B_{0}$ by
\begin{equation}
\mathrm{H}_{0}:=\underset{k,\ell }{\sum }\left\{ \Omega _{0}\right\}
_{k,\ell }a_{k}^{\ast }a_{\ell }+\left\{ B_{0}\right\} _{k,\ell }a_{k}^{\ast
}a_{\ell }^{\ast }+\left\{ \bar{B}_{0}\right\} _{k,\ell }a_{k}a_{\ell
}+C_{0}\ ,  \label{hamilton}
\end{equation}%
with $\left\{ X\right\} _{k,\ell }:=\left\langle \varphi _{k}|X\varphi
_{\ell }\right\rangle $ for all operators $X$ acting on $\mathfrak{h}$. Note
that a continuous integral $\int \mathrm{d}k\ \mathrm{d}\ell $ could also
replace the discrete sum $\sum_{k,\ell }$ in this definition since \emph{all
results also hold in the continuous case}.

We first establish the self--adjointness of $\mathrm{H}_{0}$.

\begin{proposition}[Self--adjointness of quadratic operators]
\label{lemma example 1 copy(1)}\mbox{ }\newline
Under Conditions A1--A2, the quadratic operator $\mathrm{H}_{0}$ is
essentially self--adjoint on the domain
\begin{equation}
\mathcal{D}\left( \mathrm{H}_{0}\right) :=\overset{\infty }{\underset{N=1}{%
\bigcup }}\left( \bigoplus_{n=0}^{N}\mathcal{S}_{n}\left( \mathcal{D}\left(
\Omega _{0}\right) ^{\otimes n}\right) \right) \ .
\end{equation}
\end{proposition}

\noindent Detailed proofs of self--adjointness of quadratic operators are
given in \cite[Thm 6.1]{Berezin} and, more recently and also richer in
detail, in \cite[Thm 5.3]{Bruneau-derezinski2007}. See also \cite%
{Shale1962,Ruijsenaars-1976,Ruijsenaars-1978}.

The aim of this paper is to diagonalize the quadratic Hamiltonian $\mathrm{H}%
_{0}$. To this end, it suffices to find a unitary transformation leading to
a quadratic operator of the form (\ref{hamilton}) with $B_{0}=0$, after
conjugation with this unitary. Such quadratic operators are called $\emph{N}$%
\emph{--diagonal} since they commute with the particle number operator
\begin{equation}
\mathrm{N}:=\underset{k}{\sum }a_{k}^{\ast }a_{k}\ .
\label{particle number operator}
\end{equation}%
In particular, the $\mathrm{N}$--diagonal part of the quadratic Hamiltonian $%
\mathrm{H}_{0}$ equals%
\begin{equation}
\Gamma _{0}:=\underset{k,\ell }{\sum }\left\{ \Omega _{0}\right\} _{k,\ell
}a_{k}^{\ast }a_{\ell }+C_{0}\ ,  \label{hamilton diagonal0}
\end{equation}%
because $[\Gamma _{0},\mathrm{N}]=0$, for any self--adjoint operator $\Omega
_{0}$ acting on $\mathfrak{h}$.

Note that we diagonalize semi--bounded quadratic boson operators. Indeed,
under Conditions A1--A4 (see below), Theorem \ref{flow equation thm 4
copy(1)} implies that the Hamiltonian $\mathrm{H}_{0}$ is bounded from below
by $\mathrm{H}_{0}\geq C_{0}$. See \cite[Thm 5.4 and Cor. 5.1]%
{Bruneau-derezinski2007}. It is a typical situation encountered in quantum
mechanics because the $\mathrm{N}$--diagonal part $\Gamma _{0}$ usually
corresponds to the kinetic energy operator, whereas the non--$\mathrm{N}$%
--diagonal part $\mathrm{W}_{0}$ of $\mathrm{H}_{0}$ represents interactions.

\subsection{Main Results\label{Section main result}}

In this paper, we prove the $\mathrm{N}$--diagonalization of quadratic
operators in a more general setting as before, by using the Brocket--Wegner
flow \cite{Brockett1,Wegner1} for quadratic boson operators. Theories of
non--autonomous evolution equations are crucial to define this flow. For
more details about our approach, we refer to Section \ref{Section evolution
equation_our results}, particularly Section \ref{Subsection flow 1}. Here,
we define this unitary flow by the following assertion:

\begin{theorem}[Local unitary flow on quadratic operators]
\label{flow equation thm 4}\mbox{ }\newline
Under Conditions A1--A3, there exist $T_{+}\in (0,\infty ]$, two operator
families $(\Omega _{t})_{t\in \lbrack 0,T_{+})}$ and $(B_{t})_{t\in \lbrack
0,T_{+})}$, satisfying A1--A2, and a strongly continuous family $(\mathrm{U}%
_{t})_{t\in \lbrack 0,T_{+})}$ of unitary operators acting on the boson Fock
space $\mathcal{F}_{b}$ such that
\begin{equation}
\mathrm{U}_{t}\mathrm{H}_{0}\mathrm{U}_{t}^{\ast }=\sum_{k,\ell }\left\{
\Omega _{t}\right\} _{k,\ell }a_{k}^{\ast }a_{\ell }+\mathrm{W}_{t}+C_{t}\ ,
\end{equation}%
where the non--$\mathrm{N}$--diagonal part of the quadratic operator $%
\mathrm{U}_{t}\mathrm{H}_{0}\mathrm{U}_{t}^{\ast }$ equals%
\begin{equation}
\mathrm{W}_{t}:=\sum_{k,\ell }\left\{ B_{t}\right\} _{k,\ell }a_{k}^{\ast
}a_{\ell }^{\ast }+\left\{ \bar{B}_{t}\right\} _{k,\ell }a_{k}a_{\ell }\ \
\text{and}\ \ C_{t}:=C_{0}+8\int_{0}^{t}\left\Vert B_{\tau }\right\Vert
_{2}^{2}\ \mathrm{d}\tau \ .
\end{equation}%
Furthermore, the map $t\mapsto \Vert B_{t}\Vert _{2}$ from $[0,T_{+})$ to $%
\mathbb{R}_{0}^{+}$ is monotonically decreasing, $\Omega _{t}\leq \Omega
_{0} $, $\Omega _{t}^{2}-8B_{t}\bar{B}_{t}\geq \Omega _{0}^{2}-8B_{0}\bar{B}%
_{0}$, and
\begin{equation}
\mathrm{tr}\left( \Omega _{t}^{2}-4B_{t}\bar{B}_{t}-\Omega _{0}^{2}+4B_{0}%
\bar{B}_{0}\right) =0\ .
\end{equation}%
If $\Omega _{0}B_{0}=B_{0}\Omega _{0}^{\mathrm{t}}$ then $\Omega
_{t}B_{t}=B_{t}\Omega _{t}^{\mathrm{t}}$ and
\begin{equation}
\Omega _{t}=\{\Omega _{0}^{2}-4B_{0}\bar{B}_{0}+4B_{t}\bar{B}_{t}\}^{1/2}\ .
\end{equation}
\end{theorem}

\begin{remark}
\label{remark utiles copy(1)}If A1--A2 hold, but not A3, then $\Omega _{t}$
and $B_{t}$ exist within a small time interval. However, we omit this case
to simplify our discussions, see as an example the proof of Proposition \ref%
{the Brocket--Wegner flow blowup}. In fact, A3 ensures the positivity of $%
\Omega _{t}$ and the decay of the Hilbert--Schmidt norm $\Vert B_{t}\Vert
_{2}$ for $t\in \lbrack 0,T_{+})$, whereas $\Omega _{t}$, $B_{t}$, and $%
\mathrm{U}_{t}$ exist for $t\in \lbrack 0,T_{\max })$ with $T_{\max }\in
(T_{+},\infty ]$.
\end{remark}

\noindent Theorem \ref{flow equation thm 4} follows from Theorems \ref%
{theorem important 1 -1}, \ref{theorem important 2bis}, and \ref{theorem
Ht=unitary orbite} (Section \ref{Subsection flow 2}). In particular, by
Theorem \ref{theorem important 2bis}, the unitary operator \textrm{$U$}$_{t}$
realizes a (time--dependent) Bogoliubov $\mathbf{u}$--$\mathbf{v}$
transformation:
\begin{equation}
\forall t\in \lbrack 0,T_{+}):\qquad \mathrm{U}_{t}a_{k}\mathrm{U}_{t}^{\ast
}=\sum_{\ell }\left\{ \mathbf{u}_{t}\right\} _{k,\ell }a_{\ell }+\left\{
\mathbf{v}_{t}\right\} _{k,\ell }a_{\ell }^{\ast }\ ,
\end{equation}%
with $\mathbf{u}_{t}\in \mathcal{B}\left( \mathfrak{h}\right) $ and $\mathbf{%
v}_{t}\in \mathcal{L}^{2}\left( \mathfrak{h}\right) $ satisfying $\mathbf{u}%
_{t}\mathbf{u}_{t}^{\ast }-\mathbf{v}_{t}\mathbf{v}_{t}^{\ast }=\mathbf{1}$,
$\mathbf{u}_{t}^{\ast }\mathbf{u}_{t}-\mathbf{v}_{t}^{\mathrm{t}}\mathbf{%
\bar{v}}_{t}=\mathbf{1}$, $\mathbf{u}_{t}\mathbf{v}_{t}^{\mathrm{t}}=\mathbf{%
v}_{t}\mathbf{u}_{t}^{\mathrm{t}}$, and $\mathbf{u}_{t}^{\ast }\mathbf{v}%
_{t}=\mathbf{v}_{t}^{\mathrm{t}}\mathbf{\bar{u}}_{t}$.

A necessary condition to $\mathrm{N}$--diagonalize the quadratic operator $%
\mathrm{H}_{0}$ is thus the convergence of the Hilbert--Schmidt operator $%
B_{t}\in \mathcal{L}^{2}(\mathfrak{h})$ of Theorem \ref{flow equation thm 4}
to zero in some topology, in fact, the Hilbert--Schmidt topology to be
precise. Unfortunately, we cannot exclude, a priori, the existence of a
\emph{blow--up} at a finite time, that is, the fact that the
Hilbert--Schmidt norm $\left\Vert B_{t}\right\Vert _{2}$ diverges as $%
t\nearrow T_{\max }$ with $T_{\max }\in (T_{+},\infty )$, see Lemma \ref%
{section extension gap copy(1)} for more details. Moreover, even if we
assume that $T_{+}=\infty $, we can only infer from Theorem \ref{flow
equation thm 4} the convergence of the Hilbert--Schmidt norm $\Vert
B_{t}\Vert _{2}$ as $t\rightarrow \infty $ to some positive constant $%
\mathrm{K}\in \lbrack 0,\Vert B_{0}\Vert _{2}]$. In other words, Conditions
A1--A3 are not sufficient, a priori, to obtain the convergence of $\Vert
B_{t}\Vert _{2}$ to $\mathrm{K}=0$, even if we a priori assume that $%
T_{+}=\infty $. As a consequence, we proceed by progressively strengthening
A1--A3, using the additional assumptions A4--A6\footnote{$\mathbf{1}%
>4B_{0}(\Omega _{0}^{\mathrm{t}})^{-2}\bar{B}_{0}$ means that, for any $%
\varphi \in \mathfrak{h}$, $\varphi \neq 0$, $\langle \varphi |(\mathbf{1}%
-4B_{0}(\Omega _{0}^{\mathrm{t}})^{-2}\bar{B}_{0})\varphi \rangle >0$.}
defined as follows:

\begin{itemize}
\item[A4:] $\Omega _{0}^{-1/2}B_{0}\in \mathcal{L}^{2}(\mathfrak{h})$ is a
Hilbert--Schmidt operator.

\item[A5:] $\mathbf{1}>4B_{0}(\Omega _{0}^{\mathrm{t}})^{-2}\bar{B}_{0}$ and
$\Omega _{0}^{-1-\varepsilon }B_{0}\in \mathcal{L}^{2}(\mathfrak{h})$ for
some constant $\varepsilon >0$.

\item[A6:] $\Omega _{0}\geq 4B_{0}(\Omega _{0}^{\mathrm{t}})^{-1}\bar{B}%
_{0}+\mu \mathbf{1}$ for some constant $\mu >0$.
\end{itemize}

\noindent It is easy to show that A5 yields A4, whereas A6 is stronger than
Conditions A3--A4 as it obviously implies $\Omega _{0}^{-\alpha }B_{0}\in
\mathcal{L}^{2}(\mathfrak{h})$ for any $\alpha >0$ because, in this case, $%
\Omega _{0}\geq \mu \mathbf{1}$. In particular, A6 yields A5, up to the
inequality $\mathbf{1}>4B_{0}(\Omega _{0}^{\mathrm{t}})^{-2}\bar{B}_{0}$,
and this sixth assumption should thus be seen as the most restrictive
condition. Opposed to A6, Conditions A4--A5 accommodate the \emph{most
difficult cases} where $\Omega _{0}^{-1}$ can be unbounded.

\begin{remark}
\label{Condition A5}Condition A5 yields the existence of a constant $\mathrm{%
r}>0$ such that
\begin{equation}
\mathbf{1}\geq \left( 4+\mathrm{r}\right) B_{0}(\Omega _{0}^{\mathrm{t}%
})^{-2}\bar{B}_{0}\ .  \label{A5bis}
\end{equation}%
Indeed, $\Omega _{0}^{-1-\varepsilon }B_{0}\in \mathcal{L}^{2}(\mathfrak{h})$
implies that the operator $B_{0}(\Omega _{0}^{\mathrm{t}})^{-2}\bar{B}_{0}$
is compact and thus, by combining $\mathbf{1}>4B_{0}(\Omega _{0}^{\mathrm{t}%
})^{-2}\bar{B}_{0}$ with an orthonormal basis of eigenvectors of $%
B_{0}(\Omega _{0}^{\mathrm{t}})^{-2}\bar{B}_{0}$, one directly gets (\ref%
{A5bis}). In fact, the assumptions $\Omega _{0}^{-1-\varepsilon }B_{0}\in
\mathcal{L}^{2}(\mathfrak{h})$ and (\ref{A5bis}) for some constants $\mathrm{%
r},\varepsilon >0$ are equivalent to Condition A5 and is what is really used
in all proofs or assertions invoking A5.
\end{remark}

\begin{remark}
\label{Condition A5 copy(1)}The necessity of Conditions A3--A6 is discussed
in Section \ref{section illustration1} via explicit examples that generalize
Bogoliubov's result \cite{Bogoliubov1,BruZagrebnov8}.
\end{remark}

Henceforth we assume at least A1--A4, from which we derive some partial
results (see Theorems \ref{theorem important 1 -1} (v), \ref{theorem
important 1 -2} (i), and \ref{theorem important 1 -3} (i)):

\begin{theorem}[Global unitary flow on quadratic operators]
\label{flow equation thm 4 copy(2)}\mbox{ }\newline
Conditions A1--A4 implies $T_{+}=\infty $ as well as the
square--integrability of the map $t\mapsto \Vert B_{t}\Vert _{2}$ of Theorem %
\ref{flow equation thm 4} on $\left[ 0,\infty \right) $. In this case, $%
\Omega _{t}$ and $B_{t}$ satisfy A1--A4 for all $t\geq 0$.
\end{theorem}

\noindent Since, by Lemma \ref{lemma estimates},
\begin{equation}
\Vert \mathrm{W}_{t}\left( \mathrm{N}+\mathbf{1}\right) ^{-1}\Vert _{\mathrm{%
op}}\leq (4+\sqrt{2})\Vert B_{t}\Vert _{2}\ ,
\end{equation}%
Theorem \ref{flow equation thm 4 copy(2)} asserts a \emph{quasi }$\mathrm{N}$%
\emph{--diagonalization} of $\mathrm{H}_{0}$, in the sense that the non--$%
\mathrm{N}$--diagonal part $\mathrm{W}_{t}$ of the quadratic operator $%
\mathrm{U}_{t}\mathrm{H}_{0}\mathrm{U}_{t}^{\ast }$, compared to the
particle number operator $\mathrm{N}$, tends to zero, as $t\rightarrow
\infty $. Moreover, this theorem yields the convergence in $\mathcal{L}^{1}(%
\mathfrak{h})$ of the (possibly unbounded) operators $\Omega _{t}$ to a
positive operator $\Omega _{\infty }=\Omega _{\infty }^{\ast }\geq 0$
because, for any $t\in \mathbb{R}_{0}^{+}\cup \{\infty \}$,%
\begin{equation}
\Omega _{t}=\Omega _{0}-16\int_{0}^{t}B_{\tau }\bar{B}_{\tau }\mathrm{d}\tau
\ ,
\end{equation}%
see Theorem \ref{theorem important 1 -1} (i). Similarly,
\begin{equation}
C_{\infty }:=\underset{t\rightarrow \infty }{\lim }C_{t}=C_{0}+8\int_{0}^{%
\infty }\Vert B_{\tau }\Vert _{2}^{2}\mathrm{d}\tau <\infty \ .
\end{equation}%
Observe also that the properties of the operator family $(\Omega
_{t})_{t\geq 0}$ described in Theorem \ref{flow equation thm 4} can be
extended to $t=\infty $: $\Omega _{\infty }\leq \Omega _{0}$ whereas $\Omega
_{\infty }^{2}\geq \Omega _{0}^{2}-8B_{0}\bar{B}_{0}$ and
\begin{equation}
\mathrm{tr}\left( \Omega _{\infty }^{2}-\Omega _{0}^{2}+4B_{0}\bar{B}%
_{0}\right) =0\ .
\end{equation}%
Moreover, in the specific cases where $\Omega _{0}B_{0}=B_{0}\Omega _{0}^{%
\mathrm{t}}$, the limit operator
\begin{equation}
\Omega _{\infty }=\{\Omega _{0}^{2}-4B_{0}\bar{B}_{0}\}^{1/2}
\label{omega infinity special case}
\end{equation}%
is \emph{explicit}, and the limit constant equals%
\begin{equation}
C_{\infty }=C_{0}+\frac{1}{2}\mathrm{tr}\left( \Omega _{0}-\{\Omega
_{0}^{2}-4B_{0}\bar{B}_{0}\}^{1/2}\right) <\infty \ .
\end{equation}%
The fact that the trace in the last equation is finite under Conditions
A1--A4 is not trivial. This property is proven under different assumptions
on $\Omega _{0}$ and $B_{0}$ in \cite[Lemma 8.1]{Berezin}. In the present
paper, however, the finiteness of $C_{\infty }$ is never used elsewhere, in
contrast to Berezin's method where \cite[Lemma 8.1]{Berezin} is crucial. In
fact, in our case, it is an obvious corollary of Theorem \ref{theorem
important 4bis} (i).

Using the second assertion (ii) of Theorem \ref{theorem important 4bis} we
can also make the quasi $\mathrm{N}$--diagonalization of $\mathrm{H}_{0}$
more precise by controlling the convergence of the operator family $(\mathrm{%
U}_{t}\mathrm{H}_{0}\mathrm{U}_{t}^{\ast })_{t\geq 0}$ to the well--defined $%
\mathrm{N}$--diagonal quadratic operator%
\begin{equation}
\mathrm{H}_{\infty }:=\underset{k,\ell }{\sum }\left\{ \Omega _{\infty
}\right\} _{k,\ell }a_{k}^{\ast }a_{\ell }+C_{\infty }\ .
\label{hamilton diagonalized}
\end{equation}

\begin{theorem}[Quasi $\mathrm{N}$--diagonalization of quadratic operators]
\label{flow equation thm 4 copy(1)}\mbox{ }\newline
Conditions A1--A4 imply convergence in the strong resolvent sense and as $%
t\rightarrow \infty $, of the operator family $(\mathrm{U}_{t}\mathrm{H}_{0}%
\mathrm{U}_{t}^{\ast })_{t\geq 0}$ of Theorem \ref{flow equation thm 4} to $%
\mathrm{H}_{\infty }$.
\end{theorem}

However, under Conditions A1--A4, the unitary operator $\mathrm{U}_{t}$
itself may not converge, as $t\rightarrow \infty $, in general. In
particular, it is not clear that $\mathrm{H}_{0}$ and $\mathrm{H}_{\infty }$%
\ are unitarily equivalent, unfortunately. Therefore, in order to obtain a
complete $\mathrm{N}$--diagonalization of $\mathrm{H}_{0}$, we use either A5
or A6 which separately lead to the strong convergence of the unitary
operator $\mathrm{U}_{t}$ to $\mathrm{U}_{\infty }$ and, thus, to the $%
\mathrm{N}$--diagonalization of quadratic boson operators (Theorem \ref%
{theorem important 3}).

\begin{theorem}[$\mathrm{N}$--diagonalization of quadratic operators]
\label{flow equation thm 4bis}\mbox{ }\newline
Under Conditions A1--A3 and either A5 or A6, there exists a Bogoliubov $%
\mathbf{u}$--$\mathbf{v}$ unitary transformation $\mathrm{U}_{\infty }$ on $%
\mathcal{F}_{b}$ such that $\mathrm{H}_{\infty }=\mathrm{U}_{\infty }\mathrm{%
H}_{0}\mathrm{U}_{\infty }^{\ast }$.
\end{theorem}

\noindent Observe that the assumptions of Theorem \ref{flow equation thm
4bis} are sufficient, but possibly not necessary, see Remarks \ref{Condition
A5 copy(1)} and \ref{remark conditions copy(1)}. Also, we do not obtain an
explicit form of $\Omega _{\infty }$ in the general case. However, under the
additional assumption that $\Omega _{0}B_{0}=B_{0}\Omega _{0}^{\mathrm{t}}$,
we derive the limit operator $\Omega _{\infty }$ explicitly, see (\ref{omega
infinity special case}). This is a new result and the assumptions of Theorem %
\ref{flow equation thm 4bis} are also more general as the ones previously
used, especially since they include the most difficult cases where both
operators $\Omega _{0}$ and $\Omega _{0}^{-1}$ are unbounded.

In Section \ref{Section evolution equation_our results} we describe our
method in detail, but we also refer to Section \ref{section illustration},
where an instructive and fully explicit example is given. In particular, we
show in Section \ref{section illustration} that (\ref{omega infinity special
case}) does not generally hold if $\Omega _{0}B_{0}\neq B_{0}\Omega _{0}^{%
\mathrm{t}}$.

\subsection{Historical Overview\label{Section historical overview}}

A diagonalization of quadratic boson operators has been performed by
Bogoliubov \cite{Bogoliubov1} in 1947, see also \cite[Appendix B.2]%
{BruZagrebnov8}. This result is a key ingredient of the Bogoliubov theory of
superfluidity, and it defines the borderline to what could possibly be
expected, see also \cite{Williamson} with discussions of \cite{Hormander} in
the finite dimensional case. It is restrictive, however, since $\Omega
_{0}=\Omega _{0}^{\mathrm{t}}$ and $\Omega _{0}B_{0}=B_{0}\Omega _{0}$ in
his case. In fact, it is one of the simplest examples to diagonalize since
this problem is reduced to the $\mathrm{N}$--diagonalization of simple
quadratic operators defined, for any $k\in \mathbb{Z}^{\nu \geq 1}/\{-1,1\}$%
, by $2\times 2$ real matrices $\Omega _{0,k}$ and $B_{0,k}$ (\ref%
{bogoliubov eq 1}) satisfying $\Omega _{0,k}B_{0,k}=B_{0,k}\Omega _{0,k}$
and
\begin{equation}
\Omega _{0,k}^{2}\geq 4B_{0,k}\bar{B}_{0,k}+\mu \mathbf{1}\ ,
\end{equation}%
see (\ref{bogoliubov eq 2}).

During the 1950ies and 1960ies, Friedrichs \cite[Part V]{Friedrichs} and
Berezin \cite[Theorem 8.1]{Berezin} gave a first general result under the
condition that $\Omega _{0}\in \mathcal{B}(\mathfrak{h})$ and $B_{0}\in
\mathcal{L}^{2}(\mathfrak{h})$ are both real symmetric operators satisfying
the gap condition
\begin{equation}
\Omega _{0}\pm 2B_{0}\geq \mu \mathbf{1}\ ,\mathbf{\quad }\mathrm{for\ some\
}\mu >0\ .  \label{gap equation}
\end{equation}%
To be precise, under these assumptions, they prove Theorem \ref{flow
equation thm 4bis}. These hypotheses are stronger than Conditions A1, A2,
and A6, which we use to prove Theorem \ref{flow equation thm 4bis}:

\begin{proposition}[On the Friedrichs--Berezin assumptions]
\label{lemma example 1 copy(3)}\mbox{ }\newline
Let $\Omega _{0}\in \mathcal{B}(\mathfrak{h})$ and $B_{0}\in \mathcal{L}%
^{2}\left( \mathfrak{h}\right) $ be real symmetric operators satisfying (\ref%
{gap equation}). Then $\Omega _{0}$ and $B_{0}$ satisfy Conditions A1, A2,
and A6.
\end{proposition}

\noindent \textbf{Proof.} Using (\ref{gap equation}) we obtain, for any $%
\varphi \in \mathfrak{h}$, that
\begin{equation}
\langle \varphi |\Omega _{0}\varphi \rangle =\frac{1}{2}\underset{\sigma
=\pm 1}{\sum }\langle \varphi |\left( \Omega _{0}+2\sigma B_{0}\right)
\varphi \rangle \geq \mu \mathbf{\Vert }\varphi \mathbf{\Vert }^{2}\ ,
\end{equation}%
i.e., $\Omega _{0}=\Omega _{0}^{\mathrm{t}}\geq \mu \mathbf{1}>0$. For any $%
\tilde{\mu}\in \left( 0,\mu \right) $,
\begin{equation}
\tilde{\Omega}:=\Omega _{0}-\tilde{\mu}\mathbf{1\geq }\left( \mu -\tilde{\mu}%
\right) \mathbf{1}>0
\end{equation}%
is globally invertible. By (\ref{gap equation}), observe also that
\begin{equation}
\tilde{\Omega}\geq \Omega _{0}-\mu \mathbf{1}\geq \pm 2B_{0}\ .
\end{equation}%
The last inequality together with the existence and the positivity of the
inverse operator $\tilde{\Omega}^{-1}\in \mathcal{B}(\mathfrak{h})$ for any $%
\tilde{\mu}\in \left( 0,\mu \right) $ yields
\begin{equation}
\pm 2\tilde{\Omega}^{-1/2}B_{0}\tilde{\Omega}^{-1/2}\leq \mathbf{1}\ ,
\end{equation}%
which in turn implies%
\begin{equation}
4\tilde{\Omega}^{-1/2}B_{0}\tilde{\Omega}^{-1}B_{0}\tilde{\Omega}^{-1/2}\leq
\mathbf{1}\ .  \label{prop 4-1}
\end{equation}%
From the inequalities $\Omega _{0}\geq \tilde{\Omega}\geq \left( \mu -\tilde{%
\mu}\right) \mathbf{1}$ we infer that $\Omega _{0}^{-1}\leq \tilde{\Omega}%
^{-1}$ for any $\tilde{\mu}\in \left( 0,\mu \right) $ because the map $%
X\mapsto X^{-1}$ is operator anti--monotone, for positive $X$. By (\ref{prop
4-1}), it follows that%
\begin{equation}
4B_{0}\Omega _{0}^{-1}B_{0}\leq 4B_{0}\tilde{\Omega}^{-1}B_{0}\leq \tilde{%
\Omega}:=\Omega _{0}-\tilde{\mu}\mathbf{1}\ ,
\end{equation}%
for any $\tilde{\mu}\in \left( 0,\mu \right) $. Consequently,
\begin{equation}
\Omega _{0}\geq 4B_{0}\Omega _{0}^{-1}B_{0}+\mu \mathbf{1}\geq \mu \mathbf{1}%
\ .
\end{equation}%
Since $\Omega _{0}^{\mathrm{t}}=\Omega _{0}\geq 0$ and $B_{0}=B_{0}^{\mathrm{%
t}}=\bar{B}_{0}\in \mathcal{L}^{2}\left( \mathfrak{h}\right) $, Conditions
A1--A2 and A6 hold.\hfill $\Box $

Later, in 1967, Kato and Mugibayashi \cite[Theorem 2]{Kato-Mugibayashi} have
relaxed the hypothesis (\ref{gap equation}) to accommodate the equality $%
\Omega _{0}=2B_{0}$ or $\Omega _{0}=-2B_{0}$ on some $n$--dimensional
subspace of $\mathfrak{h}$. In particular, Theorem \ref{flow equation thm
4bis} was proven in \cite{Kato-Mugibayashi}, replacing the assumption (\ref%
{gap equation}) by the two inequalities\footnote{%
Their result \cite{Kato-Mugibayashi} is also valid by substituting $-B_{0}$
for $B_{0}$ in (\ref{kato inequality}).}
\begin{equation}
\Omega _{0}+2B_{0}\geq \mu _{-}\mathbf{1\qquad }\mathrm{and}\mathbf{\qquad }%
\Omega _{0}-2B_{0}+P^{(n)}\geq \mu _{+}\mathbf{1}\ ,  \label{kato inequality}
\end{equation}%
where $\mu _{\pm }>0$ are two strictly positive constants and
\begin{equation*}
P^{(n)}:=\mathbf{1}\left[ \Omega _{0}=2B_{0}\right]
\end{equation*}%
is the projection onto the $n$--dimensional subspace of $\mathfrak{h}$ where
$\Omega _{0}=2B_{0}$.

However, after elementary manipulations, the hypotheses (\ref{kato
inequality}) yields again a gap condition and the boundedness of the inverse
operator $\Omega _{0}^{-1}\in \mathcal{B}(\mathfrak{h})$. Indeed, the
equality $\Omega _{0}\varphi =2B_{0}\varphi =0$ for some $\varphi \in
\mathfrak{h}$ would contradict the first inequality of (\ref{kato inequality}%
), as it is confirmed by the following proposition:

\begin{proposition}[On the Kato--Mugibayashi assumption]
\label{lemma example 1}\mbox{ }\newline
Let $\Omega _{0}\in \mathcal{B}(\mathfrak{h})$ and $B_{0}\in \mathcal{L}%
^{2}\left( \mathfrak{h}\right) $ be real symmetric operators satisfying the
assumption (\ref{kato inequality}). Then $\Omega _{0}$ satisfies the gap
equation
\begin{equation}
\Omega _{0}\geq \frac{\mu _{-}}{2}\mathbf{1}>0  \label{gap equationbis}
\end{equation}%
and its inverse $\Omega _{0}^{-1}\in \mathcal{B}(\mathfrak{h})$ is thus
bounded.
\end{proposition}

\noindent \textbf{Proof.} Let $\mathfrak{h}_{0}=\mathrm{Ran}P^{(n)}$ be the
finite dimensional subspace of $\mathfrak{h}=\mathfrak{h}_{0}\oplus
\mathfrak{h}_{0}^{\bot }$. For any $\varphi \in \mathfrak{h}$, there are $%
\varphi _{0}\in \mathfrak{h}_{0}$ and $\varphi _{1}\in \mathfrak{h}%
_{0}^{\bot }$ such that $\varphi =\varphi _{0}+\varphi _{1}$. By using $%
\Omega _{0}^{\mathrm{t}}=\Omega _{0}=\Omega _{0}^{\ast }$ and $B_{0}=B_{0}^{%
\mathrm{t}}=B_{0}^{\ast }$, the first inequality of (\ref{kato inequality})
applied on such $\varphi \in \mathfrak{h}$ implies that%
\begin{equation}
\langle \varphi _{0}|\Omega _{0}\varphi _{1}\rangle +\langle \varphi
_{1}|\Omega _{0}\varphi _{0}\rangle \geq \frac{\mu _{-}}{2}\Vert \varphi
\Vert ^{2}-\langle \varphi _{1}|B_{0}\varphi _{1}\rangle -\langle \varphi
_{0}|\Omega _{0}\varphi _{0}\rangle \mathbf{-}\frac{1}{2}\langle \varphi
_{1}|\Omega _{0}\varphi _{1}\rangle \ .
\end{equation}%
It follows that%
\begin{eqnarray}
\langle \varphi |\Omega _{0}\varphi \rangle &=&\langle \varphi _{1}|\Omega
_{0}\varphi _{1}\rangle +\langle \varphi _{0}|\Omega _{0}\varphi _{0}\rangle
+\langle \varphi _{0}|\Omega _{0}\varphi _{1}\rangle +\langle \varphi
_{1}|\Omega _{0}\varphi _{0}\rangle  \notag \\
&\geq &\frac{1}{2}\left( \langle \varphi _{1}|\left( \Omega
_{0}-2B_{0}\right) \varphi _{1}\rangle +\mu _{-}\Vert \varphi \Vert
^{2}\right) \ ,
\end{eqnarray}%
which, combined with the second inequality of (\ref{kato inequality}), gives
\begin{equation}
\langle \varphi |\Omega _{0}\varphi \rangle \geq \frac{1}{2}\left( \mu
_{+}\Vert \varphi _{1}\Vert ^{2}+\mu _{-}\Vert \varphi \Vert ^{2}\right)
\geq \frac{\mu _{-}}{2}\Vert \varphi \Vert ^{2}
\label{gap equation bis bis bis end}
\end{equation}%
for any $\varphi \in \mathfrak{h}$.\hfill $\Box $

\noindent Therefore, the assumption (\ref{kato inequality}) can be used to
define the Brocket--Wegner flow for quadratic boson operators as explained
in the next section. In particular, it can be shown that the condition (\ref%
{kato inequality}) is preserved by the flow for all times and, because of
Proposition \ref{lemma example 1} which yields $\Vert B_{t}\Vert _{2}=%
\mathcal{O}\left( \mathrm{e}^{-t\mu _{-}}\right) $, we can $\mathrm{N}$%
--diagonalize in this way the quadratic operator $\mathrm{H}_{0}$, as
explained in Theorem \ref{flow equation thm 4bis}. This hypothesis can even
be relaxed as, for instance, the assumption $\Omega _{0}\in \mathcal{B}(%
\mathfrak{h})$ is not necessary. This result is, however, not proven here as
it would lead to adding new assumptions without providing new conceptual
ideas.

Note that the diagonalization problem of quadratic boson Hamiltonians has
been studied by Grech and Seiringer in \cite{+} when $\Omega _{0}$ is a
positive operator with compact resolvent. This has been developed
independently of our study.

\section{Brocket--Wegner Flow for Quadratic Boson Operators\label{Section
evolution equation_our results}}

\setcounter{equation}{0}%
The novelty of our approach lies in the use of a non--autonomous evolution
equation, called the Brocket--Wegner flow \cite{Brockett1,Wegner1,bach-bru},
to diagonalize quadratic boson operators. On a formal level, it is easy to
describe, but the mathematically rigorous treatment is rather involved due
to the unboundedness of the operators we are dealing with. More
specifically, we can distinguish three sources of unboundedness in the
quadratic Hamiltonian $\mathrm{H}_{0}$ (\ref{hamilton}):

\begin{itemize}
\item The first one is related to the unboundedness of creation/annihilation
operators, and it can be controlled without much efforts by our methods.
This was not a problem for Friedrichs \cite[Part V]{Friedrichs}, Berezin
\cite{Berezin} or Kato and Mugibayashi \cite{Kato-Mugibayashi}, either.

\item The second one is more serious and corresponds to the unboundedness of
the self--adjoint operator $\Omega _{0}$, i.e., the ultraviolet divergence
features of the $\mathrm{N}$--diagonal part $\Gamma _{0}$ (\ref{hamilton
diagonal0}) of the Hamiltonian $\mathrm{H}_{0}$. This situation is already
out of the scope of previous statements \cite%
{Friedrichs,Berezin,Kato-Mugibayashi}. Only Bogoliubov's result \cite%
{Bogoliubov1,BruZagrebnov8} can still be used but in a very restricted way,
rather far from being satisfactory, see Sections \ref{Section historical
overview} and \ref{section illustration}.

\item Even worse is the infrared property\footnote{%
Absence of a spectral gap for $\Omega _{0}=\Omega _{0}^{\ast }\geq 0$.} of
the $\mathrm{N}$--diagonal operator $\Gamma _{0}$ making the inverse $\Omega
_{0}^{-1}$ of the positive operator $\Omega _{0}\geq 0$ also unbounded. This
last problem turns out to be rather non--trivial and all previous approaches
required an infrared cutoff of the form (\ref{gap equation}) or (\ref{gap
equationbis}). This infrared cutoff condition is shown to be unnecessary,
see Section \ref{Subsection flow 4}.
\end{itemize}

We start by a heuristic description of the Brocket--Wegner flow in Section %
\ref{Subsection flow 1}, which is then rigorously formulated in Section \ref%
{Subsection flow 2}. It leads to a (time--dependent) Bogoliubov $\mathbf{u}$%
--$\mathbf{v}$ transformation, which allows us to diagonalize the quadratic
Hamiltonian in the limit of infinite time, cf. Section \ref{Subsection flow
4}.

\subsection{Setup of the Brocket--Wegner Flow \label{Subsection flow 1}}

The Brocket--Wegner flow is a method proposed more than two decades ago by
Brockett \cite{Brockett1} to solve linear programming problems and
independently by Wegner \cite{Wegner1} to diagonalize Hamiltonians. It is
defined as the following (quadratically) non--linear first--order
differential equation for positive times:
\begin{equation}
\forall t\geq 0:\qquad \partial _{t}Y_{t}=\left[ Y_{t},\left[ Y_{t},X\right] %
\right] \ ,\quad Y_{t=0}:=Y_{0}\ ,  \label{flow equation}
\end{equation}%
with (possibly unbounded) operators $X$ and $Y_{0}$ acting on a Hilbert or
Banach space, and with
\begin{equation}
\lbrack Y_{t},X]:=Y_{t}X-XY_{t}
\end{equation}%
being the commutator between operators $Y_{t}$ and $X$, as usual. This flow
is closely related to non--autonomous evolution equations, see\ Section \ref%
{Section Non-autonomous evolution}. Indeed, let $U_{t,s}$ be an evolution
operator, that is, a jointly strongly continuous in $s$ and $t$ operator
family $(U_{t,s})_{t\geq s\geq 0}$ satisfying $U_{s,s}:=\mathbf{1}$ and the
cocycle property, also called Chapman--Kolmogorov property:%
\begin{equation}
\forall t\geq x\geq s\geq 0:\qquad U_{t,s}=U_{t,x}U_{x,s}\ .
\end{equation}%
Suppose that $U_{t,s}$ solves of the non--autonomous evolution equation%
\begin{equation}
\forall t\geq s\geq 0:\qquad \partial _{t}U_{t,s}=-iG_{t}U_{t,s}\ ,\quad
U_{s,s}:=\mathbf{1}\ ,  \label{flow equationbis}
\end{equation}%
with infinitesimal generator $G_{t}=i\left[ X,Y_{t}\right] $. Then the
operator
\begin{equation}
Y_{t}=U_{t,s}Y_{s}U_{t,s}^{-1}=U_{t}Y_{0}U_{t}^{-1}
\label{flow equation the solution}
\end{equation}%
is a solution of (\ref{flow equation}), where $U_{t}:=U_{t,0}$ and $%
U_{t}^{-1}$ is its right inverse. In the context of self--adjoint operators $%
X\ $and $Y_{0}$ on a Hilbert space, the Brocket--Wegner flow generates a
family $(Y_{t})_{t\geq 0}$ of mutually unitarily equivalent operators.
Furthermore, Brockett's observation is that a solution $Y_{t}$ of (\ref{flow
equation}) converges, at least for real symmetric\textit{\ }matrices $Y_{0}$
and $X$, to a symmetric matrix $Y_{\infty }$ such that $G_{\infty }=i\left[
X,Y_{\infty }\right] =0$. In other words, $Y_{\infty }$ is $X$--diagonal%
\footnote{%
An operator $Y$ is called $X$--\emph{diagonal} iff $[Y,X]=0$.}. A thorough
treatment of the foundations of the Brocket--Wegner flow can be found in
\cite{bach-bru}.

Now, by using the particle number operator $\mathrm{N}$ (\ref{particle
number operator}) and the quadratic boson operator $\mathrm{H}_{0}$ (\ref%
{hamilton}), the infinitesimal generator $\mathrm{G}_{t}$ at $t=0$ equals%
\begin{equation}
\mathrm{G}_{0}=i\left[ \mathrm{N},\mathrm{H}_{0}\right] =2i\underset{k,\ell }%
{\sum }\left\{ B_{0}\right\} _{k,\ell }a_{k}^{\ast }a_{\ell }^{\ast
}-\left\{ \bar{B}_{0}\right\} _{k,\ell }a_{k}a_{\ell }\ .
\label{generator quadratic}
\end{equation}%
In particular, $\mathrm{G}_{0}$ is again a quadratic boson operator.
Therefore, in (\ref{flow equation}) we take $Y_{0}:=\mathrm{H}_{0}$ and the
particle number operator $\mathrm{N}$ for $X$. A solution of the
Brocket--Wegner flow for this example has the following form
\begin{equation}
\mathrm{H}_{t}:=\underset{k,\ell }{\sum }\left\{ \Omega _{t}\right\}
_{k,\ell }a_{k}^{\ast }a_{\ell }+\left\{ B_{t}\right\} _{k,\ell }a_{k}^{\ast
}a_{\ell }^{\ast }+\left\{ \bar{B}_{t}\right\} _{k,\ell }a_{k}a_{\ell
}+C_{t}\ ,  \label{hamiltonbis}
\end{equation}%
with $\Omega _{t}=\Omega _{t}^{\ast }$ and $B_{t}=B_{t}^{\mathrm{t}}$ both
acting on $\mathfrak{h}$ and where $C_{t}\in \mathbb{R}$ is a real number.
It follows that the time--dependent infinitesimal generator equals
\begin{equation}
\mathrm{G}_{t}:=i\left[ \mathrm{N},\mathrm{H}_{t}\right] =2i\underset{k,\ell
}{\sum }\left\{ B_{t}\right\} _{k,\ell }a_{k}^{\ast }a_{\ell }^{\ast
}-\left\{ \bar{B}_{t}\right\} _{k,\ell }a_{k}a_{\ell }
\label{generator quadraticbis}
\end{equation}%
for any $t\geq 0$. Consequently, the strong convergence of the infinitesimal
generator $\mathrm{G}_{t}$ on some domain to $0$ heuristically means that $%
B_{t}\rightarrow 0$ in some topology. The limit operator $\mathrm{H}_{\infty
}$ of $\mathrm{H}_{t}$ is then $\mathrm{N}$--diagonal. In particular, it can
then be diagonalized by using another unitary operator acting only on the
one--particle Hilbert space $\mathfrak{h}$.

\subsection{Mathematical Foundations of our Method\label{Subsection flow 2}}

As shown in \cite[Theorem 2]{bach-bru}, the Brocket--Wegner flow (\ref{flow
equation}) has a unique solution for unbounded operators $Y_{0}$ on a
Hilbert space provided its iterated commutators with $X$ define relatively
bounded operators whose norm tends to zero, as the order increases
sufficiently fast. Until now, it is the only known result about the
well--posedness of the Brocket--Wegner flow for unbounded operators.\ In the
special case where $Y_{0}=\mathrm{H}_{0}$ (\ref{hamilton}) and $X=\mathrm{N}$
(\ref{particle number operator}), however, the infinitesimal generator $%
\mathrm{G}_{0}$ (\ref{generator quadratic}) is unbounded, and we cannot
invoke \cite[Theorem 2]{bach-bru} to ensure the existence of a solution $%
\mathrm{H}_{t}$ of (\ref{flow equation}).

Such a proof in the general case is difficult since the Brocket--Wegner flow
is a (quadratically) non--linear differential equation on operators.
However, if $\mathrm{H}_{t}$ takes the form (\ref{hamiltonbis}) and also
satisfies (\ref{flow equation}) with $Y_{0}=\mathrm{H}_{0}$ and $X=\mathrm{N}
$, then a straightforward computation shows that the operators $\Omega
_{t}=\Omega _{t}^{\ast }$ and $B_{t}=B_{t}^{\mathrm{t}}$ in (\ref%
{hamiltonbis}) satisfy\footnote{%
Recall that the operator $\Omega _{t}^{\mathrm{t}}$ denotes the transpose of
$\Omega _{t}$ and $\bar{B}_{t}$ the complex conjugate of $B_{t}$, see
Section \ref{Section bosonic quadratic operators}.}%
\begin{equation}
\forall t\geq 0:\qquad \left\{
\begin{array}{lll}
\partial _{t}\Omega _{t}=-16B_{t}\bar{B}_{t} & , & \quad \Omega
_{t=0}:=\Omega _{0}\ , \\
\partial _{t}B_{t}=-2\left( \Omega _{t}B_{t}+B_{t}\Omega _{t}^{\mathrm{t}%
}\right) & , & \quad B_{t=0}:=B_{0}\ ,%
\end{array}%
\right.  \label{flow equation-quadratic}
\end{equation}%
whereas the real number $C_{t}$ equals%
\begin{equation}
C_{t}=C_{0}+8\int_{0}^{t}\left\Vert B_{\tau }\right\Vert _{2}^{2}\mathrm{d}%
\tau \ .  \label{flow equation-quadratic-constante}
\end{equation}%
So, the alternative route is to prove the existence of a solution of the
system (\ref{flow equation-quadratic}) and then to define $\mathrm{H}_{t}$
by (\ref{hamiltonbis}) and (\ref{flow equation-quadratic-constante}). The
equations of (\ref{flow equation-quadratic}) form a system of
(quadratically) non--linear first--order differential equations. The
existence and uniqueness of a solution $(\Omega _{t},B_{t})_{t\geq 0}$ for
such a problem is also not obvious, either, especially for unbounded
operators $\Omega _{t}$, but still easier to derive since an explicit
solution of $B_{t}$ as a function of $\Omega _{t}$ can be found by using
again non--autonomous evolution equations. This is carried out in Sections %
\ref{section well posed}--\ref{section constant of motion} under Conditions
A1--A4, i.e., $\Omega _{0}\geq 0$, $B_{0}=B_{0}^{\mathrm{t}}\in \mathcal{L}%
^{2}(\mathfrak{h})$, $\Omega _{0}\geq 4B_{0}(\Omega _{0}^{\mathrm{t}})^{-1}%
\bar{B}_{0}$, $\Omega _{0}^{-1/2}B_{0}\in \mathcal{L}^{2}(\mathfrak{h})$. We
summarize these results in the following theorem about the well--posedness
of the system (\ref{flow equation-quadratic}) of differential equations:

\begin{theorem}[Existence of operators $\Omega _{t}\ $and $B_{t}$]
\label{theorem important 1 -1}\mbox{ }\newline
Under Conditions A1--A3, there are $T_{+}\in (0,\infty ]$ and a unique
family $(\Omega _{t},B_{t})_{t\in \lbrack 0,T_{+})}$ satisfying: \newline
\emph{(i)} The (possibly unbounded) operator $\Omega _{t}$ is positive and
self--adjoint. The family $(\Omega _{t}-\Omega _{0})_{t\in \lbrack
0,T_{+})}\in C^{1}[[0,T_{+});\mathcal{L}^{2}(\mathfrak{h})]$ is Lipschitz
continuous and is a solution of $\partial _{t}\Omega _{t}=-16B_{t}\bar{B}%
_{t} $ in the Hilbert--Schmidt topology. \newline
\emph{(ii)} The family $(B_{t})_{t\in \lbrack 0,T_{+})}\in C[[0,T_{+});%
\mathcal{L}^{2}(\mathfrak{h})]$ is a solution\footnote{%
The integral equation $B_{t}=B_{0}-2\underset{0}{\overset{t}{\int }}\left(
\Omega _{\tau }B_{\tau }+B_{\tau }\Omega _{\tau }^{\mathrm{t}}\right)
\mathrm{d}\tau $ is valid for any $t\in \lbrack 0,T_{+})$ on $\mathfrak{h}$.}
of the non--autonomous parabolic evolution equation $\partial
_{t}B_{t}=-2(\Omega _{t}B_{t}+B_{t}\Omega _{t}^{\mathrm{t}})$ in $\mathcal{L}%
^{2}(\mathfrak{h})$, provided $t>0$. \newline
\emph{(iii)} The constant of motion of the flow is given by
\begin{equation}
\forall t\in \lbrack 0,T_{+}):\qquad \mathrm{tr}\left( \Omega _{t}^{2}-4B_{t}%
\bar{B}_{t}-\Omega _{0}^{2}+4B_{0}\bar{B}_{0}\right) =0\ .
\end{equation}%
\emph{(iv)} Furthermore, for all $t,s\in \lbrack 0,T_{+})$,
\begin{equation}
\Omega _{t}^{2}-8B_{t}\bar{B}_{t}=\Omega _{s}^{2}-8B_{s}\bar{B}%
_{s}+32\int_{s}^{t}B_{\tau }\Omega _{\tau }^{\mathrm{t}}\bar{B}_{\tau }%
\mathrm{d}\tau \ ,
\end{equation}%
and if $\Omega _{0}B_{0}=B_{0}\Omega _{0}^{\mathrm{t}}$ then $\Omega
_{t}B_{t}=B_{t}\Omega _{t}^{\mathrm{t}}$ and $\Omega _{t}^{2}-4B_{t}\bar{B}%
_{t}=\Omega _{0}^{2}-4B_{0}\bar{B}_{0}$.\newline
\emph{(v)} If additionally Condition A4 holds, that is, $\Omega
_{0}^{-1/2}B_{0}\in \mathcal{L}^{2}(\mathfrak{h})$, then $T_{+}=\infty $.
\end{theorem}

The system (\ref{flow equation-quadratic}) is therefore well--posed, at
least for small times. In Lemmata \ref{existence_flow_5 copy(1)}, \ref{new
lemma 2 copy(2)}, and \ref{new lemma 2 copy(1)}, we give additional
properties on the flow and we summarize them in the theorem below.

\begin{theorem}[Inequalities conserved by the flow]
\label{theorem important 1 -2}\mbox{ }\newline
Assume Conditions A1--A4 and that $(\Omega _{t},B_{t})_{t\geq 0}$ is the
solution of (\ref{flow equation-quadratic}).\newline
\emph{(i)} If at initial time $t=0$,
\begin{equation}
\Omega _{0}^{-1/2}B_{0}\in \mathcal{L}^{2}(\mathfrak{h})\quad \text{and}%
\quad \Omega _{0}\geq (4+\mathrm{r})B_{0}(\Omega _{0}^{\mathrm{t}})^{-1}\bar{%
B}_{0}+\mu \mathbf{1}  \label{condition 1}
\end{equation}%
for some $\mu ,\mathrm{r}\geq 0$, then these assumptions are conserved for
all times:%
\begin{equation}
\forall t\geq 0:\quad \Omega _{t}^{-1/2}B_{t}\in \mathcal{L}^{2}(\mathfrak{h}%
)\quad \text{and}\quad \Omega _{t}\geq (4+\mathrm{r})B_{t}(\Omega _{t}^{%
\mathrm{t}})^{-1}\bar{B}_{t}+\mu \mathbf{1}\ .
\end{equation}%
\emph{(ii)} If at initial time $t=0$,
\begin{equation}
\Omega _{0}^{-1}B_{0}\in \mathcal{L}^{2}(\mathfrak{h})\in \mathcal{L}^{2}(%
\mathfrak{h})\quad \text{and}\quad \mathbf{1}\geq \left( 4+\mathrm{r}\right)
B_{0}(\Omega _{0}^{\mathrm{t}})^{-2}\bar{B}_{0}  \label{condition 2bis}
\end{equation}%
for some $\mathrm{r}\geq 0$, then these assumptions are conserved for all
times:%
\begin{equation}
\forall t\geq 0:\quad \Omega _{t}^{-1}B_{t}\in \mathcal{L}^{2}(\mathfrak{h}%
)\ ,\ \Omega _{t}\geq 4B_{t}(\Omega _{t}^{\mathrm{t}})^{-1}\bar{B}_{t}\ ,\
\mathbf{1}\geq \left( 4+\mathrm{r}\right) B_{t}(\Omega _{t}^{\mathrm{t}%
})^{-2}\bar{B}_{t}\ .
\end{equation}
\end{theorem}

\begin{remark}
\label{remark relax condition}Theorem \ref{theorem important 1 -2} also
holds if the operators $\Omega _{t}$, $B_{t}(\Omega _{t}^{\mathrm{t}})^{-1}%
\bar{B}_{t}$, and $B_{t}(\Omega _{t}^{\mathrm{t}})^{-2}\bar{B}_{t}$ are
replaced by $\Omega _{t}^{\mathrm{t}}$, $\bar{B}_{t}\Omega _{t}^{-1}B_{t}$,
and $\bar{B}_{t}\Omega _{t}^{-2}B_{t}$, respectively. Furthermore, strict
inequalities in A3, (\ref{condition 1}) and (\ref{condition 2bis}) ,
respectively, are also preserved for all times $t>0$.
\end{remark}

The inequalities of A3, (\ref{condition 1}) and (\ref{condition 2bis}),
which are conserved by the flow (\ref{flow equation-quadratic}), are crucial
to derive the behavior at infinite time of the solution $(\Omega
_{t},B_{t})_{t\geq 0}$ of (\ref{flow equation-quadratic}), see Section \ref%
{Section Conserved inequalities}.

The Hamiltonian $\mathrm{H}_{t}$ defined by (\ref{hamiltonbis}), with $%
(\Omega _{t},B_{t})_{t\geq 0}$ solution of (\ref{flow equation-quadratic})
and $C_{t}\ $defined by (\ref{flow equation-quadratic-constante}), should
satisfy the Brocket--Wegner flow (\ref{flow equation}) with $Y_{0}=\mathrm{H}%
_{0}$ (\ref{hamilton}) and $X=\mathrm{N}$ (\ref{particle number operator}).
Note, however, that we still do not know whether $\mathrm{H}_{t}$ belongs to
the unitary orbit of $\mathrm{H}_{0}$, i.e., whether (\ref{flow equation the
solution}) is satisfied in this case. Consequently, one has to prove the
existence of the unitary propagator\footnote{%
It is a (strongly continuous) evolution operator which is unitary for all
times.} $\mathrm{U}_{t,s}$ as a solution of (\ref{flow equationbis}) with
infinitesimal generator $\mathrm{G}_{t}$ defined by (\ref{generator
quadraticbis}).

For bounded generators $G_{t}$, this is established by standard methods
involving the Dyson series (\ref{dyson series}) as an explicit solution of (%
\ref{flow equationbis}). For unbounded $G_{t}$, the problem is more
delicate, see for instance \cite%
{Katobis,Caps,Schnaubelt1,Pazy,Neidhardt-zagrebnov} and the corresponding
references cited therein. This problem is solved in Section \ref{Section
Non-autonomous evolution} by using the Yosida approximation. More precisely,
we give another proof of the well--posedness of the Cauchy problem (\ref%
{flow equationbis}) in the hyperbolic case in terms of standard, sufficient
conditions on the generator $G_{t}$, see Conditions B1--B3 in Section \ref%
{Section Non-autonomous evolution}. Since the generator $\mathrm{G}_{t}\ $(%
\ref{generator quadraticbis}) is self--adjoint, B1 is directly satisfied,
see\ (\ref{stability condition in Hilbert space}). But B2--B3 require at
least one auxiliary closed operator $\Theta $. In fact, we use the particle
number operator $\mathrm{N}$ to define the auxiliary self--adjoint operator $%
\Theta $ by setting $\Theta =\mathrm{N}+\mathbf{1}$. This proof is laid out
in detail in Section \ref{Section proof theroem important 2bis}, see the
introduction of Section \ref{Section technical proofs copy(1)} about A1--A4
and Lemmata \ref{lemma estimates}--\ref{lemma uv-Bogoliubov transformation}.
The assertion is the following:

\begin{theorem}[Existence and uniqueness of the operator $\mathrm{U}_{t,s}$]

\label{theorem important 2bis}\mbox{ }\newline
Under Conditions A1--A3, there is a unique unitary propagator $\mathrm{U}%
_{t,s}$ satisfying, for any $s\in \lbrack 0,T_{+})$ and $t\in \lbrack
s,T_{+})$, the following properties:\newline
\emph{(i)} $\mathrm{U}_{t,s}$ conserves the domain $\mathcal{D}\left(
\mathrm{N}\right) $ and is the strong solution on $\mathcal{D}\left( \mathrm{%
N}\right) $ of the non--autonomous evolution equations%
\begin{equation}
\forall s\in \lbrack 0,T_{+}),\ t\in \lbrack s,T_{+}):\qquad \left\{
\begin{array}{llll}
\partial _{t}\mathrm{U}_{t,s}=-i\mathrm{G}_{t}\mathrm{U}_{t,s} & , & \mathrm{%
U}_{s,s}:=\mathbf{1} & . \\
\partial _{s}\mathrm{U}_{t,s}=i\mathrm{U}_{t,s}\mathrm{G}_{s} & , & \mathrm{U%
}_{t,t}:=\mathbf{1} & .%
\end{array}%
\right.
\end{equation}%
\emph{(ii)} $\mathrm{U}_{t,s}$ realizes a Bogoliubov $\mathbf{u}$--$\mathbf{v%
}$ transformation: there are $\mathbf{u}_{t,s}\in \mathcal{B}(\mathfrak{h})$
and $\mathbf{v}_{t,s}\in \mathcal{L}^{2}(\mathfrak{h})$ such that
\begin{eqnarray}
\mathbf{u}_{t,s}\mathbf{u}_{t,s}^{\ast }-\mathbf{v}_{t,s}\mathbf{v}%
_{t,s}^{\ast } &=&\mathbf{1}\ ,\text{\qquad }\mathbf{u}_{t,s}\mathbf{v}%
_{t,s}^{\mathrm{t}}=\mathbf{v}_{t,s}\mathbf{u}_{t,s}^{\mathrm{t}}\ ,
\label{bog1} \\
\mathbf{u}_{t,s}^{\ast }\mathbf{u}_{t,s}-\mathbf{v}_{t,s}^{\mathrm{t}}%
\mathbf{\bar{v}}_{t,s} &=&\mathbf{1}\ ,\qquad \mathbf{u}_{t,s}^{\ast }%
\mathbf{v}_{t,s}=\mathbf{v}_{t,s}^{\mathrm{t}}\mathbf{\bar{u}}_{t,s}\ ,
\label{bog2}
\end{eqnarray}%
and, on the domain $\mathcal{D}(\mathrm{N}^{1/2})$,%
\begin{equation}
\forall k\in \mathbb{N},\ s,t\in \lbrack 0,T_{+}),\ t\geq s:\mathrm{U}%
_{t,s}a_{s,k}\mathrm{U}_{t,s}^{\ast }=\underset{\ell }{\sum }\left\{ \mathbf{%
u}_{t,s}\right\} _{k,\ell }a_{s,\ell }+\left\{ \mathbf{v}_{t,s}\right\}
_{k,\ell }a_{s,\ell }^{\ast }\ ,
\end{equation}%
where $a_{s,k}:=\mathrm{U}_{s}a_{k}\mathrm{U}_{s}^{\ast }$ with $\mathrm{U}%
_{s}:=\mathrm{U}_{s,0}$.
\end{theorem}

\begin{remark}
For the reader's convenience, we explain in Section \ref{section bogoliubov}
that there exists a self--adjoint quadratic boson operator $\mathbb{Q}_{t,s}=%
\mathbb{Q}_{t,s}^{\ast }$ such that $\mathrm{U}_{t,s}=\exp (i\mathbb{Q}%
_{t,s})$ for the finite dimensional case (Theorem \ref{flow equation thm 5
copy(1)}). We refer to \cite{Bruneau-derezinski2007} for the general case.
\end{remark}

\begin{remark}
The family $(\mathrm{U}_{t,s})_{t\geq s\geq 0}$ can naturally be extended to
all $s,t\in \mathbb{R}_{0}^{+}$. For more details, see Theorem \ref{theorem
Ht=unitary orbite copy(1)}.
\end{remark}

It remains to establish the link between $\Omega _{t}$, $B_{t}$, $C_{t}$ and
the operator $\mathrm{U}_{t}\mathrm{H}_{0}\mathrm{U}_{t}^{\ast }$ in terms
of (\ref{hamiltonbis}), where we recall that $\mathrm{U}_{t}:=\mathrm{U}%
_{t,0}$. This is done in Section \ref{section unitarly equi H}, see the
introduction of Section \ref{Section technical proofs copy(1)} about A1--A4
and Lemma \ref{lemma resolvent 3}. It proves the following theorem:

\begin{theorem}[Unitarly equivalence between $\mathrm{H}_{s}$ and $\mathrm{H}%
_{t}$]
\label{theorem Ht=unitary orbite}\mbox{ }\newline
Under Conditions A1--A3, the self--adjoint operator $\mathrm{H}_{t}$ defined
by (\ref{hamiltonbis}) equals $\mathrm{H}_{t}=\mathrm{U}_{t,s}\mathrm{H}_{s}%
\mathrm{U}_{t,s}^{\ast }$, for all $s\in \lbrack 0,T_{+})$ and $t\in \lbrack
s,T_{+})$.
\end{theorem}

\subsection{Asymptotic Properties of the Brocket--Wegner Flow\label%
{Subsection flow 4}}

We devote this subsection to the study of the limits of the time--dependent
quadratic operator $\mathrm{H}_{t}$ and the unitary propagator $\mathrm{U}%
_{t,s}$ as $t\rightarrow \infty $, which require an analysis of $\Omega _{t}$%
, $B_{t}$, $C_{t}$, $\mathbf{u}_{t,s}$, and $\mathbf{v}_{t,s}$ in the limit
of infinite time. All asymptotics of $\Omega _{t},\ C_{t}$, $\mathbf{u}%
_{t,s} $, and $\mathbf{v}_{t,s}$ depend on the behavior of the
Hilbert--Schmidt norm $\Vert B_{t}\Vert _{2}$ for large times. Therefore, we
first summarize in the next theorem a few possible scenarios for the
asymptotic behavior of the map $t\mapsto \Vert B_{t}\Vert _{2}$, which are
taken from Lemmata \ref{lemma integrability copy(1)}, \ref{lemma
integrability copy(2)} and Corollary \ref{lemma integrability}.

\begin{theorem}[Integrability properties of the flow]
\label{theorem important 1 -3}\mbox{ }\newline
Assume Conditions A1--A4.\newline
\emph{(i)} The map $t\mapsto \Vert B_{t}\Vert _{2}$ is square--integrable on
$\left[ 0,\infty \right) $. \newline
\emph{(ii)} If A5 holds, that is, $\mathbf{1}>4B_{0}(\Omega _{0}^{\mathrm{t}%
})^{-2}\bar{B}_{0}$ and $\Omega _{0}^{-1-\varepsilon }B_{0}\in \mathcal{L}%
^{2}(\mathfrak{h})$ for some $\varepsilon >0$, then the map $t\mapsto \Vert
B_{t}\Vert _{2}$ is integrable on $\left[ 0,\infty \right) $.\newline
\emph{(iii)} If A6 holds, that is, $\Omega _{0}\geq 4B_{0}(\Omega _{0}^{%
\mathrm{t}})^{-1}\bar{B}_{0}+\mu \mathbf{1}$ for some $\mu >0$, then $\Vert
B_{t}\Vert _{2}=\mathcal{O}\left( \mathrm{e}^{-2t\mu }\right) $ decays
exponentially to zero in the limit $t\rightarrow \infty $.
\end{theorem}

\begin{remark}
More precise asymptotics are given in Section \ref{Section Conserved
inequalities}: See Corollaries \ref{lemma integrability copy(3)} and \ref%
{lemma integrability} as well as the discussions after Lemma \ref{lemma
integrability copy(4)}.
\end{remark}

\begin{remark}
Under the Kato--Mugibayashi assumption (\ref{kato inequality}), it can be
shown from Proposition \ref{lemma example 1} that $\Vert B_{t}\Vert _{2}=%
\mathcal{O}\left( \mathrm{e}^{-t\mu _{-}}\right) $, similar to Theorem \ref%
{theorem important 1 -3} (iii).
\end{remark}

By Theorem \ref{theorem important 1 -3} (i), the Hilbert--Schmidt norm $%
\Vert B_{t}\Vert _{2}$ is square--integrable on $\left[ 0,\infty \right) $
provided A1--A4 hold. From Section \ref{Section theorem important 4bis}
(Lemmata \ref{lemma omega t infinity}--\ref{lemma omega t infinity2} and \ref%
{lemma convergence in strong resolvent sence of H}), this assertion implies
the existence of the limits of $\Omega _{t}$, $C_{t}$, and $\mathrm{H}_{t}$%
\textbf{, }as $t\rightarrow \infty $:

\begin{theorem}[Limits of $\Omega _{t}$, $C_{t}$, and $\mathrm{H}_{t}\mathbf{%
\ }$as $t\rightarrow \infty $]
\label{theorem important 4bis}\mbox{ }\newline
Assume Conditions A1--A4.\newline
\emph{(i)} The operator $(\Omega _{0}-\Omega _{t})\in \mathcal{L}^{1}(%
\mathfrak{h})$ converges in trace--norm to $(\Omega _{0}-\Omega _{\infty })$%
, where
\begin{equation}
\Omega _{\infty }:=\Omega _{0}-16\int_{0}^{\infty }B_{\tau }\bar{B}_{\tau }%
\mathrm{d}\tau =\Omega _{\infty }^{\ast }\geq 0
\end{equation}%
on $\mathcal{D}\left( \Omega _{0}\right) $, and $2C_{\infty }=\mathrm{tr}%
(\Omega _{0}-\Omega _{\infty })+2C_{0}<\infty $. Furthermore,%
\begin{equation}
\Omega _{\infty }^{2}=\Omega _{0}^{2}-8B_{0}\bar{B}_{0}+32\int_{0}^{\infty
}B_{\tau }\Omega _{\tau }^{\mathrm{t}}\bar{B}_{\tau }\mathrm{d}\tau
\end{equation}%
with $\mathrm{tr}\left( \Omega _{\infty }^{2}-\Omega _{0}^{2}+4B_{0}\bar{B}%
_{0}\right) =0$. If $\Omega _{0}B_{0}=B_{0}\Omega _{0}^{\mathrm{t}}$ then
\begin{equation}
\Omega _{\infty }=\{\Omega _{0}^{2}-4B_{0}\bar{B}_{0}\}^{1/2}\ .
\end{equation}%
\emph{(ii)} The operator $\mathrm{H}_{t}$ converges in the strong resolvent
sense to
\begin{equation}
\mathrm{H}_{\infty }:=\underset{k,\ell }{\sum }\{\Omega _{\infty }\}_{k,\ell
}a_{k}^{\ast }a_{\ell }+C_{\infty }\ .
\end{equation}
\end{theorem}

To obtain the limits of $\mathbf{u}_{t,s}$, $\mathbf{v}_{t,s}$, and $\mathrm{%
U}_{t,s}$, as $t\rightarrow \infty $, the square--integrability of the
function $t\mapsto \Vert B_{t}\Vert _{2}$ is not sufficient. In fact, we
need its integrability which is ensured by either A5 or A6, see Theorem \ref%
{theorem important 1 -3} (ii)--(iii). Indeed, from Theorem \ref{theorem
important 1 -3} and Lemmata \ref{lemma unitary asymptotics 1}--\ref{lemma
convergence in strong resolvent sence of H copy(1)} we infer the existence
of the bounded operators $\mathbf{u}_{t,s}$, $\mathbf{v}_{t,s}$, and $%
\mathrm{U}_{t,s}$ for all $t\in \left[ s,\infty \right] $:

\begin{theorem}[Limits of $\mathbf{u}_{t,s}$, $\mathbf{v}_{t,s}$, and $%
\mathrm{U}_{t,s}$ as $t\rightarrow \infty $]
\label{theorem important 3}\mbox{ }\newline
Assume Conditions A1--A3 and either A5 or A6.\newline
\emph{(i)} $\mathbf{u}_{t,s}$ and $\mathbf{v}_{t,s}$ converge in $\mathcal{L}%
^{2}(\mathfrak{h})$ to $\mathbf{u}_{\infty ,s}\in \mathcal{B}(\mathfrak{h})$
and $\mathbf{v}_{\infty ,s}\in \mathcal{L}^{2}(\mathfrak{h})$ respectively%
\footnote{%
See (\ref{generalized Bog transf 2bis}) and (\ref{generalized Bog transf
2bisbis}) with $t=\infty $.}, with $\mathbf{u}_{\infty ,\infty }=\mathbf{1}$%
, $\mathbf{v}_{\infty ,\infty }=0$, and%
\begin{eqnarray}
\mathbf{u}_{\infty ,s}\mathbf{u}_{\infty ,s}^{\ast }-\mathbf{v}_{\infty ,s}%
\mathbf{v}_{\infty ,s}^{\ast } &=&\mathbf{1}\ ,\text{\qquad }\mathbf{u}%
_{\infty ,s}\mathbf{v}_{\infty ,s}^{\mathrm{t}}=\mathbf{v}_{\infty ,s}%
\mathbf{u}_{\infty ,s}^{\mathrm{t}}\ , \\
\mathbf{u}_{\infty ,s}^{\ast }\mathbf{u}_{\infty ,s}-\mathbf{v}_{\infty ,s}^{%
\mathrm{t}}\mathbf{\bar{v}}_{\infty ,s} &=&\mathbf{1}\ ,\qquad \mathbf{u}%
_{\infty ,s}^{\ast }\mathbf{v}_{\infty ,s}=\mathbf{v}_{\infty ,s}^{\mathrm{t}%
}\mathbf{\bar{u}}_{\infty ,s}\ .
\end{eqnarray}%
\emph{(ii)} $\mathrm{U}_{t,s}$ (resp. $\mathrm{U}_{t,s}^{\ast }$) converges
strongly to a unitary operator $\mathrm{U}_{\infty ,s}$ (resp. $\mathrm{U}%
_{\infty ,s}^{\ast }$) which is strongly continuous in $s$ and satisfies $%
\partial _{s}\mathrm{U}_{\infty ,s}=i\mathrm{U}_{\infty ,s}\mathrm{G}_{s}$
on $\mathcal{D}\left( \mathrm{N}\right) $, $\mathrm{U}_{\infty ,s}=\mathrm{U}%
_{\infty ,x}\mathrm{U}_{x,s}$ for any $x\geq s\geq 0$, and $\mathrm{U}%
_{\infty ,\infty }:=\underset{s\rightarrow \infty }{\lim }\mathrm{U}_{\infty
,s}=\mathbf{1}$ in the strong topology. \newline
\emph{(iii)} For any $s\in \mathbb{R}_{0}^{+}\cup \left\{ \infty \right\} $,
$\mathrm{U}_{\infty ,s}$ realizes a Bogoliubov $\mathbf{u}$--$\mathbf{v}$
transformation:%
\begin{equation}
\forall k\in \mathbb{N},\ \forall s\in \mathbb{R}_{0}^{+}\cup \left\{ \infty
\right\} :\ \mathrm{U}_{\infty ,s}a_{s,k}\mathrm{U}_{\infty ,s}^{\ast }=%
\underset{\ell }{\sum }\left\{ \mathbf{u}_{\infty ,s}\right\} _{k,\ell
}a_{s,\ell }+\left\{ \mathbf{v}_{\infty ,s}\right\} _{k,\ell }a_{s,\ell
}^{\ast }
\end{equation}%
with $a_{s,k}:=\mathrm{U}_{s}a_{k}\mathrm{U}_{s}^{\ast }$. \newline
\emph{(iv)} For any $s\in \mathbb{R}_{0}^{+}\cup \left\{ \infty \right\} $,
the unitary operator $\mathrm{U}_{\infty ,s}$ realizes a $\mathrm{N}$%
--diagonalization of the quadratic boson operator $\mathrm{H}_{s}$ as $%
\mathrm{H}_{\infty }=\mathrm{U}_{\infty ,s}\mathrm{H}_{s}\mathrm{U}_{\infty
,s}^{\ast }$.
\end{theorem}

\noindent In other words, we have $\mathrm{N}$--diagonalized the quadratic
operator $\mathrm{H}_{0}$ (\ref{hamilton}) under the assumptions of Theorem %
\ref{theorem important 3}.

\begin{remark}
\label{remark conditions copy(1)}Theorem \ref{theorem important 3} only
depends on the integrability of the map $t\mapsto \Vert B_{t}\Vert _{2}$ on $%
\left[ 0,\infty \right) $. This property is ensured by A5 or A6, or the
Kato--Mugibayashi assumption (\ref{kato inequality}).
\end{remark}

\begin{remark}
\label{remark extension of unitary}Theorem \ref{theorem important 3}
(ii)--(iii) means that we can continuously extend the definition of the
unitary propagator $\mathrm{U}_{t,s}$ to $\mathbb{R}_{0}^{+}\cup \left\{
\infty \right\} $ by setting $\mathrm{G}_{\infty }:=0$.
\end{remark}

\section{Illustration of the Method\label{section illustration}}

\setcounter{equation}{0}%
In this section we apply the Brocket--Wegner flow on Bogoliubov's example,
which is the simplest quadratic boson operator one can study. This example
is a crucial ingredient of his celebrated microscopic theory of
superfluidity \cite{Bogoliubov1} for liquid $^{4}$He as its diagonalization
by the Bogoliubov $\mathbf{u}$--$\mathbf{v}$ transformation shows a
Landau--type excitation spectrum, see also \cite%
{BruZagrebnov8,Adams-Bru1,Adams-Bru2,bru3} for more details. In this simple
case, the flow can explicitly be computed and we then generalize
Bogoliubov's result by relaxing his gap condition (\ref{bogoliubov eq 2}).
All this study is the subject of Section \ref{section illustration1}. In
Section \ref{section illustration2} we use this example to show a \emph{%
blow--up} of a solution of the Brocket--Wegner flow. This second subsection
is thus a strong warning on the possible pathological behavior of the
Brocket--Wegner flow for unbounded operators.

\subsection{The Brocket--Wegner Flow on Bogoliubov's Example\label{section
illustration1}}

We illustrate first our method on the simplest quadratic boson operator
which is of the form
\begin{equation}
\mathrm{H}_{0}=\omega _{-}a_{-}^{\ast }a_{-}+\omega _{+}a_{+}^{\ast }a_{+}+%
\mathrm{b}a_{+}^{\ast }a_{-}^{\ast }+\mathrm{b}a_{-}a_{+}\ .
\label{quadratic exemple}
\end{equation}%
Here,%
\begin{equation}
\Omega _{0}=\left(
\begin{array}{ll}
\omega _{-} & 0 \\
0 & \omega _{+}%
\end{array}%
\right) \qquad \mathrm{and}\qquad B_{0}=\left(
\begin{array}{ll}
0 & \mathrm{b} \\
\mathrm{b} & 0%
\end{array}%
\right)  \label{bogoliubov eq 1}
\end{equation}%
with strictly positive $\omega _{-},\omega _{+}\in \mathbb{R}^{+}$ and $%
\mathrm{b}\in \mathbb{R}$. We assume without loss of generality that $\omega
_{+}\geq \omega _{-}$.

For any positive real $\omega _{-}$ and $\omega _{+}$, the Brocket--Wegner
flow (\ref{flow equation}) when $Y_{0}=\mathrm{H}_{0}$ and
\begin{equation}
X=\mathrm{N}=a_{-}^{\ast }a_{-}+a_{+}^{\ast }a_{+}
\label{operator number easy}
\end{equation}%
yields a time--dependant quadratic operator%
\begin{equation}
\mathrm{H}_{t}=\omega _{-,t}a_{-}^{\ast }a_{-}+\omega _{+,t}a_{+}^{\ast
}a_{+}+b_{t}a_{+}^{\ast }a_{-}^{\ast }+b_{t}a_{-}a_{+}\ .
\label{quadratic time dep exemple}
\end{equation}%
More precisely,
\begin{equation}
\Omega _{t}=\left(
\begin{array}{ll}
\omega _{-,t} & 0 \\
0 & \omega _{+,t}%
\end{array}%
\right) \quad \mathrm{and}\quad B_{t}=\left(
\begin{array}{ll}
0 & b_{t} \\
b_{t} & 0%
\end{array}%
\right)
\end{equation}%
are real symmetric $2\times 2$--matrices with matrix elements satisfying the
differential equations (\ref{flow equation-quadratic}), that is in this
elementary example,%
\begin{equation}
\forall t\geq 0:\quad \left\{
\begin{array}{llll}
\partial _{t}\omega _{-,t}=-16b_{t}^{2} & , & \quad \omega _{-,0}:=\omega
_{-} & . \\
\partial _{t}\omega _{+,t}=-16b_{t}^{2} & , & \quad \omega _{+,0}:=\omega
_{+} & . \\
\partial _{t}b_{t}=-2\left( \omega _{-,t}+\omega _{+,t}\right) b_{t} & , &
\quad b_{0}:=\mathrm{b} & .%
\end{array}%
\right.  \label{flow equation-quadratic-simplified}
\end{equation}%
This system of differential equations can \emph{explicitly} be solved under
the assumption that $\Omega _{0}\geq 4B_{0}(\Omega _{0}^{\mathrm{t}})^{-1}%
\bar{B}_{0}$ (cf. A3), i.e., $\omega _{+}\omega _{-}\geq 4\mathrm{b}^{2}$.

\begin{remark}
\label{remark conditions copy(3)}In this example, $\Omega _{0}^{2}\geq 4B_{0}%
\bar{B}_{0}$ is equivalent to $\omega _{\pm }^{2}\geq 4\mathrm{b}^{2}$,
which is different from $\omega _{+}\omega _{-}\geq 4\mathrm{b}^{2}$, unless
$[\Omega _{0},B_{0}]=0$, i.e., $\omega _{+}=\omega _{-}$. In fact, the
condition $\Omega _{0}^{2}\geq 4B_{0}\bar{B}_{0}$, together with A1--A2,
yields A3 but the converse does not hold: If $\omega _{-}=1$, $\omega _{+}=2$
and $\mathrm{b}\in (1/2,1/\sqrt{2})$ then A1--A3 hold true but $\Omega
_{0}^{2}\geq 4B_{0}\bar{B}_{0}$ is false.
\end{remark}

\noindent We start this explicit computation with the case $\omega
_{+}\omega _{-}=4\mathrm{b}^{2}$.

\begin{proposition}[Example of a flow when $\protect\omega _{+}\protect%
\omega _{-}=4\mathrm{b}^{2}$]
\label{lemma example 1 copy(2)}\mbox{ }\newline
Let $\Omega _{0}$ and $B_{0}$ be the real symmetric $2\times 2$--matrices (%
\ref{bogoliubov eq 1}). If
\begin{equation}
\omega _{+}\omega _{-}=4\mathrm{b}^{2}\quad \text{and}\quad \delta :=\omega
_{+}-\omega _{-}\geq 0\ ,  \label{condition simple 1}
\end{equation}%
then a solution of (\ref{flow equation-quadratic-simplified}) is given, for
all $t\geq 0$, by either%
\begin{equation}
\omega _{-,t}=\dfrac{\omega _{-}\delta }{\omega _{+}e^{4\delta t}-\omega _{-}%
}\ ,\ \omega _{+,t}=\dfrac{\omega _{+}\delta }{\omega _{+}-\omega
_{-}e^{-4\delta t}}\ ,\ b_{t}^{2}=\frac{\mathrm{b}^{2}\delta
^{2}e^{-4t\delta }}{\left( \omega _{+}-\omega _{-}e^{-4t\delta }\right) ^{2}}%
\ ,
\end{equation}%
provided $\delta >0$ is strictly positive, or by%
\begin{equation}
\omega _{-,t}=\dfrac{\omega _{-}}{4t\omega _{-}+1}\ ,\ \omega _{+,t}=\dfrac{%
\omega _{+}}{4t\omega _{+}+1}\ ,\ b_{t}^{2}=\dfrac{\omega _{+}^{2}}{4\left(
4t\omega _{+}+1\right) ^{2}}\ ,
\end{equation}%
in case that $\delta =0$.
\end{proposition}

\noindent \textbf{Proof.} The constant of motion given in Theorem \ref%
{theorem important 1 -1} (iii) implies that%
\begin{equation}
\omega _{-,t}^{2}+\omega _{+,t}^{2}-8b_{t}^{2}=\omega _{-}^{2}+\omega
_{+}^{2}-8\mathrm{b}^{2}\ .  \label{illustration eq 0}
\end{equation}%
Together with (\ref{flow equation-quadratic-simplified}) it yields the
differential equation%
\begin{equation}
\partial _{t}\omega _{+,t}=2(\omega _{-}^{2}+\omega _{+}^{2}-8\mathrm{b}%
^{2}-\omega _{-,t}^{2}-\omega _{+,t}^{2})\ .
\label{flow equation-quadratic-simplified 1}
\end{equation}%
Since the system (\ref{flow equation-quadratic-simplified}) of differential
equations imposes the equality%
\begin{equation}
\omega _{-,t}=\omega _{+,t}+\omega _{-}-\omega _{+}\ ,
\label{illustration eq 00}
\end{equation}%
we infer from (\ref{flow equation-quadratic-simplified 1}) that%
\begin{equation}
\partial _{t}\omega _{+,t}=4\left( \omega _{-}\omega _{+}-4\mathrm{b}%
^{2}\right) -4\left( \omega _{+,t}\left( \omega _{-}-\omega _{+}\right)
+\omega _{+,t}^{2}\right) \ .
\label{flow equation-quadratic-simplified 1bis}
\end{equation}%
The proposition then follows from (\ref{condition simple 1}) together with
elementary computations using (\ref{illustration eq 0}) and (\ref%
{illustration eq 00})--(\ref{flow equation-quadratic-simplified 1bis}%
).\hfill $\Box $

Proposition \ref{lemma example 1 copy(2)} means that the quadratic boson
operator $\mathrm{H}_{t}$ (\ref{quadratic time dep exemple}) converges in
the strong resolvent sense to $\mathrm{H}_{\infty }=\delta a_{+}^{\ast
}a_{+} $, because the matrix element $b_{t}$ is square--integrable at
infinity, see Lemma \ref{lemma convergence in strong resolvent sence of H}.
If $\delta =0$ then the quadratic operator $\mathrm{H}_{0}$ defined by (\ref%
{quadratic exemple}) is obviously not unitarily equivalent to $\mathrm{H}%
_{\infty }=0$. This is confirmed by the non--integrability of the function $%
b_{t}$ at infinity. If $\delta >0$ then the matrix element $b_{t}$ is
integrable on $\left[ 0,\infty \right) $ and, by Lemma \ref{lemma
convergence in strong resolvent sence of H copy(1)}, $\mathrm{H}_{0}$ and $%
\mathrm{H}_{\infty }$ are unitarily equivalent.

Next we give the explicit solution of the system (\ref{flow
equation-quadratic-simplified}) of differential equations under the
assumption that $\omega _{+}\omega _{-}>4\mathrm{b}^{2}$.

\begin{proposition}[Example of a flow when $\protect\omega _{+}\protect%
\omega _{-}>4\mathrm{b}^{2}$]
\label{lemma example 2 copy(1)}\mbox{ }\newline
Let $\Omega _{0}$ and $B_{0}$ be the real symmetric $2\times 2$--matrices (%
\ref{bogoliubov eq 1}). If $\omega _{+}\omega _{-}>4\mathrm{b}^{2}$ then a
solution of (\ref{flow equation-quadratic-simplified}) is given, for all $%
t\geq 0$, by%
\begin{equation}
\omega _{-,t}=\mathrm{h}_{t}\left( -\delta \right) \ ,\ \omega _{+,t}=%
\mathrm{h}_{t}\left( \delta \right) \ ,\ b_{t}^{2}=\frac{4\varsigma ^{2}%
\mathrm{b}^{2}\mathrm{e}^{-4\varsigma t}}{\left( \sigma +\varsigma +\left(
\varsigma -\sigma \right) \mathrm{e}^{-4\varsigma t}\right) ^{2}}\ ,
\end{equation}%
where%
\begin{equation}
\mathrm{h}_{t}\left( \delta \right) :=\frac{\left( \varsigma +\delta \right)
\left( \varsigma +\sigma \right) +\left( \varsigma -\delta \right) \left(
\sigma -\varsigma \right) \mathrm{e}^{-4\varsigma t}}{2\left( \varsigma
+\sigma +\left( \varsigma -\sigma \right) \mathrm{e}^{-4\varsigma t}\right) }
\end{equation}%
and
\begin{equation}
\delta :=\omega _{+}-\omega _{-},\ \sigma :=\omega _{+}+\omega _{-},\
\varsigma :=\{\sigma ^{2}-16\mathrm{b}^{2}\}^{1/2}>0\ .
\end{equation}%
In particular, in the limit $t\rightarrow \infty $, the functions $\omega
_{\pm ,t}$ and $b_{t}$ converge exponentially to $\left( \varsigma \pm
\delta \right) /2$ and $0$, respectively.
\end{proposition}

\noindent \textbf{Proof.} The proof is obtained by a combination of
Equations (\ref{illustration eq 0}), (\ref{illustration eq 00}), and (\ref%
{flow equation-quadratic-simplified 1bis}) with direct computations using $%
\omega _{+}\omega _{-}>4\mathrm{b}^{2}$, which implies $\varsigma >0$. We
omit the details.\hfill $\Box $

The explicit solution given in Proposition \ref{lemma example 2 copy(1)}
obviously yields a solution $\mathrm{H}_{t}$ (\ref{quadratic time dep
exemple}) of the Brocket--Wegner flow. In this case, the matrix element $%
b_{t}$ is integrable at infinity and, by Lemma \ref{lemma convergence in
strong resolvent sence of H copy(1)}, the quadratic boson operators $\mathrm{%
H}_{0}$ and
\begin{equation}
\mathrm{H}_{\infty }=\frac{1}{2}\left\{ \left( \varsigma +\delta \right)
a_{+}^{\ast }a_{+}+\left( \varsigma -\delta \right) a_{-}^{\ast
}a_{-}\right\}
\end{equation}%
are unitarily equivalent.

To compare with Bogoliubov's result \cite{Bogoliubov1,BruZagrebnov8}, we
extend our example to the quadratic boson operator
\begin{equation}
\mathrm{H}_{0}=\underset{[k]\in \mathbb{Z}^{\nu }/\{-1,1\}}{\sum }\omega
_{-k}a_{-k}^{\ast }a_{-k}+\omega _{k}a_{k}^{\ast }a_{k}+\mathrm{b}%
_{k}a_{k}^{\ast }a_{-k}^{\ast }+\mathrm{b}_{k}a_{-k}a_{k}\ ,
\label{quadratic example}
\end{equation}%
where the positive self--adjoint operator $\Omega _{0}$ and the
Hilbert--Schmidt operator $B_{0}=B_{0}^{\mathrm{t}}\in \mathcal{L}^{2}\left(
\mathfrak{h}\right) $ can be written as direct sums
\begin{equation}
\Omega _{0}=\underset{[k]\in \mathbb{Z}^{\nu }/\{-1,1\}}{\bigoplus }\left(
\begin{array}{ll}
\omega _{-k} & 0 \\
0 & \omega _{k}%
\end{array}%
\right) \ ,\quad B_{0}=\underset{[k]\in \mathbb{Z}^{\nu }/\{-1,1\}}{%
\bigoplus }\left(
\begin{array}{ll}
0 & \mathrm{b}_{k} \\
\mathrm{b}_{k} & 0%
\end{array}%
\right)  \label{definition of omega 0 et b0 exemple}
\end{equation}%
of real symmetric $2\times 2$--matrices, at fixed dimension $\nu \in \mathbb{%
N}$. Here, $\{-1,1\}$ are the point reflections defined on $\mathbb{Z}^{\nu
} $ by $k\rightarrow \pm k$ and $\mathbb{Z}^{\nu }/\{-1,1\}$ is by
definition the set $\{[k]:=\{k,-k\}\}_{k\in \mathbb{Z}^{\nu }}$ of
equivalence classes. Note also that $\omega _{k}$ and $\mathrm{b}_{k}$ are
real numbers such that $\Omega _{0}=\Omega _{0}^{\ast }>0$, i.e.,
\begin{equation}
\forall \lbrack k]\in \mathbb{Z}^{\nu }/\{-1,1\}:\qquad \omega _{\pm k}\in
\mathbb{R}^{+}\ ,  \label{A1 easy}
\end{equation}%
$B_{0}\in \mathcal{L}^{2}\left( \mathfrak{h}\right) $, i.e.,%
\begin{equation}
\left\Vert B_{0}\right\Vert _{2}^{2}=\underset{[k]\in \mathbb{Z}^{\nu
}/\{-1,1\}}{\sum }\mathrm{b}_{k}^{2}<\infty \ ,  \label{norm of B}
\end{equation}%
and $\Omega _{0}\geq 4B_{0}(\Omega _{0}^{\mathrm{t}})^{-1}\bar{B}_{0}$,
i.e.,
\begin{equation}
\forall \lbrack k]\in \mathbb{Z}^{\nu }/\{-1,1\}:\qquad \omega _{k}\omega
_{-k}\geq 4\mathrm{b}_{k}^{2}\ .  \label{A3 easy}
\end{equation}

The case where $\omega _{k}=\omega _{-k}$ for $[k]\in \mathbb{Z}^{\nu
}/\{-1,1\}$, i.e., $\left[ \Omega _{0},B_{0}\right] =0$, was solved by
Bogoliubov in \cite{Bogoliubov1} through the so--called Bogoliubov $\mathbf{u%
}$--$\mathbf{v}$ transformation under the assumptions (\ref{A1 easy})--(\ref%
{norm of B}) and $\Omega _{0}^{2}\geq 4B_{0}\bar{B}_{0}+\mu _{B}\mathbf{1}$
for some constant $\mu _{B}>0$, i.e.,
\begin{equation}
\forall \lbrack k]\in \mathbb{Z}^{\nu }/\{-1,1\}:\qquad \omega
_{k}^{2}=\omega _{-k}^{2}\geq 4\mathrm{b}_{k}^{2}+\mu _{B}>0\ ,
\label{bogoliubov eq 2}
\end{equation}%
Indeed, in Bogoliubov's theory of superfluidity, the sum is over the set $%
(2\pi \mathbb{Z}^{3}/L)/\{-1,1\}$ with $L>0$, $\mathrm{b}_{k}=\hat{\varphi}%
(k)=\hat{\varphi}(-k)$ is the Fourier transform of an absolutely integrable
(real) two--body interaction potential $\varphi (x)=\varphi (\Vert x\Vert )$%
, $x\in \mathbb{R}^{3}$, and $\omega _{k}=\hbar ^{2}k^{2}/2m$ is the
one--particle energy spectrum in the modes $k\in (2\pi \mathbb{Z}^{3}/L)$ of
free bosons of mass $m$ enclosed in a cubic box $\Lambda =L\times L\times
L\subset \mathbb{R}^{3}$ of side length $L$. ($\hbar $ is the Planck
constant divided by $2\pi $.) See \cite[Appendix B.2]{BruZagrebnov8} for
more details.

Note that A6, i.e., $\Omega _{0}\geq 4B_{0}(\Omega _{0}^{\mathrm{t}})^{-1}%
\bar{B}_{0}+\mu \mathbf{1}$ for some constant $\mu >0$, corresponds in this
example to
\begin{equation}
\forall \lbrack k]\in \mathbb{Z}^{\nu }/\{-1,1\}:\qquad \omega _{-k}\omega
_{k}\geq 4\mathrm{b}_{1}^{2}+\mu \omega _{\pm k}\ .  \label{A6 easy}
\end{equation}%
When $\omega _{k}=\omega _{-k}$, Condition A6 is thus equivalent to (\ref%
{bogoliubov eq 2}): On the one hand, (\ref{A6 easy}) implies $\omega _{\pm
k}\geq \mu $ for all $[k]\in \mathbb{Z}^{\nu }/\{-1,1\}$ and (\ref%
{bogoliubov eq 2}) with $\mu _{B}=\mu ^{2}$. On the other hand, since $\mu
x+4\mathrm{b}_{1}^{2}=o(x^{2})$, as $x\rightarrow \infty $, we can assume
without loss of generality that $\omega _{k}\leq \omega _{\max }<\infty $
for all $[k]\in \mathbb{Z}^{\nu }/\{-1,1\}$ and (\ref{bogoliubov eq 2})
yields in this case (\ref{A6 easy}) with $\mu \leq \mu _{B}\omega _{\max
}^{-1}$.

We generalize Bogoliubov's result for $\mu _{B}=0$ in (\ref{bogoliubov eq 2}%
) by using Proposition \ref{lemma example 2 copy(1)} because the $\mathrm{N}$%
--diagonalization of $\mathrm{H}_{0}$ can be done on each $2\times 2$--block
separately. Indeed, the time--dependent quadratic operator $\mathrm{H}_{t}$
computed from the Brocket--Wegner flow (\ref{flow equation}) ($Y_{0}$ being
equal to (\ref{quadratic example}) and $X=\mathrm{N}$) equals
\begin{equation}
\mathrm{H}_{t}=\underset{[k]\in \mathbb{Z}^{\nu }/\{-1,1\}}{\sum }\omega
_{-k,t}a_{-k}^{\ast }a_{-k}+\omega _{k,t}a_{k}^{\ast
}a_{k}+b_{k,t}a_{k}^{\ast }a_{-k}^{\ast }+b_{k,t}a_{-k}a_{k}\ .
\label{quadratic time dep exemplebis}
\end{equation}%
In particular,
\begin{equation}
\Omega _{t}=\underset{[k]\in \mathbb{Z}^{\nu }/\{-1,1\}}{\bigoplus }\left(
\begin{array}{ll}
\omega _{-k,t} & 0 \\
0 & \omega _{k,t}%
\end{array}%
\right) \quad \mathrm{and}\quad B_{t}=\underset{[k]\in \mathbb{Z}^{\nu
}/\{-1,1\}}{\bigoplus }\left(
\begin{array}{ll}
0 & b_{k,t} \\
b_{k,t} & 0%
\end{array}%
\right) \ ,  \label{sol example}
\end{equation}%
where the matrix elements $\omega _{\pm k,t}$ and $b_{k,t}$ are solutions of
the system (\ref{flow equation-quadratic-simplified}) of differential
equations with initial values $\omega _{\pm k,0}=\omega _{\pm k}\in \mathbb{R%
}^{+}$ and $b_{k,0}=\mathrm{b}_{k}\in \mathbb{R}$ for all $[k]\in \mathbb{Z}%
^{\nu }/\{-1,1\}$.

To analyze the $\mathrm{N}$--diagonalization of $\mathrm{H}_{0}$, the
asymptotics of the Hilbert--Schmidt norm $\Vert B_{t}\Vert _{2}$ is pivotal
as explained in Section \ref{Subsection flow 4}. In fact, by Theorem \ref%
{theorem important 1 -3} (i), $t\mapsto \Vert B_{t}\Vert _{2}$ is
square--integrable on $\left[ 0,\infty \right) $ provided A4 holds, that is,
\begin{equation}
\Vert \Omega _{0}^{-1/2}B_{0}\Vert _{2}^{2}=\underset{[k]\in \mathbb{Z}^{\nu
}/\{-1,1\}}{\sum }\mathrm{b}_{k}^{2}\left( \frac{1}{\omega _{k}}+\frac{1}{%
\omega _{-k}}\right) <\infty \ .  \label{A4 easy}
\end{equation}%
Together with Theorem \ref{theorem important 4bis} (ii) and Proposition \ref%
{lemma example 2 copy(1)}, the quadratic boson operator $\mathrm{H}_{t}$
defined by (\ref{quadratic time dep exemplebis}) converges in the strong
resolvent sense to
\begin{eqnarray}
\mathrm{H}_{\infty } &=&\frac{1}{2}\underset{[k]\in \mathbb{Z}^{\nu
}/\{-1,1\}}{\sum }\left( \omega _{-k}-\omega _{k}\right) a_{-k}^{\ast
}a_{-k}+\left( \omega _{k}-\omega _{-k}\right) a_{k}^{\ast }a_{k}  \notag \\
&&+\{\left( \omega _{k}+\omega _{-k}\right) ^{2}-16\mathrm{b}%
_{k}^{2}\}^{1/2}\left( a_{k}^{\ast }a_{k}+a_{-k}^{\ast }a_{-k}\right) \ ,
\label{limit quadratic operator}
\end{eqnarray}%
under the assumptions (\ref{A1 easy})--(\ref{A3 easy}) and (\ref{A4 easy}).

Observe that the limit operator $\Omega _{\infty }$ is equal in this
elementary example to%
\begin{equation}
\Omega _{\infty }=S_{0,-}+(S_{0,+}^{2}-4B_{0}\bar{B}_{0})^{1/2}
\label{omega infinity exemple}
\end{equation}%
with $S_{0,\pm }:=(\Omega _{0}\pm \hat{\Omega}_{0})/2$ and
\begin{equation}
\hat{\Omega}_{0}:=\underset{[k]\in \mathbb{Z}^{\nu }/\{-1,1\}}{\bigoplus }%
\left(
\begin{array}{ll}
\omega _{k} & 0 \\
0 & \omega _{-k}%
\end{array}%
\right) \ .  \label{omega infinity exemplebis}
\end{equation}%
(The operator $\hat{\Omega}_{0}$ satisfies the equality $\hat{\Omega}%
_{0}B_{0}=B_{0}\Omega _{0}$.) It explicitly shows that $\Omega _{\infty }$
is generally not equal to
\begin{equation}
(\Omega _{0}^{2}-4B_{0}\bar{B}_{0})^{1/2}\ ,
\end{equation}%
when $\Omega _{0}B_{0}\neq B_{0}\Omega _{0}^{\mathrm{t}}$. See also Theorem %
\ref{lemma constante of motion} (ii). Also, the explicit form (\ref{omega
infinity exemple})--(\ref{omega infinity exemplebis}) cannot be generalized.
It is only due to peculiar properties of the operators $\Omega _{0}$ and $%
B_{0}$ chosen in (\ref{definition of omega 0 et b0 exemple}). Indeed, from (%
\ref{flow equation-quadratic}) combined with direct computations using (\ref%
{definition of omega 0 et b0 exemple}),
\begin{equation}
\forall t\geq 0:\quad \left\{
\begin{array}{llll}
\partial _{t}S_{t,-}=0 & , & \quad S_{t=0,-}:=S_{0,-} & . \\
\partial _{t}S_{t,+}=-16B_{t}\bar{B}_{t} & , & \quad S_{t=0,+}:=S_{0,+} & .
\\
\partial _{t}B_{t}=-2\left( S_{t,+}B_{t}+B_{t}S_{t,+}^{\mathrm{t}}\right) & ,
& \quad B_{t=0}:=B_{0} & .%
\end{array}%
\right.
\end{equation}%
In this special example, $S_{0,+}B_{0}=B_{0}S_{0,+}^{\mathrm{t}}$.
Therefore, if (\ref{A1 easy})--(\ref{A3 easy}) and (\ref{A4 easy}) hold,
then we directly obtain the explicit form (\ref{omega infinity exemple})--(%
\ref{omega infinity exemplebis}) of $\Omega _{\infty }$ by applying Theorem %
\ref{theorem important 4bis} (i) and substituting $S_{0,+}$ for $\Omega _{0}$%
.

Finally, Theorem \ref{theorem important 3} (iv) shows that the quadratic
boson operators $\mathrm{H}_{0}$ (\ref{quadratic example}) and $\mathrm{H}%
_{\infty }$ (\ref{limit quadratic operator}) are unitarily equivalent
provided that either A5 or A6 is satisfied. In the case where A5 holds, that
is, $\mathbf{1}>4B_{0}(\Omega _{0}^{\mathrm{t}})^{-2}\bar{B}_{0}$, i.e.,%
\begin{equation}
\forall \lbrack k]\in \mathbb{Z}^{\nu }/\{-1,1\}:\qquad \omega _{k}^{2}>4%
\mathrm{b}_{k}^{2}\ ,\ \omega _{-k}^{2}>4\mathrm{b}_{k}^{2}\ ,
\label{A5 ineq}
\end{equation}%
and $\Omega _{0}^{-1-\varepsilon }B_{0}\in \mathcal{L}^{2}(\mathfrak{h})$
for some constant $\varepsilon >0$, i.e.,%
\begin{equation}
\Vert \Omega _{0}^{-1-\varepsilon }B_{0}\Vert _{2}^{2}=\underset{[k]\in
\mathbb{Z}^{\nu }/\{-1,1\}}{\sum }\mathrm{b}_{k}^{2}\left( \frac{1}{\omega
_{k}^{2+2\varepsilon }}+\frac{1}{\omega _{-k}^{2+2\varepsilon }}\right)
<\infty \ ,  \label{A5 easy}
\end{equation}%
this result is already out of the scope of other studies because all
previous approaches required an infrared cutoff of the form (\ref{gap
equation}) or (\ref{kato inequality}), see Section \ref{Section historical
overview}. This example generalizes Bogoliubov's result \cite%
{Bogoliubov1,BruZagrebnov8} for $\mu =0$ in (\ref{bogoliubov eq 2}).

On the other hand, this example can also be used to study the mathematical
necessity of Condition A5, that is, (\ref{A5 ineq})--(\ref{A5 easy}). To
this end, we study A5 with respect to the integrability of the map $t\mapsto
\Vert B_{t}\Vert _{2}$ on $\left[ 0,\infty \right) $ that is \emph{crucial}
to prove Theorem \ref{theorem important 3}, see Remark \ref{remark
conditions copy(1)}.

We first show below that the property $\Omega _{0}^{-1}B_{0}\in \mathcal{L}%
^{2}(\mathfrak{h})$ is \emph{pivotal}:

\begin{proposition}[$\Omega _{0}^{-1}B_{0}\in \mathcal{L}^{2}(\mathfrak{h})$
as a pivotal condition]
\label{lemma example 2}\mbox{ }\newline
There is a choice $(\Omega _{0},B_{0})$ defined by (\ref{definition of omega
0 et b0 exemple}), for $\nu =1$ and satisfying A1--A4, (\ref{A5 ineq}), and $%
\Omega _{0}^{-1+\epsilon }B_{0}\in \mathcal{L}^{2}(\mathfrak{h})$, for any $%
\epsilon >0$, such that $\Vert B_{t}\Vert _{2}\geq \mathcal{O}\left(
t^{-1}\right) $\textbf{, }as $t\rightarrow \infty $. In particular, the map $%
t\mapsto \Vert B_{t}\Vert _{2}$ is not integrable on $\left[ 0,\infty
\right) $. Here, $(\Omega _{t},B_{t})_{t\geq 0}$ is the solution of (\ref%
{flow equation-quadratic}).
\end{proposition}

\noindent \textbf{Proof.} In (\ref{definition of omega 0 et b0 exemple})
take $\mathrm{b}_{k}:=1/2^{k}$ and $\omega _{-k}:=\omega _{k}:=2\sqrt{2}%
\mathrm{b}_{k}$, for $k\in \mathbb{N}$, and $\mathrm{b}_{0}:=\omega _{0}:=0$%
. Then, $\Omega _{0}$ obviously satisfies A1. Moreover,%
\begin{equation}
\left\Vert B_{0}\right\Vert _{2}^{2}=\underset{k=1}{\overset{\infty }{\sum }}%
\frac{1}{2^{2k-1}}=\frac{2}{3}
\end{equation}%
and $B_{0}=B_{0}^{\mathrm{t}}\in \mathcal{L}^{2}(\mathfrak{h})$, see
Condition A2. By (\ref{A3 easy}) and (\ref{A5 ineq}), one gets A3. Finally,
by (\ref{A5 easy}), for any $\epsilon >0$,
\begin{equation}
\left\Vert \Omega _{0}^{-1+\epsilon }B_{0}\right\Vert _{2}^{2}=\frac{%
2^{3\epsilon }}{4}\underset{k=1}{\overset{\infty }{\sum }}\frac{1}{%
2^{2\epsilon k}}=\frac{2^{3\epsilon }}{4\left( 2^{2\epsilon }-1\right) }%
<\infty \ ,
\end{equation}%
i.e., $\Omega _{0}^{-1+\epsilon }B_{0}\in \mathcal{L}^{2}(\mathfrak{h})$. By
(\ref{sol example}) and Proposition \ref{lemma example 2 copy(1)},
straightforward computations show from (\ref{norm of B}) that%
\begin{equation}
\left\Vert B_{t}\right\Vert _{2}^{2}\geq \left( 12-8\sqrt{2}\right) \frac{%
\mathrm{e}^{-\frac{16}{2^{k}}t}}{2^{2k}}\ ,
\end{equation}%
for any $k\in \mathbb{N}$, discarding all terms in the sum but one. In
particular, choosing $k\in \mathbb{N}$ such that $2^{k-1}\leq 16t<2^{k}$ one
gets that, as $t\rightarrow \infty $,
\begin{equation*}
\left\Vert B_{t}\right\Vert _{2}^{2}\geq \left( 12-8\sqrt{2}\right) \frac{%
\mathrm{e}^{-\frac{16}{2^{k}}t}}{2^{2k}}=\mathcal{O}\left( t^{-2}\right) \ .
\end{equation*}%
\hfill $\Box $

Therefore, Proposition \ref{lemma example 2} shows that the condition $%
\Omega _{0}^{-1-\varepsilon }B_{0}\in \mathcal{L}^{2}(\mathfrak{h})$ for
some constant $\varepsilon >0$ in A5 is almost optimal to get the
integrability of $t\mapsto \Vert B_{t}\Vert _{2}$ on $\left[ 0,\infty
\right) $. The borderline is given by the condition $\Omega
_{0}^{-1}B_{0}\in \mathcal{L}^{2}(\mathfrak{h})$ for which the behavior of $%
\Vert B_{t}\Vert _{2}$, as $t\rightarrow \infty $, is unclear: $t\mapsto
\Vert B_{t}\Vert _{2}$ may or may not be integrable -- both cases can occur.

We now study the inequality $\mathbf{1}>4B_{0}(\Omega _{0}^{\mathrm{t}})^{-2}%
\bar{B}_{0}$ of A5 with respect to the integrability of the Hilbert--Schmidt
norm $\Vert B_{t}\Vert _{2}$ on $\left[ 0,\infty \right) $:

\begin{proposition}[$\mathbf{1}>4B_{0}(\Omega _{0}^{\mathrm{t}})^{-2}\bar{B}%
_{0}$ and integrability of $\Vert B_{t}\Vert _{2}$]
\label{lemma example 2 copy(2)}\mbox{ }\newline
There is a choice $(\Omega _{0},B_{0})$ defined by (\ref{definition of omega
0 et b0 exemple}), for $\nu =1$ and satisfying A1--A4, $\Omega
_{0}^{-2}B_{0}\in \mathcal{L}^{2}(\mathfrak{h})$, but not (\ref{A5 ineq}),
such that the map $t\mapsto \Vert B_{t}\Vert _{2}$ is integrable on $\left[
0,\infty \right) $. Here, $(\Omega _{t},B_{t})_{t\geq 0}$ is the solution of
(\ref{flow equation-quadratic}).
\end{proposition}

\noindent \textbf{Proof.} In (\ref{definition of omega 0 et b0 exemple})
take $\omega _{-1}:=1$, $\omega _{1}:=2$ and $\mathrm{b}_{1}\in (1/2,1/\sqrt{%
2})$, while for $k\in \mathbb{N}\backslash \{1\}$, choose $\mathrm{b}%
_{k}:=1/2^{k}$, $\omega _{-k}:=\omega _{k}:=(3/4)^{k}$. Similar to the proof
of Proposition \ref{lemma example 2}, Conditions A1--A3 and $\Omega
_{0}^{-2}B_{0}\in \mathcal{L}^{2}(\mathfrak{h})$ are clearly satisfied for
this choice. In particular, A3 for $k=1$ holds because $\mathrm{b}%
_{1}^{2}<1/2$, see (\ref{A3 easy}). Inequality (\ref{A5 ineq}) requires for $%
k=1$ that $1-\mathrm{b}_{1}^{2}>0$ and $1-4\mathrm{b}_{1}^{2}>0$. Therefore,
if $\mathrm{b}_{1}>1/2$ then (\ref{A5 ineq}) does \emph{not} hold.

We analyze the flow (\ref{flow equation-quadratic-simplified}) for $k=1$ and
$k\in \mathbb{N}\backslash \{1\}$ separately. For $k=1$, we study the real
symmetric $2\times 2$--matrices
\begin{equation}
\Omega _{0}^{(1)}=\left(
\begin{array}{ll}
\omega _{-1} & 0 \\
0 & \omega _{1}%
\end{array}%
\right) \qquad \mathrm{and}\qquad B_{0}^{(1)}=\left(
\begin{array}{ll}
0 & \mathrm{b}_{1} \\
\mathrm{b}_{1} & 0%
\end{array}%
\right) \ .
\end{equation}%
$(\Omega _{0}^{(1)},B_{0}^{(1)})$ satisfies A1--A3, and A6, see in
particular (\ref{A6 easy}) for $k=1$ and $\mu <1-2\mathrm{b}_{1}^{2}$, where
$\mathrm{b}_{1}^{2}<1/2$. The other cases $k\in \mathbb{N}\backslash \{1\}$
also defines operators $(\Omega _{0}^{(2)},B_{0}^{(2)})$, similar to (\ref%
{sol example}), that satisfy A1--A3 and A5. The assertion then follows from
Theorem \ref{theorem important 1 -3} (ii)--(iii). We omit the details.\hfill
$\Box $

By Proposition \ref{lemma example 2 copy(2)}, Inequality (\ref{A5 ineq}) of
A5 is not necessary to get the integrability of $\Vert B_{t}\Vert _{2}$ on $%
\left[ 0,\infty \right) $, but it is partially replaced by A6 (cf. (\ref{A6
easy})) in the example given in the proof of Proposition \ref{lemma example
2 copy(2)}. Indeed, the example above is very special and \emph{trivial} in
some sense because it is an independent combination of two flows defined
from two couples of operators $(\Omega _{0}^{(1)},B_{0}^{(1)})$ and $(\Omega
_{0}^{(2)},B_{0}^{(2)})$, respectively: $(\Omega _{0}^{(1)},B_{0}^{(1)})$\
does \emph{not} satisfy
\begin{equation}
\mathbf{1}>4B_{0}^{(1)}((\Omega _{0}^{(1)})^{\mathrm{t}})^{-2}\bar{B}%
_{0}^{(1)}\ ,
\end{equation}%
but Condition A6. The second one $(\Omega _{0}^{(2)},B_{0}^{(2)})$ satisfies
A5 but not A6. Then, the example of Proposition \ref{lemma example 2 copy(2)}
is constructed by using
\begin{equation}
\Omega _{0}=\Omega _{0}^{(1)}\oplus \Omega _{0}^{(2)}\quad \mathrm{and}\quad
B_{0}=B_{0}^{(1)}\oplus B_{0}^{(2)}  \label{break}
\end{equation}%
because neither A5 nor A6 is satisfied for this choice of operators $(\Omega
_{0},B_{0})$. In fact, if one imposes A1--A3 and $\Omega _{0}^{-1}B_{0}\in
\mathcal{L}^{2}(\mathfrak{h})$, then in the example (\ref{quadratic example}%
) there is \emph{no other way} than the method explained above to break (\ref%
{A5 ineq}) while keeping the integrability of $\Vert B_{t}\Vert _{2}$ on $%
\left[ 0,\infty \right) $.

Up to trivial cases presented above, the inequality $\mathbf{1}%
>4B_{0}(\Omega _{0}^{\mathrm{t}})^{-2}\bar{B}_{0}$ really appears in a
natural way in our proofs and it is not a technical artefact. Moreover,
similar to Proposition \ref{lemma example 2}, it is easy to see that
Condition A4 is optimal with respect to the square--integrability of the map
$t\mapsto \Vert B_{t}\Vert _{2}$ on $\left[ 0,\infty \right) $. See also
Propositions \ref{lemma example 1 copy(2)}--\ref{lemma example 2 copy(1)}
and the proof of Proposition \ref{the Brocket--Wegner flow blowup} that
illustrate A3.

\subsection{Blow--up of the Brocket--Wegner Flow\label{section illustration2}%
}

For bounded operators, the Brocket--Wegner flow (\ref{flow equation}), that
is,
\begin{equation}
\forall t\geq 0:\qquad \partial _{t}Y_{t}=\left[ Y_{t},\left[ Y_{t},X\right] %
\right] \ ,\quad Y_{t=0}:=Y_{0}\ ,  \label{flow equationbisbis}
\end{equation}%
has a unique \emph{global} solution $(Y_{t})_{t\geq 0}$. This assertion is
\cite[Theorem 1]{bach-bru}:

\begin{theorem}[Global existence of the Brocket--Wegner flow]
\label{thm-2a.1}\mbox{ }\newline
Let $(\mathcal{X},\Vert \cdot \Vert _{\mathcal{X}})$ be a Banach subalgebra
of the Banach algebra $\mathcal{B}\left( \mathcal{H}\right) \supset \mathcal{%
X}$ of bounded operators on a separable Hilbert space $\mathcal{H}$ such
that $\Vert \cdot \Vert _{\mathcal{X}}$ is a unitarily invariant norm.
Suppose that $Y_{0}=Y_{0}^{\ast },X=X^{\ast }\in \mathcal{X}$ are two
self--adjoint operators such that $X\geq 0$. Then the Brocket--Wegner flow
has a unique solution $(Y_{t})_{t\geq 0}\in C^{\infty }[\mathbb{R}_{0}^{+};%
\mathcal{X}]$ and $Y_{t}$ is unitarily equivalent to $Y_{0}$ for all $t>0$.
\end{theorem}

\noindent For unbounded operators, the well--posedness of the
Brocket--Wegner flow is much more delicate and in \cite[Theorem 2]{bach-bru}
we used a Nash--Moser type of estimate to show that the Brocket--Wegner flow
has a unique, smooth \emph{local} unbounded solution $(Y_{t})_{t\in \lbrack
0,T_{\ast }]}$ under some restricted conditions on iterated commutators.
Here $T_{\ast }<\infty $ is finite and the existence of such a solution for
larger times $t>T_{\ast }$ is, a priori, not excluded in this case.
Nevertheless, \cite[Theorem 2]{bach-bru} suggests that there is generally
\emph{no global} solution of (\ref{flow equationbisbis}) for unbounded
operators, by opposition to the bounded case. This is confirmed by using the
quadratic boson operator of the previous subsection:

\begin{proposition}[Blow--up of the Brocket--Wegner flow]
\label{the Brocket--Wegner flow blowup}\mbox{ }\newline
There exist two unbounded self--adjoint operators $Y_{0}=Y_{0}^{\ast }$ and $%
X\geq 0$ acting on a separable Hilbert space such that the Brocket--Wegner
flow has a (unbounded) local solution $(Y_{t})_{t\in \left[ 0,T_{\max
}\right) }$ with domain $\mathcal{D}\left( Y_{t}\right) =\mathcal{D}\left(
Y_{0}\right) $ and whose norm $\Vert Y_{t}\varphi \Vert $, $\varphi \in
\mathcal{D}\left( Y_{0}\right) $, diverges at a finite time $T_{\max
}<\infty $.
\end{proposition}

\begin{remark}
Note that in this example the one--particle Hilbert space $\mathfrak{h}$ is
isomorphic to $\mathbb{C}^{2}$.
\end{remark}

\noindent \textbf{Proof.} Take the unbounded self--adjoint operator $%
Y_{0}=H_{0}$ (\ref{quadratic exemple}) with $\omega _{+}=\omega _{-}=0$ and $%
\mathrm{b}>0$, that is, a quadratic boson operator satisfying A1--A2 but not
A3. As before, the unbounded positive operator $X$ is the particle number
operator $X=\mathrm{N}\geq 0$, see (\ref{operator number easy}). The
separable Hilbert space is obviously the boson Fock space $\mathcal{F}_{b}$
defined by (\ref{Fock}). In this case, it is easy to verify that $%
Y_{t}=H_{t} $ (\ref{quadratic time dep exemple}) is a solution of the
Brocket--Wegner flow, provided the three matrix coefficients $\omega _{\pm
,t}$ and $b_{t}$ solve the system (\ref{flow equation-quadratic-simplified})
of differential equations. If $\omega _{+}=\omega _{-}=0$ and $\mathrm{b}>0$
then $b_{t}$ is solution of the initial value problem%
\begin{equation}
\forall t\geq 0:\qquad \partial _{t}b_{t}=64b_{t}\int_{0}^{t}b_{\tau }^{2}%
\mathrm{d}\tau \ ,\quad b_{0}:=\mathrm{b}\ .
\end{equation}%
It is easy to check that the solution of this ODE is
\begin{equation}
\forall t\in \lbrack 0,T_{\max }):\qquad b_{t}=\mathrm{b}\sqrt{\tan
^{2}\left( 8\mathrm{b}t\right) +1}\ ,
\end{equation}%
where $T_{\max }=\pi /(16\mathrm{b})$. By (\ref{flow
equation-quadratic-simplified}), it follows that
\begin{equation}
\forall t\in \lbrack 0,T_{\max }):\qquad \omega _{-,t}=\omega
_{+,t}=-16\int_{0}^{t}b_{\tau }^{2}\mathrm{d}\tau =-2\mathrm{b}\tan \left( 8%
\mathrm{b}t\right) \ ,
\end{equation}%
whereas
\begin{equation}
\underset{t\nearrow T_{\max }}{\lim }\left\vert \omega _{-,t}\right\vert =%
\underset{t\nearrow T_{\max }}{\lim }\left\vert \omega _{+,t}\right\vert =%
\underset{t\nearrow T_{\max }}{\lim }b_{t}=\infty \ .
\end{equation}%
In other words, the Brocket--Wegner flow has a (unbounded) solution $%
(H_{t})_{t\in \left[ 0,T_{\max }\right) }$ defined by (\ref{quadratic time
dep exemple}), which \emph{blows up} on its domain in the limit $t\nearrow
T_{\max }$.\hfill $\Box $

Therefore, this rather elementary example using a quadratic boson operator
shows a pathological behavior of the Brocket--Wegner flow and we thus has to
be really careful while using this flow for unbounded operators. More
generally, Proposition \ref{the Brocket--Wegner flow blowup} provides a
strong warning which should discourage us from performing sloppy
manipulations with the Brocket--Wegner flow, for instance by producing
perturbation series under different ansätze on unbounded Hamiltonians.

\section{Technical Proofs on the One--Particle Hilbert Space\label{Section
technical proofs}}

\setcounter{equation}{0}%
Together with Section \ref{Section technical proofs copy(1)}, it is the
heart of our work. To be precise, here we give the following proofs: the
proofs of Theorems \ref{theorem important 1 -1} and \ref{theorem important 1
-2} (i) in Sections \ref{section well posed}--\ref{section constant of
motion} as well as the proofs of Theorems \ref{theorem important 1 -2} (ii)
and \ref{theorem important 1 -3} in Section \ref{Section Conserved
inequalities}. All these proofs are broken up into several lemmata or
theorems, which often yield information beyond the contents of the above
theorems.

\subsection{Well--Posedness of the Flow\label{section well posed}}

Observe that a solution $B_{t}$ of (\ref{flow equation-quadratic}) can be
written as a function of $\Omega _{t}$ by using a non--autonomous evolution
equation. Indeed, if we assume the existence of a (bounded positive)
evolution family $(W_{t,s})_{t\geq s\geq 0}$ acting on $\mathfrak{h}$ and
solving the non--autonomous evolution equation
\begin{equation}
\forall t>s\geq 0:\qquad \partial _{t}W_{t,s}=-2\Omega _{t}W_{t,s}\ ,\quad
W_{s,s}:=\mathbf{1}\ ,  \label{B_initial value problem}
\end{equation}%
then, by using the notation $W_{t}:=W_{t,0}$, the operator
\begin{equation}
B_{t}=B_{t}^{\mathrm{t}}=W_{t}B_{0}W_{t}^{\mathrm{t}}
\label{B_explicit_solution}
\end{equation}%
is a solution of the second differential equation of (\ref{flow
equation-quadratic}). Consequently, we prove below the existence,
uniqueness, and strong continuity of the bounded operator family $(\Delta
_{t})_{t\geq 0}$ solution of the initial value problem
\begin{equation}
\forall t\geq 0:\qquad \partial _{t}\Delta _{t}=16W_{t}B_{0}W_{t}^{\mathrm{t}%
}\left( W_{t}^{\mathrm{t}}\right) ^{\ast }\bar{B}_{0}W_{t}^{\ast }\ ,\quad
\Delta _{0}:=0\ ,  \label{flow equation-quadratic delta}
\end{equation}%
with $W_{t}:=W_{t,0}$ satisfying (\ref{B_initial value problem}) for $\Omega
_{t}:=\Omega _{0}-\Delta _{t}$.

For this purpose, we first need to define $W_{t,s}$ for any uniformly
bounded operator family $(\Delta _{t})_{t\geq 0}$. Then, we can use the
contraction mapping principle, first for small times $t\in \left[ 0,T_{0}%
\right] \ $($T_{0}>0$). Next, we prove the positivity of the operator $%
\Omega _{t}$ for small times, the uniqueness of a solution of (\ref{flow
equation-quadratic delta}), and the so--called \emph{blow--up alternative}
of the flow in order to extend -- via \emph{an a priori estimate} -- the
domain of existence $\left[ 0,T_{0}\right] $ to the positive real line $%
\mathbb{R}_{0}^{+}$ provided A4 holds. The proofs of Theorems \ref{theorem
important 1 -1}--\ref{theorem important 1 -2} are broken up into several
lemmata or theorems. But first, we need some additional notation.

For any fixed $T>0$, we define $\mathfrak{C}:=C[[0,T];\mathcal{B}(\mathfrak{h%
})]$ to be the Banach space of strongly continuous maps $t\mapsto \Delta
_{t} $\ from $\left[ 0,T\right] \subseteq \mathbb{R}_{0}^{+}$ to the set $%
\mathcal{B}\left( \mathfrak{h}\right) $ of bounded operators acting on a
separable, complex Hilbert space $\mathfrak{h}$. The Banach space $\mathcal{B%
}\left( \mathfrak{h}\right) $ is equipped with the usual operator norm $%
\left\Vert \cdot \right\Vert _{\mathrm{op}}$, whereas the norm on $\mathfrak{%
C}$ is defined by
\begin{equation}
\left\Vert \Delta \right\Vert _{\infty }:=\underset{t\in \left[ 0,T\right] }{%
\sup }\left\Vert \Delta _{t}\right\Vert _{\mathrm{op}}\ .
\label{norm for contraction mapping principle}
\end{equation}%
Also, by $\mathbf{B}_{r}\left( X\right) \subset \mathfrak{C}$ we denote the
open ball of radius $r>0$ centered at $X:=(X_{t})_{t\in \left[ 0,T\right]
}\in \mathfrak{C}$, and below we take $(\Delta _{t})_{t\in \left[ 0,T\right]
}\in \mathbf{B}_{r}\left( 0\right) $ for some finite constant $r>0$.

To define the operator $W_{t,s}$ the existence of a solution of (\ref%
{B_initial value problem}) cannot be deduced from Section \ref{Section
Non-autonomous evolution} or from existence theorems of parabolic evolution
equations. In both cases, we would have to add additional assumptions on $%
\Delta _{t}$, for instance some regularity conditions to apply results on
parabolic equations.\ However, our problem is much easier to analyze since
the operator family $(\Delta _{t})_{t\in \left[ 0,T\right] }\ $is uniformly
bounded in norm and $\Omega _{0}\geq 0$. Indeed, for any $t\geq s\geq 0$, we
explicitly define the operator $W_{t,s}$ by the series%
\begin{eqnarray}
W_{t,s} &\equiv &W_{t,s}\left( \Delta \right) :=\mathrm{e}^{-2\left(
t-s\right) \Omega _{0}}  \label{definition of W} \\
&&+\underset{n=1}{\overset{\infty }{\sum }}2^{n}\int_{s}^{t}\mathrm{d}\tau
_{1}\cdots \int_{s}^{\tau _{n-1}}\mathrm{d}\tau _{n}\mathrm{e}^{-2\left(
t-\tau _{1}\right) \Omega _{0}}\left( \underset{j=1}{\overset{n}{\prod }}%
\Delta _{\tau _{j}}\mathrm{e}^{-2\left( \tau _{j}-\tau _{j+1}\right) \Omega
_{0}}\right) \ ,  \notag
\end{eqnarray}%
where $\tau _{n+1}:=s$ inside the product. In other words, we have chosen $%
W_{t,s}$ to be the unique solution of the integral equation $W=\mathcal{T}%
\left( W\right) $, with%
\begin{equation}
\left[ \mathcal{T}\left( W\right) \right] _{t,s}=\mathrm{e}^{-2\left(
t-s\right) \Omega _{0}}+2\int_{s}^{t}\mathrm{e}^{-2\left( t-\tau \right)
\Omega _{0}}\Delta _{\tau }W_{\tau ,s}\mathrm{d}\tau \ ,
\label{definition of W 2}
\end{equation}%
which follows from a standard contraction mapping principle argument. It is
important to remark that the operator $W_{t,s}$ is well--defined by (\ref%
{definition of W}) since%
\begin{equation}
\max \Big\{\left\Vert W_{t,s}\right\Vert _{\mathrm{op}},\left\Vert W_{t,s}^{%
\mathrm{t}}\right\Vert _{\mathrm{op}}\Big\}\leq \mathrm{e}^{2r\left(
t-s\right) }\ ,  \label{upper bound for W}
\end{equation}%
by $\Vert \Delta _{t}\Vert _{\mathrm{op}}=\Vert \Delta _{t}^{\mathrm{t}%
}\Vert _{\mathrm{op}}\leq r$ and $\Omega _{0}\geq 0$. In fact, it is
standard to prove that $W_{t,s}$ is an evolution operator:

\begin{lemma}[Properties of the operator $W_{t,s}$]
\label{lemma existence 1}\mbox{ }\newline
Assume that $\Omega _{0}=\Omega _{0}^{\ast }\geq 0$ and $(\Delta _{t})_{t\in %
\left[ 0,T\right] }\in \mathfrak{C}$ for some $T>0$. Then, for any $s,x,t\in %
\left[ 0,T\right] $ so that $t\geq x\geq s$:\newline
\emph{(i)} $W_{t,s}$ satisfies the cocycle property\ $W_{t,x}W_{x,s}=W_{t,s}$%
. \newline
\emph{(ii)} $W_{t,s}$ is jointly strongly continuous in $s$ and $t$. \newline
Similar properties as (i)--(ii) also hold for $W_{t,s}^{\ast }$, $W_{t,s}^{%
\mathrm{t}}$, and $\left( W_{t,s}^{\mathrm{t}}\right) ^{\ast }$.
\end{lemma}

\noindent \textbf{Proof.} Note that there exists $r>0$ such that $(\Delta
_{t})_{t\in \left[ 0,T\right] }\in \mathbf{B}_{r}\left( 0\right) $.\smallskip

\underline{(i):} Observing that%
\begin{equation}
W_{t,s}=\left[ \mathcal{T}\left( W\right) \right] _{t,s}=\mathrm{e}%
^{-2\left( t-x\right) \Omega _{0}}W_{x,s}+2\int_{x}^{t}\mathrm{e}^{-2\left(
t-\tau \right) \Omega _{0}}\Delta _{\tau }W_{\tau ,s}\mathrm{d}\tau
\end{equation}%
for $t\geq x\geq s\geq 0$, the cocycle property $W_{t,x}W_{x,s}=W_{t,s}\ $is
proven by multiplying the series (\ref{definition of W}).\smallskip

\underline{(ii):} Its strong continuity is standard to verify. Indeed, for
any $t^{\prime }\geq t\geq s$, the properties (\ref{definition of W 2})--(%
\ref{upper bound for W}), combined with the cocycle property, the uniform
bound $\Vert \Delta _{t}\Vert _{\mathrm{op}}\leq r$, and $\Omega _{0}\geq 0$%
, imply after some elementary computations that, for all $\varphi \in
\mathfrak{h}$,%
\begin{equation}
\left\Vert \left\{ W_{t^{\prime },s}-W_{t,s}\right\} \varphi \right\Vert
\leq \left\Vert \left\{ \mathbf{1}-\mathrm{e}^{-2(t^{\prime }-t)\Omega
_{0}}\right\} W_{t,s}\varphi \right\Vert +\left( \mathrm{e}^{2r(t^{\prime
}-s)}-\mathrm{e}^{2r(t-s)}\right) \left\Vert \varphi \right\Vert \ .
\label{strong continuity of W}
\end{equation}%
For any $\alpha >0$, the semigroup $\mathrm{e}^{2\alpha \Omega _{0}}$ has a
dense domain because $\Omega _{0}=\Omega _{0}^{\ast }$. Hence, since $\Omega
_{0}\geq 0\ $and
\begin{eqnarray}
\left\Vert \left\{ \mathbf{1}-\mathrm{e}^{-2(t^{\prime }-t)\Omega
_{0}}\right\} \mathrm{e}^{-2\alpha \Omega _{0}}\right\Vert _{\mathrm{op}}
&\leq &\underset{\omega \geq 0}{\sup }\left\{ \left( 1-\mathrm{e}%
^{-2(t^{\prime }-t)\omega }\right) \mathrm{e}^{-2\alpha \omega }\right\}
\notag \\
&\leq &\left\vert t^{\prime }-t\right\vert \alpha ^{-1},
\label{inequality idiote bis}
\end{eqnarray}%
then, for any $\delta >0$, there is $\psi \in \mathcal{D}\left( \mathrm{e}%
^{2\alpha \Omega _{0}}\right) $ such that $\Vert W_{t,s}\varphi -\psi \Vert
\leq \delta $ and
\begin{equation}
\left\Vert \left\{ \mathbf{1}-\mathrm{e}^{-2(t^{\prime }-t)\Omega
_{0}}\right\} W_{t,s}\varphi \right\Vert \leq \left\vert t^{\prime
}-t\right\vert \alpha ^{-1}\left\Vert \mathrm{e}^{2\alpha \Omega _{0}}\psi
\right\Vert +2\delta \ .
\end{equation}%
In other words, from (\ref{strong continuity of W}) we get the limits%
\begin{equation}
\underset{t^{\prime }\rightarrow t}{\lim }\left\Vert \left\{ W_{t^{\prime
},s}-W_{t,s}\right\} \varphi \right\Vert =\underset{t\rightarrow t^{\prime }}%
{\lim }\left\Vert \left\{ W_{t^{\prime },s}-W_{t,s}\right\} \varphi
\right\Vert =0  \label{strong continuity of W 1}
\end{equation}%
for all $t^{\prime }\geq t\geq s$.\hfill $\Box $

Note that we do not need to know whether the evolution family $%
(W_{t,s})_{t\geq s\geq 0}$ solves the non--autonomous evolution equation (%
\ref{B_initial value problem}) to prove the existence of a solution $(\Delta
_{t})_{t\in \left[ 0,T\right] }$ of (\ref{flow equation-quadratic delta})
for sufficiently small times $T>0$:

\begin{lemma}[Existence of $\Omega _{t}$ for small times]
\label{lemma existence 2}\mbox{ }\newline
Assume $\Omega _{0}=\Omega _{0}^{\ast }\geq 0$ and $B_{0}=B_{0}^{\mathrm{t}%
}\in \mathcal{L}^{2}\left( \mathfrak{h}\right) $. Then, for
\begin{equation}
T_{0}:=(128\Vert B_{0}\Vert _{2})^{-1}\ ,  \label{T_not}
\end{equation}%
there exists a unique strongly continuous operator family $(\Omega
_{t})_{t\in \lbrack 0,T_{0}]}$ that is a strong solution in $\mathfrak{h}$
of the initial value problem
\begin{equation}
\forall t\in \lbrack 0,T_{0}]:\qquad \partial _{t}\Omega _{t}=-16B_{t}\bar{B}%
_{t}\ ,\quad \Omega _{t=0}=\Omega _{0}\ ,  \label{initial value problem}
\end{equation}%
with $(B_{t})_{t\in \lbrack 0,T_{0}]}\subset \mathcal{L}^{2}\left( \mathfrak{%
h}\right) $ defined by (\ref{B_explicit_solution}) and (\ref{definition of W}%
). Furthermore, $(\Omega _{t})_{t\in \lbrack 0,T_{0}]}$ is self--adjoint,
with domain $\mathcal{D}\left( \Omega _{0}\right) $, and Lipschitz
continuous in the norm topology.
\end{lemma}

\noindent \textbf{Proof.} Observe that the initial value problem (\ref{flow
equation-quadratic delta}) can be rewritten as the integral equation
\begin{equation}
\forall t\geq 0:\quad \Delta _{t}=\mathfrak{T}\left( \Delta \right)
_{t}=\left\{ \mathfrak{T}\left( \Delta \right) \right\} _{t}^{\ast
}:=16\int_{0}^{t}W_{\tau }B_{0}W_{\tau }^{\mathrm{t}}\left( W_{\tau }^{%
\mathrm{t}}\right) ^{\ast }\bar{B}_{0}W_{\tau }^{\ast }\mathrm{d}\tau
\label{flow equation definition of f(delta)}
\end{equation}%
with $W_{t}:=W_{t,0}$ defined by (\ref{definition of W}). If $r>0$ and $%
\Delta =\Delta ^{\ast }\in \mathbf{B}_{r}\left( 0\right) $ then, thanks to (%
\ref{upper bound for W}), we obtain
\begin{equation}
\left\Vert \mathfrak{T}\left( \Delta \right) \right\Vert _{\infty }\leq
16\int_{0}^{T}\left\Vert W_{\tau }B_{0}W_{\tau }^{\mathrm{t}}\left( W_{\tau
}^{\mathrm{t}}\right) ^{\ast }\bar{B}_{0}W_{\tau }^{\ast }\right\Vert _{%
\mathrm{op}}\mathrm{d}\tau \leq \frac{2}{r}\left( \mathrm{e}^{8rT}-1\right)
\Vert B_{0}\Vert _{2}^{2}\ .  \label{toto1}
\end{equation}%
Now, let $\Delta ^{\left( 1\right) },\Delta ^{\left( 2\right) }\in \mathbf{B}%
_{r}\left( 0\right) $ which define via (\ref{definition of W}) two evolution
operators $W_{t}^{\left( 1\right) }:=W_{t,0}^{\left( 1\right) }$ and $%
W_{t}^{\left( 2\right) }:=W_{t,0}^{\left( 2\right) }$. Again from (\ref%
{upper bound for W}),
\begin{equation}
\Vert \mathfrak{T}(\Delta ^{\left( 1\right) })-\mathfrak{T}(\Delta ^{\left(
2\right) })\Vert _{\infty }\leq \frac{16}{3r}\left( \mathrm{e}%
^{6rT}-1\right) \Vert B_{0}\Vert _{2}^{2}\Big\{\Vert \Lambda \Vert _{\infty
}+\Vert \Lambda ^{\mathrm{t}}\Vert _{\infty }\Big\}  \label{condition new 0}
\end{equation}%
with $\Lambda _{t}:=W_{t}^{\left( 1\right) }-W_{t}^{\left( 2\right) }$ for
any $t\in \lbrack 0,T]$. Since the operator $W_{t,s}$ solves the integral
equation $W=\mathcal{T}\left( W\right) $ (cf. (\ref{definition of W 2})),
\begin{equation}
\Lambda _{t}=2\int_{0}^{t}\mathrm{e}^{-2\left( t-\tau \right) \Omega
_{0}}\left\{ \Delta _{\tau }^{(1)}\Lambda _{\tau }+(\Delta _{\tau
}^{(1)}-\Delta _{\tau }^{(2)})W_{\tau }^{\left( 2\right) }\right\} \mathrm{d}%
\tau \ ,
\end{equation}%
which, together with (\ref{upper bound for W}), implies that
\begin{equation}
\Vert \Lambda \Vert _{\infty }\leq 2rT\Vert \Lambda \Vert _{\infty }+\frac{1%
}{r}\left( \mathrm{e}^{2rT}-1\right) \Vert \Delta ^{\left( 1\right) }-\Delta
^{\left( 2\right) }\Vert _{\infty }\ .
\end{equation}%
That is equivalent when $2rT<1$ to
\begin{equation}
\Vert \Lambda \Vert _{\infty }\leq \frac{\mathrm{e}^{2rT}-1}{r\left(
1-2rT\right) }\Vert \Delta ^{\left( 1\right) }-\Delta ^{\left( 2\right)
}\Vert _{\infty }\ .
\end{equation}%
Inserting this and a similar bound for $\left\Vert \Lambda ^{\mathrm{t}%
}\right\Vert _{\infty }$ in (\ref{condition new 0}), we obtain%
\begin{equation}
\Vert \mathfrak{T}(\Delta ^{\left( 1\right) })-\mathfrak{T}(\Delta ^{\left(
2\right) })\Vert _{\infty }\leq \frac{32\left( \mathrm{e}^{6rT}-1\right)
\left( \mathrm{e}^{2rT}-1\right) }{3r^{2}\left( 1-2rT\right) }\Vert
B_{0}\Vert _{2}^{2}\ \Vert \Delta ^{\left( 1\right) }-\Delta ^{\left(
2\right) }\Vert _{\infty }\ .  \label{toto2}
\end{equation}%
Therefore, upon choosing
\begin{equation}
r=r_{0}:=\sqrt{32}\Vert B_{0}\Vert _{2}\qquad \mathrm{and}\qquad
T=T_{0}:=(128\Vert B_{0}\Vert _{2})^{-1}\ ,  \label{flow equation T}
\end{equation}%
we observe that%
\begin{equation}
\left\Vert \mathfrak{T}\left( \Delta \right) \right\Vert _{\infty }\leq
\frac{\sqrt{2}}{4}(\mathrm{e}^{\frac{\sqrt{2}}{4}}-1)\Vert B_{0}\Vert
_{2}<r_{0}\ ,  \label{flow equation condition (i)}
\end{equation}%
by (\ref{toto1}), whereas we infer from (\ref{toto2}) that
\begin{equation}
\Vert \mathfrak{T}(\Delta ^{\left( 1\right) })-\mathfrak{T}(\Delta ^{\left(
2\right) })\Vert _{\infty }<\frac{1}{2}\Vert \Delta ^{\left( 1\right)
}-\Delta ^{\left( 2\right) }\Vert _{\infty }\ .
\end{equation}%
In other words, the map%
\begin{equation}
\mathfrak{T}:\overline{\mathbf{B}_{r_{0}}\left( 0\right) }\rightarrow
\mathbf{B}_{r_{0}}\left( 0\right) \subset \overline{\mathbf{B}_{r_{0}}\left(
0\right) }
\end{equation}%
is a contraction. Consequently, the contraction mapping principle on the
closed set defined by $\overline{\mathbf{B}_{r_{0}}\left( 0\right) }\subset
\mathfrak{C}$, which is equipped with the norm $\left\Vert \cdot \right\Vert
_{\infty }$, yields a unique fixed point $\Delta =\mathfrak{T}\left( \Delta
\right) $ with $\Delta =\Delta ^{\ast }\in \overline{\mathbf{B}%
_{r_{0}}\left( 0\right) }$, for $t\leq T_{0}$. In other words, the lemma
follows by defining $\Omega _{t}:=\Omega _{0}-\Delta _{t}=\Omega _{t}^{\ast
} $ for $t\leq T_{0}$. Indeed, since $\Vert \Delta _{t}\Vert _{\mathrm{op}%
}\leq r_{0}$, it is clear that $\mathcal{D}\left( \Omega _{t}\right) =%
\mathcal{D}\left( \Omega _{0}\right) $ and, thanks to (\ref{upper bound for
W}),%
\begin{equation}
\Vert \Omega _{t}-\Omega _{s}\Vert _{\mathrm{op}}\leq 16\Vert B_{0}\Vert
_{2}^{2}\mathrm{e}^{\frac{\sqrt{2}}{4}}|t-s|  \label{delta lipschitz}
\end{equation}%
for any $t,s\in \left[ 0,T_{0}\right] $, i.e., the family $(\Omega
_{t})_{t\in \lbrack 0,T_{0}]}$ is Lipschitz continuous in the norm topology
for any $t\in \left[ 0,T_{0}\right] $.\hfill $\Box $

Now we observe that the evolution family $(W_{t,s})_{t\geq s\geq 0}$ solves
the non--autonomous evolution equation (\ref{B_initial value problem}),
provided $(\Delta _{t})_{t\in \left[ 0,T_{0}\right] }$ is a solution of (\ref%
{flow equation definition of f(delta)}). From now, we set
\begin{equation}
\forall t\in \lbrack 0,T_{0}]:\qquad \Delta _{t}\equiv 16\int_{0}^{t}B_{\tau
}\bar{B}_{\tau }\mathrm{d}\tau \quad \mathrm{and}\quad \Omega _{t}:=\Omega
_{0}-\Delta _{t}\ ,  \label{definition de delta}
\end{equation}%
(cf. Lemma \ref{lemma existence 2}) and prove additional properties on $%
(W_{t,s})_{t\geq s\geq 0}$ in the next lemma:

\begin{lemma}[$W_{t,s}$ as solution of a parabolic evolution equation]
\label{lemma existence 1 copy(1)}\mbox{ }\newline
Assume $\Omega _{0}=\Omega _{0}^{\ast }\geq 0$ and $B_{0}=B_{0}^{\mathrm{t}%
}\in \mathcal{L}^{2}\left( \mathfrak{h}\right) $, define $T_{0}>0$ by (\ref%
{T_not}) and let $(\Omega _{t})_{t\in \left[ 0,T_{0}\right] }$ be the unique
solution of (\ref{initial value problem}). \newline
\emph{(i) }$\Omega _{t}W_{t,s}\in \mathcal{B}\left( \mathfrak{h}\right) $ is
a bounded operator provided $t>s$:%
\begin{equation}
\forall s\in \lbrack 0,T_{0}),\ t\in (s,T_{0}]:\qquad \left\Vert \Omega
_{t}W_{t,s}\right\Vert _{\mathrm{op}}\leq \mathrm{C}_{0}+\mathrm{D}%
_{0}\left( t-s\right) ^{-1}  \label{eq diff 3}
\end{equation}%
with \textrm{C}$_{0},\mathrm{D}_{0}<\infty $. Moreover, $(\Omega
_{t}W_{t,s})_{t>s\geq 0}$ is strongly continuous in $t>s$.\newline
\emph{(ii)} The evolution family $(W_{t,s})_{t\geq s\geq 0}$ is the solution
of the non--autonomous evolution equations%
\begin{equation}
\left\{
\begin{array}{llllll}
\forall s\in \lbrack 0,T_{0}),\ t\in (s,T_{0}] & : & \partial
_{t}W_{t,s}=-2\Omega _{t}W_{t,s} & , & W_{s,s}:=\mathbf{1} & . \\
\forall t\in (0,T_{0}],\ s\in \lbrack 0,t] & : & \partial
_{s}W_{t,s}=2W_{t,s}\Omega _{s} & , & W_{t,t}:=\mathbf{1} & .%
\end{array}%
\right.  \label{eq diff 4}
\end{equation}%
The derivatives with respect to $t$ and $s$ are in the strong sense in $%
\mathfrak{h}$ and $\mathcal{D}\left( \Omega _{0}\right) $, respectively.%
\newline
\emph{(iii)} For any $s\in \lbrack 0,T_{0})$ and every $\epsilon \in \lbrack
0,T_{0}-s)$, the map $t\mapsto W_{t,s}$ is Lipschitz continuous on $%
[s+\epsilon ,T_{0}]$ in the norm topology:
\begin{equation}
\forall t,t^{\prime }\in \lbrack s+\epsilon ,T_{0}]:\qquad \left\Vert
W_{t^{\prime },s}-W_{t,s}\right\Vert _{\mathrm{op}}\leq 2\left( \mathrm{C}%
_{0}+\mathrm{D}_{0}\epsilon ^{-1}\right) \left\vert t^{\prime }-t\right\vert
\ .
\end{equation}%
Similar properties as (i)--(iii) also hold for $W_{t,s}^{\ast }$, $W_{t,s}^{%
\mathrm{t}}$, and $(W_{t,s}^{\mathrm{t}})^{\ast }$.
\end{lemma}

\noindent \textbf{Proof.} The operator $\Omega _{t}$ has domain $\mathcal{D}%
\left( \Omega _{t}\right) =\mathcal{D}\left( \Omega _{0}\right) $ and is
continuous in the norm topology for $t\leq T_{0}$, with $T_{0}$ defined by (%
\ref{T_not}), see (\ref{delta lipschitz}). Because $\Omega _{0}\geq 0$,
there is $m>-\infty $ such that $\Omega _{t}\geq m$ and $-2\Omega _{t}$ is
the generator of an analytic semigroup $(\mathrm{e}^{-2\alpha \Omega
_{t}})_{\alpha \geq 0}$ for any fixed $t\in \left[ 0,T_{0}\right] $.
Consequently, the Cauchy problem (\ref{eq diff 4}) is a parabolic evolution
equation. In particular, existence and uniqueness of its solution is
standard in this case, see, e.g., \cite[p. 407, Theorem 2.2]{Schnaubelt1} or
\cite[Chap. 5, Theorem 6.1]{Pazy}. Nevertheless, we give here a complete
proof as our case is a specific parabolic evolution equation which is easier
to study.

Before starting the proof, note that, for any positive operator $X=X^{\ast
}\geq 0$, $\beta >0$, and $\alpha \geq 0$, the operator $X^{\alpha }\mathrm{e%
}^{-\beta X}$ is bounded in operator norm by%
\begin{equation}
\Vert X^{\alpha }\mathrm{e}^{-\beta X}\Vert _{\mathrm{op}}\leq \ \underset{%
x\geq 0}{\sup }\left\{ x^{\alpha }\mathrm{e}^{-\beta x}\right\} =\left(
\frac{\alpha }{\mathrm{e}\beta }\right) ^{\alpha }\ .
\label{petit inequality new}
\end{equation}%
Since $\Omega _{0}=\Omega _{0}^{\ast }\geq 0$, we then conclude that%
\begin{equation}
\forall t>s\geq 0:\qquad \Vert \Omega _{0}\mathrm{e}^{-2\left( t-s\right)
\Omega _{0}}\Vert _{\mathrm{op}}\leq \frac{1}{2\mathrm{e}\left( t-s\right) }%
\ .  \label{petit inequality newbisbis}
\end{equation}%
Moreover, $\Omega _{0}\geq 0$ is the generator of an analytic semigroup for $%
t\in \left[ 0,T_{0}\right] $. In particular, for any $t^{\prime }\geq t>s$,
\begin{equation}
\forall t^{\prime }\geq t>s\geq 0:\qquad \left\Vert \mathrm{e}^{-2\left(
t^{\prime }-s\right) \Omega _{0}}-\mathrm{e}^{-2\left( t-s\right) \Omega
_{0}}\right\Vert _{\mathrm{op}}\leq \left( t^{\prime }-t\right) \left(
t-s\right) ^{-1}\ ,  \label{interpolation1}
\end{equation}%
using (\ref{inequality idiote bis}) with $\alpha =t-s$. Meanwhile, as $%
\Omega _{0}\geq 0$,%
\begin{equation}
\left\Vert \mathrm{e}^{-2\left( t^{\prime }-s\right) \Omega _{0}}-\mathrm{e}%
^{-2\left( t-s\right) \Omega _{0}}\right\Vert _{\mathrm{op}}\leq 2\ .
\label{interpolation2}
\end{equation}%
As consequence, we can interpolate the estimates (\ref{interpolation1}) and (%
\ref{interpolation2}) to get
\begin{equation}
\left\Vert \mathrm{e}^{-2\left( t^{\prime }-s\right) \Omega _{0}}-\mathrm{e}%
^{-2\left( t-s\right) \Omega _{0}}\right\Vert _{\mathrm{op}}\leq 2^{1-\nu
}\left( t^{\prime }-t\right) ^{\nu }\left( t-s\right) ^{-\nu }
\label{interpolation3}
\end{equation}%
for any $t^{\prime }\geq t>s$ and $\nu \in \left[ 0,1\right] $. We then
conclude these preliminary remarks on the semigroup $\left\{ \mathrm{e}%
^{-2\left( t-s\right) \Omega _{0}}\right\} _{t\geq s}$ by observing that,
for any $t^{\prime }>t>s$,%
\begin{eqnarray}
\left\Vert \Omega _{0}\left( \mathrm{e}^{-2\left( t^{\prime }-s\right)
\Omega _{0}}-\mathrm{e}^{-2\left( t-s\right) \Omega _{0}}\right) \right\Vert
_{\mathrm{op}} &\leq &2\int_{t-s}^{t^{\prime }-s}\left\Vert \Omega _{0}^{2}%
\mathrm{e}^{-2\tau \Omega _{0}}\right\Vert _{\mathrm{op}}\mathrm{d}\tau
\notag \\
&\leq &\frac{\left( t^{\prime }-t\right) }{\mathrm{e}^{2}\left( t^{\prime
}-s\right) \left( t-s\right) }\ ,  \label{interpolation3bis}
\end{eqnarray}%
using (\ref{petit inequality new}) with $X=\Omega _{0}$, $\alpha =2$, and $%
\beta =2\tau $. Now, we are in position to prove the lemma by following
similar arguments as in \cite[p. 157-159]{Pazy}.

\noindent Recall that the operator family $(\Delta _{t})_{t\in \left[ 0,T_{0}%
\right] }\subset \mathbf{B}_{r}\left( 0\right) $ is Lipschitz continuous in
the norm topology. See, e.g., (\ref{delta lipschitz})--(\ref{definition de
delta}). By using two times (\ref{interpolation3}) together with $\Omega
_{0}\geq 0$, $\Delta =\Delta ^{\ast }\in \mathbf{B}_{r}\left( 0\right) $, (%
\ref{definition of W 2}) and (\ref{upper bound for W}), we then deduce, for
any $t^{\prime }\geq t>s$ and $\nu \in \lbrack 0,1)$, that%
\begin{eqnarray}
\left\Vert W_{t^{\prime },s}-W_{t,s}\right\Vert _{\mathrm{op}} &\leq
&2^{1-\nu }\left( t^{\prime }-t\right) ^{\nu }\left( t-s\right) ^{-\nu
}+2r\left( t^{\prime }-t\right) \mathrm{e}^{2r\left( t^{\prime }-s\right) }
\notag \\
&&+\frac{2^{2-\nu }r}{1-\nu }\mathrm{e}^{2r\left( t-s\right) }\left(
t^{\prime }-t\right) ^{\nu }\left( t-s\right) ^{1-\nu }\ .
\label{interpolation4}
\end{eqnarray}%
Now, for any $s\in \lbrack 0,T_{0}]$ and sufficiently small $\epsilon >0$,
let
\begin{equation}
\forall t\in (s+\epsilon ,T_{0}]:\qquad \mathcal{W}_{t,s}^{(\epsilon
)}:=2\int_{s}^{t-\epsilon }\mathrm{e}^{-2\left( t-\tau \right) \Omega
_{0}}\Delta _{\tau }W_{\tau ,s}\mathrm{d}\tau \ .
\end{equation}%
Recall that $-2\Omega _{0}\leq 0$ is the generator of an analytic semigroup $%
(\mathrm{e}^{-2\alpha \Omega _{0}})_{\alpha \geq 0}$. Consequently, as $%
\epsilon \rightarrow 0$, $\mathcal{W}_{t,s}^{\left( \epsilon \right) }$
strongly converges to $\mathcal{W}_{t,s}^{(0)}$ and $\mathcal{W}%
_{t,s}^{(\epsilon )}$ is strongly differentiable with respect to $%
t>s+\epsilon $ with derivative%
\begin{equation}
\partial _{t}\mathcal{W}_{t,s}^{(\epsilon )}=2\mathrm{e}^{-2\epsilon \Omega
_{0}}\Delta _{t-\epsilon }W_{t-\epsilon ,s}-4\int_{s}^{t-\epsilon }\Omega
_{0}\mathrm{e}^{-2\left( t-\tau \right) \Omega _{0}}\Delta _{\tau }W_{\tau
,s}\mathrm{d}\tau  \label{interpolation4bis}
\end{equation}%
for any $t\in (s+\epsilon ,T_{0}]$. Therefore, using again $\Omega _{0}\geq
0 $, $\Delta =\Delta ^{\ast }\in \mathbf{B}_{r}\left( 0\right) $, and (\ref%
{upper bound for W}) together with
\begin{eqnarray}
-4\int_{s}^{t-\epsilon }\Omega _{0}\mathrm{e}^{-2\left( t-\tau \right)
\Omega _{0}}\Delta _{\tau }W_{\tau ,s}\mathrm{d}\tau
&=&-4\int_{s}^{t-\epsilon }\Omega _{0}\mathrm{e}^{-2\left( t-\tau \right)
\Omega _{0}}\left( \Delta _{\tau }W_{\tau ,s}-\Delta _{t}W_{t,s}\right)
\mathrm{d}\tau  \notag \\
&&-4\int_{s}^{t-\epsilon }\Omega _{0}\mathrm{e}^{-2\left( t-\tau \right)
\Omega _{0}}\Delta _{t}W_{t,s}\mathrm{d}\tau  \notag \\
&=&-4\int_{s}^{t-\epsilon }\Omega _{0}\mathrm{e}^{-2\left( t-\tau \right)
\Omega _{0}}\left( \Delta _{\tau }W_{\tau ,s}-\Delta _{t}W_{t,s}\right)
\mathrm{d}\tau  \notag \\
&&+2\left( \mathrm{e}^{-2\left( t-s\right) \Omega _{0}}-\mathrm{e}%
^{-2\epsilon \Omega _{0}}\right) \Delta _{t}W_{t,s}\ ,
\end{eqnarray}%
we find that
\begin{eqnarray}
\Vert \partial _{t}\mathcal{W}_{t,s}^{(\epsilon )}\Vert _{\mathrm{op}} &\leq
&2r\mathrm{e}^{2r\left( t-s\right) }+4r\mathrm{e}^{2r\left( t-s\right) }
\label{interpolation5} \\
&&+4\int_{s}^{t-\epsilon }\Vert \Omega _{0}\mathrm{e}^{-2\left( t-\tau
\right) \Omega _{0}}\Vert _{\mathrm{op}}\Vert \Delta _{\tau }W_{\tau
,s}-\Delta _{t}W_{t,s}\Vert _{\mathrm{op}}\mathrm{d}\tau \ .  \notag
\end{eqnarray}%
We use now (\ref{upper bound for W}), (\ref{petit inequality newbisbis}), (%
\ref{interpolation4}) for $\nu \in (0,1)$ and the Lipschitz continuity of
the operator family $(\Delta _{t})_{t\in \left[ 0,T_{0}\right] }$ to get
\begin{equation}
\int_{s}^{t-\epsilon }\Vert \Omega _{0}\mathrm{e}^{-2\left( t-\tau \right)
\Omega _{0}}\Vert _{\mathrm{op}}\Vert \Delta _{\tau }W_{\tau ,s}-\Delta
_{t}W_{t,s}\Vert _{\mathrm{op}}\mathrm{d}\tau \leq \mathrm{C}<\infty
\end{equation}%
for some constant $\mathrm{C}\in (0,\infty )$ which is independent of $s\in %
\left[ 0,T_{0}\right] $, $\epsilon >0$, and $t\in (s+\epsilon ,T_{0}]$. As a
consequence, by (\ref{interpolation5}), we arrive at the inequality
\begin{equation}
\Vert \partial _{t}\mathcal{W}_{t,s}^{(\epsilon )}\Vert _{\mathrm{op}}\leq
\mathrm{C}<\infty  \label{interpolation6}
\end{equation}%
for some constant $\mathrm{C}\in (0,\infty )$ not depending on $s\in \left[
0,T_{0}\right] $, $\epsilon >0$, and $t\in (s+\epsilon ,T_{0}]$. Moreover,
as we prove (\ref{interpolation6}), one can verify that $\partial _{t}%
\mathcal{W}_{t,s}^{(\epsilon )}$ strongly converges to
\begin{equation}
\mathcal{V}_{t,s}:=\partial _{t}\mathcal{W}_{t,s}^{(\epsilon )}|_{\epsilon
=0}=2\Delta _{t}W_{t,s}-4\int_{s}^{t}\Omega _{0}\mathrm{e}^{-2\left( t-\tau
\right) \Omega _{0}}\Delta _{\tau }W_{\tau ,s}\mathrm{d}\tau \ ,
\label{interpolation6+1}
\end{equation}%
as $\epsilon \rightarrow 0$, for all $s\in \left[ 0,T_{0}\right] $ and $t\in
(s,T_{0}]$, and%
\begin{equation}
\Vert \mathcal{V}_{t,s}\Vert _{\mathrm{op}}\leq \mathrm{C}<\infty \ ,
\label{petit inequality newbisbisbis}
\end{equation}%
by (\ref{interpolation6}). The operator $\mathcal{V}_{t,s}$ is also strongly
continuous in $t>s$, see (\ref{interpolation6+1}). Since $\mathcal{W}%
_{t,s}^{(\epsilon )}$ strongly converges to $\mathcal{W}_{t,s}^{(0)}$, we
can take the limit $\epsilon \rightarrow 0$ in%
\begin{equation}
\mathcal{W}_{t^{\prime },s}^{\left( \epsilon \right) }-\mathcal{W}%
_{t,s}^{\left( \epsilon \right) }=\int_{t}^{t^{\prime }}\partial _{\tau }%
\mathcal{W}_{\tau ,s}^{(\epsilon )}\mathrm{d}\tau
\end{equation}%
to arrive at the equality
\begin{equation}
\mathcal{W}_{t^{\prime },s}^{\left( 0\right) }-\mathcal{W}_{t,s}^{\left(
0\right) }=\int_{t}^{t^{\prime }}\mathcal{V}_{\tau ,s}\mathrm{d}\tau \ ,
\end{equation}%
with $s\in \left[ 0,T_{0}\right] $ and $t\in (s,T_{0}]$. Because $\mathcal{V}%
_{t,s}$ is strongly continuous in $t>s$, it follows that
\begin{equation}
\forall t\in (s,T_{0}]:\qquad \mathcal{V}_{t,s}=\partial _{t}\mathcal{W}%
_{t,s}^{(0)}\ ,
\end{equation}%
i.e., $\mathcal{W}_{t,s}^{(0)}$\ is strongly differential with respect to $%
t>s$. Since the operator $W_{t,s}$ solves the integral equation
\begin{equation}
W_{t,s}=\left[ \mathcal{T}\left( W\right) \right] _{t,s}=\mathrm{e}%
^{-2\left( t-s\right) \Omega _{0}}+\mathcal{W}_{t,s}^{(0)}
\label{definition of W 2bis}
\end{equation}%
(cf. (\ref{definition of W 2})), $W_{t,s}$ is strongly differential with
respect to $t>s$ with derivative%
\begin{equation}
\forall t\in (s,T_{0}]:\qquad \partial _{t}W_{t,s}=-2\Omega _{0}\mathrm{e}%
^{-2\left( t-s\right) \Omega _{0}}+\mathcal{V}_{t,s}\ .
\end{equation}%
In particular, by (\ref{petit inequality newbisbis}) and (\ref{petit
inequality newbisbisbis}),
\begin{equation}
\forall t>s\geq 0:\qquad \Vert \partial _{t}W_{t,s}\Vert _{\mathrm{op}}\leq
\mathrm{e}^{-1}\left( t-s\right) ^{-1}+\mathrm{C}<\infty \ .
\label{definition of W 2bisbis}
\end{equation}%
Finally, for any $s\in \lbrack 0,T_{0}]$ and sufficiently small $\epsilon >0$%
, we define the bounded operator
\begin{equation}
\forall t\in (s+\epsilon ,T_{0}]:\qquad W_{t,s}^{(\epsilon )}:=\mathrm{e}%
^{-2\left( t-s\right) \Omega _{0}}+\mathcal{W}_{t,s}^{(\epsilon )}\ .
\end{equation}%
Using (\ref{interpolation4bis}) and the fact that $W_{t,s}^{(\epsilon )}%
\mathfrak{h}\subset \mathcal{D}\left( \Omega _{0}\right) $ for $t>s+\epsilon
$ and $\epsilon >0$ we then observe that
\begin{equation}
\partial _{t}W_{t,s}^{(\epsilon )}+2\Omega _{t}W_{t,s}^{(\epsilon
)}=-2\Delta _{t}\mathrm{e}^{-2\left( t-s\right) \Omega _{0}}-2\Delta _{t}%
\mathcal{W}_{t,s}^{(\epsilon )}+2\mathrm{e}^{-2\epsilon \Omega _{0}}\Delta
_{t-\epsilon }W_{t-\epsilon ,s}\ ,
\end{equation}%
where we recall that $\Omega _{t}:=\Omega _{0}-\Delta _{t}$. In the limit $%
\epsilon \rightarrow 0$, the right hand side strongly converges to zero
because of (\ref{definition of W 2bis}). On the other hand, $\partial
_{t}W_{t,s}^{(\epsilon )}$ strongly converges to $\partial _{t}W_{t,s}=%
\mathcal{V}_{t,s}$, as $\epsilon \rightarrow 0$. Therefore, the operator $%
2\Omega _{t}W_{t,s}^{(\epsilon )}$ converges strongly, as $\epsilon
\rightarrow 0$, whereas $W_{t,s}^{(\epsilon )}$ converges strongly to $%
W_{t,s}$ when $\epsilon \rightarrow 0$. As $\Omega _{t}$ is a closed
operator, it follows that $W_{t,s}\mathfrak{h}\subset \mathcal{D}\left(
\Omega _{0}\right) $ and
\begin{equation}
\forall t\in (s,T_{0}]:\qquad \partial _{t}W_{t,s}=-2\Omega _{0}\mathrm{e}%
^{-2\left( t-s\right) \Omega _{0}}+\mathcal{V}_{t,s}=-2\Omega _{t}W_{t,s}\ .
\label{diff equ W}
\end{equation}%
On the other hand, the assertion
\begin{equation}
\forall t\in (0,T_{0}],\ s\in \lbrack 0,t]:\qquad \partial
_{s}W_{t,s}=2W_{t,s}\Omega _{s}\ ,  \label{diff equ W+1}
\end{equation}%
directly follows from the norm continuity of the family $(\Omega _{t})_{t\in
\lbrack 0,T_{0}]}$ together with the equality%
\begin{eqnarray}
W_{t,s} &=&\mathrm{e}^{-2\left( t-s\right) \Omega _{0}}+\underset{n=1}{%
\overset{\infty }{\sum }}2^{n}\int_{s}^{t}\mathrm{d}\tau _{1}\cdots
\int_{\tau _{n-1}}^{t}\mathrm{d}\tau _{n}  \notag \\
&&\left( \underset{j=1}{\overset{n}{\prod }}\mathrm{e}^{-2\left( \tau
_{n+2-j}-\tau _{n+1-j}\right) \Omega _{0}}\Delta _{\tau _{n+1-j}}\right)
\mathrm{e}^{-2\left( \tau _{1}-s\right) \Omega _{0}}\ ,
\label{eq aup adjoint}
\end{eqnarray}%
with $\Delta _{t}:=\Omega _{0}-\Omega _{t}\in \mathcal{B}\left( \mathfrak{h}%
\right) $ and $\tau _{n+1}:=t$. This last equation can easily be derived
from (\ref{definition of W}) and Fubini's theorem.

Therefore, by (\ref{definition of W 2bisbis}) and (\ref{diff equ W})--(\ref%
{diff equ W+1}), we have proven Assertion (ii) as well as the upper bound of
(i). Additionally, (\ref{petit inequality newbisbis}), (\ref{petit
inequality newbisbisbis}) and (\ref{diff equ W}) yield (\ref{eq diff 3}) and
$(\Omega _{t}W_{t,s})_{t>s\geq 0}$ is strongly continuous in $t>s$ because
of (\ref{interpolation3bis}), (\ref{diff equ W}) and the strong continuity
of the operator $\mathcal{V}_{t,s}$ with respect to $t>s$. Uniqueness of the
solution of (\ref{diff equ W}) (or (\ref{eq diff 4})) is standard to verify.
We omit the details. This concludes the proof of (i)--(ii), which in turn
imply (iii).

Note that similar properties as (i)--(iii) also hold for $W_{t,s}^{\ast }$, $%
W_{t,s}^{\mathrm{t}}$, and $(W_{t,s}^{\mathrm{t}})^{\ast }$. For instance,
let us consider the operator family $(W_{t,s}^{\ast })_{t\geq s\geq 0}$.
Since $\Omega _{0}=\Omega _{0}^{\ast }$ and $\Delta _{t}=\Delta _{t}^{\ast }$
for any $t\in \lbrack 0,T_{0}]$ (see (\ref{definition de delta})), one
verifies from (\ref{eq aup adjoint}) that $W_{t,s}^{\ast }$ has a
representation in term of a series constructed from the integral equation%
\begin{equation}
W_{t,s}^{\ast }=\mathrm{e}^{-2\left( t-s\right) \Omega _{0}}+2\int_{s}^{t}%
\mathrm{e}^{-2\left( \tau -s\right) \Omega _{0}}\Delta _{\tau }W_{t,\tau
}^{\ast }\mathrm{d}\tau \ ,  \label{eq aup adjoint2}
\end{equation}%
for any $t\in \lbrack 0,T_{0}]$ and $s\in \lbrack 0,t]$. Therefore, similar
properties as (i)--(iii) hold for $W_{t,s}^{\ast }$ which follow in a
similar way as for $W_{t,s}$, but interchanging the role of $s$ and $t$
(compare (\ref{eq aup adjoint2}) to (\ref{definition of W 2})). Analogous
observations can be done for $W_{t,s}^{\mathrm{t}}$ and $(W_{t,s}^{\mathrm{t}%
})^{\ast }$ with $(\Omega _{0}^{\mathrm{t}},\Delta _{t}^{\mathrm{t}})$
replacing $(\Omega _{0},\Delta _{t})$. We omit the details.\hfill $\Box $

\begin{remark}
\label{remark utiles copy(2)}One can verify that $(\Omega
_{t}W_{t,s})_{t\geq s+\epsilon \geq 0}$ is also Hölder continuous in the
norm topology for every $\epsilon >0$. See for instance similar arguments
done to prove \cite[Chap. 5, Theorem 6.9.]{Pazy}.
\end{remark}

By Lemmata \ref{lemma existence 1} (ii) and \ref{lemma existence 1 copy(1)},
the bounded operator $B_{t}=B_{t}^{\mathrm{t}}\in \mathcal{B}\left(
\mathfrak{h}\right) $ of Lemma \ref{lemma existence 2} satisfies (\ref{flow
equation-quadratic}), that is:

\begin{corollary}[Properties of $B_{t}$ for small times]
\label{section unicity of B}\mbox{ }\newline
Assume $\Omega _{0}=\Omega _{0}^{\ast }\geq 0$ and $B_{0}=B_{0}^{\mathrm{t}%
}\in \mathcal{L}^{2}\left( \mathfrak{h}\right) $. Then, for $%
T_{0}:=(128\Vert B_{0}\Vert _{2})^{-1}$, the bounded operator family $%
(B_{t})_{t\in \lbrack 0,T_{0}]}$, defined by (\ref{B_explicit_solution}) and
(\ref{definition of W}), is strongly continuous and satisfies%
\begin{equation}
B_{t}=B_{0}-2\int_{0}^{t}\left( \Omega _{\tau }B_{\tau }+B_{\tau }\Omega
_{\tau }^{\mathrm{t}}\right) \mathrm{d}\tau \ .
\end{equation}%
Moreover, $(B_{t})_{t\in \lbrack 0,T_{0}]}\in \mathfrak{C}\ $is the unique
strong solution on the domain $\mathcal{D}\left( \Omega _{0}^{\mathrm{t}%
}\right) $ of $\partial _{t}B_{t}=-2(\Omega _{t}B_{t}+B_{t}\Omega _{t}^{%
\mathrm{t}})$ for $t\in (0,T_{0}]$ as $B_{t>0}\mathfrak{h}\subseteq \mathcal{%
D}(\Omega _{0})$. $(B_{t})_{t\in (0,T_{0}]}$ is also locally Lipschitz
continuous in the norm topology.
\end{corollary}

\noindent \textbf{Proof.} The only non--trivial statement to prove is the
uniqueness of a strong solution of the differential equation
\begin{equation}
\forall t\in (0,T_{0}]:\qquad \partial _{t}B_{t}=-2(\Omega
_{t}B_{t}+B_{t}\Omega _{t}^{\mathrm{t}})\ ,\quad B_{t=0}=B_{0}\ ,
\label{uniquness new1}
\end{equation}%
on $\mathcal{D}\left( \Omega _{0}^{\mathrm{t}}\right) $. To this end, let $(%
\tilde{B}_{t})_{t\in \lbrack 0,T_{0}]}\subset \mathcal{B}\left( \mathfrak{h}%
\right) $ be a strongly continuous family of bounded operators obeying (\ref%
{uniquness new1}) on $\mathcal{D}\left( \Omega _{0}^{\mathrm{t}}\right) $.
It means, in particular, that $\tilde{B}_{t}\mathcal{D}\left( \Omega _{0}^{%
\mathrm{t}}\right) \subset \mathcal{D}\left( \Omega _{0}\right) $, whereas $%
W_{t,s}\mathfrak{h}\subseteq \mathcal{D}(\Omega _{0})$ for any $t>s$ (Lemma %
\ref{lemma existence 1 copy(1)} (i)). Therefore, using Lemma \ref{lemma
existence 1 copy(1)} (ii) we observe that
\begin{equation}
\forall t\in (0,T_{0}],\ s\in (0,t):\qquad \partial _{s}\{W_{t,s}\tilde{B}%
_{s}W_{t,s}^{\mathrm{t}}\}=0\ .
\end{equation}%
In other words,
\begin{equation}
\forall t\in (0,T_{0}],\ s\in (0,t):\qquad \tilde{B}_{t}=W_{t,s}\tilde{B}%
_{s}W_{t,s}^{\mathrm{t}}\ .
\end{equation}%
By passing to the limit $s\rightarrow 0^{+}$ (cf. Lemma \ref{lemma existence
1 copy(1)} (iii)), it follows that
\begin{equation}
\forall t\in (0,T_{0}]:\qquad \tilde{B}_{t}=W_{t}B_{0}W_{t}^{\mathrm{t}%
}=B_{t}\ ,
\end{equation}%
see (\ref{B_explicit_solution}). \hfill $\Box $

We now prove that Condition A6 is preserved under the dynamics. This
preliminary result is crucial to get the positivity of operators $\Omega
_{t} $ under A3.

\begin{lemma}[Conservation of Condition A6]
\label{strict posivity marker copy(1)}\mbox{ }\newline
Let $\Omega _{0}=\Omega _{0}^{\ast }\geq 0$ and $B_{0}=B_{0}^{\mathrm{t}}\in
\mathcal{L}^{2}\left( \mathfrak{h}\right) $ such that $\Omega _{0}$ is
invertible on $\mathrm{Ran}B_{0}$ and $\Omega _{0}\geq 4B_{0}(\Omega _{0}^{%
\mathrm{t}})^{-1}\bar{B}_{0}+\mu \mathbf{1}$ for some $\mu >0$. Then, the
operator family $(\Omega _{t})_{t\in \lbrack 0,T_{0}]}$ of Lemma \ref{lemma
existence 2} satisfies the operator inequality
\begin{equation}
\forall t\in \lbrack 0,T_{0}]:\qquad \Omega _{t}\geq 4B_{t}(\Omega _{t}^{%
\mathrm{t}})^{-1}\bar{B}_{t}+\mu \mathbf{1}\ .
\end{equation}
\end{lemma}

\noindent \textbf{Proof.} From Lemma \ref{lemma existence 2} and Corollary %
\ref{section unicity of B}, there is a strong solution $(\Omega
_{t},B_{t})_{t\in \lbrack 0,T_{0}]}$ of (\ref{flow equation-quadratic}). Let
\begin{equation}
\varkappa :=\sup \left\{ T\in \left[ 0,T_{0}\right] \
\Big |%
\ \forall t\in \left[ 0,T\right] :\quad \Omega _{t}^{\mathrm{t}}\geq \frac{%
\mu }{2}\ \mathbf{1}\right\} >0  \label{positivity 1}
\end{equation}%
and observe, by Lemma \ref{lemma existence 1 copy(1)} (ii), that
\begin{equation}
\partial _{t}\left\{ W_{t,s}^{\ast }W_{t,s}\right\} =-4W_{t,s}^{\ast }\Omega
_{t}W_{t,s}\leq 0\ ,  \label{petit inequalitybis}
\end{equation}%
for all $s\in (0,\varkappa )$ and $t\in (s,\varkappa )$. Integrating this we
obtain $W_{t,s}^{\ast }W_{t,s}\leq \mathbf{1}$ for $s\in (0,\varkappa )$ and
$t\in \left[ s,\varkappa \right) $. In the same way, $W_{t,s}^{\mathrm{t}%
}\left( W_{t,s}^{\mathrm{t}}\right) ^{\ast }\leq \mathbf{1}$ since $\Omega
_{t}\geq 0$ implies $\Omega _{t}^{\mathrm{t}}\geq 0$. In fact, the
inequality $\Omega _{x}\geq 0$ for all $x\in \left[ s,t\right] \subset \left[
0,\varkappa \right) $ yields%
\begin{equation}
\max \Big\{\left\Vert W_{t,s}\right\Vert _{\mathrm{op}},\left\Vert W_{t,s}^{%
\mathrm{t}}\right\Vert _{\mathrm{op}}\Big\}\leq 1\ .
\label{petit inequalitybisbis}
\end{equation}

For all $t\in \lbrack 0,\varkappa )$, let us consider two important
operators:%
\begin{equation}
\mathfrak{B}_{t}:=B_{t}\left( \Omega _{t}^{\mathrm{t}}\right) ^{-1}\bar{B}%
_{t}\mathrm{\qquad and\qquad }\mathfrak{D}_{t}:=\Omega _{t}-4\mathfrak{B}%
_{t}\ .  \label{new differential equation 0}
\end{equation}%
Note that $\mathfrak{B}_{t}$ is bounded for all $t\in \lbrack 0,\varkappa )$
and $\mu >0$ because (\ref{B_explicit_solution}) and (\ref{petit
inequalitybisbis}) imply that%
\begin{equation}
\forall t\in \lbrack 0,\varkappa ):\qquad \Vert \mathfrak{B}_{t}\Vert _{%
\mathrm{op}}\leq 2\left\Vert B_{0}\right\Vert _{2}^{2}\mu ^{-1}<\infty \ .
\label{upper bound toto0}
\end{equation}%
Additionally, by Lemma \ref{lemma existence 2} and Corollary \ref{section
unicity of B}, on the domain $\mathcal{D}(\Omega _{0})$ we have%
\begin{equation}
\forall t\in \left( 0,T_{0}\right] :\qquad \partial _{t}\mathfrak{B}%
_{t}=-2\left( \mathfrak{B}_{t}\mathfrak{D}_{t}+\mathfrak{D}_{t}\mathfrak{B}%
_{t}\right) -4B_{t}\bar{B}_{t}\ .  \label{diff eq B}
\end{equation}%
This derivative is justified by combining Lemma \ref{lemma existence 2} and
Corollary \ref{section unicity of B} with the upper bound%
\begin{eqnarray}
&&\left\Vert \left( \epsilon ^{-1}(\mathfrak{B}_{t+\epsilon }-\mathfrak{B}%
_{t})-\partial _{t}\{B_{t}\}(\Omega _{t}^{\mathrm{t}})^{-1}\bar{B}%
_{t}\right. \right.  \notag \\
&&\left. \left. -B_{t}\partial _{t}\{(\Omega _{t}^{\mathrm{t}})^{-1}\}\bar{B}%
_{t}-B_{t}(\Omega _{t}^{\mathrm{t}})^{-1}\partial _{t}\{\bar{B}_{t}\}\right)
\varphi \right\Vert  \notag \\
&\leq &2\left\Vert B_{0}\right\Vert _{2}\mu ^{-1}\left\Vert \left( \epsilon
^{-1}(\bar{B}_{t+\epsilon }-\bar{B}_{t})-\partial _{t}\{\bar{B}_{t}\}\right)
\varphi \right\Vert +128\left\Vert B_{0}\right\Vert _{2}^{3}\mu
^{-2}\left\Vert \bar{B}_{t+\epsilon }-\bar{B}_{t}\right\Vert _{\mathrm{op}}
\notag \\
&&+2\mu ^{-1}\left\Vert B_{t+\epsilon }-B_{t}\right\Vert _{\mathrm{op}%
}\left\Vert \epsilon ^{-1}(\bar{B}_{t+\epsilon }-\bar{B}_{t})\varphi
\right\Vert  \notag \\
&&+\left\Vert B_{0}\right\Vert _{2}\left\Vert \left( \epsilon ^{-1}((\Omega
_{t+\epsilon }^{\mathrm{t}})^{-1}-(\Omega _{t}^{\mathrm{t}})^{-1})-\partial
_{t}\{(\Omega _{t}^{\mathrm{t}})^{-1}\}\right) \bar{B}_{t}\varphi \right\Vert
\notag \\
&&+\left\Vert \left( \epsilon ^{-1}(B_{t+\epsilon }-B_{t})-\partial
_{t}\{B_{t}\}\right) (\Omega _{t}^{\mathrm{t}})^{-1}\bar{B}_{t}\varphi
\right\Vert \ ,
\end{eqnarray}%
for any $\varphi \in \mathcal{D}(\Omega _{0})$, $t\in \left( 0,T_{0}\right] $%
, and sufficiently small $|\epsilon |>0$. Note that it is easy to show that $%
\left( \Omega _{t}^{\mathrm{t}}\right) ^{-1}\bar{B}_{t}\mathfrak{h}\subset
\mathcal{D}\left( \Omega _{0}\right) $. Therefore, by Lemma \ref{lemma
existence 2} and (\ref{diff eq B}), we observe that $\mathfrak{D}_{t}$ is
the strong solution on the domain $\mathcal{D}(\Omega _{0})$ of the initial
value problem%
\begin{equation}
\forall t\in \left( 0,T_{0}\right] :\qquad \partial _{t}\mathfrak{D}%
_{t}=8\left( \mathfrak{B}_{t}\mathfrak{D}_{t}+\mathfrak{D}_{t}\mathfrak{B}%
_{t}\right) \ ,\quad \mathfrak{D}_{0}:=\Omega _{0}-4\mathfrak{B}_{0}\ .
\label{new differential equation 1}
\end{equation}

Let the operator $\mathfrak{V}_{t,s}$ be the strong solution in $\mathcal{B}%
\left( \mathfrak{h}\right) $ of the non--autonomous evolution equation%
\begin{equation}
\forall s,t\in \lbrack 0,\varkappa ):\qquad \partial _{t}\mathfrak{V}_{t,s}=8%
\mathfrak{B}_{t}\mathfrak{V}_{t,s}\ ,\quad \mathfrak{V}_{s,s}:=\mathbf{1}\ .
\label{new differential equation 2}
\end{equation}%
By (\ref{upper bound toto0}), the operator $\mathfrak{B}_{t}$ is bounded for
any $t\in \lbrack 0,\varkappa )$ and the evolution operator $\mathfrak{V}%
_{t,s}$ is of course well--defined, for any $s,t\in \lbrack 0,\varkappa )$,
by the Dyson series%
\begin{equation}
\mathfrak{V}_{t,s}:=\mathbf{1}+\underset{n=1}{\overset{\infty }{\sum }}%
8^{n}\int_{s}^{t}\mathrm{d}\tau _{1}\cdots \int_{s}^{\tau _{n-1}}\mathrm{d}%
\tau _{n}\underset{j=1}{\overset{n}{\prod }}\mathfrak{B}_{\tau _{j}}\ ,
\end{equation}%
whose operator norm is bounded from above, for any $\mu >0$, by
\begin{equation}
\Vert \mathfrak{V}_{t,s}\Vert _{\mathrm{op}}\leq \exp \left\{ 16\left\Vert
B_{0}\right\Vert _{2}^{2}\mu ^{-1}\left\vert t-s\right\vert \right\} <\infty
\ .  \label{upper bound toto}
\end{equation}%
In particular, $\mathfrak{V}_{t,s}$ satisfies the cocycle property and its
adjoint is a strong solution in $\mathcal{B}\left( \mathfrak{h}\right) $ of
the non--autonomous evolution equation
\begin{equation}
\forall s,t\in \lbrack 0,\varkappa ):\qquad \partial _{t}\mathfrak{V}%
_{t,s}^{\ast }=8\mathfrak{V}_{t,s}^{\ast }\mathfrak{B}_{t}\ ,\quad \mathfrak{%
V}_{s,s}^{\ast }:=\mathbf{1}\ ,
\end{equation}%
as $\mathfrak{B}_{t}^{\ast }=\mathfrak{B}_{t}$ is self--adjoint.
Furthermore, $\mathfrak{V}_{t,s}$ and $\mathfrak{V}_{t,s}^{\ast }$ satisfy
the non--autonomous evolution equations%
\begin{equation}
\forall s,t\in \lbrack 0,\varkappa ):\qquad \partial _{s}\mathfrak{V}%
_{t,s}=-8\mathfrak{V}_{t,s}\mathfrak{B}_{s}\ ,\quad \mathfrak{V}_{t,t}:=%
\mathbf{1}\ ,  \label{reverse ev eq0}
\end{equation}%
and%
\begin{equation}
\forall s,t\in \lbrack 0,\varkappa ):\qquad \partial _{s}\mathfrak{V}%
_{t,s}^{\ast }=-8\mathfrak{B}_{s}\mathfrak{V}_{t,s}^{\ast }\ ,\quad
\mathfrak{V}_{t,t}:=\mathbf{1}\ ,  \label{reverse ev eq}
\end{equation}%
respectively. In fact, for any $\epsilon >0$ the bounded operator family $(%
\mathfrak{B}_{t})_{t\in \lbrack \epsilon ,\varkappa )}$ is Lipschitz
continuous in the norm topology, by Lemma \ref{lemma existence 2} and
Corollary \ref{section unicity of B}. A detailed proof of basic properties
of $(\mathfrak{V}_{t,s})_{_{t\in \lbrack 0,\varkappa )}}$ is thus given for
instance by \cite[Chap. 5, Thm 5.2]{Pazy}. In particular, $(\mathfrak{V}%
_{t,s})_{_{t\in \lbrack \epsilon ,\varkappa )}}$ is also Lipschitz norm
continuous. We additionally observe that $\mathfrak{V}_{t,s}^{\ast }$
conserves the domain $\mathcal{D}\left( \Omega _{0}\right) $, i.e.,
\begin{equation}
\forall s,t\in (0,\varkappa ):\qquad \mathfrak{V}_{t,s}^{\ast }\mathcal{D}%
\left( \Omega _{0}\right) \subset \mathcal{D}\left( \Omega _{0}\right) \ .
\label{holds}
\end{equation}%
Indeed, by (\ref{reverse ev eq}), the evolution operator $\mathfrak{V}%
_{t,s}^{\ast }$ satisfies the integral equation
\begin{equation}
\forall s,t\in \lbrack 0,\varkappa ):\qquad \mathfrak{V}_{t,s}^{\ast }:=%
\mathbf{1}+8\int_{s}^{t}\mathfrak{B}_{\tau }\mathfrak{V}_{t,\tau }^{\ast }%
\mathrm{d}\tau  \label{holdsbis}
\end{equation}%
from which we deduce that, for any $s,t\in (0,\varkappa )$,
\begin{eqnarray}
\Omega _{0}\left( \mathfrak{V}_{t,s}^{\ast }-\mathbf{1}\right) &=&8\Omega
_{0}\int_{s}^{t}\left( \Omega _{0}+\mathbf{1}\right) ^{-1}\left( \Omega _{0}+%
\mathbf{1}\right) \mathfrak{B}_{\tau }\mathfrak{V}_{t,\tau }^{\ast }\mathrm{d%
}\tau  \notag \\
&=&8\frac{\Omega _{0}}{\Omega _{0}+\mathbf{1}}\int_{s}^{t}\left( \Omega _{0}+%
\mathbf{1}\right) \mathfrak{B}_{\tau }\mathfrak{V}_{t,\tau }^{\ast }\mathrm{d%
}\tau \ ,  \label{eq flow necessary}
\end{eqnarray}%
because the closed operator $\left( \Omega _{0}+\mathbf{1}\right) ^{-1}\in
\mathcal{B}\left( \mathfrak{h}\right) $ is bounded, the integrands are
continuous (see in particular Lemma \ref{lemma existence 1 copy(1)} (i)) and
one only has Riemann integrals. It follows that, for all $s,t\in
(0,\varkappa )$ and $\mu >0$,%
\begin{eqnarray}
\left\Vert \Omega _{0}\left( \mathfrak{V}_{t,s}^{\ast }-\mathbf{1}\right)
\right\Vert _{\mathrm{op}} &\leq &8\int_{\min \left\{ s,t\right\} }^{\max
\left\{ s,t\right\} }\left\Vert \mathfrak{B}_{\tau }\mathfrak{V}_{t,\tau
}^{\ast }\right\Vert _{\mathrm{op}}\mathrm{d}\tau  \notag \\
&&+8\int_{\min \left\{ s,t\right\} }^{\max \left\{ s,t\right\} }\left\Vert
\Delta _{\tau }\mathfrak{B}_{\tau }\mathfrak{V}_{t,\tau }^{\ast }\right\Vert
_{\mathrm{op}}\mathrm{d}\tau  \notag \\
&&+8\int_{\min \left\{ s,t\right\} }^{\max \left\{ s,t\right\} }\left\Vert
\Omega _{\tau }\mathfrak{B}_{\tau }\mathfrak{V}_{t,\tau }^{\ast }\right\Vert
_{\mathrm{op}}\mathrm{d}\tau  \notag \\
&<&\infty \ ,
\end{eqnarray}%
using (\ref{B_explicit_solution}) together with Lemma \ref{lemma existence 1
copy(1)} (i), (\ref{petit inequalitybisbis}), and the same upper bound on
the operator norm $\Vert \mathfrak{V}_{s,t}^{\ast }\Vert _{\mathrm{op}}$ as
the one (\ref{upper bound toto}) on $\Vert \mathfrak{V}_{t,s}\Vert _{\mathrm{%
op}}$. Therefore, (\ref{holds}) holds and the (possibly unbounded) operator $%
\mathfrak{V}_{t,s}\mathfrak{D}_{s}\mathfrak{V}_{t,s}^{\ast }$ is
well--defined on the domain $\mathcal{D}\left( \Omega _{0}\right) $ whenever
$s,t\in (0,\varkappa )$. Moreover, using (\ref{new differential equation 1})
and (\ref{reverse ev eq0})--(\ref{reverse ev eq}) one verifies that its time
derivative on $\mathcal{D}\left( \Omega _{0}\right) $ vanishes, i.e.,
\begin{equation}
\forall s,t\in (0,\varkappa ),\ \varphi \in \mathcal{D}\left( \Omega
_{0}\right) :\qquad \partial _{s}\{\mathfrak{V}_{t,s}\mathfrak{D}_{s}%
\mathfrak{V}_{t,s}^{\ast }\}\varphi =0\ .  \label{infer}
\end{equation}%
To prove (\ref{infer}), one needs to know on the domain $\mathcal{D}\left(
\Omega _{0}\right) $ that%
\begin{equation}
\partial _{s}\{\mathfrak{D}_{s}\mathfrak{V}_{t,s}^{\ast }\}=\partial _{s}\{%
\mathfrak{D}_{s}\}\mathfrak{V}_{t,s}^{\ast }+\mathfrak{D}_{s}\partial _{s}\{%
\mathfrak{V}_{t,s}^{\ast }\}\ ,  \label{infer+1}
\end{equation}%
as well as%
\begin{equation}
\partial _{s}\{\mathfrak{V}_{t,s}\mathfrak{D}_{s}\mathfrak{V}_{t,s}^{\ast
}\}=\partial _{s}\{\mathfrak{V}_{t,s}\}\mathfrak{D}_{s}\mathfrak{V}%
_{t,s}^{\ast }+\mathfrak{V}_{t,s}\partial _{s}\{\mathfrak{D}_{s}\mathfrak{V}%
_{t,s}^{\ast }\}\ .  \label{infer+2}
\end{equation}%
To prove (\ref{infer+1}), take $s\in (0,\varkappa )$, some sufficiently
small parameter $\left\vert \epsilon \right\vert >0$, and $\varphi \in
\mathcal{D}\left( \Omega _{0}\right) $. Then, observe that%
\begin{eqnarray}
&&\left\Vert \left( \epsilon ^{-1}(\mathfrak{D}_{s+\epsilon }\mathfrak{V}%
_{t,s+\epsilon }^{\ast }-\mathfrak{D}_{s}\mathfrak{V}_{t,s}^{\ast
})-\partial _{s}\{\mathfrak{D}_{s}\}\mathfrak{V}_{t,s}^{\ast }-\mathfrak{D}%
_{s}\partial _{s}\{\mathfrak{V}_{t,s}^{\ast }\}\right) \varphi \right\Vert
\notag \\
&\leq &\left\Vert \left( \epsilon ^{-1}(\mathfrak{D}_{s+\epsilon }-\mathfrak{%
D}_{s})-\partial _{s}\mathfrak{D}_{s}\right) \mathfrak{V}_{t,s}^{\ast
}\varphi \right\Vert +\left\Vert (\mathfrak{D}_{s+\epsilon }-\mathfrak{D}%
_{s})\epsilon ^{-1}(\mathfrak{V}_{t,s+\epsilon }^{\ast }-\mathfrak{V}%
_{t,s}^{\ast })\varphi \right\Vert  \notag \\
&&+\left\Vert \mathfrak{D}_{s}\left( \epsilon ^{-1}(\mathfrak{V}%
_{t,s+\epsilon }^{\ast }-\mathfrak{V}_{t,s}^{\ast })-\partial _{s}\mathfrak{V%
}_{t,s}^{\ast }\right) \varphi \right\Vert \ .  \label{infer+2+1}
\end{eqnarray}%
By (\ref{new differential equation 1}) and (\ref{holds}), for any $s\in
(0,\varkappa )$ and $\varphi \in \mathcal{D}\left( \Omega _{0}\right) $,
\begin{equation}
\underset{\epsilon \rightarrow 0}{\lim }\left\Vert \left( \epsilon ^{-1}(%
\mathfrak{D}_{s+\epsilon }-\mathfrak{D}_{s})-\partial _{s}\mathfrak{D}%
_{s}\right) \mathfrak{V}_{t,s}^{\ast }\varphi \right\Vert =0\ .
\label{infer+2+2}
\end{equation}%
Recall meanwhile that, for every $\delta \in (0,\varkappa )$, $(\mathfrak{B}%
_{t})_{t\in \lbrack \delta ,\varkappa )}$ and $(\Omega _{t})_{t\in \lbrack
0,T_{0})}$ are both (Lipschitz) norm continuous, by Lemma \ref{lemma
existence 2} and Corollary \ref{section unicity of B}. Therefore, for any $%
s\in (0,\varkappa )$ and $\varphi \in \mathcal{D}\left( \Omega _{0}\right) $%
,
\begin{equation}
\underset{\epsilon \rightarrow 0}{\lim }\left\Vert (\mathfrak{D}_{s+\epsilon
}-\mathfrak{D}_{s})\epsilon ^{-1}(\mathfrak{V}_{t,s+\epsilon }^{\ast }-%
\mathfrak{V}_{t,s}^{\ast })\varphi \right\Vert =0\ ,  \label{infer+2+3}
\end{equation}%
using also (\ref{new differential equation 0})--(\ref{upper bound toto0})
and (\ref{reverse ev eq}). Similar to Equation (\ref{eq flow necessary}), we
get from (\ref{holdsbis}) that
\begin{equation}
\Omega _{0}\left( \mathfrak{V}_{t,s+\epsilon }^{\ast }-\mathfrak{V}%
_{t,s}^{\ast }\right) =-\frac{8\Omega _{0}}{\Omega _{0}+\mathbf{1}}%
\int_{s}^{s+\epsilon }\left( \Omega _{0}+\mathbf{1}\right) \mathfrak{B}%
_{\tau }\mathfrak{V}_{t,\tau }^{\ast }\mathrm{d}\tau \ ,
\end{equation}%
which, by Lemma \ref{lemma existence 2}, implies that
\begin{eqnarray}
\Omega _{0}\left( \mathfrak{V}_{t,s+\epsilon }^{\ast }-\mathfrak{V}%
_{t,s}^{\ast }\right) &=&-\frac{8\Omega _{0}}{\Omega _{0}+\mathbf{1}}%
\int_{s}^{s+\epsilon }\Omega _{\tau }\mathfrak{B}_{\tau }\mathfrak{V}%
_{t,\tau }^{\ast }\mathrm{d}\tau  \notag \\
&&-\frac{8\Omega _{0}}{\Omega _{0}+\mathbf{1}}\int_{s}^{s+\epsilon }\left(
\Delta _{\tau }+\mathbf{1}\right) \mathfrak{B}_{\tau }\mathfrak{V}_{t,\tau
}^{\ast }\mathrm{d}\tau \ .  \label{infer+3-1}
\end{eqnarray}%
Since, by (\ref{B_explicit_solution}) and (\ref{new differential equation 0}%
),%
\begin{equation}
\forall t\in \lbrack 0,\varkappa ):\quad \mathfrak{B}_{t}=W_{t}\mathfrak{%
\hat{B}}_{t}\ ,\quad \mathfrak{\hat{B}}_{t}:=B_{0}W_{t}^{\mathrm{t}}\left(
\Omega _{t}^{\mathrm{t}}\right) ^{-1}\left( W_{t}^{\mathrm{t}}\right) ^{\ast
}\bar{B}_{0}W_{t}^{\ast }\ ,
\end{equation}%
note that, for any $s\in (0,\varkappa )$ and $\varphi \in \mathcal{D}\left(
\Omega _{0}\right) $,%
\begin{eqnarray}
&&\left\Vert \epsilon ^{-1}\int_{s}^{s+\epsilon }\left( \Omega _{\tau }%
\mathfrak{B}_{\tau }\mathfrak{V}_{t,\tau }^{\ast }-\Omega _{s}\mathfrak{B}%
_{s}\mathfrak{V}_{t,s}^{\ast }\right) \varphi \mathrm{d}\tau \right\Vert
\notag \\
&\leq &\epsilon ^{-1}\int_{s}^{s+\epsilon }\left\Vert (\Omega _{\tau
}W_{\tau }-\Omega _{s}W_{s})\mathfrak{\hat{B}}_{s}\mathfrak{V}_{t,s}^{\ast
}\varphi \right\Vert \mathrm{d}\tau  \notag \\
&&+\epsilon ^{-1}\int_{s}^{s+\epsilon }\left\Vert \Omega _{\tau }W_{\tau }(%
\mathfrak{\hat{B}}_{\tau }-\mathfrak{\hat{B}}_{s})\mathfrak{V}_{t,s}^{\ast
}\varphi \right\Vert \mathrm{d}\tau \ .  \label{infer+3}
\end{eqnarray}%
Therefore, we invoke Lemmata \ref{lemma existence 2}--\ref{lemma existence 1
copy(1)} and Corollary \ref{section unicity of B} to deduce from (\ref{new
differential equation 0}), (\ref{reverse ev eq}), (\ref{infer+3-1}), and (%
\ref{infer+3}) that, for any $s\in (0,\varkappa )$ and $\varphi \in \mathcal{%
D}\left( \Omega _{0}\right) $,
\begin{equation}
\underset{\epsilon \rightarrow 0}{\lim }\left\Vert \mathfrak{D}_{s}\left(
\epsilon ^{-1}(\mathfrak{V}_{t,s+\epsilon }^{\ast }-\mathfrak{V}_{t,s}^{\ast
})-\partial _{s}\mathfrak{V}_{t,s}^{\ast }\right) \varphi \right\Vert =0\ .
\label{infer+4}
\end{equation}%
Thus, (\ref{infer+1}) results from (\ref{infer+2+1})--(\ref{infer+2+3}) and (%
\ref{infer+4}). To prove (\ref{infer+2}), take again $s\in (0,\varkappa )$,
some sufficiently small $\left\vert \epsilon \right\vert >0$, and $\varphi
\in \mathcal{D}\left( \Omega _{0}\right) $, and consider the upper bound%
\begin{eqnarray}
&&\left\Vert \left( \epsilon ^{-1}(\mathfrak{V}_{t,s+\epsilon }\mathfrak{D}%
_{s+\epsilon }\mathfrak{V}_{t,s+\epsilon }^{\ast }-\mathfrak{V}_{t,s}%
\mathfrak{D}_{s}\mathfrak{V}_{t,s}^{\ast })\right. \right.  \notag \\
&&\left. \left. -\partial _{s}\{\mathfrak{V}_{t,s}\}\mathfrak{D}_{s}%
\mathfrak{V}_{t,s}^{\ast }-\mathfrak{V}_{t,s}\partial _{s}\{\mathfrak{D}_{s}%
\mathfrak{V}_{t,s}^{\ast }\}\right) \varphi \right\Vert  \notag \\
&\leq &\left\Vert \mathfrak{V}_{t,s}\right\Vert _{\mathrm{op}}\left\Vert
\left( \epsilon ^{-1}(\mathfrak{D}_{s+\epsilon }\mathfrak{V}_{t,s+\epsilon
}^{\ast }-\mathfrak{D}_{s}\mathfrak{V}_{t,s}^{\ast })-\partial _{s}\{%
\mathfrak{D}_{s}\mathfrak{V}_{t,s}^{\ast }\}\right) \varphi \right\Vert
\notag \\
&&+\left\Vert \mathfrak{V}_{t,s+\epsilon }-\mathfrak{V}_{t,s}\right\Vert _{%
\mathrm{op}}\left\Vert \epsilon ^{-1}(\mathfrak{D}_{s+\epsilon }\mathfrak{V}%
_{t,s+\epsilon }^{\ast }-\mathfrak{D}_{s}\mathfrak{V}_{t,s}^{\ast })\varphi
\right\Vert  \notag \\
&&+\left\Vert \left( \epsilon ^{-1}(\mathfrak{V}_{t,s+\epsilon }-\mathfrak{V}%
_{t,s})-\partial _{s}\mathfrak{V}_{t,s}\right) \mathfrak{D}_{s}\mathfrak{V}%
_{t,s}^{\ast }\varphi \right\Vert \ .  \label{infer+5}
\end{eqnarray}%
In the limit $\epsilon \rightarrow 0$ the three terms of the upper bound
vanish because of (\ref{infer+1}), the norm continuity of $(\mathfrak{V}%
_{t,s})_{t,s\in \lbrack \epsilon ,\varkappa )}\subset \mathcal{B}\left(
\mathfrak{h}\right) $ for $\epsilon >0$ and (\ref{reverse ev eq0}).

We thus obtain (\ref{infer+1}) and (\ref{infer+2}). In other words, (\ref%
{infer}) holds and implies the equality
\begin{equation}
\forall s,t\in (0,\varkappa ):\qquad \mathfrak{D}_{t}=\mathfrak{V}_{t,s}%
\mathfrak{D}_{s}\mathfrak{V}_{t,s}^{\ast }\ .
\label{new differential equation 3bis}
\end{equation}%
Since from Lemma \ref{lemma existence 2}
\begin{equation}
\forall s\in \lbrack 0,\varkappa ):\qquad \Omega _{s}=\Omega
_{0}-16\int_{0}^{s}B_{\tau }\bar{B}_{\tau }\mathrm{d}\tau \ ,
\label{toto encore}
\end{equation}%
we get, for any $s,t\in (0,\varkappa )$, that%
\begin{equation}
\mathfrak{D}_{t}=\mathfrak{V}_{t,s}\mathfrak{D}_{0}\mathfrak{V}_{t,s}^{\ast
}+4\mathfrak{V}_{t,s}\mathfrak{B}_{0}\mathfrak{V}_{t,s}^{\ast }-4\mathfrak{V}%
_{t,s}\mathfrak{B}_{s}\mathfrak{V}_{t,s}^{\ast }-16\int_{0}^{s}\mathfrak{V}%
_{t,s}B_{\tau }\bar{B}_{\tau }\mathfrak{V}_{t,s}^{\ast }\mathrm{d}\tau \ ,
\label{new differential equation 3bis+1}
\end{equation}%
where the interchange of the Riemann integral on $[0,s]$ with $\mathfrak{V}%
_{t,s}$ is justified by the fact that $\mathfrak{V}_{t,s}\in \mathcal{B}%
\left( \mathfrak{h}\right) $. As, by assumption,%
\begin{equation}
\Omega _{0}\geq 4B_{0}(\Omega _{0}^{\mathrm{t}})^{-1}\bar{B}_{0}+\mu \mathbf{%
1}\ ,
\end{equation}%
i.e., $\mathfrak{D}_{0}\geq \mu \mathbf{1}$, it follows that%
\begin{equation}
\mathfrak{D}_{t}\geq \mu \mathfrak{V}_{t,s}\mathfrak{V}_{t,s}^{\ast }+4%
\mathfrak{V}_{t,s}\mathfrak{B}_{0}\mathfrak{V}_{t,s}^{\ast }-4\mathfrak{V}%
_{t,s}\mathfrak{B}_{s}\mathfrak{V}_{t,s}^{\ast }-16\int_{0}^{s}\mathfrak{V}%
_{t,s}B_{\tau }\bar{B}_{\tau }\mathfrak{V}_{t,s}^{\ast }\mathrm{d}\tau \ ,
\label{new differential equation 3bisbis}
\end{equation}%
for any $s,t\in (0,\varkappa )$.

We proceed by taking the limit $s\rightarrow 0^{+}$ in (\ref{new
differential equation 3bisbis}). First, for all $\varphi \in \mathfrak{h}$, $%
s\in \left[ 0,\varkappa \right) $, and $\mu >0$,
\begin{equation}
\underset{s\rightarrow 0^{+}}{\lim }\langle \varphi |\left( \mathfrak{B}_{s}-%
\mathfrak{B}_{0}\right) \varphi \rangle =0\ ,
\label{new differential equation 3bisbistoto}
\end{equation}%
because the operator family $(\bar{B}_{t})_{t\in \lbrack 0,T_{0}]}$ is
strongly continuous (Corollary \ref{section unicity of B}) and for any $s\in %
\left[ 0,\varkappa \right) $,
\begin{equation}
\langle \varphi |\left( \mathfrak{B}_{s}-\mathfrak{B}_{0}\right) \varphi
\rangle \leq \frac{4}{\mu }\left\Vert B_{0}\right\Vert _{2}\left\Vert \left(
\bar{B}_{s}-\bar{B}_{0}\right) \varphi \right\Vert +\frac{64s}{\mu ^{2}}%
\left\Vert B_{0}\right\Vert _{2}^{4}\ ,
\end{equation}%
see (\ref{B_explicit_solution}), (\ref{petit inequalitybisbis}), and (\ref%
{toto encore}). Since the bounded operator family $(\mathfrak{V}%
_{t,s})_{s\in \left[ 0,\varkappa \right) }$ is strongly continuous, it is
then easy to prove, for $\mu >0$, that
\begin{equation}
\forall t\in \lbrack 0,\varkappa ):\qquad \mathfrak{D}_{t}\geq \mu \mathfrak{%
V}_{t,0}\mathfrak{V}_{t,0}^{\ast }\ ,  \label{new differential equation 3}
\end{equation}%
by passing to the limit $s\rightarrow 0^{+}$ in (\ref{new differential
equation 3bisbis}) with the help of (\ref{upper bound toto0}) and (\ref{new
differential equation 3bisbistoto}).

From (\ref{reverse ev eq0})--(\ref{reverse ev eq}) combined with $\mathfrak{B%
}_{t}=\mathfrak{B}_{t}^{\ast }\geq 0$, for $t\in \left[ 0,\varkappa \right) $%
, we observe that
\begin{equation}
\partial _{s}\left\{ \mathfrak{V}_{t,s}\mathfrak{V}_{t,s}^{\ast }\right\}
=-16\mathfrak{V}_{t,s}\mathfrak{B}_{s}\mathfrak{V}_{t,s}^{\ast }\leq 0%
\mathbf{\ ,}
\end{equation}%
which yields in this case the inequality
\begin{equation}
\forall t\in \lbrack 0,\varkappa ):\qquad \mathfrak{V}_{t,0}\mathfrak{V}%
_{t,0}^{\ast }\geq \mathbf{1}\ .
\end{equation}%
Therefore, we use (\ref{new differential equation 0}) and (\ref{new
differential equation 3}), and obtain
\begin{equation}
\forall t\in \lbrack 0,\varkappa ):\qquad \mathfrak{D}_{t}:=\Omega
_{t}-4B_{t}\left( \Omega _{t}^{\mathrm{t}}\right) ^{-1}\bar{B}_{t}\geq \mu
\mathbf{1}\ ,  \label{new differential equation 4}
\end{equation}%
provided that $\mu >0$. In particular, for any $t\in \left[ 0,\varkappa
\right) $ we have $\Omega _{t}\geq \mu \mathbf{1}$. This implies $\Omega
_{\varkappa }\geq \mu \mathbf{1}$ by norm continuity of $(\Omega _{t})_{t\in
\lbrack 0,T_{0}]}$ (Lemma \ref{lemma existence 2}), and hence
\begin{equation}
\varkappa =T_{0}\ .  \label{positivity 1fin}
\end{equation}%
It is thus easy to see that (\ref{new differential equation 4}) holds for
any $t\in \lbrack 0,T_{0}]$, i.e., for $t=\varkappa =T_{0}$ included. The
latter proves the lemma.\hfill $\Box $

Observe that the (possibly unbounded) operator $\Omega _{t}$ is bounded from
below as already mentioned in the proof of Lemma \ref{lemma existence 1
copy(1)}. If Condition A3 also holds then, by using Lemmata \ref{lemma
existence 2}, \ref{strict posivity marker copy(1)} and Corollary \ref%
{section unicity of B}, we prove next that $\Omega _{t}$ is a positive
operator:

\begin{lemma}[Positivity of the operator $\Omega _{t}$\ for small times]
\label{strict posivity marker}\mbox{ }\newline
Let $\Omega _{0}=\Omega _{0}^{\ast }\geq 0$ and $B_{0}=B_{0}^{\mathrm{t}}\in
\mathcal{L}^{2}\left( \mathfrak{h}\right) $ such that $\Omega _{0}$ is
invertible on $\mathrm{Ran}B_{0}$ and $\Omega _{0}\geq 4B_{0}(\Omega _{0}^{%
\mathrm{t}})^{-1}\bar{B}_{0}$. Let $(\Omega _{t})_{t\in \lbrack 0,T_{0}]}$
be the operator family of Lemma \ref{lemma existence 2}. Then, for all $t\in
\lbrack 0,T_{0}]$, $\Omega _{t}\geq 0$ is a positive operator.
\end{lemma}

\noindent \textbf{Proof.} Let us consider the general case where it is
assumed that
\begin{equation}
\Omega _{0}\geq 4B_{0}(\Omega _{0}^{\mathrm{t}})^{-1}\bar{B}_{0}\ .
\end{equation}%
Pick an arbitrary real parameter $\mu >0$. From Lemma \ref{lemma existence 2}
and Corollary \ref{section unicity of B}, there is a strong solution $%
(\Omega _{t,\mu },B_{t,\mu })_{t\in \lbrack 0,T_{0}]}$ of (\ref{flow
equation-quadratic}) with initial values $\Omega _{0,\mu }:=\Omega _{0}+\mu
\mathbf{1}$ and $B_{0,\mu }=B_{0}$. Note that $r_{0}$ and $T_{0}$ are
defined by (\ref{flow equation T}) and so, they do not depend on $\mu >0$.
Now, we perform the limit $\mu \rightarrow 0^{+}$ in order to prove that $%
\Omega _{t}\geq 0$ in all cases. In fact, by using (\ref{flow
equation-quadratic}) with initial values $\Omega _{0,\mu }=\Omega _{0}+\mu
\mathbf{1}$ and $B_{0,\mu }=B_{0}$, for $\mu \geq 0$ and $t\in \left[ 0,T_{0}%
\right] $, one gets that
\begin{equation}
\left\Vert \Omega _{t,\mu }-\Omega _{t}\right\Vert _{\mathrm{op}}\leq \mu
+16\int_{0}^{t}\left\Vert B_{\tau ,\mu }\bar{B}_{\tau ,\mu }-B_{\tau }\bar{B}%
_{\tau }\right\Vert _{\mathrm{op}}\mathrm{d}\tau \ .
\end{equation}%
Here,
\begin{equation}
B_{t,\mu }:=W_{t}\left( \mu \right) B_{0}W_{t}^{\mathrm{t}}\left( \mu
\right) \ ,
\end{equation}%
where the evolution operator $W_{t}\left( \mu \right) :=W_{t,0}\left( \mu
\right) $ is the strong solution of (\ref{B_initial value problem}) with $%
\Omega _{t,\mu }$ replacing $\Omega _{t}$. Therefore, using (\ref{upper
bound for W}) we proceed from the previous inequality to obtain
\begin{equation}
\left\Vert \Omega _{t,\mu }-\Omega _{t}\right\Vert _{\mathrm{op}}\leq \mu +64%
\mathrm{e}^{6r_{0}T_{0}}\Vert B_{0}\Vert _{2}^{2}\int_{0}^{T_{0}}\left\Vert
W_{\tau }\left( \mu \right) -W_{\tau }\right\Vert _{\mathrm{op}}\mathrm{d}%
\tau \ .  \label{norm continuity as function of mu}
\end{equation}%
For any $t\in \left[ 0,T_{0}\right] $, recall that $W_{t,s}$ is the unique
solution of the integral equation $W=\mathcal{T}\left( W\right) $, with $%
\mathcal{T}$ defined by (\ref{definition of W 2}). Since%
\begin{equation}
\left\Vert \mathrm{e}^{-2\alpha \Omega _{0}}-\mathrm{e}^{-2\alpha \left(
\Omega _{0}+\mu \right) }\right\Vert _{\mathrm{op}}\leq \left\vert 1-\mathrm{%
e}^{-2\alpha \mu }\right\vert
\end{equation}%
for any $\alpha ,\mu >0$, we combine $W=\mathcal{T}\left( W\right) $ with (%
\ref{flow equation T}) and the estimate (\ref{petit inequalitybisbis})
applied to $W_{t}\left( \mu \right) $ in order to deduce that%
\begin{eqnarray}
\underset{t\in \left[ 0,T_{0}\right] }{\sup }\left\Vert W_{t}\left( \mu
\right) -W_{t}\right\Vert _{\mathrm{op}} &\leq &\left\vert 1-\mathrm{e}%
^{-2T_{0}\mu }\right\vert \left( 1+2T_{0}^{2}\Vert B_{0}\Vert _{2}^{2}\right)
\\
&&+2T_{0}\ \underset{t\in \left[ 0,T_{0}\right] }{\sup }\left\Vert \Omega
_{t,\mu }-\Omega _{t}\right\Vert _{\mathrm{op}}  \notag \\
&&+2r_{0}T_{0}\ \underset{t\in \left[ 0,T_{0}\right] }{\sup }\left\Vert
W_{t}\left( \mu \right) -W_{t}\right\Vert _{\mathrm{op}}\ .  \notag
\end{eqnarray}%
By (\ref{flow equation T}), note that $2r_{0}T_{0}<1/2$ and so, for any $%
t\in \left[ 0,T_{0}\right] $,%
\begin{eqnarray}
\underset{t\in \left[ 0,T_{0}\right] }{\sup }\left\Vert W_{t}\left( \mu
\right) -W_{t}\right\Vert _{\mathrm{op}} &\leq &2\left\vert 1-\mathrm{e}%
^{-2T_{0}\mu }\right\vert \left( 1+2T_{0}^{2}\Vert B_{0}\Vert _{2}^{2}\right)
\notag \\
&&+4T_{0}\underset{t\in \left[ 0,T_{0}\right] }{\sup }\left\Vert \Omega
_{t,\mu }-\Omega _{t}\right\Vert _{\mathrm{op}}\ .
\label{norm continuity as function of mubis}
\end{eqnarray}%
Consequently, we infer from (\ref{flow equation T}), (\ref{norm continuity
as function of mu}), and (\ref{norm continuity as function of mubis}) that%
\begin{equation}
\underset{t\in \left[ 0,T_{0}\right] }{\sup }\left\Vert \Omega _{t,\mu
}-\Omega _{t}\right\Vert _{\mathrm{op}}\leq \mu +2\left\vert 1-\mathrm{e}%
^{-2T_{0}\mu }\right\vert \ \Vert B_{0}\Vert _{2}+\frac{1}{2}\ \underset{%
t\in \left[ 0,T_{0}\right] }{\sup }\left\Vert \Omega _{t,\mu }-\Omega
_{t}\right\Vert _{\mathrm{op}}\ ,
\end{equation}%
i.e.,
\begin{equation}
\underset{t\in \left[ 0,T_{0}\right] }{\sup }\left\Vert \Omega _{t,\mu
}-\Omega _{t}\right\Vert _{\mathrm{op}}\leq 2\mu +4\left\vert 1-\mathrm{e}%
^{-2T_{0}\mu }\right\vert \ \Vert B_{0}\Vert _{2}\ .
\end{equation}%
In particular, for any $t\in \left[ 0,T_{0}\right] $,
\begin{equation}
\underset{\mu \rightarrow 0^{+}}{\lim }\left\{ \underset{t\in \left[ 0,T_{0}%
\right] }{\sup }\left\Vert \Omega _{t,\mu }-\Omega _{t}\right\Vert _{\mathrm{%
op}}\right\} =0\ ,  \label{new differential equation 4bisbis}
\end{equation}%
which implies, for all $\varphi \in \mathfrak{h}$, that
\begin{equation}
\forall t\in \lbrack 0,T_{0}]:\qquad \left( \varphi ,\Omega _{t}\varphi
\right) =\underset{\mu \rightarrow 0^{+}}{\lim }\left( \varphi ,\Omega
_{t,\mu }\varphi \right) \geq 0\ ,  \label{positivity 1finfin}
\end{equation}%
because of Lemma \ref{strict posivity marker copy(1)} applied to the
operator family $(\Omega _{t,\mu })_{t\in \lbrack 0,T_{0}]}$. In other
words, the operator $\Omega _{t}$ is positive for any $t\in \left[ 0,T_{0}%
\right] $.\hfill $\Box $

By Lemma \ref{lemma existence 2}, there is a unique solution $(\Delta
_{t})_{t\in \lbrack 0,T_{0}]}\in \mathfrak{C}$ of the initial value problem (%
\ref{flow equation-quadratic delta}) for small times and it is thus natural
to define the (possibly infinite) maximal time for which such a solution
exists, that is,
\begin{equation}
T_{\max }:=\sup \left\{ T\geq 0\
\Big |%
\ \exists \mathrm{\mathrm{\ a}\ solution\ }(\Delta _{t})_{t\in \lbrack
0,T]}\in \mathfrak{C}\ \mathrm{of\ (\ref{flow equation-quadratic delta})}%
\right\} \in (0,\infty ]\ .  \label{Tmax}
\end{equation}%
As already mentioned, Lemma \ref{lemma existence 2} gives an explicit lower
bound
\begin{equation}
T_{\max }\geq T_{0}:=(128\Vert B_{0}\Vert _{2})^{-1}>0  \label{Tmaxbis}
\end{equation}%
on this maximal time. Then, we can clearly extend Lemma \ref{lemma existence
2} and Corollary \ref{section unicity of B} to all times $t\in \lbrack
0,T_{\max })$:

\begin{theorem}[Local existence of $(\Omega _{t},B_{t})$]
\label{section extension}\mbox{ }\newline
Assume $\Omega _{0}=\Omega _{0}^{\ast }\geq 0$ and $B_{0}=B_{0}^{\mathrm{t}%
}\in \mathcal{L}^{2}\left( \mathfrak{h}\right) $. Then there exists an
operator family $(\Omega _{t},B_{t})_{t\in \lbrack 0,T_{\max })}$ satisfying:%
\newline
\emph{(i)} $(\Omega _{t})_{t\in \lbrack 0,T_{\max })}$ is a family of
self--adjoint operators with all the same domain $\mathcal{D}\left( \Omega
_{0}\right) $. Moreover, it satisfies the initial value problem%
\begin{equation}
\forall t\in \lbrack 0,T_{\max }):\qquad \partial _{t}\Omega _{t}=-16B_{t}%
\bar{B}_{t}\ ,\quad \Omega _{t=0}=\Omega _{0}\ ,
\end{equation}%
in the strong topology and it is Lipschitz continuous in the norm topology
on any compact set $\left[ 0,T\right] $ with $T\in (0,T_{\max })$.\newline
\emph{(ii)} $(B_{t})_{t\in \lbrack 0,T_{\max })}=(B_{t}^{\mathrm{t}})_{t\in
\lbrack 0,T_{\max })}$, defined by (\ref{B_explicit_solution}) and (\ref%
{definition of W}), is a family of Hilbert--Schmidt operators. It is
strongly continuous and%
\begin{equation}
\forall t\in \lbrack 0,T_{\max }):\qquad B_{t}=B_{0}-2\int_{0}^{t}\left(
\Omega _{\tau }B_{\tau }+B_{\tau }\Omega _{\tau }^{\mathrm{t}}\right)
\mathrm{d}\tau \ ,
\end{equation}%
with $B_{t>0}\mathfrak{h}\subseteq \mathcal{D}(\Omega _{0})$. Furthermore, $%
(B_{t})_{t\in (0,T_{\max })}$ is locally Lipschitz norm continuous.
\end{theorem}

\noindent \textbf{Proof.} (i) The first assertion trivially follows from the
definition (\ref{Tmax}) of $T_{\max }$ as $\Omega _{t}:=\Omega _{0}-\Delta
_{t}$ for all $t\in \lbrack 0,T_{\max })$. Only the Lipschitz continuity of $%
(\Omega _{t})_{t\in \lbrack 0,T_{0}]}$ must be proven. In fact, for any $%
T\in \lbrack 0,T_{\max })$,
\begin{equation}
r=\Vert \Delta \Vert _{\infty }:=\underset{t\in \left[ 0,T\right] }{\sup }%
\left\Vert \Delta _{t}\right\Vert _{\mathrm{op}}<\infty  \label{rplus}
\end{equation}%
is obviously finite as $(\Delta _{t})_{t\in \lbrack 0,T]}\in \mathfrak{C}$.
Therefore, $\mathcal{D}\left( \Omega _{t}\right) =\mathcal{D}\left( \Omega
_{0}\right) $ and, thanks to (\ref{upper bound for W}),%
\begin{equation}
\Vert \Omega _{t}-\Omega _{s}\Vert _{\mathrm{op}}\leq 16\Vert B_{0}\Vert
_{2}^{2}\mathrm{e}^{8r(t-s)}|t-s|  \label{rplusbis}
\end{equation}%
for any $t,s\in \left[ 0,T\right] $ and $T\in \lbrack 0,T_{\max })$, i.e.,
the operator family $(\Omega _{t})_{t\in \lbrack 0,T_{\max })}$ is Lipschitz
continuous on $\left[ 0,T\right] $.

(ii) The second assertion results from Lemma \ref{lemma existence 1 copy(1)}
which can clearly be extended to all $t,s\in \left[ 0,T\right] $ and $T\in
\lbrack 0,T_{\max })$ because of (\ref{rplusbis}). Note that $B_{t}\in
\mathcal{L}^{2}\left( \mathfrak{h}\right) $ is clearly a Hilbert--Schmidt
operator as
\begin{equation}
\Vert B_{t}\Vert _{2}\leq \mathrm{e}^{8rt}\Vert B_{0}\Vert _{2}<\infty
\end{equation}%
for any $t\in \left[ 0,T\right] $ and $T\in \lbrack 0,T_{\max })$, because
of (\ref{B_explicit_solution}), (\ref{upper bound for W}), and (\ref{rplus}%
). \hfill $\Box $

This theorem says nothing about the uniqueness of the solutions $(\Omega
_{t})_{t\in \lbrack 0,T_{\max })}$ and $(B_{t})_{t\in \lbrack 0,T_{\max })}$%
. In our next lemma, we show a partial result, that is, the existence of a
unique solution $(\Delta _{t})_{t\in \lbrack 0,T_{\max })}$ of\ (\ref{flow
equation-quadratic delta}):

\begin{lemma}[Uniqueness of the operator family $(\Omega _{t})_{t\in \lbrack
0,T_{\max })}$]
\label{section extension copy(3)}\mbox{ }\newline
Assume $\Omega _{0}=\Omega _{0}^{\ast }\geq 0$ and $B_{0}=B_{0}^{\mathrm{t}%
}\in \mathcal{L}^{2}\left( \mathfrak{h}\right) $. Then there exists a unique
operator family $(\Omega _{t})_{t\in \lbrack 0,T_{\max })}$, with all the
same domain $\mathcal{D}\left( \Omega _{0}\right) $, solving in the strong
topology the initial value problem%
\begin{equation}
\forall t\in \lbrack 0,T_{\max }):\qquad \partial _{t}\Omega _{t}=-16B_{t}%
\bar{B}_{t}\ ,\quad \Omega _{t=0}=\Omega _{0}\ ,
\label{flow equation definition of f(delta)bis}
\end{equation}%
where $(B_{t})_{t\in \lbrack 0,T_{\max })}\subset \mathcal{L}^{2}\left(
\mathfrak{h}\right) $ is defined by (\ref{B_explicit_solution}) and (\ref%
{definition of W}).
\end{lemma}

\noindent \textbf{Proof.} For any $T\in (0,T_{\max })$, let $\Delta ^{\left(
1\right) },\Delta ^{\left( 2\right) }\in \mathfrak{C}$ be two solutions of
the initial value problem (\ref{flow equation-quadratic delta}), which also
define via (\ref{definition of W}) two evolution operators $W_{t}^{\left(
1\right) }:=W_{t,0}^{\left( 1\right) }$ and $W_{t}^{\left( 2\right)
}:=W_{t,0}^{\left( 2\right) }$. Pick
\begin{equation}
r=\sup \left\{ \Vert \Delta ^{\left( 1\right) }\Vert _{\infty },\Vert \Delta
^{\left( 2\right) }\Vert _{\infty }\right\} <\infty \ .
\end{equation}%
If
\begin{equation}
0<\delta \leq \min \left\{ \frac{1}{6r},\frac{1}{50\Vert B_{0}\Vert _{2}}%
,T_{\max }\right\} \ ,
\end{equation}%
then we do similar estimates as those used to prove (\ref{norm continuity as
function of mu}) and (\ref{norm continuity as function of mubis}), and
deduce the inequality
\begin{eqnarray}
\underset{t\in \left[ 0,\delta \right] }{\sup }\Vert \Delta ^{\left(
1\right) }-\Delta ^{\left( 2\right) }\Vert _{\mathrm{op}} &\leq &\frac{128%
\mathrm{e}^{6r\delta }}{\left( 1-2r\delta \right) }\delta ^{2}\Vert
B_{0}\Vert _{2}^{2}\underset{t\in \left[ 0,\delta \right] }{\sup }\Vert
\Delta ^{\left( 1\right) }-\Delta ^{\left( 2\right) }\Vert _{\mathrm{op}}
\notag \\
&<&\frac{1}{2}\underset{t\in \left[ 0,\delta \right] }{\sup }\Vert \Delta
^{\left( 1\right) }-\Delta ^{\left( 2\right) }\Vert _{\mathrm{op}}\ ,
\end{eqnarray}%
i.e., $\Delta _{t}^{\left( 1\right) }=\Delta _{t}^{\left( 2\right) }$ for
all $t\in \lbrack 0,\delta ]$. Now, we can shift the starting point from $%
t=0 $ to any fixed $x\in (0,\delta ]$ to deduce that $\Delta _{t}^{\left(
1\right) }=\Delta _{t}^{\left( 2\right) }$ for any $t\in \left[ x,x+\delta %
\right] \cap \left[ 0,T\right] $. In other words, by recursively using these
arguments, there is a unique solution $(\Delta _{t})_{t\in \lbrack 0,T]}$ in
$\mathfrak{C}$ of the initial value problem\ (\ref{flow equation-quadratic
delta}) for any $T\in (0,T_{\max })$. This concludes the proof of the lemma
as $\Omega _{t}:=\Omega _{0}-\Delta _{t}$ for all $t\in \lbrack 0,T_{\max })$%
. \hfill $\Box $

To diagonalize quadratic boson operators, we need the existence of the
operator family $(\Omega _{t},B_{t})_{t\geq 0}$ for all times, i.e., we want
to prove that $T_{\max }=\infty $ is infinite, see (\ref{Tmax}).
Unfortunately, \emph{a blow--up is generally not excluded}, that is, the
Hilbert--Schmidt norm $\left\Vert B_{t}\right\Vert _{2}$ may diverge in a
finite time. Indeed, if $T_{\max }<\infty $ then the continuous map $%
t\mapsto \Vert B_{t}\Vert _{2}$ is unbounded on the set $[0,T_{\max })$:

\begin{lemma}[The blow--up alternative]
\label{section extension gap copy(1)}\mbox{ }\newline
Assume $\Omega _{0}=\Omega _{0}^{\ast }\geq 0$ and $B_{0}=B_{0}^{\mathrm{t}%
}\in \mathcal{L}^{2}\left( \mathfrak{h}\right) $. Then, either $T_{\max
}=\infty $ and we have a global solution $(\Omega _{t})_{t\geq 0}$\ of\ (\ref%
{flow equation definition of f(delta)bis}) or $T_{\max }<\infty $ and
\begin{equation}
\underset{t\nearrow T_{\max }}{\lim }\left\Vert B_{t}\right\Vert _{2}=\infty
\ .
\end{equation}
\end{lemma}

\noindent \textbf{Proof.} By contradiction, assume the finiteness of $%
T_{\max }<\infty $ and the existence of a sequence $\left\{ t_{n}\right\}
_{n\in \mathbb{N}}\subset \lbrack 0,T_{\max })$ converging to $T_{\max }$
such that
\begin{equation}
\kappa :=\underset{n\in \mathbb{N}}{\sup }\left\Vert B_{t_{n}}\right\Vert
_{2}<\infty \ .  \label{toto3bisassumption}
\end{equation}%
Consider the integral equation%
\begin{equation}
\mathfrak{T}_{t_{n}}\left( X\right) _{t}=\left\{ \mathfrak{T}_{t_{n}}\left(
X\right) \right\} _{t}^{\ast }:=16\int_{t_{n}}^{t}W_{\tau ,t_{n}}B_{t_{n}}^{%
\mathrm{t}}W_{\tau ,t_{n}}^{\mathrm{t}}\left( W_{\tau ,t_{n}}^{\mathrm{t}%
}\right) ^{\ast }\bar{B}_{t_{n}}W_{\tau ,t_{n}}^{\ast }\mathrm{d}\tau
\label{toto5}
\end{equation}%
for $t\geq t_{n}$, where the evolution operator $W_{t,t_{n}}$ is defined by (%
\ref{definition of W}) with $(X_{t})_{t\in \lbrack t_{n},t_{n}+T]}\in
\mathfrak{C}$ replacing $(\Delta _{t})_{t\in \lbrack 0,T]}$. Obviously, the
proof of Lemma \ref{lemma existence 2} can also be performed for any
starting point $t_{n}\in \lbrack 0,T_{\max })$\ and by (\ref%
{toto3bisassumption}), there is a solution $(X_{t})_{t\in \lbrack
t_{n},t_{n}+T]}\in \mathbf{B}_{r}\left( 0\right) $ of $\mathfrak{T}%
_{t_{n}}\left( X\right) =X$, where the real positive parameters%
\begin{equation}
r=\sqrt{32}\kappa >0\qquad \mathrm{and}\qquad T=(128\kappa )^{-1}>0
\end{equation}%
do not depend on $n\in \mathbb{N}$. Using the cocycle property (Lemma \ref%
{lemma existence 1} (i)) as well as (\ref{B_explicit_solution}), one can
thus check that
\begin{equation}
\forall t\in \lbrack t_{n},t_{n}+T]:\qquad X_{t}+\Delta _{t_{n}}=\mathfrak{T}%
\left( X+\Delta _{t_{n}}\right) _{t}
\end{equation}%
for any $n\in \mathbb{N}$, whereas we have the equality
\begin{equation}
\forall t\in \lbrack t_{n},t_{n}+T]\cap \lbrack 0,T_{\max }):\qquad \Delta
_{t}=X_{t}+\Delta _{t_{n}}\ ,
\end{equation}%
by uniqueness of the solution of $\Delta =\mathfrak{T}\left( \Delta \right) $
for any $t\in \lbrack 0,T_{\max })$, see Lemma \ref{section extension
copy(3)}. Because of (\ref{Tmax}), these two last assertions cannot hold
when $t_{n}+T>T_{\max }$ for any sufficiently large $n\in \mathbb{N}$. We
thus conclude that either $T_{\max }=\infty $ or $\kappa =\infty $. \hfill $%
\Box $

A sufficient condition to prevent from having a blow--up is given by the gap
condition, i.e., Condition A6. Indeed, we can in this case extend the domain
of existence of the operator family $(\Omega _{t},B_{t})_{t\in \lbrack
0,T_{0}]}$ to any time $t\geq 0$.

\begin{lemma}[Global existence of $(\Omega _{t},B_{t})$ under the gap
condition]
\label{section extension gap}Let $\Omega _{0}=\Omega _{0}^{\ast }\geq 0$ and
$B_{0}=B_{0}^{\mathrm{t}}\in \mathcal{L}^{2}\left( \mathfrak{h}\right) $
such that $\Omega _{0}$ is invertible on $\mathrm{Ran}B_{0}$ and $\Omega
_{0}\geq 4B_{0}(\Omega _{0}^{\mathrm{t}})^{-1}\bar{B}_{0}+\mu \mathbf{1}$,
for some $\mu >0$. Then $T_{\max }=\infty $ and%
\begin{equation}
\forall t\geq 0:\qquad \Omega _{t}\geq 4B_{t}(\Omega _{t}^{\mathrm{t}})^{-1}%
\bar{B}_{t}+\mu \mathbf{1}\ .
\end{equation}
\end{lemma}

\noindent \textbf{Proof.} The arguments from (\ref{positivity 1}) to (\ref%
{positivity 1fin}) show that the gap equation holds on $[0,T_{\max })$:%
\begin{equation}
\forall t\in \lbrack 0,T_{\max }):\qquad \Omega _{t}:=\Omega _{0}-\Delta
_{t}\geq 4B_{t}(\Omega _{t}^{\mathrm{t}})^{-1}\bar{B}_{t}+\mu \mathbf{1}\ ,
\end{equation}%
In particular, $\Omega _{t}\geq 0$ is a positive operator for all $t\in
\lbrack 0,T_{\max })$. This property yields (\ref{petit inequalitybisbis})
from which we infer the global estimate%
\begin{equation}
\forall t\in \lbrack 0,T_{\max }):\qquad \left\Vert B_{t}\right\Vert
_{2}\leq \left\Vert B_{0}\right\Vert _{2}<\infty \ ,  \label{toto3bis}
\end{equation}%
see (\ref{B_explicit_solution}). Using Lemma \ref{section extension gap
copy(1)} we deduce that $T_{\max }=\infty $.\hfill $\Box $

If $\Omega _{0}\geq 0$ does not satisfy the gap equation then the absence of
a \emph{blow--up} is far from being clear. Another weaker, sufficient
condition to prevent from having this behavior is given by Condition A4,
that is, $\Omega _{0}^{-1/2}B_{0}\in \mathcal{L}^{2}(\mathfrak{h})$ is a
Hilbert--Schmidt operator. However, this proof is rather non--trivial and we
need an \emph{a priori estimate} to control the Hilbert--Schmidt norm $%
\left\Vert B_{t}\right\Vert _{2}$. To this end, assume Conditions A1--A3,
define the time
\begin{equation}
T_{+}:=\sup \left\{ T\in \lbrack 0,T_{\max })\
\Big |%
\ \forall t\in \left[ 0,T\right] :\quad \Omega _{t}\geq 0\right\} \in
(0,\infty ]  \label{temps xi}
\end{equation}%
and note that
\begin{equation}
T_{+}\geq T_{0}:=(128\Vert B_{0}\Vert _{2})^{-1}>0\ ,  \label{temps xibis}
\end{equation}%
by Lemma \ref{strict posivity marker}. Furthermore, we observe that (\ref%
{petit inequalitybisbis}) is clearly satisfied for any $t\in \lbrack
0,T_{+}) $, that is,%
\begin{equation}
\forall t\in \lbrack 0,T_{+}):\qquad \max \Big\{\left\Vert
W_{t,s}\right\Vert _{\mathrm{op}},\left\Vert W_{t,s}^{\mathrm{t}}\right\Vert
_{\mathrm{op}}\Big\}\leq 1\ ,  \label{new upper bound for W}
\end{equation}%
which in turn implies the a priori estimate
\begin{equation}
\forall s\in \lbrack 0,T_{+}),\ t\in \lbrack s,T_{+}):\qquad \left\Vert
B_{t}\right\Vert _{2}\leq \left\Vert B_{s}\right\Vert _{2}\ ,
\label{a priori estimate}
\end{equation}%
see (\ref{B_explicit_solution}). We now proceed by proving few crucial
properties which hold for all $t\in \lbrack 0,T_{+})$.

First, we aim at establishing the differential equation
\begin{equation}
\forall t\in (0,T_{+}):\qquad \partial _{t}B_{t}=-2\left( \Omega
_{t}B_{t}+B_{t}\Omega _{t}^{\mathrm{t}}\right) \ ,\quad B_{t=0}:=B_{0}\ ,
\label{Hilbert--Schmidt topology}
\end{equation}%
to hold in the Hilbert--Schmidt topology. This fact is important for our
study on the boson Fock space in\ Section \ref{Section technical proofs
copy(1)}. To show this, we need to prove that both $\Omega _{t}B_{t}$ and $%
B_{t}\Omega _{t}^{\mathrm{t}}$ are Hilbert--Schmidt operators. This is a
direct consequence of (\ref{B_explicit_solution}) and Lemma \ref{lemma
existence 1 copy(1)} (i):

\begin{lemma}[$\Omega _{t}B_{t}$ and $B_{t}\Omega _{t}^{\mathrm{t}}$ as
Hilbert--Schmidt operators]
\label{section extension copy(1)}\mbox{ }\newline
Let $\Omega _{0}=\Omega _{0}^{\ast }\geq 0$ and $B_{0}=B_{0}^{\mathrm{t}}\in
\mathcal{L}^{2}\left( \mathfrak{h}\right) $ such that $\Omega _{0}$ is
invertible on $\mathrm{Ran}B_{0}$ and $\Omega _{0}\geq 4B_{0}(\Omega _{0}^{%
\mathrm{t}})^{-1}\bar{B}_{0}$. Then, for any $t\in (0,T_{+})$, both $\Omega
_{t}B_{t}$ and $B_{t}\Omega _{t}^{\mathrm{t}}$ are Hilbert--Schmidt
operators and $\Vert \Omega _{t}B_{t}\Vert _{2}=\Vert B_{t}\Omega _{t}^{%
\mathrm{t}}\Vert _{2}<\infty $.
\end{lemma}

\noindent \textbf{Proof.} By using (\ref{B_explicit_solution}), (\ref{new
upper bound for W}), Lemma \ref{lemma existence 1 copy(1)} (i) extended to
all $t,s\in \left[ 0,T\right] $ for any $T\in \lbrack 0,T_{\max })$, and the
cyclicity of the trace (cf. Lemma \ref{lemma existence 3 copy(3)} (i)), we
deduce that
\begin{equation}
\forall t\in (0,T_{+}):\quad \Vert \Omega _{t}B_{t}\Vert _{2}\leq \left\Vert
\Omega _{t}W_{t}\right\Vert _{\mathrm{op}}\Vert B_{0}\Vert _{2}\leq \mathrm{%
\tilde{C}}_{0}+\frac{\mathrm{\tilde{D}}_{0}}{t}<\infty
\end{equation}%
with $\mathrm{\tilde{C}}_{0},\mathrm{\tilde{D}}_{0}<\infty $. Moreover, note
that
\begin{equation}
\mathrm{tr}(B_{t}^{\ast }\Omega _{t}^{2}B_{t})=\mathrm{tr}(\bar{B}_{t}\Omega
_{t}^{2}B_{t})=\mathrm{tr}(B_{t}(\Omega _{t}^{\mathrm{t}})^{2}\bar{B}_{t})=%
\mathrm{tr}(B_{t}\left( \Omega _{t}^{\mathrm{t}}\right) ^{2}B_{t}^{\ast
})<\infty \ ,
\end{equation}%
because of Lemma \ref{lemma existence 3 copy(2)} and $B_{t}^{\mathrm{t}%
}=B_{t}$. From the cyclicity of the trace (Lemma \ref{lemma existence 3
copy(3)} (iii)), it follows that
\begin{equation}
\forall t\in (0,T_{+}):\qquad \Vert B_{t}\Omega _{t}^{\mathrm{t}}\Vert
_{2}=\Vert \Omega _{t}^{\mathrm{t}}B_{t}^{\ast }\Vert _{2}=\Vert \Omega
_{t}B_{t}\Vert _{2}<\infty \ .
\end{equation}%
\hfill $\Box $

As a consequence, since $B_{t}\in \mathcal{L}^{2}\left( \mathfrak{h}\right) $
is Hilbert--Schmidt, one can use the isomorphism $\mathcal{J}$, defined in
Section \ref{section hilbert schmidt}, between $\mathcal{L}^{2}(\mathfrak{h}%
) $ and $\mathfrak{h}\otimes \mathfrak{h}^{\ast }$. In this space, $\left(
\Omega _{t}B_{t}+B_{t}\Omega _{t}^{\mathrm{t}}\right) _{t>0}$ is seen as
vectors of $\mathcal{L}^{2}(\mathfrak{h})$, instead of operators acting on
the one--particle Hilbert space $\mathfrak{h}$. In particular, in the
Hilbert space $\mathfrak{h}\otimes \mathfrak{h}^{\ast }$,%
\begin{equation}
\mathcal{J}\left( \Omega _{t}X+X\Omega _{t}^{\mathrm{t}}\right) =\mathbf{%
\Omega }_{t}\mathcal{J}\left( X\right) \ ,
\end{equation}%
provided $\mathcal{J}\left( X\right) \in \mathcal{D}(\mathbf{\Omega }_{t})$,
where, for all $t\in (0,T_{+})$,
\begin{equation}
\mathbf{\Omega }_{t}:=\Omega _{t}\otimes \mathbf{1}+\mathbf{1}\otimes \Omega
_{t}^{\mathrm{t}}\ .
\end{equation}%
The differential equation (\ref{Hilbert--Schmidt topology}) is, in $\mathcal{%
L}^{2}(\mathfrak{h})$ or $\mathfrak{h}\otimes \mathfrak{h}^{\ast }$, a
non--autonomous \emph{parabolic} evolution equation as
\begin{equation}
\forall t\in (0,T_{+}):\qquad \partial _{t}\{\mathcal{J}\left( B_{t}\right)
\}=-2\mathbf{\Omega }_{t}\mathcal{J}\left( B_{t}\right) \ ,\quad \mathcal{J}%
\left( B_{t=0}\right) :=\mathcal{J}\left( B_{0}\right) \ .  \label{eq iso}
\end{equation}

Therefore, to establish (\ref{Hilbert--Schmidt topology}) in $\mathcal{L}%
^{2}(\mathfrak{h})$, or (\ref{eq iso}) in $\mathfrak{h}\otimes \mathfrak{h}%
^{\ast }$, we proceed by strengthening the regularity properties satisfied
by the operators $\Omega _{t}$\ and $B_{t}$:

\begin{theorem}[Continuity properties of $\Omega _{t}$\ and $B_{t}$]
\label{lemma existence 3}\mbox{ }\newline
Let $\Omega _{0}=\Omega _{0}^{\ast }\geq 0$ and $B_{0}=B_{0}^{\mathrm{t}}\in
\mathcal{L}^{2}\left( \mathfrak{h}\right) $ such that $\Omega _{0}$ is
invertible on $\mathrm{Ran}B_{0}$ and $\Omega _{0}\geq 4B_{0}(\Omega _{0}^{%
\mathrm{t}})^{-1}\bar{B}_{0}$. \newline
\emph{(i)} The operator family $(\Omega _{t})_{t\in \lbrack 0,T_{+})}$\ is
globally Lipschitz continuous in $\mathcal{L}^{1}(\mathfrak{h})$ and $%
\mathcal{L}^{2}(\mathfrak{h})$.\newline
\emph{(ii)} The Hilbert--Schmidt operator family $(B_{t})_{t\in \lbrack
0,T_{+})}$ is continuous in $\mathcal{L}^{2}(\mathfrak{h})$ and even locally
Lipschitz continuous in $\mathcal{L}^{2}(\mathfrak{h})$ on $(0,T_{+})$.%
\newline
\emph{(iii)} The Hilbert--Schmidt operator families $(\Omega
_{t}B_{t})_{t\in (0,T_{+})}$ and $(B_{t}\Omega _{t}^{\mathrm{t}})_{t\in
(0,T_{+})}$ are continuous in $\mathcal{L}^{2}(\mathfrak{h})$.
\end{theorem}

\noindent \textbf{Proof.} \underline{(i):} For all $s\in \lbrack 0,T_{+})$
and $t\in \lbrack s,T_{+})$,
\begin{equation}
\left\Vert \Omega _{t}-\Omega _{s}\right\Vert _{2}\leq 16\left( t-s\right)
\left\Vert B_{0}\right\Vert _{2}^{2}\ .
\end{equation}%
The latter means that $(\Omega _{t})_{t\geq 0}$ is globally Lipschitz
continuous in the Hilbert--Schmidt topology. The same trivially holds in $%
\mathcal{L}^{1}(\mathfrak{h})$.\smallskip

\noindent \underline{(ii):} By using (\ref{B_explicit_solution}), that is,
\begin{equation}
B_{t}:=W_{t}B_{0}W_{t}^{\mathrm{t}}=W_{t,s}B_{s}W_{t,s}^{\mathrm{t}}\ ,
\label{B_explicit_solution2}
\end{equation}%
we have, for all $s\in \lbrack 0,T_{+})$ and $t\in \lbrack s,T_{+})$ that%
\begin{equation}
\Vert B_{t}-B_{s}\Vert _{2}\leq \Vert \left( W_{t,s}-\mathbf{1}\right)
B_{s}\Vert _{2}+\Vert W_{t,s}B_{s}\left( W_{t,s}^{\mathrm{t}}-\mathbf{1}%
\right) \Vert _{2}\ .  \label{continuity lebegue plus0}
\end{equation}%
Therefore, we need to analyze both terms of this upper bound. Starting with
the first one we observe that
\begin{equation}
\Vert \left( W_{t,s}-\mathbf{1}\right) B_{s}\Vert _{2}^{2}=\mathrm{tr}(\bar{B%
}_{s}\left( W_{t,s}-\mathbf{1}\right) ^{\ast }\left( W_{t,s}-\mathbf{1}%
\right) B_{s})=\underset{n=1}{\overset{\infty }{\sum }}h_{t,s}(n)\ ,
\end{equation}%
where
\begin{equation}
h_{t,s}\left( n\right) :=\Vert \left( W_{t,s}-\mathbf{1}\right) B_{s}\varphi
_{n}\Vert ^{2}
\end{equation}%
and $\{\varphi _{n}\}_{n=1}^{\infty }\subset \mathfrak{h}$ is any
orthonormal basis of the Hilbert space $\mathfrak{h}$. Since, thanks to (\ref%
{new upper bound for W}), $\Vert W_{t,s}\Vert _{\mathrm{op}}\leq 1$, the
coefficient $h_{t,s}\left( n\right) $ is uniformly bounded in $t$ by%
\begin{equation}
h_{t,s}\left( n\right) \leq 4\Vert B_{s}\varphi _{n}\Vert ^{2}
\end{equation}%
and, as $B_{s}\in \mathcal{L}^{2}\left( \mathfrak{h}\right) $ is a
Hilbert--Schmidt operator,
\begin{equation}
\underset{n=1}{\overset{\infty }{\sum }}4\Vert B_{s}\varphi _{n}\Vert
^{2}=4\Vert B_{s}\Vert _{2}^{2}<\infty \ .
\end{equation}%
Therefore, by Lebesgue's dominated convergence theorem, we obtain that
\begin{equation}
\underset{t\rightarrow s^{+}}{\lim }\Vert (W_{t,s}-\mathbf{1})B_{s}\Vert
_{2}^{2}=\underset{n=1}{\overset{\infty }{\sum }}\left( \underset{%
t\rightarrow s^{+}}{\lim }h_{t,s}(n)\right) \ ,
\label{continuity lebegue plus}
\end{equation}%
provided $h_{t,s}(n)$ has a limit for all $n\in \mathbb{N}$, as $%
t\rightarrow s^{+}$. By Lemma \ref{lemma existence 1} (ii), the operator $%
W_{t,s}$ is strongly continuous in $t$ with $W_{s,s}=\mathbf{1}$. Hence, for
all $n\in \mathbb{N}$,
\begin{equation}
\underset{t\rightarrow s^{+}}{\lim }h_{t,s}\left( n\right) =0\ ,
\end{equation}%
which, together with (\ref{continuity lebegue plus}), implies that
\begin{equation}
\underset{t\rightarrow s^{+}}{\lim }\Vert \left( W_{t,s}-\mathbf{1}\right)
B_{s}\Vert _{2}^{2}=0\ .  \label{continuity lebegue plus1}
\end{equation}%
The limit
\begin{equation}
\underset{t\rightarrow s^{+}}{\lim }\Vert W_{t,s}B_{s}\left( W_{t,s}^{%
\mathrm{t}}-\mathbf{1}\right) \Vert _{2}^{2}=0
\label{continuity lebegue plus2}
\end{equation}%
is obtained in the same way as (\ref{continuity lebegue plus1}) because,
using $\Vert W_{t,s}\Vert _{\mathrm{op}}\leq 1$ and the cyclicity of the
trace (Lemma \ref{lemma existence 3 copy(3)} (i)),
\begin{equation}
\Vert W_{t,s}B_{s}(W_{t,s}^{\mathrm{t}}-\mathbf{1})\Vert _{2}^{2}\leq \Vert
((W_{t,s}^{\mathrm{t}})^{\ast }-\mathbf{1})\bar{B}_{s}\Vert _{2}^{2}
\label{continuity lebegue plus01}
\end{equation}%
and $\left( W_{t,s}^{\mathrm{t}}\right) ^{\ast }$ is also strongly
continuous in $t\geq s\geq 0$ with $\left( W_{s,s}^{\mathrm{t}}\right)
^{\ast }=\mathbf{1}$ and $\Vert W_{t,s}^{\mathrm{t}}\Vert _{\mathrm{op}}\leq
1$.

Consequently, the limits (\ref{continuity lebegue plus1}) and (\ref%
{continuity lebegue plus2}), together with the upper bound (\ref{continuity
lebegue plus0}), yield the right continuity in $\mathcal{L}^{2}(\mathfrak{h}%
) $ of the Hilbert--Schmidt operator family $(B_{t})_{t\in \lbrack 0,T_{+})}$%
. Furthermore, since the evolution operators $W_{t,s}$ and $W_{t,s}^{\mathrm{%
t}}$ are also strongly continuous in $s\leq t$, the left continuity in $%
\mathcal{L}^{2}(\mathfrak{h})$ of $(B_{t})_{t\in \lbrack 0,T_{+})}$ is
verified in the same way. In other words, the Hilbert--Schmidt operator
family $(B_{t})_{t\in \lbrack 0,T_{+})}$ is continuous in $\mathcal{L}^{2}(%
\mathfrak{h})$.

We prove now that $B_{t}$ is locally Lipschitz continuous in $\mathcal{L}%
^{2}(\mathfrak{h})$ for $t\in (0,T_{+})$. Using the integral equation of
Theorem \ref{section extension} (ii) we directly derive the inequality%
\begin{equation}
\Vert B_{t}-B_{s}\Vert _{2}\leq 2\int_{s}^{t}\left( \Vert \Omega _{\tau
}B_{\tau }\Vert _{2}+\Vert B_{t}\Omega _{t}^{\mathrm{t}}\Vert _{2}\right)
\mathrm{d}\tau \ .  \label{eq cyclicity unbounded1bis-changer}
\end{equation}%
From Lemma \ref{section extension copy(1)} both $\Omega _{t}B_{t}$ and $%
B_{t}\Omega _{t}^{\mathrm{t}}$ are Hilbert--Schmidt operators for any $t\in
(0,T_{+})$ and so, the operator family $(B_{t})_{t\in (0,T_{+})}$ is locally
Lipschitz continuous in the Hilbert--Schmidt topology.\smallskip

\noindent \underline{(iii):} Finally, we can again use (\ref%
{B_explicit_solution2}) to obtain that%
\begin{eqnarray}
\Vert \Omega _{t}B_{t}-\Omega _{s}B_{s}\Vert _{2} &\leq &\Vert \Omega
_{t}W_{t}\Vert _{\mathrm{op}}\Vert B_{0}W_{s}^{\mathrm{t}}\left( W_{t,s}^{%
\mathrm{t}}-\mathbf{1}\right) \Vert _{2}  \notag \\
&&+\Vert \left( \Omega _{t}W_{t}-\Omega _{s}W_{s}\right) B_{0}W_{s}^{\mathrm{%
t}}\Vert _{2}\ ,  \label{inequality autrebis}
\end{eqnarray}%
for all $s\in \lbrack 0,T_{+})$ and $t\in \lbrack s,T_{+})$. By Lemma \ref%
{lemma existence 1 copy(1)} (i) extended to all $t,s\in \left[ 0,T\right] $
for any $T\in \lbrack 0,T_{\max })$, the bounded operator family $(\Omega
_{t}W_{t})_{t>0}$ is strongly continuous. Therefore, similar to (\ref%
{continuity lebegue plus1}) and (\ref{continuity lebegue plus2}), we can use
Lebesgue's dominated convergence theorem to deduce that, for all $s\in
(0,T_{+})$,%
\begin{equation}
\underset{t\rightarrow s^{+}}{\lim }\Vert \left( \Omega _{t}W_{t}-\Omega
_{s}W_{s}\right) B_{0}W_{s}^{\mathrm{t}}\Vert _{2}=0
\end{equation}%
whereas
\begin{equation}
\underset{t\rightarrow s^{+}}{\lim }\left\{ \Vert \Omega _{t}W_{t}\Vert _{%
\mathrm{op}}\Vert B_{0}W_{s}^{\mathrm{t}}\left( W_{t,s}^{\mathrm{t}}-\mathbf{%
1}\right) \Vert _{2}\right\} =0\ .
\end{equation}%
Therefore, by (\ref{inequality autrebis}), we arrive at the right continuity
in $\mathcal{L}^{2}(\mathfrak{h})$ of the Hilbert--Schmidt operator family $%
(\Omega _{t}B_{t})_{t\in (0,T_{+})}$. The left continuity in $\mathcal{L}%
^{2}(\mathfrak{h})$ of $(\Omega _{t}B_{t})_{t\in (0,T_{+})}$ is proven in
the same way. Furthermore, using the cyclicity of the trace (Lemma \ref%
{lemma existence 3 copy(3)}), Lemma \ref{lemma existence 3 copy(2)}, $B_{t}^{%
\mathrm{t}}=B_{t}$, and Lemma \ref{section extension copy(1)}, one shows
that
\begin{eqnarray}
\Vert B_{t}\Omega _{t}^{\mathrm{t}}-B_{s}\Omega _{s}^{\mathrm{t}}\Vert
_{2}^{2} &=&\Vert B_{t}\Omega _{t}^{\mathrm{t}}\Vert _{2}^{2}+\Vert
B_{s}\Omega _{s}^{\mathrm{t}}\Vert _{2}^{2}-\mathrm{tr}(\Omega _{s}^{\mathrm{%
t}}\bar{B}_{s}B_{t}\Omega _{t}^{\mathrm{t}})-\mathrm{tr}(\Omega _{t}^{%
\mathrm{t}}\bar{B}_{t}B_{s}\Omega _{s}^{\mathrm{t}})  \notag \\
&=&\Vert \Omega _{t}B_{t}\Vert _{2}^{2}+\Vert \Omega _{s}B_{s}\Vert _{2}^{2}-%
\mathrm{tr}(\Omega _{t}B_{t}\bar{B}_{s}\Omega _{s})-\mathrm{tr}(\Omega
_{s}B_{s}\bar{B}_{t}\Omega _{t})  \notag \\
&=&\Vert \Omega _{t}B_{t}\Vert _{2}^{2}+\Vert \Omega _{s}B_{s}\Vert _{2}^{2}-%
\mathrm{tr}(\bar{B}_{s}\Omega _{s}\Omega _{t}B_{t})-\mathrm{tr}(\bar{B}%
_{t}\Omega _{t}\Omega _{s}B_{s})  \notag \\
&=&\Vert \Omega _{t}B_{t}-\Omega _{s}B_{s}\Vert _{2}^{2}
\end{eqnarray}%
for any $s\in (0,T_{+})$ and $t\in \lbrack s,T_{+})$. In other words, the
family $(B_{t}\Omega _{t}^{\mathrm{t}})_{t\in (0,T_{+})}$ of
Hilbert--Schmidt operators is also continuous in $\mathcal{L}^{2}(\mathfrak{h%
})$.\hfill $\Box $

As a consequence, the differential equation (\ref{Hilbert--Schmidt topology}%
) holds true in the Hilbert--Schmidt topology:

\begin{corollary}[Well--posedness of the flow on $\mathcal{L}^{2}\left(
\mathfrak{h}\right) $]
\label{Corollary existence}\mbox{ }\newline
Let $\Omega _{0}=\Omega _{0}^{\ast }\geq 0$ and $B_{0}=B_{0}^{\mathrm{t}}\in
\mathcal{L}^{2}\left( \mathfrak{h}\right) $ such that $\Omega _{0}$ is
invertible on $\mathrm{Ran}B_{0}$ and $\Omega _{0}\geq 4B_{0}(\Omega _{0}^{%
\mathrm{t}})^{-1}\bar{B}_{0}$. Then the operator family $(\Omega _{t}-\Omega
_{0},B_{t})_{t\in \lbrack 0,T_{+})}\in C[[0,T_{+});\mathcal{L}^{2}(\mathfrak{%
h})\oplus \mathcal{L}^{2}(\mathfrak{h})]$ is the unique solution of the
system of differential equations%
\begin{equation}
\left\{
\begin{array}{lllll}
\forall t\in \lbrack 0,T_{+}) & : & \partial _{t}\Omega _{t}=-16B_{t}\bar{B}%
_{t} & , & \Omega _{t=0}=\Omega _{0}\ , \\
\forall t\in (0,T_{+}) & : & \partial _{t}B_{t}=-2\left( \Omega
_{t}B_{t}+B_{t}\Omega _{t}^{\mathrm{t}}\right) & , & B_{t=0}=B_{0}\ ,%
\end{array}%
\right.  \label{toto dif eq}
\end{equation}%
both in $\mathcal{L}^{2}(\mathfrak{h})$, i.e., in the Hilbert--Schmidt
topology.
\end{corollary}

\noindent \textbf{Proof.} We use Theorem \ref{section extension} and the a
priori estimate (\ref{a priori estimate}) to deduce, for all $t\in \lbrack
0,T_{+})$ and $\delta \in \lbrack -t,T_{+}-t)$, the two inequalities%
\begin{eqnarray}
\left\Vert \delta ^{-1}\left\{ \Omega _{t+\delta }-\Omega _{t}\right\}
+16B_{t}\bar{B}_{t}\right\Vert _{2} &\leq &32\delta ^{-1}\int_{t}^{t+\delta
}\left\Vert B_{t}-B_{\tau }\right\Vert _{2}\left\Vert B_{t}\right\Vert _{2}\
\mathrm{d}\tau  \notag \\
&&  \label{strengthening 1}
\end{eqnarray}%
and
\begin{eqnarray}
&&\left\Vert \delta ^{-1}\left\{ B_{t+\delta }-B_{t}\right\} +2\Omega
_{t}B_{t}+2B_{t}\Omega _{t}^{\mathrm{t}}\right\Vert _{2}
\label{strengthening 2} \\
&\leq &2\delta ^{-1}\int_{t}^{t+\delta }\left\{ \left\Vert \Omega
_{t}B_{t}-\Omega _{\tau }B_{\tau }\right\Vert _{2}+\left\Vert B_{t}\Omega
_{t}^{\mathrm{t}}-B_{\tau }\Omega _{\tau }^{\mathrm{t}}\right\Vert
_{2}\right\} \ \mathrm{d}\tau \ .  \notag
\end{eqnarray}%
Therefore, from (\ref{a priori estimate}) and Theorem \ref{lemma existence 3}
(ii)--(iii) combined with the inequalities (\ref{strengthening 1})--(\ref%
{strengthening 2}) in the limit $\delta \rightarrow 0$, we conclude that $%
(\Omega _{t}-\Omega _{0},B_{t})_{t\in \lbrack 0,T_{+})}$ is a solution of
the differential equations stated in the corollary.

Uniqueness of the family $(\Omega _{t},B_{t})_{t\in \lbrack 0,T_{+})}$
defined in Theorem \ref{section extension} is then a direct corollary of
Lemmata \ref{section extension copy(3)} and \ref{lemma existence 3 copy(1)}:
As soon as $B_{t}$ is defined by (\ref{B_explicit_solution}), the operator $%
\Omega _{t}$ is unique, see Lemma \ref{section extension copy(3)}.
Uniqueness of $B_{t}\in \mathcal{L}^{2}\left( \mathfrak{h}\right) $ solution
of the integral equation of Theorem \ref{section extension} (ii) directly
follows from Lemma \ref{lemma existence 3 copy(1)} for $T=T_{+}$, $\alpha =2$%
, $Z_{0}=\Omega _{0}=Z_{0}^{\ast }$, and
\begin{equation}
Q_{t}=-16\int_{0}^{t}B_{\tau }\bar{B}_{\tau }\mathrm{d}\tau =Q_{t}^{\ast }\ .
\label{Q_t}
\end{equation}%
Indeed, the self--adjoint operator $Q_{t}\in \mathcal{B}\left( \mathfrak{h}%
\right) $ is bounded for any $t\in \lbrack 0,T_{+})$, as the a priori
estimate (\ref{a priori estimate}) yields
\begin{equation}
\left\Vert Q_{t}\right\Vert _{2}\leq 16t\left\Vert B_{0}\right\Vert
_{2}<\infty  \label{Q_tbis}
\end{equation}%
for all $t\in \lbrack 0,T_{+})$. Note that the uniqueness can also be
directly deduced from Corollary \ref{section unicity of B}\ extended to $%
[0,T_{+})$.

Therefore, assume that some family $(\tilde{\Omega}_{t}-\Omega _{0},\tilde{B}%
_{t})_{t\in \lbrack 0,T_{+})}\in C[[0,T_{+});\mathcal{L}^{2}(\mathfrak{h}%
)\oplus \mathcal{L}^{2}(\mathfrak{h})]$ solves (\ref{toto dif eq}). Then, $%
\tilde{B}_{t}$ is equal to (\ref{B_explicit_solution}) with $\tilde{\Omega}%
_{t}$ replacing $\Omega _{t}$, by Lemma \ref{lemma existence 3 copy(1)}. As
a consequence, $(\tilde{\Omega}_{t}-\Omega _{0})_{t\in \lbrack 0,T_{+})}\in
C[[0,T_{+});\mathcal{L}^{2}(\mathfrak{h})]$ solves (\ref{flow
equation-quadratic delta}) and $\tilde{\Omega}_{t}=\Omega _{t}$ for all $%
t\in \lbrack 0,T_{+})$, which in turn implies\ $\tilde{B}_{t}=B_{t}$ for all
$t\in \lbrack 0,T_{+})$.\hfill $\Box $

We now come back to the problem of the global existence of $(\Omega
_{t},B_{t})_{t\geq 0}$. We would like to prevent from having any blow--up,
which means that $T_{\max }=\infty $, see (\ref{Tmax}). In the proof of
Lemma \ref{section extension gap}, the gap condition, i.e.,%
\begin{equation}
\Omega _{0}\geq 4B_{0}(\Omega _{0}^{\mathrm{t}})^{-1}\bar{B}_{0}+\mu \mathbf{%
1}\ ,\quad \mu >0\ ,
\end{equation}%
was crucial and is preserved for all times $t\in \lbrack 0,T_{\max })$ in
order to prove next that $T_{\max }=\infty $. However, when $\mu =0$, the
inequality
\begin{equation}
\Omega _{0}\geq 4B_{0}(\Omega _{0}^{\mathrm{t}})^{-1}\bar{B}_{0}
\end{equation}%
is not necessarily preserved for all times $t\in \lbrack 0,T_{\max })$.

\begin{remark}
\label{flow equation-quadratic remark 3}From Theorem \ref{lemma constante of
motion} (i) combined with $\Omega _{t}^{\mathrm{t}}\geq 0$,%
\begin{equation}
\Omega _{0}^{2}\geq 8B_{0}\bar{B}_{0}\Longrightarrow \Omega _{t}^{2}\geq
8B_{t}\bar{B}_{t}\ ,
\end{equation}%
which implies $\Omega _{t}\geq 8B_{t}(\Omega _{t}^{\mathrm{t}})^{-1}\bar{B}%
_{t}$ for all $t\in \lbrack 0,T_{+})$. However, this proof of the first
statement (i) of Theorem \ref{theorem important 1 -2} only works for this
special example. Note that one can also prove in this specific case that $%
\Omega _{t}\geq \frac{\Omega _{0}}{2t\Omega _{0}+\mathbf{1}}$ for any $t\in
\lbrack 0,T_{+})$.
\end{remark}

At first sight, one could use similar arguments performed from (\ref{new
differential equation 1}) to (\ref{new differential equation 4}) for $\mu =0$%
. Indeed, these arguments, which use the system (\ref{flow
equation-quadratic}), suggest that%
\begin{equation}
\mathfrak{D}_{t}:=\Omega _{t}-4\mathfrak{B}_{t}=\mathfrak{V}_{t}\mathfrak{D}%
_{0}\mathfrak{V}_{t}^{\ast }\geq 0\ ,  \label{new system 1}
\end{equation}%
where
\begin{equation}
\mathfrak{B}_{t}:=B_{t}(\Omega _{t}^{\mathrm{t}})^{-1}\bar{B}_{t}=\mathfrak{B%
}_{t}^{\ast }\geq 0
\end{equation}%
and $\mathfrak{V}_{t}:=\mathfrak{V}_{t,0}$ is the strong solution of the
non--autonomous evolution equation (\ref{new differential equation 2}), that
is,
\begin{equation}
\forall s,t\in \lbrack 0,T_{+}):\qquad \partial _{t}\mathfrak{V}_{t,s}=8%
\mathfrak{B}_{t}\mathfrak{V}_{t,s}\ ,\quad \mathfrak{V}_{s,s}:=\mathbf{1}\ .
\label{new system 1bis}
\end{equation}

Unfortunately, since $(\Omega _{t}^{\mathrm{t}})^{-1}$ is possibly
unbounded, it is not clear that the evolution operator $\mathfrak{V}_{t,s}$
is well--defined. For instance, the boundedness of $\mathfrak{B}_{0}\in
\mathcal{B}\left( \mathfrak{h}\right) $ does not imply, a priori, the
boundedness of the operator $\mathfrak{B}_{t}$ for all times $t\in \lbrack
0,T_{+})$. However, if $\mathfrak{B}_{0}\in \mathcal{L}^{1}(\mathfrak{h})$
is not only bounded but also trace--class, i.e., $\Omega _{0}^{-1/2}B_{0}\in
\mathcal{L}^{2}(\mathfrak{h})$ (Condition A4), then we show below that the
operator $\mathfrak{B}_{t}\in \mathcal{L}^{1}(\mathfrak{h})$ stays
trace--class for all times $t\in \lbrack 0,T_{+})$ and Inequality (\ref{new
system 1}) can thus be justified.

\begin{lemma}[$\mathfrak{B}_{t}\in \mathcal{L}^{1}(\mathfrak{h})$ and
positivity of $\mathfrak{D}_{t}\geq 0$]
\label{existence_flow_5 copy(1)}\mbox{ }\newline
Assume $\Omega _{0}=\Omega _{0}^{\ast }\geq 0$, $B_{0}=B_{0}^{\mathrm{t}}\in
\mathcal{L}^{2}\left( \mathfrak{h}\right) $, $\Omega _{0}\geq 4B_{0}(\Omega
_{0}^{\mathrm{t}})^{-1}\bar{B}_{0}$, $\Omega _{0}^{-1/2}B_{0}\in \mathcal{L}%
^{2}(\mathfrak{h})$, and let $(\Omega _{t},B_{t})_{t\in \lbrack 0,T_{+})}$
be the solution of (\ref{flow equation-quadratic}). \newline
\emph{(i)} The map $t\mapsto \Vert \mathfrak{B}_{t}\Vert _{1}$ is
monotonically decreasing. In particular, $\Vert \mathfrak{B}_{t}\Vert
_{1}\leq \Vert \mathfrak{B}_{0}\Vert _{1}$ for all $t\in \lbrack 0,T_{+})$.
\newline
\emph{(ii)} The operator $\mathfrak{D}_{t}\geq 0$ is positive for all $t\in
\lbrack 0,T_{+})$.
\end{lemma}

\noindent \textbf{Proof.} First, observe that $\Omega _{0}^{-1/2}B_{0}\in
\mathcal{L}^{2}(\mathfrak{h})$ implies that $(\Omega _{0}^{\mathrm{t}%
})^{-1/2}\bar{B}_{0}\in \mathcal{L}^{2}(\mathfrak{h})$, by Lemma \ref{lemma
existence 3 copy(2)} and $B_{0}=B_{0}^{\mathrm{t}}$. In particular, $%
\mathfrak{B}_{0}\in \mathcal{L}^{1}(\mathfrak{h})$ as
\begin{equation}
\mathfrak{b}_{0}:=\Vert \mathfrak{B}_{0}\Vert _{1}=\Vert \Omega
_{0}^{-1/2}B_{0}\Vert _{2}^{2}<\infty \ .
\end{equation}

Let $\mu >0$, $T\in (0,T_{+})$, and set
\begin{equation}
\forall t\in \left[ 0,T\right] :\qquad \mathfrak{B}_{t,\mu }:=B_{t}\left(
\Omega _{t}^{\mathrm{t}}+\mu \mathbf{1}\right) ^{-1}\bar{B}_{t}\geq 0\ .
\label{eq-2}
\end{equation}%
It is a bounded operator for all $t\in \lbrack 0,T_{+})$ as $\mu >0$ and $%
\Omega _{t}^{\mathrm{t}}\geq 0$ is a positive operator by definition of $%
T_{+}>0$, see (\ref{temps xi}). In fact, $\mathfrak{B}_{t,\mu }\in \mathcal{L%
}^{1}(\mathfrak{h})$ is trace--class. Therefore, we introduce the function $%
q_{\mu }$ defined on $[0,T_{+})$ by
\begin{equation}
q_{\mu }(t):=\mathrm{tr}\{\mathfrak{B}_{t,\mu }\}=\Vert \mathfrak{B}_{t,\mu
}\Vert _{1}\ .  \label{eq-3}
\end{equation}%
Using Corollary \ref{Corollary existence}, Lemma \ref{lemma existence 3
copy(4)}, the cyclicity of the trace (Lemma \ref{lemma existence 3 copy(3)}
(i)), $\mathfrak{B}_{t,\mu }\in \mathcal{L}^{1}(\mathfrak{h})$ with $%
\mathfrak{B}_{t,\mu }\geq 0$, and the positivity of the self--adjoint
operators $\Omega _{t},\Omega _{t}^{\mathrm{t}}\geq 0$ for all $t\in \lbrack
0,T_{+})$, we observe that its derivative is well--defined for any strictly
positive time $t\in (0,T_{+})$ and satisfies:
\begin{eqnarray}
\partial _{t}q_{\mu }(t) &=&16\mathrm{tr}\left\{ B_{t}\left( \Omega _{t}^{%
\mathrm{t}}+\mu \mathbf{1}\right) ^{-1}\bar{B}_{t}B_{t}\left( \Omega _{t}^{%
\mathrm{t}}+\mu \mathbf{1}\right) ^{-1}\bar{B}_{t}\right\}  \label{eq-3bis}
\\[0.01in]
&&-4\mathrm{tr}\left\{ \Omega _{t}B_{t}\left( \Omega _{t}^{\mathrm{t}}+\mu
\mathbf{1}\right) ^{-1}\bar{B}_{t}+B_{t}\Omega _{t}^{\mathrm{t}}\left(
\Omega _{t}^{\mathrm{t}}+\mu \mathbf{1}\right) ^{-1}\bar{B}_{t}\right\}
\notag \\[0.01in]
&\leq &16\mathrm{tr}\{\mathfrak{B}_{t,\mu }^{2}\}\leq 16\,q_{\mu }^{2}(t)\ .
\end{eqnarray}%
By majorisation, we thus obtain the inequality
\begin{equation}
q_{\mu }(t)\leq \frac{1}{q_{\mu }^{-1}(s)-16\left( t-s\right) }\leq 2q_{\mu
}(s)\ ,  \label{eq-5}
\end{equation}%
provided that the times $s\in (0,T_{+})$ and $t\in \lbrack s,T_{+})$ satisfy
the inequality
\begin{equation}
t-s\leq \frac{1}{32q_{\mu }(s)}\ .  \label{eq-6}
\end{equation}%
Now, using the resolvent identity
\begin{equation}
\left( X+\mathbf{1}\right) ^{-1}-\left( Y+\mathbf{1}\right) ^{-1}=\left( X+%
\mathbf{1}\right) ^{-1}\left( Y-X\right) \left( Y+\mathbf{1}\right) ^{-1}
\label{V.197}
\end{equation}%
for any positive operator $X,Y\geq 0$, as well as the Cauchy--Schwarz
inequality, the cyclicity of the trace (Lemma \ref{lemma existence 3 copy(3)}
(i)), the positivity of the operator $\Omega _{s}^{\mathrm{t}}\geq 0$ for
all $s\in \lbrack 0,T_{+})$ and the a priori estimate (\ref{a priori
estimate}), note that%
\begin{eqnarray}
\left\vert q_{\mu }(t)-q_{\mu }(s)\right\vert &\leq &\Vert \mathfrak{B}%
_{t,\mu }-\mathfrak{B}_{s,\mu }\Vert _{1}  \notag \\
&\leq &\mu ^{-2}\Vert B_{s}\Vert _{2}^{2}\ \Vert \Omega _{t}^{\mathrm{t}%
}-\Omega _{s}^{\mathrm{t}}\Vert _{1}  \notag \\
&&+2\mu ^{-1}\Vert B_{s}\Vert _{2}\ \Vert B_{t}-B_{s}\Vert _{2}
\label{continuity1}
\end{eqnarray}%
for all $\mu >0$ and $s,t\in \lbrack 0,T_{+})$. In particular, thanks to
Theorem \ref{lemma existence 3} (i)--(ii), the function $q_{\mu }$ is
continuous on the whole interval $[0,T_{+})$ and in particular at zero:%
\begin{equation}
\underset{s\rightarrow 0^{+}}{\lim }\ q_{\mu }(s)=q_{\mu }(0)\leq q_{0}(0)=%
\mathfrak{b}_{0}:=\Vert \mathfrak{B}_{0}\Vert _{1}\ .  \label{eq-7}
\end{equation}%
We thus combine (\ref{eq-7}) with (\ref{eq-5}) and (\ref{eq-6}) to arrive at
the inequality
\begin{equation}
\forall \mu >0,\ \forall t\in \left[ 0,T\right] :\qquad \Vert \mathfrak{B}%
_{t,\mu }\Vert _{\mathrm{op}}\ \leq \ q_{\mu }(t)\ \leq \ 2\mathfrak{b}_{0}\
,  \label{eq-8}
\end{equation}%
provided that $T<1/(32\ \mathfrak{b}_{0})$. In particular, since, by (\ref%
{continuity1}) and Theorem \ref{lemma existence 3} (i)--(ii), the
trace--class operator family $(\mathfrak{B}_{t,\mu })_{t\in \left[ 0,T\right]
}$ is (at least) norm continuous ($T<1/(32\ \mathfrak{b}_{0})$), the
evolution operator defined by the Dyson series
\begin{equation}
\forall s,t\in \lbrack 0,T]\ :\qquad \mathfrak{W}_{t,s}\ :=\mathbf{1}%
+\sum_{n=1}^{\infty }8^{n}\int_{s}^{t}\mathrm{d}\tau _{1}\cdots
\int_{s}^{\tau _{n-1}}\mathrm{d}\tau _{n}\;\mathfrak{B}_{\tau _{1},\mu
}\cdots \mathfrak{B}_{\tau _{n},\mu }  \label{eq-9}
\end{equation}%
is the unique, bounded and norm continuous solution of the non--autonomous
evolution equation
\begin{equation}
\forall s,t\in \lbrack 0,T]:\qquad \partial _{t}\mathfrak{W}_{t,s}=8\,%
\mathfrak{B}_{t,\mu }\,\mathfrak{W}_{t,s}\ ,\quad \mathfrak{W}_{s,s}=\mathbf{%
1}\ .  \label{eq-10}
\end{equation}%
Next, we set
\begin{equation}
\mathfrak{D}_{t,\mu }:=\Omega _{t}-4\mathfrak{B}_{t,\mu }  \label{eq-11}
\end{equation}%
on the interval $[0,T]$ and observe that, for all strictly positive times $%
t\in (0,T]$,%
\begin{equation}
\partial _{t}\mathfrak{D}_{t,\mu }=8\mathfrak{B}_{t,\mu }\big(\mathfrak{D}%
_{t,\mu }-\mu \mathbf{1}\big)+8\big(\mathfrak{D}_{t,\mu }-\mu \mathbf{1}\big)%
\mathfrak{B}_{t,\mu }  \label{eq-12}
\end{equation}%
on the domain $\mathcal{D}\left( \Omega _{0}\right) $, with positive initial
value
\begin{equation}
\mathfrak{D}_{0,\mu }\ :=\Omega _{0}-4B_{0}\left( \Omega _{0}^{\mathrm{t}%
}+\mu \right) ^{-1}\bar{B}_{0}\geq 0\ .  \label{eq-13}
\end{equation}%
The proof of (\ref{eq-12}) uses exactly the same arguments as those proving (%
\ref{new differential equation 1}). Using (\ref{eq-10}) and also the
arguments justifying (\ref{new differential equation 3bis}) we remark that
\begin{equation}
\forall s,t\in (0,T]:\qquad \mathfrak{D}_{t,\mu }-\mu \mathbf{1}=\mathfrak{W}%
_{t,s}\big(\mathfrak{D}_{s,\mu }-\mu \mathbf{1}\big)\mathfrak{W}_{t,s}^{\ast
}\ ,  \label{eq-14}
\end{equation}%
from which we obtain, for $s,t\in (0,T]$, the inequality%
\begin{eqnarray}
\mathfrak{D}_{t,\mu } &\geq &-\mu \left( \mathfrak{W}_{t,s}\mathfrak{W}%
_{t,s}^{\ast }-\mathbf{1}\right) +4\mathfrak{W}_{t,s}\mathfrak{B}_{0,\mu }%
\mathfrak{W}_{t,s}^{\ast }-4\mathfrak{W}_{t,s}\mathfrak{B}_{s,\mu }\mathfrak{%
W}_{t,s}^{\ast }  \notag \\
&&-16\int_{0}^{s}\mathfrak{W}_{t,s}B_{\tau }\bar{B}_{\tau }\mathfrak{W}%
_{t,s}^{\ast }\mathrm{d}\tau ,
\end{eqnarray}%
using (\ref{toto encore}). As it is similarly done to prove (\ref{new
differential equation 3}), we take the limit $s\rightarrow 0$ in this last
inequality to arrive at the following assertion:%
\begin{equation}
\forall t\in \left[ 0,T\right] :\qquad \mathfrak{D}_{t,\mu }\geq -\mu \left(
\mathfrak{W}_{t,0}\mathfrak{W}_{t,0}^{\ast }-\mathbf{1}\right) \ ,
\label{eq-15}
\end{equation}%
with strictly positive arbitrary parameter $\mu >0$. Now, since, by (\ref%
{eq-10}),
\begin{equation}
\mathfrak{W}_{t,s}\mathfrak{W}_{t,s}^{\ast }-\mathbf{1}=16\int_{s}^{t}%
\mathfrak{W}_{t,\tau }\mathfrak{B}_{\tau ,\mu }\mathfrak{W}_{t,\tau }^{\ast }
\label{eq-16}
\end{equation}%
for any $s,t\in \left[ 0,T\right] $, one gets the upper bound%
\begin{equation}
\Vert \mathfrak{W}_{t,s}\mathfrak{W}_{t,s}^{\ast }-\mathbf{1}\Vert _{{%
\mathrm{op}}}\leq 16t\,\sup_{s\leq \tau \leq t}\left\{ \Vert \mathfrak{B}%
_{\tau ,\mu }\Vert _{{\mathrm{op}}}\right\} \left( \sup_{s\leq \tau \leq
t}\left\{ \Vert \mathfrak{W}_{t,\tau }\mathfrak{W}_{t,\tau }^{\ast }-\mathbf{%
1}\Vert _{{\mathrm{op}}}\right\} +1\right) \ .  \label{eq-17bis}
\end{equation}%
Introducing the function
\begin{equation}
w_{\mu }:=\sup_{0\leq \tau \leq T}\Vert \mathfrak{W}_{T,\tau }\mathfrak{W}%
_{T,\tau }^{\ast }-\mathbf{1}\Vert _{{\mathrm{op}}}
\end{equation}%
and using the upper bound%
\begin{equation}
\sup_{0\leq \tau \leq T}\left\{ \Vert \mathfrak{B}_{\tau ,\mu }\Vert _{{%
\mathrm{op}}}\right\} \leq 2\mathfrak{b}_{0}
\end{equation}%
(cf. (\ref{eq-8})), we infer from (\ref{eq-17bis}) that
\begin{equation}
w_{\mu }\ \leq \frac{32T\mathfrak{b}_{0}}{1-32T\mathfrak{b}_{0}}=:\mathrm{K}%
_{T}<\infty  \label{eq-18}
\end{equation}%
because $T<1/(32\ \mathfrak{b}_{0})$. Inserting (\ref{eq-18}) into (\ref%
{eq-15}), we arrive at the following inequality:%
\begin{equation}
\forall t\in \left[ 0,T\right] :\qquad \mathfrak{D}_{t,\mu }\geq -\mu
\mathrm{K}_{T}\ \mathbf{1}\ .  \label{eq-20}
\end{equation}%
Since $\Omega _{t}\geq \mathfrak{D}_{t,\mu }$, this last statement reads
\begin{equation}
\forall \mu >0,\ \forall t\in \left[ 0,T\right] :\qquad \Omega _{t}\geq
\mathfrak{D}_{t,\mu }\geq -\mu \mathrm{K}_{T}\ \mathbf{1}\ .  \label{eq-22}
\end{equation}%
Therefore, it remains to perform the limit $\mu \rightarrow 0^{+}$ in (\ref%
{eq-22}). Note first that $\ker (\Omega _{t}^{\mathrm{t}})\cap \mathrm{Ran}%
(B_{t})=\emptyset $, by (\ref{eq-2})--(\ref{eq-3}), (\ref{eq-5}), and (\ref%
{eq-7}). Since the net $(y_{\mu })_{\mu >0}$ of bounded real functions
\begin{equation}
x\mapsto y_{\mu }\left( x\right) =\frac{1}{x+\mu }
\end{equation}%
from $\mathbb{R}_{0}^{+}$ to $\mathbb{R}_{0}^{+}$ is monotonically
decreasing in $\mu $, the spectral theorem applied to the positive
self--adjoint operator $\Omega _{t}^{\mathrm{t}}\geq 0$ for $t\in \lbrack
0,T_{+})$ and the monotone convergence theorem both yield the limit
\begin{equation}
\lim_{\mu \searrow 0}(\varphi ,\mathfrak{B}_{t,\mu }\varphi )=(\varphi ,%
\mathfrak{B}_{t}\varphi )\in \lbrack 0,2\mathfrak{b}_{0}]  \label{eq-23-1}
\end{equation}%
for all $\varphi \in \mathfrak{h}$. It follows that $\mathfrak{B}%
_{t}^{1/2}\in \mathcal{B}\left( \mathfrak{h}\right) $, i.e., $\mathfrak{B}%
_{t}\in \mathcal{B}\left( \mathfrak{h}\right) $, and, by (\ref{eq-22}),
\begin{equation}
\forall t\in \left[ 0,T\right] :\qquad \mathfrak{D}_{t}:=\Omega _{t}-4%
\mathfrak{B}_{t}\geq 0\ ,  \label{eq-23}
\end{equation}%
provided that $T\in \lbrack 0,T_{+})$ satisfies $T<1/(32\ \mathfrak{b}_{0})$.

In fact, since the functional
\begin{equation}
\varphi \mapsto z_{\mu }\left( \varphi \right) =(\varphi ,\mathfrak{B}%
_{t,\mu }\varphi )
\end{equation}%
from $\mathfrak{h}$ to $\mathbb{R}_{0}^{+}$ is monotonically decreasing in $%
\mu $, we obtain that
\begin{equation}
\lim_{\mu \rightarrow 0^{+}}\Vert \mathfrak{B}_{t,\mu }\Vert _{1}=\Vert
\mathfrak{B}_{t}\Vert _{1}\in \lbrack 0,2\mathfrak{b}_{0}]\ ,  \label{eq-24}
\end{equation}%
by applying again the monotone convergence theorem together with (\ref{eq-8}%
) and (\ref{eq-23-1}).

Now, from (\ref{eq-3bis}) and knowing that $\Omega _{t}\geq 4\mathfrak{B}%
_{t} $ for all $t\in \left[ 0,T\right] $, we remark that
\begin{equation}
\partial _{t}q_{\mu }(t)\leq -4\mathrm{tr}\left\{ B_{t}\Omega _{t}^{\mathrm{t%
}}\left( \Omega _{t}^{\mathrm{t}}+\mu \mathbf{1}\right) ^{-1}\bar{B}%
_{t}\right\} \leq 0
\end{equation}%
for any strictly positive time $t\in \left( 0,T\right] $. It follows that $%
\Vert \mathfrak{B}_{t,\mu }\Vert _{1}\leq \Vert \mathfrak{B}_{s,\mu }\Vert
_{1}<\infty $ for all $s\in \left( 0,T\right] $ and $t\in \lbrack s,T]$,
which, by (\ref{eq-7}), can be extended by continuity to $s=0$. In
particular, through (\ref{eq-24}), $\Vert \mathfrak{B}_{t}\Vert _{1}\leq
\Vert \mathfrak{B}_{s}\Vert _{1}<\infty $ for any $s\in \left[ 0,T\right] $
and $t\in \lbrack s,T]$. As a consequence,
\begin{equation}
T<1/(32\ \mathfrak{b}_{0})\leq 1/(32\ \mathfrak{b}_{t})
\end{equation}%
for any $t\in \lbrack 0,T]$. We can thus shift the starting point from $t=0$
to any fixed $x\in \left( 0,T\right) $ and use the same arguments. In
particular, by using an induction argument, we deduce that $\Omega _{t}\geq 4%
\mathfrak{B}_{t}$ and $\Vert \mathfrak{B}_{t}\Vert _{1}\leq \Vert \mathfrak{B%
}_{s}\Vert _{1}<\infty $ for all $s\in \lbrack 0,T_{+})$ and $t\in \lbrack
s,T_{+})$. \hfill $\Box $

We are now in position to prove that Condition A4, that is, $\Omega
_{0}^{-1/2}B_{0}\in \mathcal{L}^{2}(\mathfrak{h})$, prevents from having a
blow--up. In particular, in this case $T_{+}=T_{\max }=\infty $, see (\ref%
{Tmax}) and (\ref{temps xi}).

\begin{theorem}[Global existence of $(\Omega _{t},B_{t})$]
\label{section extension copy(4)}\mbox{ }\newline
Assume Conditions A1--A4, that is, $\Omega _{0}=\Omega _{0}^{\ast }\geq 0$, $%
B_{0}=B_{0}^{\mathrm{t}}\in \mathcal{L}^{2}\left( \mathfrak{h}\right) $, $%
\Omega _{0}\geq 4B_{0}(\Omega _{0}^{\mathrm{t}})^{-1}\bar{B}_{0}$, and $%
\Omega _{0}^{-1/2}B_{0}\in \mathcal{L}^{2}(\mathfrak{h})$. Then $%
T_{+}=T_{\max }=\infty $ and
\begin{equation}
\forall t\geq 0:\qquad \Omega _{t}\geq 4B_{t}(\Omega _{t}^{\mathrm{t}})^{-1}%
\bar{B}_{t}\ ,\qquad \Omega _{t}^{-1/2}B_{t}\in \mathcal{L}^{2}(\mathfrak{h}%
)\ .
\end{equation}
\end{theorem}

\noindent \textbf{Proof.} For any $x\geq 0$ and $\mu >0$, we consider the
integral equation $X_{\mu }=\mathfrak{T}_{x}\left( X_{\mu }\right) $, where $%
\Omega _{x,\mu }:=\Omega _{x}+\mu \mathbf{1}$ and $B_{x,\mu }=B_{x}$. By
Lemma \ref{section extension gap}, there is a (unique) global solution $%
(X_{t,\mu })_{t\in \lbrack x,\infty )}$ of such integral equation for any $%
x\in \lbrack 0,T_{+})$ and $\mu >0$ satisfying
\begin{equation}
\forall x\in \mathbb{R}_{0}^{+},t\in \lbrack x,\infty ):\qquad \Omega
_{t,\mu }=\Omega _{x}+\mu \mathbf{1}-X_{t,\mu }\geq \mu \mathbf{1}\ ,
\label{toto4}
\end{equation}%
because
\begin{equation}
\forall t\in \lbrack 0,T_{+}):\qquad \Omega _{t}\geq 4B_{t}(\Omega _{t}^{%
\mathrm{t}})^{-1}\bar{B}_{t}\ ,
\end{equation}%
see Lemma \ref{existence_flow_5 copy(1)} (ii). On the other hand, the
solution $(\Delta _{t})_{t\in \lbrack 0,T]}\in \mathfrak{C}$\ of\ the
initial value problem (\ref{flow equation-quadratic delta}) satisfies%
\begin{equation}
\forall t\in \lbrack x,T):\qquad \Delta _{t}-\Delta _{x}=\mathfrak{T}%
_{x}\left( \Delta -\Delta _{x}\right) _{t}\ ,  \label{toto3-1}
\end{equation}%
for all $x\in \lbrack 0,T_{\max })$ and $T\in \lbrack x,T_{\max })$.
Therefore, using (\ref{a priori estimate}) together with similar arguments
as those used to prove Lemma \ref{strict posivity marker} we show that the
bounded operator $X_{t,\mu }$ converges in norm to $(\Delta _{t}-\Delta
_{x}) $ as $\mu \rightarrow 0^{+}$, uniformly for $t\in \lbrack
x,x+T_{0}]\cap \lbrack 0,T_{\max })$ whenever $x\in \lbrack 0,T_{+})$. By (%
\ref{toto4}), it follows that $\Omega _{t}\geq 0$ for all $t\in \lbrack
x,x+T_{0}]\cap \lbrack 0,T_{\max })$ and any $x\in \lbrack 0,T_{+})$. The
positive time $T_{0}$ does not depend on $x\in \lbrack 0,T_{+})$ and, by (%
\ref{temps xi}), we conclude that $T_{+}=T_{\max }$. The latter yields
\begin{equation}
\forall t\in \lbrack 0,T_{\max }):\quad \Omega _{t}:=\Omega _{0}-\Delta
_{t}\geq 0\ ,\ \left\Vert B_{t}\right\Vert _{2}\leq \left\Vert
B_{0}\right\Vert _{2}<\infty \ ,
\end{equation}%
where $(\Delta _{t})_{t\in \lbrack 0,T]}$ is the solution in $\mathfrak{C}$
of\ the initial value problem (\ref{flow equation-quadratic delta}) for any $%
T\in (0,T_{\max })$. In particular, by Lemma \ref{section extension gap
copy(1)}, we arrive at the assertion $T_{\max }=\infty $. Moreover, $\Omega
_{t}^{-1/2}B_{t}\in \mathcal{L}^{2}(\mathfrak{h})$ because of Lemma \ref%
{existence_flow_5 copy(1)} (i) and
\begin{equation}
\forall t\geq 0:\qquad \Vert \mathfrak{B}_{t}\Vert _{1}=\Vert \Omega
_{t}^{-1/2}B_{t}\Vert _{2}^{2}\ .
\end{equation}%
\hfill $\Box $

We finally conclude this subsection by some observations on the flow. Assume
Conditions A1--A4. By (\ref{flow equation-quadratic}), we note that,
formally,%
\begin{equation}
\forall t>0:\qquad \left\{
\begin{array}{ll}
\partial _{t}\mathfrak{D}_{t}=8\left( \mathfrak{B}_{t}\mathfrak{D}_{t}+%
\mathfrak{D}_{t}\mathfrak{B}_{t}\right) & , \\
\partial _{t}\mathfrak{B}_{t}=-2\left( \mathfrak{B}_{t}\mathfrak{D}_{t}+%
\mathfrak{D}_{t}\mathfrak{B}_{t}\right) -4B_{t}\bar{B}_{t} & ,%
\end{array}%
\right.
\end{equation}%
with $\mathfrak{B}_{t}\in \mathcal{L}^{1}(\mathfrak{h})$, whereas $\mathfrak{%
D}_{t}:=\Omega _{t}-4\mathfrak{B}_{t}\geq 0$ is the sum of a positive
operator $\Omega _{0}$ and an operator which is bounded for all $t\geq 0$,
by Lemma \ref{existence_flow_5 copy(1)}. By Lemma \ref{strict posivity
marker copy(1)} and Theorem \ref{lemma existence 3}, the trace--class
operator family $(\mathfrak{B}_{t})_{t\geq 0}$ is clearly norm continuous
when A6 holds and this property should persist under A1--A4. Since $(\Omega
_{t})_{t\geq 0}$\ is at least norm continuous (Theorem \ref{lemma existence
3}), it should exist two evolution operators $\mathfrak{U}_{t,s}$ and $%
\mathfrak{V}_{t,s}$ which are the strong solution in $\mathcal{B}\left(
\mathfrak{h}\right) $ of
\begin{equation}
\forall t>s\geq 0:\qquad \partial _{t}\mathfrak{U}_{t,s}=-2\mathfrak{D}_{t}%
\mathfrak{U}_{t,s}\ ,\quad \mathfrak{U}_{s,s}:=\mathbf{1}\ ,
\end{equation}%
and (\ref{new system 1bis}) (here $T_{+}=\infty $), respectively. In
particular,
\begin{equation}
\forall t\geq s\geq 0:\qquad \mathfrak{D}_{t}=\mathfrak{V}_{t,s}\mathfrak{D}%
_{s}\mathfrak{V}_{t,s}^{\ast }\geq 0\ ,
\end{equation}%
as heuristically explained before Lemma \ref{existence_flow_5 copy(1)} (cf. (%
\ref{new system 1})), whereas we should have
\begin{equation}
\forall t\geq s\geq 0:\qquad \mathfrak{B}_{t}=\mathfrak{U}_{t,s}\mathfrak{B}%
_{s}\mathfrak{U}_{t,s}^{\ast }-4\int_{s}^{t}\mathfrak{U}_{t,\tau }B_{\tau }%
\bar{B}_{\tau }\mathfrak{U}_{t,\tau }^{\ast }\mathrm{d}\tau \ .
\label{Bt tilde}
\end{equation}%
This equation may be proven by using the operator families $(\mathfrak{B}%
_{t,\mu })_{t\geq 0}$ and $(\mathfrak{D}_{t,\mu })_{t\geq 0}$ respectively
defined, for any arbitrarily constant $\mu >0$, by (\ref{eq-2}) and (\ref%
{eq-11}), together with the limit $\mu \rightarrow 0^{+}$.

Another interesting observation on the flow is the conservation for all
times of all inequalities of the type
\begin{equation}
\Omega _{0}\geq (4+\mathrm{r})B_{0}(\Omega _{0}^{\mathrm{t}})^{-1}\bar{B}%
_{0}+\mu \mathbf{1=}(4+\mathrm{r})\mathfrak{B}_{0}+\mu \mathbf{1}\ ,
\label{generalized A3-A6}
\end{equation}%
where $\mu ,\mathrm{r}\geq 0$ are two nonnegative constants. This is already
proven when $\mathrm{r}=0$\ and $\mu \geq 0$, see Lemmata \ref{section
extension gap} and Theorem \ref{section extension copy(4)}, but this
property can now be generalized to all $\mu ,\mathrm{r}\geq 0$:

\begin{lemma}[Conservation by the flow of $\Omega _{0}\geq (4+\mathrm{r})%
\mathfrak{B}_{0}+\protect\mu \mathbf{1}$]
\label{new lemma 2 copy(2)}\mbox{ }\newline
Assume Conditions A1--A2 and A4. If (\ref{generalized A3-A6}) holds for some
$\mu ,\mathrm{r}\geq 0$, then%
\begin{equation}
\forall t\geq 0:\qquad \Omega _{t}\geq (4+\mathrm{r})B_{t}(\Omega _{t}^{%
\mathrm{t}})^{-1}\bar{B}_{t}+\mu \mathbf{1}\ .  \label{asserted estimate}
\end{equation}
\end{lemma}

\noindent \textbf{Proof.} First note that
\begin{equation}
\Omega _{0}\geq (4+\mathrm{r})B_{0}(\Omega _{0}^{\mathrm{t}})^{-1}\bar{B}%
_{0}+\mu \mathbf{1}\geq B_{0}(\Omega _{0}^{\mathrm{t}})^{-1}\bar{B}_{0}
\end{equation}%
and, hence, by Theorem \ref{section extension copy(4)}, there exists a
unique solution $(\Omega _{t},B_{t})_{t\geq 0}$ of (\ref{flow
equation-quadratic}) obeying
\begin{equation}
\forall t\geq 0:\qquad \Omega _{t}\geq \mathfrak{B}_{t}:=4B_{t}(\Omega _{t}^{%
\mathrm{t}})^{-1}\bar{B}_{t}\ ,
\end{equation}%
see also Theorem \ref{section extension} for more details. Moreover, $%
\mathfrak{B}_{t}$ is a positive operator with trace uniformly bounded in $%
t\geq 0$, see Lemma \ref{existence_flow_5 copy(1)}.

Now assume that $\mu >0$. Then, by Theorem \ref{lemma existence 3}, the
bounded operator family $(\mathfrak{B}_{t})_{t\geq 0}$\ is norm continuous.
(This property should also hold for $\mu =0$, but it is not necessary for
this proof.) Then, we use $\mathfrak{B}_{t}$\ to define a norm continuous
evolution operator by the non--autonomous evolution equation
\begin{equation}
\forall t,s\in \mathbb{R}_{0}^{+}:\qquad \partial _{t}\widetilde{\mathfrak{V}%
}_{t,s}=2(4+\mathrm{r})\mathfrak{B}_{t}\widetilde{\mathfrak{V}}_{t,s}\
,\quad \widetilde{\mathfrak{V}}_{s,s}:=\mathbf{1}\ .
\label{new system 2-avec r}
\end{equation}%
Since $\mathfrak{B}_{t}$ is bounded, uniformly in $t\geq 0$, $\widetilde{%
\mathfrak{V}}_{t,s}$ is bounded. Indeed,
\begin{equation}
\Vert \widetilde{\mathfrak{V}}_{t,s}\Vert _{\mathrm{op}}^{2}\leq \exp
\left\{ 2\left( 4+\mathrm{r}\right) \left\vert t-s\right\vert \Vert
\mathfrak{B}_{0}\Vert _{1}\right\} \ .
\end{equation}%
Furthermore,
\begin{equation}
\forall t,s\in \mathbb{R}_{0}^{+}:\qquad \partial _{s}\widetilde{\mathfrak{V}%
}_{t,s}=-2(4+\mathrm{r})\widetilde{\mathfrak{V}}_{t,s}\mathfrak{B}_{s}\
,\quad \widetilde{\mathfrak{V}}_{t,t}:=\mathbf{1}\ .
\label{new system 2-avec r+1}
\end{equation}%
We set
\begin{equation}
\widetilde{\mathfrak{D}}_{t}:=\Omega _{t}-(4+\mathrm{r})\mathfrak{B}_{t}=%
\mathfrak{D}_{t}-\mathrm{r}\mathfrak{B}_{t}
\end{equation}%
and observe that, for all $t>0$,
\begin{equation}
\partial _{t}\widetilde{\mathfrak{D}}_{t}=2(4+\mathrm{r})(\mathfrak{B}_{t}%
\widetilde{\mathfrak{D}}_{t}+\widetilde{\mathfrak{D}}_{t}\mathfrak{B}_{t})+4%
\mathrm{r}B_{t}\bar{B}_{t}+4\mathrm{r}(4+\mathrm{r})\mathfrak{B}_{t}^{2}\ .
\label{eq toto1}
\end{equation}%
We can use the arguments justifying (\ref{new differential equation 3bis})
to prove that%
\begin{equation}
\forall t,s\in \mathbb{R}^{+}:\quad \widetilde{\mathfrak{D}}_{t}=\widetilde{%
\mathfrak{V}}_{t,s}\widetilde{\mathfrak{D}}_{s}\widetilde{\mathfrak{V}}%
_{t,s}^{\ast }+\int_{s}^{t}\widetilde{\mathfrak{V}}_{t,\tau }\left( 4\mathrm{%
r}B_{\tau }\bar{B}_{\tau }+4\mathrm{r}(4+\mathrm{r})\mathfrak{B}_{\tau
}^{2}\right) \widetilde{\mathfrak{V}}_{t,\tau }^{\ast }\mathrm{d}\tau \ .
\end{equation}%
The proof of this equality clearly uses (\ref{new system 2-avec r+1}) and (%
\ref{eq toto1}), but we omit the details and directly conclude that%
\begin{equation}
\forall t\geq s>0:\qquad \widetilde{\mathfrak{D}}_{t}\geq \widetilde{%
\mathfrak{V}}_{t,s}\widetilde{\mathfrak{D}}_{s}\widetilde{\mathfrak{V}}%
_{t,s}^{\ast }\ .
\end{equation}%
By (\ref{generalized A3-A6}), it follows that%
\begin{equation}
\widetilde{\mathfrak{D}}_{t}\geq \mu \widetilde{\mathfrak{V}}_{t,s}%
\widetilde{\mathfrak{V}}_{t,s}^{\ast }+(4+\mathrm{r})\widetilde{\mathfrak{V}}%
_{t,s}\mathfrak{B}_{0}\widetilde{\mathfrak{V}}_{t,s}^{\ast }-(4+\mathrm{r})%
\widetilde{\mathfrak{V}}_{t,s}\mathfrak{B}_{s}\widetilde{\mathfrak{V}}%
_{t,s}^{\ast }-16\int_{0}^{s}\widetilde{\mathfrak{V}}_{t,s}B_{\tau }\bar{B}%
_{\tau }\widetilde{\mathfrak{V}}_{t,s}^{\ast }\mathrm{d}\tau
\label{s limit con}
\end{equation}%
for all $t\geq s>0$. Using the norm continuity of the bounded operator
families $(\mathfrak{B}_{t})_{t\geq 0}$ and $(\widetilde{\mathfrak{V}}%
_{t,s})_{t,s\in \mathbb{R}^{+}}$, we take the limit $s\rightarrow 0^{+}$ in (%
\ref{s limit con}) to obtain that%
\begin{equation}
\forall t\geq 0:\qquad \widetilde{\mathfrak{D}}_{t}\geq \mu \widetilde{%
\mathfrak{V}}_{t,0}\widetilde{\mathfrak{V}}_{t,0}^{\ast }\ .
\label{eq toto2}
\end{equation}%
By (\ref{new system 2-avec r+1}), we note that%
\begin{equation}
\partial _{s}\left\{ \widetilde{\mathfrak{V}}_{t,s}\widetilde{\mathfrak{V}}%
_{t,s}^{\ast }\right\} =-2(4+\mathrm{r})\widetilde{\mathfrak{V}}_{t,s}^{\ast
}\mathfrak{B}_{t}\widetilde{\mathfrak{V}}_{t,s}\leq 0\ ,
\end{equation}%
i.e., $\widetilde{\mathfrak{V}}_{t,s}\widetilde{\mathfrak{V}}_{t,s}^{\ast
}\geq \mathbf{1}$ for all $t\geq s\geq 0$. Using this together with (\ref{eq
toto2}) we then deduce that $\widetilde{\mathfrak{D}}_{t}\geq \mu \mathbf{1}$%
, which is the asserted estimate (\ref{asserted estimate}) for $\mu >0$.

Now, if $\mu =0$ then one uses the proof of Lemma \ref{existence_flow_5
copy(1)} by using the operator $\mathfrak{B}_{t,\mu }$ defined, for any
arbitrarily constant $\mu >0$, by (\ref{eq-2}) together with the limit $\mu
\rightarrow 0^{+}$. We omit the details, especially since similar arguments
are also performed many times for other operators, see the proofs of Lemmata %
\ref{new lemma 2}, \ref{new lemma 2 copy(1)}, \ref{lemma integrability
copy(6)}, and \ref{lemma integrability copy(4)} where this strategy is
always used. \hfill $\Box $

Therefore, Theorem \ref{section extension copy(4)} and Lemma \ref{new lemma
2 copy(2)} yield Assertion (i) of Theorem \ref{theorem important 1 -2}.

\subsection{Constants of Motion\label{section constant of motion}}

Theorem \ref{lemma existence 3} and Corollary \ref{Corollary existence}
together with Theorem \ref{section extension copy(4)} already imply
Assertions (i)--(ii) and (v) of Theorem \ref{theorem important 1 -1}. It
remains to prove the third (iii) and fourth (iv) ones, which are carried out
in this subsection through four steps.

First, we relate the operator family $(\Omega _{t})_{t\geq 0}$ to a
\textquotedblleft commutator\textquotedblright\ defined by\
\begin{equation}
K_{t}:=\Omega _{t}B_{t}-B_{t}\Omega _{t}^{\mathrm{t}}=-K_{t}^{\mathrm{t}}\ .
\label{definition de K}
\end{equation}%
This bounded (linear) operator is well--defined for any strictly positive
time $t>0$ on the whole Hilbert space $\mathfrak{h}$ because both $\Omega
_{t}B_{t}$ and $B_{t}\Omega _{t}^{\mathrm{t}}$ are Hilbert--Schmidt
operators, by Lemma \ref{section extension copy(1)}. The relation between
the operator families $(K_{t})_{t>0}$ and $(\Omega _{t})_{t\geq 0}$ is
explained in Theorem \ref{lemma constante of motion}. It yields a \emph{%
constant of motion}, see Theorem \ref{lemma constante of motion} (iii).

Secondly, if one assumes (for simplicity) the condition
\begin{equation}
\Omega _{0}B_{0}(\Omega _{0}^{\mathrm{t}}+\mathbf{1})^{-1}\in \mathcal{B}%
\left( \mathfrak{h}\right) \ ,  \label{Condition sup}
\end{equation}%
then $K_{0}$ is well--defined on the domain $\mathcal{D}\left( \Omega _{0}^{%
\mathrm{t}}\right) $ and the operator family $(K_{t})_{t\geq 0}$ can be
computed like the operator family $(B_{t})_{t\geq 0}$, see Equation (\ref%
{B_explicit_solution}). This study requires a preliminary step (Lemma \ref%
{lemma constante of motion copy(1)}) and it is concluded by Lemma \ref{lemma
constante of motion copy(2)}.

Finally, combining all these results one directly deduces a simple
expression for the operator family $(\Omega _{t})_{t\geq 0}$ under the
condition $K_{0}=0$. The latter is Corollary \ref{lemma constante of motion
copy(3)}, which expresses \emph{two} constants of motion in this special
case.

So, we start by deriving a first constant of motion of the flow under
Conditions A1--A3:

\begin{theorem}[$\Omega _{t}^{2}$ and constant of motion]
\label{lemma constante of motion}\mbox{ }\newline
Assume Conditions A1--A3. Let $t,s\in \lbrack 0,T_{+})$ with $T_{+}$ defined
by (\ref{temps xi}) and $(\Omega _{t},B_{t})_{t\in \lbrack 0,T_{+})}$ be
defined via Theorem \ref{section extension}. Then the following statements
hold true:\newline
\emph{(i)} Domains: $\mathcal{D}(\Omega _{t}^{2})=\mathcal{D}(\Omega
_{s}^{2})$ and%
\begin{equation}
\Omega _{t}^{2}-8B_{t}\bar{B}_{t}=\Omega _{s}^{2}-8B_{s}\bar{B}%
_{s}+32\int_{s}^{t}B_{\tau }\Omega _{\tau }^{\mathrm{t}}\bar{B}_{\tau }%
\mathrm{d}\tau \ .  \label{(i) chiant}
\end{equation}%
\emph{(ii)} $(\Omega _{t}^{2})_{t\geq 0}$ and the \textquotedblleft
commutator\textquotedblright\ family $(K_{t})_{t>0}$ (see (\ref{definition
de K})) satisfy:
\begin{equation}
\Omega _{t}^{2}-4B_{t}\bar{B}_{t}=\Omega _{s}^{2}-4B_{s}\bar{B}%
_{s}+8\int_{s}^{t}\left( B_{\tau }\bar{K}_{\tau }+K_{\tau }^{\mathrm{t}}\bar{%
B}_{\tau }\right) \mathrm{d}\tau \ .
\end{equation}%
\emph{(iii)} Constant of motion of the flow:%
\begin{equation}
\mathrm{tr}\left( \Omega _{t}^{2}-4B_{t}\bar{B}_{t}-\Omega _{s}^{2}+4B_{s}%
\bar{B}_{s}\right) =0\ .
\end{equation}
\end{theorem}

\noindent \textbf{Proof.} Assume for convenience that $T_{+}=\infty $%
.\smallskip

\underline{(i):} For any $t>0$, Corollary \ref{Corollary existence} yields
\begin{equation}
\partial _{t}\left\{ B_{t}\bar{B}_{t}\right\} =-2\left( \Omega _{t}B_{t}\bar{%
B}_{t}+B_{t}\bar{B}_{t}\Omega _{t}\right) -4B_{t}\Omega _{t}^{\mathrm{t}}%
\bar{B}_{t}  \label{eq sup 0}
\end{equation}%
and%
\begin{equation}
\partial _{t}\left\{ \Omega _{t}^{2}-4B_{t}\bar{B}_{t}\right\} =8\left(
B_{t}\Omega _{t}^{\mathrm{t}}-\Omega _{t}B_{t}\right) \bar{B}%
_{t}+8B_{t}\left( \Omega _{t}^{\mathrm{t}}\bar{B}_{t}-\bar{B}_{t}\Omega
_{t}\right) \ ,  \label{eq sup 1}
\end{equation}%
in the strong sense in $\mathcal{D}(\Omega _{0})$. To compute this last
derivative, note that one invokes similar arguments as those used to prove (%
\ref{infer+4}). We omit the details. In particular, thanks to (\ref{eq sup 0}%
),
\begin{equation}
2\int_{s}^{t}\left( \Omega _{\tau }B_{\tau }\bar{B}_{\tau }+B_{\tau }\bar{B}%
_{\tau }\Omega _{\tau }\right) \mathrm{d}\tau =B_{s}\bar{B}_{s}-B_{t}\bar{B}%
_{t}-4\int_{s}^{t}B_{\tau }\Omega _{\tau }^{\mathrm{t}}\bar{B}_{\tau }%
\mathrm{d}\tau  \label{eq important}
\end{equation}%
for any $t\geq s>0$, which is combined with (\ref{eq sup 1}) to yield the
first statement (i) for strictly positive times $t\geq s>0$:
\begin{equation}
\forall t\geq s>0:\quad \Omega _{t}^{2}-8B_{t}\bar{B}_{t}=\Omega
_{s}^{2}-8B_{s}\bar{B}_{s}+32\int_{s}^{t}B_{\tau }\Omega _{\tau }^{\mathrm{t}%
}\bar{B}_{\tau }\mathrm{d}\tau \ ,  \label{(i) chiantbis}
\end{equation}%
on the domain $\mathcal{D}(\Omega _{t}^{2})=\mathcal{D}(\Omega
_{s}^{2})\subseteq \mathcal{D}(\Omega _{0}^{2})$. So, it remains to take the
limit $s\rightarrow 0^{+}$ in this equality and to show also that $\mathcal{D%
}(\Omega _{t}^{2})=\mathcal{D}(\Omega _{0}^{2})$, for any $t\geq 0$.

We start with the last integral by observing that the limit
\begin{equation}
\int_{0}^{t}B_{\tau }\Omega _{\tau }^{\mathrm{t}}\bar{B}_{\tau }\mathrm{d}%
\tau :=\ \underset{s\rightarrow 0^{+}}{\lim }\int_{s}^{t}B_{\tau }\Omega
_{\tau }^{\mathrm{t}}\bar{B}_{\tau }\mathrm{d}\tau  \label{operator}
\end{equation}%
defines a bounded operator. Indeed, for any $x,t>0$, let%
\begin{equation}
A_{x,t}:=W_{x}B_{0}W_{t}^{\mathrm{t}}\Omega _{t}^{\mathrm{t}}\left( W_{t}^{%
\mathrm{t}}\right) ^{\ast }\bar{B}_{0}W_{x}^{\ast }=A_{x,t}^{\ast }\ .
\end{equation}%
For any $t>0$, the operator $W_{t}^{\mathrm{t}}\Omega _{t}^{\mathrm{t}%
}\left( W_{t}^{\mathrm{t}}\right) ^{\ast }$ is bounded (cf. Lemma \ref{lemma
existence 1 copy(1)} (i)) and positive, i.e., $W_{t}^{\mathrm{t}}\Omega
_{t}^{\mathrm{t}}\left( W_{t}^{\mathrm{t}}\right) ^{\ast }\geq 0$. Since $%
B_{0}\in \mathcal{L}^{2}(\mathfrak{h})$, $A_{x,t}\in \mathcal{L}^{1}(%
\mathfrak{h})$ is trace--class and, by cyclicity of the trace (Lemma \ref%
{lemma existence 3 copy(3)} (i)),
\begin{equation}
\mathrm{tr}\left( A_{x,t}\right) =\mathrm{tr}\left( \left( W_{t}^{\mathrm{t}%
}\Omega _{t}^{\mathrm{t}}\left( W_{t}^{\mathrm{t}}\right) ^{\ast }\right)
^{1/2}\bar{B}_{0}W_{x}^{\ast }W_{x}B_{0}\left( W_{t}^{\mathrm{t}}\Omega
_{t}^{\mathrm{t}}\left( W_{t}^{\mathrm{t}}\right) ^{\ast }\right)
^{1/2}\right) \ ,  \label{derivative utile2}
\end{equation}%
for any $x,t>0$. We thus combine the derivative (\ref{petit inequalitybis})
extended to all $t>s\geq 0$ and (\ref{derivative utile2}) with Lemma \ref%
{lemma existence 3 copy(4)} to arrive at $\partial _{x}\left\{ \mathrm{tr}%
\left( A_{x,t}\right) \right\} \leq 0$ for all $x,t>0$, i.e.,
\begin{equation}
\forall t\geq x>0:\qquad \mathrm{tr}\left( B_{t}\Omega _{t}^{\mathrm{t}}\bar{%
B}_{t}\right) \leq \mathrm{tr}\left( W_{x}B_{0}W_{t}^{\mathrm{t}}\Omega
_{t}^{\mathrm{t}}\left( W_{t}^{\mathrm{t}}\right) ^{\ast }\bar{B}%
_{0}W_{x}^{\ast }\right) \ .  \label{derivative utile3}
\end{equation}%
Since $W_{t}^{\mathrm{t}}\Omega _{t}^{\mathrm{t}}\left( W_{t}^{\mathrm{t}%
}\right) ^{\ast }\in \mathcal{B}(\mathfrak{h})$ for all $t>0$, the upper
bound of this last estimate is continuous at $x=0$, similar to (\ref%
{continuity lebegue plus1}). Therefore, (\ref{derivative utile3}) also holds
for $x=0$ and we conclude that, for any $t>0$ and $s\in \lbrack 0,t]$,
\begin{equation}
\int_{s}^{t}\mathrm{tr}\left( B_{\tau }\Omega _{\tau }^{\mathrm{t}}\bar{B}%
_{\tau }\right) \mathrm{d}\tau \leq \int_{s}^{t}\mathrm{tr}\left(
B_{0}W_{\tau }^{\mathrm{t}}\Omega _{\tau }^{\mathrm{t}}\left( W_{\tau }^{%
\mathrm{t}}\right) ^{\ast }\bar{B}_{0}\right) \mathrm{d}\tau \ .
\label{fubiniplus0}
\end{equation}%
We note that, for any $t>0$ and $s\in \lbrack 0,t]$,
\begin{equation}
\forall \varphi \in \mathfrak{h}:\quad \int_{s}^{t}%
\big{\langle}%
\varphi
\big{|}%
W_{\tau }^{\mathrm{t}}\Omega _{\tau }^{\mathrm{t}}\left( W_{\tau }^{\mathrm{t%
}}\right) ^{\ast }\varphi
\big{\rangle}%
\mathrm{d}\tau =\frac{1}{4}%
\big{\langle}%
\varphi
\big{|}%
(W_{s}^{\mathrm{t}}(W_{s}^{\mathrm{t}})^{\ast }-W_{t}^{\mathrm{t}}(W_{t}^{%
\mathrm{t}})^{\ast })\varphi
\big{\rangle}%
\ ,  \label{fubiniplus1}
\end{equation}%
because $W_{t}^{\mathrm{t}}$ and $\left( W_{t}^{\mathrm{t}}\right) ^{\ast }$
are strongly continuous on $[0,\infty )$ and%
\begin{equation}
\forall t>0:\qquad \partial _{t}\left\{ W_{t}^{\mathrm{t}}\left( W_{t}^{%
\mathrm{t}}\right) ^{\ast }\right\} =-4W_{t}^{\mathrm{t}}\Omega _{t}^{%
\mathrm{t}}\left( W_{t}^{\mathrm{t}}\right) ^{\ast }\leq 0\ ,
\end{equation}%
see Lemma \ref{lemma existence 1 copy(1)} (ii) applied to $W_{t}^{\mathrm{t}%
} $ and $\left( W_{t}^{\mathrm{t}}\right) ^{\ast }$, and extended to all $%
t>s\geq 0$. Using (\ref{fubiniplus1}) and taking any orthonormal basis $%
\{g_{k}\}_{k=1}^{\infty }\subseteq \mathfrak{h}$ we now remark that
\begin{eqnarray}
\underset{k=1}{\overset{\infty }{\sum }}\int_{s}^{t}%
\big{\langle}%
\bar{B}_{0}g_{k}%
\big{|}%
W_{\tau }^{\mathrm{t}}\Omega _{\tau }^{\mathrm{t}}\left( W_{\tau }^{\mathrm{t%
}}\right) ^{\ast }\bar{B}_{0}g_{k}%
\big{\rangle}%
\ \mathrm{d}\tau &=&\frac{1}{4}\mathrm{tr}\left( B_{0}(W_{s}^{\mathrm{t}%
}(W_{s}^{\mathrm{t}})^{\ast }-W_{t}^{\mathrm{t}}(W_{t}^{\mathrm{t}})^{\ast })%
\bar{B}_{0}\right)  \notag \\
&\leq &\frac{1}{4}\left\Vert B_{0}\right\Vert _{2}^{2}<\infty \ ,
\end{eqnarray}%
because of (\ref{new upper bound for W}). Therefore, for any $t>0$ and $s\in
\lbrack 0,t]$, we invoke Fubini's theorem to exchange the trace with the
integral in the upper bound of (\ref{fubiniplus0}) and get
\begin{equation}
\int_{0}^{t}\mathrm{tr}\left( B_{\tau }\Omega _{\tau }^{\mathrm{t}}\bar{B}%
_{\tau }\right) \mathrm{d}\tau \leq \frac{1}{4}\mathrm{tr}\left(
B_{0}(W_{s}^{\mathrm{t}}(W_{s}^{\mathrm{t}})^{\ast }-W_{t}^{\mathrm{t}%
}(W_{t}^{\mathrm{t}})^{\ast })\bar{B}_{0}\right) \leq \frac{1}{4}\left\Vert
B_{0}\right\Vert _{2}^{2}<\infty \ .  \label{V.192}
\end{equation}%
As $\Omega _{t}^{\mathrm{t}}\geq 0$, we can then use a second time Fubini's
theorem to infer from (\ref{V.192}) that
\begin{eqnarray}
\left\Vert \int_{s}^{t}B_{\tau }\Omega _{\tau }^{\mathrm{t}}\bar{B}_{\tau }%
\mathrm{d}\tau \right\Vert _{1} &=&\mathrm{tr}\left( \int_{s}^{t}B_{\tau
}\Omega _{\tau }^{\mathrm{t}}\bar{B}_{\tau }\mathrm{d}\tau \right)
=\int_{s}^{t}\mathrm{tr}\left( B_{\tau }\Omega _{\tau }^{\mathrm{t}}\bar{B}%
_{\tau }\right) \mathrm{d}\tau  \notag \\
&\leq &\frac{1}{4}\mathrm{tr}\left( B_{0}(W_{s}^{\mathrm{t}}(W_{s}^{\mathrm{t%
}})^{\ast }-W_{t}^{\mathrm{t}}(W_{t}^{\mathrm{t}})^{\ast })\bar{B}_{0}\right)
\label{operator limit0} \\
&\leq &\frac{1}{4}\left\Vert B_{0}\right\Vert _{2}^{2}<\infty \ ,  \notag
\end{eqnarray}%
for any $t>0$ and $s\in \lbrack 0,t]$. In other words, the positive operator
(\ref{operator}) belongs to $\mathcal{L}^{1}\left( \mathfrak{h}\right) $ and
it is straightforward to verify that%
\begin{equation}
\underset{t\rightarrow 0^{+}}{\lim }\left\Vert \int_{0}^{t}B_{\tau }\Omega
_{\tau }^{\mathrm{t}}\bar{B}_{\tau }\mathrm{d}\tau \right\Vert _{1}=0\ ,
\label{operator limit}
\end{equation}%
because of (\ref{operator limit0}) and the strong continuity of $W_{t}^{%
\mathrm{t}}$ and $\left( W_{t}^{\mathrm{t}}\right) ^{\ast }$ on $[0,\infty )$%
, see Lemma \ref{lemma existence 1} (ii) applied to $W_{t}^{\mathrm{t}}$ and
$\left( W_{t}^{\mathrm{t}}\right) ^{\ast }$.

We study now the (time--independent) domain $\mathcal{D}(\Omega _{t}^{2})$.
First, by Corollary \ref{Corollary existence}, recall again that, for any $%
t\geq 0$,
\begin{equation}
\Omega _{t}:=\Omega _{0}-\Delta _{t}=\Omega _{0}-16\int_{0}^{t}B_{\tau }\bar{%
B}_{\tau }\mathrm{d}\tau \geq 0\ .  \label{again eq}
\end{equation}%
Then, a formal calculation implies that, for any $t\geq 0$,%
\begin{equation}
\Omega _{t}^{2}=\Omega _{0}^{2}-16\Omega _{0}\int_{0}^{t}B_{\tau }\bar{B}%
_{\tau }\mathrm{d}\tau -16\int_{0}^{t}B_{\tau }\bar{B}_{\tau }\Omega _{0}%
\mathrm{d}\tau +16^{2}\left( \int_{0}^{t}B_{\tau }\bar{B}_{\tau }\mathrm{d}%
\tau \right) ^{2}\ .  \label{again eq+1}
\end{equation}%
We observe that%
\begin{equation}
\int_{0}^{t}B_{\tau }\Omega _{\tau }^{\mathrm{t}}\bar{B}_{\tau }\mathrm{d}%
\tau \qquad \mathrm{and}\qquad \int_{0}^{t}B_{\tau }\bar{B}_{\tau }\Omega
_{\tau }\left( \Omega _{0}+\mathbf{1}\right) ^{-1}\mathrm{d}\tau
\label{operator1}
\end{equation}%
define trace--class operators because of (\ref{operator limit0}) and
\begin{equation}
\left\Vert \int_{s}^{t}B_{\tau }\bar{B}_{\tau }\Omega _{\tau }\left( \Omega
_{0}+\mathbf{1}\right) ^{-1}\mathrm{d}\tau \right\Vert _{1}\leq \left(
t-s\right) \left\Vert B_{0}\right\Vert _{2}^{2}+16\left( t-s\right)
^{2}\left\Vert B_{0}\right\Vert _{2}^{4}<\infty  \label{ine debile plus}
\end{equation}%
for any $t\geq s\geq 0$, by (\ref{a priori estimate}) and (\ref{again eq}).
Via (\ref{eq important}) and Theorem \ref{lemma existence 3} (ii), it
follows that%
\begin{equation}
\int_{0}^{t}\Omega _{\tau }B_{\tau }\bar{B}_{\tau }\left( \Omega _{0}+%
\mathbf{1}\right) ^{-1}\mathrm{d}\tau :=\ \underset{s\rightarrow 0^{+}}{\lim
}\int_{s}^{t}\Omega _{\tau }B_{\tau }\bar{B}_{\tau }\left( \Omega _{0}+%
\mathbf{1}\right) ^{-1}\mathrm{d}\tau  \label{operator2}
\end{equation}%
defines also a trace--class operator.

On the other hand, we can interchange the operator $\Omega _{0}$ and the
first integral of (\ref{again eq+1}) -- just as we did to prove (\ref{eq
flow necessary}) -- and use again (\ref{again eq}) to obtain that, for any $%
t\geq s>0$,
\begin{eqnarray}
\Omega _{0}\int_{s}^{t}B_{\tau }\bar{B}_{\tau }\mathrm{d}\tau &=&\frac{%
\Omega _{0}}{\Omega _{0}+\mathbf{1}}\int_{s}^{t}\Omega _{\tau }B_{\tau }\bar{%
B}_{\tau }\mathrm{d}\tau +\frac{\Omega _{0}}{\Omega _{0}+\mathbf{1}}%
\int_{s}^{t}B_{\tau }\bar{B}_{\tau }\mathrm{d}\tau  \notag \\
&&+\frac{16\Omega _{0}}{\Omega _{0}+\mathbf{1}}\int_{s}^{t}\mathrm{d}\tau
_{1}\int_{s}^{\tau _{1}}\mathrm{d}\tau _{2}B_{\tau _{2}}\bar{B}_{\tau
_{2}}B_{\tau _{1}}\bar{B}_{\tau _{1}}\ .  \label{operartor3}
\end{eqnarray}%
It follows from (\ref{operator2}) that, for any $t>0$,
\begin{equation}
\left( \underset{s\rightarrow 0^{+}}{\lim }\Omega _{0}\int_{s}^{t}B_{\tau }%
\bar{B}_{\tau }\left( \Omega _{0}+\mathbf{1}\right) ^{-1}\mathrm{d}\tau
\right) \in \mathcal{L}^{1}(\mathfrak{h})\ .  \label{operator4}
\end{equation}%
Here, the limit is in the strong sense in $\mathfrak{h}$. Since, for any $%
t>0 $,
\begin{equation}
\underset{s\rightarrow 0^{+}}{\lim }\int_{s}^{t}B_{\tau }\bar{B}_{\tau
}\left( \Omega _{0}+\mathbf{1}\right) ^{-1}\mathrm{d}\tau
=\int_{0}^{t}B_{\tau }\bar{B}_{\tau }\left( \Omega _{0}+\mathbf{1}\right)
^{-1}\mathrm{d}\tau
\end{equation}%
(in the strong topology) and $\Omega _{0}$ is a closed operator, we infer
from (\ref{operator4}) that
\begin{equation}
\Omega _{0}\int_{0}^{t}B_{\tau }\bar{B}_{\tau }\left( \Omega _{0}+\mathbf{1}%
\right) ^{-1}\mathrm{d}\tau =\underset{s\rightarrow 0^{+}}{\lim }\Omega
_{0}\int_{s}^{t}B_{\tau }\bar{B}_{\tau }\left( \Omega _{0}+\mathbf{1}\right)
^{-1}\mathrm{d}\tau  \label{operator5}
\end{equation}%
defines also a trace--class operator for any $t\geq 0$. It is in particular
bounded and we deduce from (\ref{again eq+1}) that $\mathcal{D}(\Omega
_{t}^{2})=\mathcal{D}(\Omega _{0}^{2})$ for all $t\geq 0$.

We proceed by analyzing the convergence of the operator $\Omega _{s}^{2}$ to
$\Omega _{0}^{2}$, as $s\rightarrow 0^{+}$. So, by using
\begin{equation}
2\left( \Omega _{s}^{2}-\Omega _{0}^{2}\right) =\left( \Omega _{s}-\Omega
_{0}\right) \left( \Omega _{s}+\Omega _{0}\right) +\left( \Omega _{s}+\Omega
_{0}\right) \left( \Omega _{s}-\Omega _{0}\right)  \label{equality utile}
\end{equation}%
and the resolvent identity (\ref{V.197}) (as $\mathcal{D}(\Omega _{t}^{2})=%
\mathcal{D}(\Omega _{0}^{2})$) together with%
\begin{equation}
\left\Vert \left( X+\mathbf{1}\right) \left( X^{2}+\mathbf{1}\right)
^{-1}\right\Vert _{\mathrm{op}}\leq \ \underset{x\geq 0}{\sup }\left( \frac{%
x+1}{x^{2}+1}\right) <2  \label{equality utilebis}
\end{equation}%
for any positive operator $X\geq 0$, one gets%
\begin{eqnarray}
\left\Vert \left( \Omega _{s}^{2}+\mathbf{1}\right) ^{-1}-\left( \Omega
_{0}^{2}+\mathbf{1}\right) ^{-1}\right\Vert _{\mathrm{op}} &\leq
&4\left\Vert \left( \Omega _{s}+\mathbf{1}\right) ^{-1}\left( \Omega
_{s}^{2}-\Omega _{0}^{2}\right) \left( \Omega _{0}+\mathbf{1}\right)
^{-1}\right\Vert _{\mathrm{op}}  \notag \\
&\leq &4\left( 2+\left\Vert \Delta _{s}\right\Vert _{2}\right) \left\Vert
\Omega _{s}-\Omega _{0}\right\Vert _{\mathrm{op}}\ .  \label{chiante1}
\end{eqnarray}%
The operator family $(\Omega _{t})_{t\geq 0}$ is continuous, at least in the
norm topology (Theorem \ref{section extension} (i)). So, we conclude from (%
\ref{chiante1}) that $\Omega _{s}^{2}$ converges in the norm resolvent sense
to $\Omega _{0}^{2}$, as $s\rightarrow 0^{+}$.

As a consequence, we arrive at Assertion (i) of the lemma by passing to the
limit $s\rightarrow 0^{+}$ in (\ref{(i) chiantbis}) with the help of (\ref%
{operator}) and the continuity of the Hilbert--Schmidt operator family $%
(B_{t})_{t\geq 0}$, see, e.g., Theorem \ref{lemma existence 3}
(ii).\smallskip

\underline{(ii):} By using (\ref{definition de K}) and integrating (\ref{eq
sup 1}), we find that, for any $t\geq s>0$,%
\begin{equation}
\Omega _{t}^{2}-4B_{t}\bar{B}_{t}=\Omega _{s}^{2}-4B_{s}\bar{B}%
_{s}+8\int_{s}^{t}\left( B_{\tau }\bar{K}_{\tau }+K_{\tau }^{\mathrm{t}}\bar{%
B}_{\tau }\right) \mathrm{d}\tau  \label{eq K1}
\end{equation}%
on the domain $\mathcal{D}(\Omega _{0}^{2})\subset \mathcal{D}(\Omega _{0})$%
. Therefore, since $\Omega _{s}^{2}$ converges in the norm resolvent sense
to $\Omega _{0}^{2}$, as $s\rightarrow 0^{+}$, and $(B_{t})_{t\geq 0}$ is
continuous in $\mathcal{L}^{2}(\mathfrak{h})$ (Theorem \ref{lemma existence
3} (ii)), one verifies Equation (\ref{eq K1}) for $s=0$ on the domain $%
\mathcal{D}(\Omega _{0}^{2})\subset \mathcal{D}(\Omega _{0})$ by elementary
computations using (\ref{definition de K}) and the fact that the operators
in (\ref{operator1}) and (\ref{operator2}) are all bounded.\smallskip

\underline{(iii):} From the cyclicity of the trace (Lemma \ref{lemma
existence 3 copy(3)}) combined with $\mathrm{tr}\left( X^{\mathrm{t}}\right)
=\mathrm{tr}\left( X\right) $, $\mathrm{tr}(\overline{X})=\overline{\mathrm{%
tr}(X)}$ (Lemma \ref{lemma existence 3 copy(2)}) and $B_{t}=B_{t}^{\mathrm{t}%
}\in \mathcal{L}^{2}\left( \mathfrak{h}\right) $, note that
\begin{equation}
\mathrm{tr}\left( \Omega _{t}B_{t}\bar{B}_{t}\right) =\mathrm{tr}\left( \bar{%
B}_{t}\Omega _{t}B_{t}\right) =\mathrm{tr}\left( B_{t}\Omega _{t}^{\mathrm{t}%
}\bar{B}_{t}\right) =\mathrm{tr}\left( B_{t}\bar{B}_{t}\Omega _{t}\right) \ .
\label{inequality trivial}
\end{equation}%
Moreover, because the positive operator (\ref{operator}) belongs to $%
\mathcal{L}^{1}\left( \mathfrak{h}\right) $\ and satisfies (\ref{operator
limit0}), we deduce from the first assertion (i) that%
\begin{equation}
\mathrm{tr}\left( \Omega _{t}^{2}-\Omega _{s}^{2}\right) =8\mathrm{tr}\left(
B_{t}\bar{B}_{t}\right) -8\mathrm{tr}\left( B_{s}\bar{B}_{s}\right)
+32\int_{s}^{t}\mathrm{tr}\left( B_{\tau }\Omega _{\tau }^{\mathrm{t}}\bar{B}%
_{\tau }\right) \mathrm{d}\tau  \label{dormir}
\end{equation}%
for all $t\geq s\geq 0$. In particular, by continuity in $\mathcal{L}^{2}(%
\mathfrak{h})$ of $(B_{t})_{t\geq 0}$,
\begin{equation}
\underset{s\rightarrow 0^{+}}{\lim }\mathrm{tr}\left( \Omega _{t}^{2}-\Omega
_{s}^{2}\right) =\mathrm{tr}\left( \Omega _{t}^{2}-\Omega _{0}^{2}\right) \ .
\label{dormir2}
\end{equation}%
Now, we use (\ref{eq sup 0}) together with Lemma \ref{lemma existence 3
copy(4)} and (\ref{inequality trivial})--(\ref{dormir}) to deduce that
\begin{equation}
\forall t\geq s>0:\qquad \mathrm{tr}\left( \Omega _{t}^{2}-\Omega
_{s}^{2}\right) =4\left( \Vert B_{t}\Vert _{2}^{2}-\Vert B_{s}\Vert
_{2}^{2}\right) \ .
\end{equation}%
The latter can be extended by continuity to $s=0$ because of the continuity
in $\mathcal{L}^{2}(\mathfrak{h})$ of $(B_{t})_{t\geq 0}$ and (\ref{dormir2}%
). \hfill $\Box $

Under Condition (\ref{Condition sup}), we aim to obtain a simple expression
for the operator family $(K_{t})_{t\geq 0}$, similar to (\ref%
{B_explicit_solution}) for the operator family $(B_{t})_{t\geq 0}$. To this
end, we need first to study the operators
\begin{equation}
\mathfrak{\Delta }_{t}:=\left( \Omega _{0}+\mathbf{1}\right) \Delta
_{t}\left( \Omega _{0}+\mathbf{1}\right) ^{-1}\quad \mathrm{and}\quad
\boldsymbol{V}_{t,s}:=\left( \Omega _{0}+\mathbf{1}\right) W_{t,s}\left(
\Omega _{0}+\mathbf{1}\right) ^{-1}\ ,  \label{evolution operato bis}
\end{equation}%
for any $t\in \lbrack 0,T_{+})$ and $s\in \lbrack 0,t]$, where $(\Delta
_{t})_{t\in \lbrack 0,T_{+})}$ is defined by (\ref{definition de delta})
extended to $[0,T_{+})$.

\begin{lemma}[$\boldsymbol{V}_{t,s}$ as an evolution operator]
\label{lemma constante of motion copy(1)}\mbox{ }\newline
Assume Conditions A1--A3. Then, for any $s,x,t\in \lbrack 0,T_{+})$ so that $%
t\geq x\geq s$:\newline
\emph{(i)} $(\mathfrak{\Delta }_{t})_{t\geq 0}\in C[[0,T_{+});\mathcal{L}%
^{1}(\mathfrak{h})]$ and $(\boldsymbol{V}_{t,s})_{t\geq s\geq 0}\subset
\mathcal{B}(\mathfrak{h})$. \newline
\emph{(ii)} $\boldsymbol{V}_{t,s}$ satisfies the cocycle property\ $%
\boldsymbol{V}_{t,x}\boldsymbol{V}_{x,s}=\boldsymbol{V}_{t,s}$. \newline
\emph{(iii)} $\boldsymbol{V}_{t,s}$ is jointly strongly continuous in $s$
and $t$.\newline
\emph{(iv)} The evolution family $(\boldsymbol{V}_{t,s})_{t\geq s\geq 0}$ is
the solution of the non--autonomous evolution equations%
\begin{equation}
\left\{
\begin{array}{llllll}
\forall s\in \lbrack 0,T_{+}),\ t\in (s,T_{+}) & : & \partial _{t}%
\boldsymbol{V}_{t,s}=-2\left( \Omega _{0}-\mathfrak{\Delta }_{t}\right)
\boldsymbol{V}_{t,s} & , & \boldsymbol{V}_{s,s}:=\mathbf{1} & . \\
\forall t\in (0,T_{+}),\ s\in \lbrack 0,t] & : & \partial _{s}\boldsymbol{V}%
_{t,s}=2\boldsymbol{V}_{t,s}\left( \Omega _{0}-\mathfrak{\Delta }_{s}\right)
& , & \boldsymbol{V}_{t,t}:=\mathbf{1} & .%
\end{array}%
\right.
\end{equation}%
The derivatives with respect to $t$ and $s$ are in the strong sense in $%
\mathfrak{h}$ and $\mathcal{D}\left( \Omega _{0}\right) $, respectively.
\newline
Similar properties as (i)--(iv) also hold for the bounded operators $%
\boldsymbol{V}_{t,s}^{\mathrm{t}}$ and $\mathfrak{\Delta }_{t}^{\mathrm{t}}$.
\end{lemma}

\noindent \textbf{Proof.} Without loss of generality assume for convenience
that $T_{+}=\infty $. Since $B_{t>0}\mathfrak{h}\subseteq \mathcal{D}(\Omega
_{0})$, we use (\ref{again eq}) and interchange the operator $(\Omega _{0}+%
\mathbf{1})$ and the integral (cf. (\ref{operator2})--(\ref{operator5})) to
arrive at
\begin{eqnarray}
\mathfrak{\Delta }_{t} &=&16\int_{0}^{t}\Omega _{\tau }B_{\tau }\bar{B}%
_{\tau }\left( \Omega _{0}+\mathbf{1}\right) ^{-1}\mathrm{d}\tau
+16\int_{0}^{t}B_{\tau }\bar{B}_{\tau }\left( \Omega _{0}+\mathbf{1}\right)
^{-1}\mathrm{d}\tau  \notag \\
&&+16^{2}\int_{0}^{t}\mathrm{d}\tau _{1}\int_{0}^{\tau _{1}}\mathrm{d}\tau
_{2}\ B_{_{\tau _{2}}}\bar{B}_{\tau _{2}}B_{\tau _{1}}\bar{B}_{\tau
_{1}}\left( \Omega _{0}+\mathbf{1}\right) ^{-1}\ .
\label{eq generator corrige}
\end{eqnarray}%
On the domain $\mathcal{D}(\Omega _{0})$, Equation (\ref{eq important})
holds for $s=0$. Indeed, $(B_{t})_{t\geq 0}\in C[\mathbb{R}_{0}^{+};\mathcal{%
L}^{2}(\mathfrak{h})]$ (Theorem \ref{lemma existence 3} (ii)) and, in the
strong limit $s\rightarrow 0^{+}$, both terms%
\begin{equation}
\int_{0}^{t}B_{\tau }\Omega _{\tau }^{\mathrm{t}}\bar{B}_{\tau }\mathrm{d}%
\tau \qquad \mathrm{and}\qquad \int_{0}^{t}B_{\tau }\bar{B}_{\tau }\Omega
_{\tau }\left( \Omega _{0}+\mathbf{1}\right) ^{-1}\mathrm{d}\tau
\label{operators chiants}
\end{equation}%
define bounded operators that are continuous in $\mathcal{L}^{1}(\mathfrak{h}%
)$ for all $t\geq 0$: To obtain the boundedness and continuity of the first
operator in (\ref{operators chiants}), one uses (\ref{operator limit0})
together with Lebesgue's dominated convergence theorem and Lemma \ref{lemma
existence 1} (ii) applied to $W_{t,s}^{\mathrm{t}}$ and $\left( W_{t,s}^{%
\mathrm{t}}\right) ^{\ast }$. To get the boundedness and continuity of the
second operator in (\ref{operators chiants}), use (\ref{ine debile plus}).
Therefore,%
\begin{equation}
\int_{0}^{t}\Omega _{\tau }B_{\tau }\bar{B}_{\tau }\mathrm{d}\tau =B_{0}\bar{%
B}_{0}-B_{t}\bar{B}_{t}-4\int_{0}^{t}B_{\tau }\Omega _{\tau }^{\mathrm{t}}%
\bar{B}_{\tau }\mathrm{d}\tau -2\int_{0}^{t}B_{\tau }\bar{B}_{\tau }\Omega
_{\tau }\mathrm{d}\tau
\end{equation}%
on the domain $\mathcal{D}(\Omega _{0})$. By (\ref{eq generator corrige})
and Theorem \ref{lemma existence 3} (ii), it follows that $(\mathfrak{\Delta
}_{t})_{t\geq 0}\in C[\mathbb{R}_{0}^{+};\mathcal{L}^{1}(\mathfrak{h})]$.

Now, for any $t\geq s\geq 0$,
\begin{eqnarray}
\boldsymbol{V}_{t,s} &=&\mathrm{e}^{-2\left( t-s\right) \Omega
_{0}}+2\int_{s}^{t}\left( \Omega _{0}+\mathbf{1}\right) \mathrm{e}^{-2\left(
t-\tau \right) \Omega _{0}}\Delta _{\tau }W_{\tau ,s}\left( \Omega _{0}+%
\mathbf{1}\right) ^{-1}\mathrm{d}\tau  \notag \\
&=&\mathrm{e}^{-2\left( t-s\right) \Omega _{0}}+2\int_{s}^{t}\mathrm{e}%
^{-2\left( t-\tau \right) \Omega _{0}}\mathfrak{\Delta }_{\tau }\boldsymbol{V%
}_{\tau ,s}\mathrm{d}\tau \ ,  \label{operator transf}
\end{eqnarray}%
because $W=\mathcal{T}\left( W\right) $ with $\mathcal{T}\left( W\right) $
defined by (\ref{definition of W 2}). Note that we again interchange the
operator $\left( \Omega _{0}+\mathbf{1}\right) $ and the integral in (\ref%
{operator transf}). This is justified by using once more the closedness of
the operator $\Omega _{0}$ as follows: For any $t>s\geq 0$ and sufficiently
small $\epsilon >0$,
\begin{eqnarray}
\Omega _{0}\int_{s}^{t-\epsilon }\mathrm{e}^{-2\left( t-\tau \right) \Omega
_{0}}\Delta _{\tau }W_{\tau ,s}\mathrm{d}\tau &=&\Omega _{0}\mathrm{e}%
^{-\epsilon \Omega _{0}}\int_{s}^{t-\epsilon }\mathrm{e}^{\left( -2\left(
t-\tau \right) +\epsilon \right) \Omega _{0}}\Delta _{\tau }W_{\tau ,s}%
\mathrm{d}\tau  \notag \\
&=&\int_{s}^{t-\epsilon }\Omega _{0}\mathrm{e}^{-2\left( t-\tau \right)
\Omega _{0}}\Delta _{\tau }W_{\tau ,s}\mathrm{d}\tau  \label{closure1}
\end{eqnarray}%
is a bounded operator. It remains to take the limit $\epsilon \rightarrow
0^{+}$ in the strong topology. On the one hand,%
\begin{equation}
\underset{\epsilon \rightarrow 0^{+}}{\lim }\int_{s}^{t-\epsilon }\mathrm{e}%
^{-2\left( t-\tau \right) \Omega _{0}}\Delta _{\tau }W_{\tau ,s}\mathrm{d}%
\tau =\int_{s}^{t}\mathrm{e}^{-2\left( t-\tau \right) \Omega _{0}}\Delta
_{\tau }W_{\tau ,s}\mathrm{d}\tau  \label{closure2}
\end{equation}%
for any $t>s\geq 0$, in the strong sense in $\mathfrak{h}$. On the other
hand, the strong limit%
\begin{equation}
\underset{\epsilon \rightarrow 0^{+}}{\lim }\int_{s}^{t-\epsilon }\Omega _{0}%
\mathrm{e}^{-2\left( t-\tau \right) \Omega _{0}}\Delta _{\tau }W_{\tau ,s}%
\mathrm{d}\tau =\int_{s}^{t}\Omega _{0}\mathrm{e}^{-2\left( t-\tau \right)
\Omega _{0}}\Delta _{\tau }W_{\tau ,s}\mathrm{d}\tau  \label{closure3}
\end{equation}%
defines again a bounded operator for any $t>s\geq 0$, see (\ref%
{interpolation4bis}) and (\ref{interpolation6+1})--(\ref{petit inequality
newbisbisbis}). Therefore, by combining (\ref{closure1})--(\ref{closure3})
with the fact that $\Omega _{0}$ is a closed operator we obtain%
\begin{equation}
\Omega _{0}\int_{s}^{t}\mathrm{e}^{-2\left( t-\tau \right) \Omega
_{0}}\Delta _{\tau }W_{\tau ,s}\mathrm{d}\tau =\int_{s}^{t}\Omega _{0}%
\mathrm{e}^{-2\left( t-\tau \right) \Omega _{0}}\Delta _{\tau }W_{\tau ,s}%
\mathrm{d}\tau
\end{equation}%
for any $t\geq s\geq 0$.

This last assertion justifies Equation (\ref{operator transf}), which leads
to $(\boldsymbol{V}_{t,s})_{t\geq s\geq 0}\subset \mathcal{B}(\mathfrak{h})$%
, because $(\mathfrak{\Delta }_{t})_{t\geq 0}\in C[\mathbb{R}_{0}^{+};%
\mathcal{L}^{1}(\mathfrak{h})]$. In other words, we obtain Assertion (i),
which combined with (\ref{operator transf}), implies that the statements of
Lemmata \ref{lemma existence 1} and \ref{lemma existence 1 copy(1)} hold
true for the bounded evolution operator $(\boldsymbol{V}_{t,s})_{t\geq s\geq
0}$. One gets in particular Assertions (ii)--(iv). Like for Lemmata \ref%
{lemma existence 1} and \ref{lemma existence 1 copy(1)}, similar properties
as (i)--(iv) also hold for $(\mathfrak{\Delta }_{t}^{\mathrm{t}})_{t\geq 0}$
and $(\boldsymbol{V}_{t,s}^{\mathrm{t}})_{t\geq s\geq 0}$. \hfill $\Box $

We are now in position to get a simple expression for the operator family $%
(K_{t})_{t\geq 0}$, similar to (\ref{B_explicit_solution}) for $%
(B_{t})_{t\geq 0}$:

\begin{lemma}[Explicit expression for the \textquotedblleft
commutator\textquotedblright ]
\label{lemma constante of motion copy(2)}\mbox{ }\newline
Assume Conditions A1--A3 and (\ref{Condition sup}). Then,%
\begin{equation}
\forall t,s\in \lbrack 0,T_{+}):\qquad K_{t}=W_{t,s}K_{s}W_{t,s}^{\mathrm{t}%
}\ .
\end{equation}
\end{lemma}

\noindent \textbf{Proof.} Without loss of generality assume for convenience
that $T_{+}=\infty $. By Corollary \ref{Corollary existence}, for any $t>0$,%
\begin{equation}
\partial _{t}\left\{ B_{t}\Omega _{t}^{\mathrm{t}}\right\} =-2(\Omega
_{t}B_{t})\Omega _{t}^{\mathrm{t}}-2(B_{t}\Omega _{t}^{\mathrm{t}})\Omega
_{t}^{\mathrm{t}}-16B_{t}\bar{B}_{t}B_{t}\ ,  \label{K derivative1}
\end{equation}%
in the strong sense in $\mathcal{D}(\Omega _{0}^{\mathrm{t}})$. Note indeed
that both $\Omega _{t}B_{t}$ and $B_{t}\Omega _{t}^{\mathrm{t}}$ are
Hilbert--Schmidt operators, by Lemma \ref{section extension copy(1)}. On the
other hand, by using (\ref{B_explicit_solution}) and the evolution operator $%
(\boldsymbol{V}_{t,s})_{t\geq s\geq 0}$ with the notation $\boldsymbol{V}%
_{t}:=\boldsymbol{V}_{t,0}$ (see (\ref{evolution operato bis}) and Lemma \ref%
{lemma constante of motion copy(1)}),%
\begin{equation}
\Omega _{t}B_{t}(\Omega _{0}^{\mathrm{t}}+\mathbf{1})^{-1}=\Omega
_{t}(\Omega _{0}+\mathbf{1})^{-1}\boldsymbol{V}_{t}(\Omega _{0}+\mathbf{1}%
)B_{0}(\Omega _{0}^{\mathrm{t}}+\mathbf{1})^{-1}\boldsymbol{V}_{t}^{\mathrm{t%
}}\ ,  \label{operator derivative}
\end{equation}%
for all $t\geq 0$. Since, by (\ref{a priori estimate}) and (\ref{again eq}),%
\begin{equation}
\left\Vert \Omega _{t}(\Omega _{0}+\mathbf{1})^{-1}\right\Vert _{\mathrm{op}%
}\leq 1+16t\left\Vert B_{0}\right\Vert _{2}^{2}<\infty \ ,
\label{toto inequality}
\end{equation}%
and (\ref{Condition sup}) holds, we can compute the derivative of the
operator (\ref{operator derivative}) and infer from Corollary \ref{Corollary
existence}, Lemma \ref{lemma constante of motion copy(1)}, and (\ref%
{evolution operato bis}) that, for any $t>0$,
\begin{equation}
\partial _{t}\left\{ \Omega _{t}B_{t}\right\} =-2\Omega
_{t}^{2}B_{t}-2\Omega _{t}B_{t}\Omega _{t}^{\mathrm{t}}-16B_{t}\bar{B}%
_{t}B_{t}\ ,  \label{K derivative2}
\end{equation}%
in the strong sense in $\mathcal{D}(\Omega _{0}^{\mathrm{t}})$. Indeed, $%
B_{t>0}\mathfrak{h}\subseteq \mathcal{D}(\Omega _{0})$ (Theorem \ref{section
extension} (ii)) and $(\Omega _{0}^{2}B_{t})\mathfrak{h}\subseteq \mathcal{D}%
(\Omega _{0})$ for $t>0$ because of (\ref{Condition sup}), $W_{t>0}^{\mathrm{%
t}}\mathfrak{h}\subseteq \mathcal{D}(\Omega _{0}^{\mathrm{t}})$ (Lemma \ref%
{lemma existence 1 copy(1)} (ii) extended to all $t>s\geq 0$), $\boldsymbol{V%
}_{t>0}\mathfrak{h}\subseteq \mathcal{D}(\Omega _{0})$ (Lemma \ref{lemma
constante of motion copy(1)} (iv)) and the equality
\begin{equation}
\Omega _{0}^{2}B_{t}=\Omega _{0}\boldsymbol{V}_{t}\left\{ (\Omega _{0}+%
\mathbf{1})B_{0}(\Omega _{0}^{\mathrm{t}}+\mathbf{1})^{-1}\right\} \left\{
(\Omega _{0}^{\mathrm{t}}+\mathbf{1})W_{t}^{\mathrm{t}}\right\} \ .
\end{equation}%
(In fact, $(\Omega _{t}^{2}B_{t})_{t>0}\subset \mathcal{L}^{2}(\mathfrak{h})$%
, by Lemma \ref{section extension copy(2)}.) We combine now (\ref{K
derivative1}) and (\ref{K derivative2}) with (\ref{definition de K}) to
deduce that, for any $t>0$,
\begin{equation}
\partial _{t}K_{t}=-2\left( \Omega _{t}K_{t}+K_{t}\Omega _{t}^{\mathrm{t}%
}\right) \ ,  \label{K derivative3}
\end{equation}%
in the strong sense in $\mathcal{D}(\Omega _{0}^{\mathrm{t}})$. In
particular, $K_{t>0}\mathfrak{h}\subseteq \mathcal{D}(\Omega _{0})$. Thus,
by Lemma \ref{lemma existence 1 copy(1)} applied to $W_{t,s}$ and $W_{t,s}^{%
\mathrm{t}}$ and extended to all $t>s>0$, it follows that, for any $t>s>0$,%
\begin{equation}
\partial _{s}\{W_{t,s}K_{s}W_{t,s}^{\mathrm{t}}\}=0\ ,
\end{equation}%
in the strong sense in $\mathcal{D}(\Omega _{0}^{\mathrm{t}})$. In other
words,
\begin{equation}
\forall t\geq s>0:\qquad K_{t}=W_{t,s}K_{s}W_{t,s}^{\mathrm{t}}\ ,
\label{equality idiote}
\end{equation}%
on the dense domain $\mathcal{D}(\Omega _{0}^{\mathrm{t}})$. Both operators $%
K_{t}$ and $W_{t,s}K_{s}W_{t,s}^{\mathrm{t}}$ are bounded, so we can extend
by continuity (\ref{equality idiote}) to the whole Hilbert space $\mathfrak{h%
}$. Hence, it remains to take the limit $s\rightarrow 0^{+}$ in (\ref%
{equality idiote}).

By (\ref{B_explicit_solution}), (\ref{definition de K}), and (\ref{evolution
operato bis}), for any $t\geq 0$,
\begin{eqnarray}
K_{t}(\Omega _{0}^{\mathrm{t}}+\mathbf{1})^{-1} &=&\Omega _{t}(\Omega _{0}+%
\mathbf{1})^{-1}\boldsymbol{V}_{t,s}(\Omega _{0}+\mathbf{1})B_{0}(\Omega
_{0}^{\mathrm{t}}+\mathbf{1})^{-1}\boldsymbol{V}_{t,s}^{\mathrm{t}}  \notag
\\
&&-B_{t}\Omega _{t}^{\mathrm{t}}(\Omega _{0}^{\mathrm{t}}+\mathbf{1})^{-1}\ .
\end{eqnarray}%
Therefore, using (\ref{Condition sup}), (\ref{toto inequality}), Theorem \ref%
{lemma existence 3} (i)--(ii), and Lemma \ref{lemma constante of motion
copy(1)} (cf. (i), (iii)) applied to $\boldsymbol{V}_{t,s}$ and $\boldsymbol{%
V}_{t,s}^{\mathrm{t}}$, we arrive at
\begin{equation}
\underset{t\rightarrow 0^{+}}{\lim }\left\Vert K_{t}(\Omega _{0}^{\mathrm{t}%
}+\mathbf{1})^{-1}-K_{0}(\Omega _{0}^{\mathrm{t}}+\mathbf{1}%
)^{-1}\right\Vert _{\mathrm{op}}=0\ .  \label{limit1}
\end{equation}%
Since, for any $t\geq s\geq 0$,
\begin{equation}
W_{t,s}K_{s}W_{t,s}^{\mathrm{t}}(\Omega _{0}^{\mathrm{t}}+\mathbf{1}%
)^{-1}=W_{t,s}K_{s}(\Omega _{0}^{\mathrm{t}}+\mathbf{1})^{-1}\boldsymbol{V}%
_{t,s}^{\mathrm{t}}\ ,
\end{equation}%
we infer from (\ref{upper bound for W}), (\ref{limit1}), Lemma \ref{lemma
existence 1} (ii), and Lemma \ref{lemma constante of motion copy(1)} (cf.
(i), (iii)) that
\begin{equation}
\underset{s\rightarrow 0^{+}}{\lim }\left\Vert W_{t,s}K_{s}W_{t,s}^{\mathrm{t%
}}(\Omega _{0}^{\mathrm{t}}+\mathbf{1})^{-1}-W_{t}K_{0}W_{t}(\Omega _{0}^{%
\mathrm{t}}+\mathbf{1})^{-1}\right\Vert _{\mathrm{op}}=0
\end{equation}%
and (\ref{equality idiote}) holds for $s=0$ on the dense domain $\mathcal{D}%
(\Omega _{0}^{\mathrm{t}})$. Again, by (\ref{Condition sup}), both operators
$K_{t}$ and $W_{t}K_{0}W_{t}^{\mathrm{t}}$ are bounded, so (\ref{equality
idiote}) is also satisfied at $s=0$ on the whole Hilbert space $\mathfrak{h}$%
.\hfill $\Box $

Therefore, by combining Theorem \ref{lemma constante of motion} (ii) with
Lemma \ref{lemma constante of motion copy(2)} we directly obtain a simple
expression for the one--particle Hamiltonian $\Omega _{t}$ under the
condition that $K_{0}=0$:

\begin{corollary}[Constants of motion when $K_{0}=0$]
\label{lemma constante of motion copy(3)}\mbox{ }\newline
Assume Conditions A1--A3 and $\Omega _{0}B_{0}=B_{0}\Omega _{0}^{\mathrm{t}}$%
. Then, for any $t,s\in \lbrack 0,T_{+})$,
\begin{equation}
\Omega _{t}B_{t}=B_{t}\Omega _{t}^{\mathrm{t}}\qquad \text{and}\qquad \Omega
_{t}^{2}-4B_{t}\bar{B}_{t}=\Omega _{s}^{2}-4B_{s}\bar{B}_{s}\ .
\end{equation}
\end{corollary}

This assertion concludes the proof of the third (iii) and fourth (iv)
assertions of Theorem \ref{theorem important 1 -1}: (iii) is Theorem \ref%
{lemma constante of motion} (iii), whereas (iv) corresponds to Theorem \ref%
{lemma constante of motion} (i) and Corollary \ref{lemma constante of motion
copy(3)}.

\subsection{Asymptotics Properties of the Flow\label{Section Conserved
inequalities}}

First, since the system (\ref{flow equation-quadratic}) of differential
equations holds in the Hilbert--Schmidt topology\ (Corollary \ref{Corollary
existence}), we start by giving an explicit upper bound on the
Hilbert--Schmidt norm of the operator $\Omega _{t}^{\alpha }B_{t}\in
\mathcal{L}^{2}\left( \mathfrak{h}\right) $:

\begin{lemma}[Asymptotics properties of $\Vert \Omega _{t}^{\protect\alpha %
}B_{t}\Vert _{2}$]
\label{section extension copy(2)}\mbox{ }\newline
Assume Conditions A1--A4. Then, for all strictly positive numbers $\alpha
,t>0$ and any integer $n\in \mathbb{N}$,
\begin{equation}
\Vert \Omega _{t}^{\alpha }B_{t}\Vert _{2}\leq \left( \frac{2^{n-1}\alpha }{%
\mathrm{e}t}\right) ^{\alpha }\Vert B_{0}\Vert _{2}^{2^{-n}}\Vert B_{t}\Vert
_{2}^{1-2^{-n}}\leq \left( \frac{2^{n-1}\alpha }{\mathrm{e}t}\right)
^{\alpha }\Vert B_{0}\Vert _{2}\ .
\end{equation}
\end{lemma}

\noindent \textbf{Proof.} Observe that
\begin{equation}
\Vert \Omega _{t}^{\alpha }B_{t}\Vert _{2}\leq \left\Vert \Omega
_{t}^{2\alpha }\mathrm{e}^{-4t\Omega _{t}}\right\Vert _{\mathrm{op}%
}^{1/2}\left\{ \mathrm{tr}\left( \bar{B}_{t}\mathrm{e}^{4t\Omega
_{t}}B_{t}\right) \right\} ^{1/2}\ ,  \label{norm continuity of B}
\end{equation}%
using
\begin{equation}
\mathrm{tr}\left( X^{\ast }YX\right) \leq \left\Vert Y\right\Vert _{\mathrm{%
op}}\mathrm{tr}\left( X^{\ast }X\right) \ .  \label{V.217}
\end{equation}%
Since $\Omega _{t}=\Omega _{t}^{\ast }\geq 0$, we already know that, for all
$\alpha ,t>0$,
\begin{equation}
\left\Vert \Omega _{t}^{2\alpha }\mathrm{e}^{-4t\Omega _{t}}\right\Vert _{%
\mathrm{op}}^{1/2}\leq \left( \frac{\alpha }{2\mathrm{e}t}\right) ^{\alpha
}\ ,  \label{petit inequality newbis}
\end{equation}%
see (\ref{petit inequality new}). To use (\ref{norm continuity of B}) we
thus need to prove that the trace
\begin{equation}
\mathrm{tr}\left( \bar{B}_{t}\mathrm{e}^{4t\Omega _{t}}B_{t}\right)
\end{equation}%
exists and is uniformly bounded for all times $t\geq 0$. The self--adjoint
operator $\Omega _{t}$ is possibly unbounded and we consequently study the
trace
\begin{equation}
\mathrm{tr}\left( \bar{B}_{t}\mathrm{e}^{4t\Omega _{t,\lambda }}B_{t}\right)
\end{equation}%
for $t\geq 0$ and $\lambda >0$, where the positive, bounded operator $\Omega
_{t,\lambda }=\Omega _{t,\lambda }^{\ast }\geq 0$ is the Yosida
approximation
\begin{equation}
\Omega _{t,\lambda }:=\frac{\lambda \Omega _{t}}{\lambda \mathbf{1}+\Omega
_{t}}=\lambda -\lambda ^{2}\left( \lambda \mathbf{1}+\Omega _{t}\right)
^{-1}=\Omega _{t,\lambda }^{\ast }\in \mathcal{B}(\mathfrak{h})
\label{Yosida approx omega}
\end{equation}%
of the positive self--adjoint operator $\Omega _{t}$. Since $\Omega _{t}\geq
0$ and $\Vert B_{t}\Vert _{2}\leq \Vert B_{0}\Vert _{2}<\infty $, note that $%
\Vert \Omega _{t,\lambda }\Vert _{\mathrm{op}}\leq \lambda $ and
\begin{equation}
\mathrm{tr}\left( \bar{B}_{t}\mathrm{e}^{4t\Omega _{t,\lambda }}B_{t}\right)
\leq \mathrm{e}^{4t\lambda }\Vert B_{t}\Vert _{2}^{2}\leq \mathrm{e}%
^{4t\lambda }\Vert B_{0}\Vert _{2}^{2}<\infty \ .  \label{petite limit}
\end{equation}%
We analyze now the derivative of $\mathrm{tr}\left( \bar{B}_{t}\mathrm{e}%
^{4t\Omega _{t,\lambda }}B_{t}\right) $ for any $t>0$. If $t\mapsto X_{t}\in
\mathcal{B}(\mathfrak{h})$ is strongly differentiable then $t\mapsto \exp
(X_{t})$ is strongly differentiable and its derivative equals%
\begin{equation}
\partial _{t}\left\{ \mathrm{e}^{X_{t}}\right\} =\int_{0}^{1}\mathrm{e}%
^{\left( 1-\tau \right) X_{t}}\left( \partial _{t}X_{t}\right) \mathrm{e}%
^{\tau X_{t}}\mathrm{d}\tau \ ,
\label{derivative of an exponential operators}
\end{equation}%
see \cite{Snider,Wilcox}. Since the operator family $(\Omega _{t})_{t\geq 0}$
is (at least) strongly differentiable and
\begin{equation}
\partial _{t}\Omega _{t,\lambda }=-16\lambda ^{2}\left( \lambda \mathbf{1}%
+\Omega _{t}\right) ^{-1}B_{t}\bar{B}_{t}\left( \lambda \mathbf{1}+\Omega
_{t}\right) ^{-1}
\end{equation}%
(see Theorem \ref{section extension} (i)), a straightforward computation
using (\ref{derivative of an exponential operators}) shows that%
\begin{equation}
\partial _{t}\left\{ \mathrm{e}^{4t\Omega _{t,\lambda }}\right\} =4\Omega
_{t,\lambda }\mathrm{e}^{4t\Omega _{t,\lambda }}-64t\lambda ^{2}\int_{0}^{1}%
\frac{\mathrm{e}^{4t\left( 1-\tau \right) \Omega _{t,\lambda }}}{\lambda
\mathbf{1}+\Omega _{t}}B_{t}\bar{B}_{t}\frac{\mathrm{e}^{4\tau t\Omega
_{t,\lambda }}}{\lambda \mathbf{1}+\Omega _{t}}\ \mathrm{d}\tau \ .
\label{derivative of an exponential operatorsbis}
\end{equation}%
Note that
\begin{equation}
\left\Vert \partial _{t}\left\{ \mathrm{e}^{4t\Omega _{t,\lambda }}\right\}
\right\Vert _{\mathrm{op}}\leq 4\lambda \mathrm{e}^{4t\lambda }+64t\mathrm{e}%
^{4t\lambda }\Vert B_{0}\Vert _{2}^{4}\ ,
\end{equation}%
because $\Vert \Omega _{t,\lambda }\Vert _{\mathrm{op}}\leq \lambda $ and $%
\Vert B_{t}\Vert _{2}\leq \Vert B_{0}\Vert _{2}<\infty $. Therefore, we
infer from Corollary \ref{Corollary existence}, Lemma \ref{lemma existence 3
copy(4)}, and (\ref{derivative of an exponential operatorsbis}) that , for
all $t>0$,%
\begin{eqnarray}
\partial _{t}\left\{ \mathrm{tr}\left( \bar{B}_{t}\mathrm{e}^{4t\Omega
_{t,\lambda }}B_{t}\right) \right\} &=&-2\mathrm{tr}\left( \bar{B}_{t}%
\mathrm{e}^{4t\Omega _{t,\lambda }}B_{t}\Omega _{t}^{\mathrm{t}}\right) -2%
\mathrm{tr}\left( \Omega _{t}^{\mathrm{t}}\bar{B}_{t}\mathrm{e}^{4t\Omega
_{t,\lambda }}B_{t}\right)  \notag \\
&&-64t\lambda ^{2}\int_{0}^{1}\mathrm{tr}\left( \bar{B}_{t}\frac{\mathrm{e}%
^{4t\left( 1-\tau \right) \Omega _{t,\lambda }}}{\lambda \mathbf{1}+\Omega
_{t}}B_{t}\bar{B}_{t}\frac{\mathrm{e}^{4\tau t\Omega _{t,\lambda }}}{\lambda
\mathbf{1}+\Omega _{t}}B_{t}\right) \mathrm{d}\tau  \notag \\
&&+4\mathrm{tr}\left( \bar{B}_{t}\mathrm{e}^{2t\Omega _{t,\lambda }}\left(
\Omega _{t,\lambda }-\Omega _{t}\right) \mathrm{e}^{2t\Omega _{t,\lambda
}}B_{t}\right) \ .  \label{derivative exponential2}
\end{eqnarray}%
Note that the exchange of the trace with the integral on the finite $\left[
0,1\right] $, which is performed in (\ref{derivative exponential2}), is
justified by Lemma \ref{lemma existence 3 copy(5)} as the operators
\begin{equation}
\frac{\mathrm{e}^{4t\left( 1-\tau \right) \Omega _{t,\lambda }}}{\lambda
\mathbf{1}+\Omega _{t}}\qquad \mathrm{and}\qquad \frac{\mathrm{e}^{4\tau
t\Omega _{t,\lambda }}}{\lambda \mathbf{1}+\Omega _{t}}
\end{equation}%
are bounded for any $t,\tau \geq 0$, whereas $B_{t}\in \mathcal{L}^{2}(%
\mathfrak{h})$ is a Hilbert--Schmidt operator. By cyclicity of the trace
(cf. Lemma \ref{lemma existence 3 copy(3)}),
\begin{equation}
\mathrm{tr}\left( \bar{B}_{t}\mathrm{e}^{4t\Omega _{t,\lambda }}B_{t}\Omega
_{t}^{\mathrm{t}}\right) =\mathrm{tr}\left( \Omega _{t}^{\mathrm{t}}\bar{B}%
_{t}\mathrm{e}^{4t\Omega _{t,\lambda }}B_{t}\right) =\mathrm{tr}\left(
\mathrm{e}^{2t\Omega _{t,\lambda }}B_{t}\Omega _{t}^{\mathrm{t}}\bar{B}_{t}%
\mathrm{e}^{2t\Omega _{t,\lambda }}\right) \geq 0\ ,
\label{derivative exponential2bis}
\end{equation}%
because $\Omega _{t}^{\mathrm{t}}=(\Omega _{t}^{\mathrm{t}})^{\ast }\geq 0$
and $\mathrm{e}^{2t\Omega _{t,\lambda }}B_{t}\in \mathcal{L}^{2}(\mathfrak{h}%
)$. One also has%
\begin{equation}
\Omega _{t,\lambda }-\Omega _{t}=-\frac{\Omega _{t}^{2}}{\lambda \mathbf{1}%
+\Omega _{t}}\leq 0  \label{derivative exponential2bis0}
\end{equation}%
and, by cyclicity of the trace (Lemma \ref{lemma existence 3 copy(3)} (i)),
\begin{eqnarray}
&&\mathrm{tr}\left( \bar{B}_{t}\frac{\mathrm{e}^{4t\left( 1-\tau \right)
\Omega _{t,\lambda }}}{\lambda \mathbf{1}+\Omega _{t}}B_{t}\bar{B}_{t}\frac{%
\mathrm{e}^{4\tau t\Omega _{t,\lambda }}}{\lambda \mathbf{1}+\Omega _{t}}%
B_{t}\right)  \label{derivative exponential2bisbis} \\
&=&\mathrm{tr}\left( \frac{\mathrm{e}^{2\tau t\Omega _{t,\lambda }}}{\left(
\lambda \mathbf{1}+\Omega _{t}\right) ^{1/2}}B_{t}\bar{B}_{t}\frac{\mathrm{e}%
^{4t\left( 1-\tau \right) \Omega _{t,\lambda }}}{\lambda \mathbf{1}+\Omega
_{t}}B_{t}\bar{B}_{t}\frac{\mathrm{e}^{2\tau t\Omega _{t,\lambda }}}{\left(
\lambda \mathbf{1}+\Omega _{t}\right) ^{1/2}}\right) \geq 0\ .  \notag
\end{eqnarray}%
Consequently, we infer from (\ref{derivative exponential2})--(\ref%
{derivative exponential2bisbis}) that $\partial _{t}\{\mathrm{tr}(\bar{B}_{t}%
\mathrm{e}^{4t\Omega _{t,\lambda }}B_{t})\}\leq 0$ for any $t>0$. In
particular, by (\ref{petite limit}),
\begin{equation}
\forall t\geq s>0:\quad \mathrm{tr}(\bar{B}_{t}\mathrm{e}^{4t\Omega
_{t,\lambda }}B_{t})\leq \mathrm{tr}(\bar{B}_{s}\mathrm{e}^{4s\Omega
_{s,\lambda }}B_{s})\leq \mathrm{e}^{4s\lambda }\Vert B_{0}\Vert _{2}^{2}\ ,
\end{equation}%
which in turn implies that
\begin{equation}
\forall t\geq 0,\lambda >0:\qquad \mathrm{tr}(\bar{B}_{t}\mathrm{e}%
^{4t\Omega _{t,\lambda }}B_{t})\leq \left\Vert B_{0}\right\Vert _{2}^{2}\ .
\label{derivative exponential3}
\end{equation}%
We proceed by taking the limit $\lambda \rightarrow \infty $ in (\ref%
{derivative exponential3}). Since the family $(y_{\lambda })_{\lambda >0}$
of real functions $\mathbb{R}_{0}^{+}\rightarrow \mathbb{R}_{0}^{+}$,%
\begin{equation}
x\mapsto y_{\lambda }\left( x\right) =\frac{\lambda x}{\lambda +x}
\end{equation}%
is monotonically increasing in $\lambda $, the spectral theorem applied to
the positive self--adjoint operator $\Omega _{t,\lambda }=\Omega _{t,\lambda
}^{\ast }\geq 0$ together with the monotone convergence theorem yields
\begin{equation}
\underset{\lambda \rightarrow \infty }{\lim }\langle \varphi |\mathrm{e}%
^{4t\Omega _{t,\lambda }}\varphi \rangle =\left\{
\begin{array}{lll}
\langle \varphi |\mathrm{e}^{4t\Omega _{t}}\varphi \rangle & , & \qquad
\mathrm{if\ }\varphi \in \mathcal{D}\left( \mathrm{e}^{2t\Omega _{t}}\right)
\ , \\
\infty & , & \qquad \mathrm{if\ }\varphi \notin \mathcal{D}\left( \mathrm{e}%
^{2t\Omega _{t}}\right) \ ,%
\end{array}%
\right.  \label{derivative exponential4}
\end{equation}%
for all $\varphi \in \mathfrak{h}$. It follows that $B_{t}\mathfrak{h}\in
\mathcal{D}\left( \mathrm{e}^{2t\Omega _{t}}\right) $ for all $t\geq 0$,
because of the inequalities (\ref{derivative exponential3}) and $\bar{B}_{t}%
\mathrm{e}^{4t\Omega _{t,\lambda }}B_{t}\geq 0$. In particular, thanks to
the limit (\ref{derivative exponential4}), we obtain
\begin{equation}
\underset{\lambda \rightarrow \infty }{\lim }\langle B_{t}\varphi |\mathrm{e}%
^{4t\Omega _{t,\lambda }}B_{t}\varphi \rangle =\langle B_{t}\varphi |\mathrm{%
e}^{4t\Omega _{t}}B_{t}\varphi \rangle <\infty
\label{derivative exponential5}
\end{equation}%
for all $\varphi \in \mathfrak{h}$. We finally invoke again the monotone
convergence theorem to infer from (\ref{derivative exponential3}) and (\ref%
{derivative exponential5}) that
\begin{equation}
\underset{\lambda \rightarrow \infty }{\lim }\left\{ \mathrm{tr}(\bar{B}_{t}%
\mathrm{e}^{4t\Omega _{t,\lambda }}B_{t})\right\} =\mathrm{tr}(\bar{B}_{t}%
\mathrm{e}^{4t\Omega _{t}}B_{t})\leq \left\Vert B_{0}\right\Vert _{2}^{2}
\label{petit inequality}
\end{equation}%
for all $t\geq 0$. Combining the inequality (\ref{norm continuity of B})
with the upper bounds (\ref{petit inequality newbis}) and (\ref{petit
inequality}) we arrive at the assertion%
\begin{equation}
\forall t>0:\qquad \Vert \Omega _{t}^{\alpha }B_{t}\Vert _{2}\leq \left(
\frac{\alpha }{2\mathrm{e}t}\right) ^{\alpha }\Vert B_{0}\Vert _{2}<\infty \
.  \label{derivative H_t1}
\end{equation}%
This bound can again be improved by recursively using the Cauchy--Schwarz
inequality. Indeed,
\begin{equation}
\Vert \Omega _{t}^{\alpha }B_{t}\Vert _{2}=\{\mathrm{tr}\left( \bar{B}%
_{t}\Omega _{t}^{2\alpha }B_{t}\right) \}^{1/2}\leq \Vert \Omega
_{t}^{2\alpha }B_{t}\Vert _{2}^{1/2}\Vert B_{t}\Vert _{2}^{1/2}\ .
\label{derivative H_t1+1}
\end{equation}%
We proceed by again applying the Cauchy--Schwarz inequality on $\Vert \Omega
_{t}^{2\alpha }B_{t}\Vert _{2}$ in (\ref{derivative H_t1+1})\ to obtain
\begin{equation}
\Vert \Omega _{t}^{\alpha }B_{t}\Vert _{2}\leq \Vert \Omega _{t}^{4\alpha
}B_{t}\Vert _{2}^{\frac{1}{4}}\Vert B_{t}\Vert _{2}^{\frac{1}{2}+\frac{1}{4}%
}\ .
\end{equation}%
Doing this $n$ times, we obtain%
\begin{equation}
\Vert \Omega _{t}^{\alpha }B_{t}\Vert _{2}\leq \Vert \Omega
_{t}^{2^{n}\alpha }B_{t}\Vert _{2}^{2^{-n}}\Vert B_{t}\Vert _{2}^{1-2^{-n}}\
,  \label{derivative H_t1+2}
\end{equation}%
as
\begin{equation}
\underset{j=1}{\overset{n}{\sum }}2^{-j}=1-2^{-n}\ .
\end{equation}%
We now combine (\ref{derivative H_t1}) with (\ref{derivative H_t1+2}) to
show that%
\begin{equation}
\Vert \Omega _{t}^{\alpha }B_{t}\Vert _{2}\leq \left( \frac{2^{n-1}\alpha }{%
\mathrm{e}t}\right) ^{\alpha }\Vert B_{0}\Vert _{2}^{2^{-n}}\Vert B_{t}\Vert
_{2}^{1-2^{-n}}
\end{equation}%
for any $n\in \mathbb{N}$. \hfill $\Box $

\begin{corollary}[Asymptotics properties of $\Vert B_{t}\Vert _{2}$]
\label{Corollary existence copy(1)}\mbox{ }\newline
Assume Conditions A1--A4. Then, for any $t\geq 0$, $n\in \mathbb{N}$, and $%
\alpha >0$,
\begin{equation}
\Vert B_{t}\Vert _{2}\leq \left( \frac{2^{n-1}\alpha }{\mathrm{e}t}\right) ^{%
\frac{\alpha }{1+2^{-n}}}\Vert \Omega _{t}^{-\alpha }B_{t}\Vert _{2}^{\frac{1%
}{1+2^{-n}}}\Vert B_{0}\Vert _{2}^{\frac{1}{2^{n}+1}}\ ,
\end{equation}%
provided $\Vert \Omega _{t}^{-\alpha }B_{t}\Vert _{2}<\infty $.
\end{corollary}

\noindent \textbf{Proof.} Using the Cauchy--Schwarz inequality for the trace
note that
\begin{equation}
\Vert B_{t}\Vert _{2}^{2}=\mathrm{tr}\left( \bar{B}_{t}\Omega _{t}^{\alpha
}\Omega _{t}^{-\alpha }B_{t}\right) \leq \Vert \Omega _{t}^{-\alpha
}B_{t}\Vert _{2}\Vert \Omega _{t}^{\alpha }B_{t}\Vert _{2}
\end{equation}%
for any $\alpha >0$. Therefore, by Lemma \ref{section extension copy(2)}, we
find that, for any $n\in \mathbb{N}$,%
\begin{equation}
\Vert B_{t}\Vert _{2}^{2}\leq \left( \frac{2^{n-1}\alpha }{\mathrm{e}t}%
\right) ^{\alpha }\Vert \Omega _{t}^{-\alpha }B_{t}\Vert _{2}\Vert
B_{0}\Vert _{2}^{2^{-n}}\Vert B_{t}\Vert _{2}^{1-2^{-n}}
\end{equation}%
from which we deduce the assertion. \hfill $\Box $

Therefore, since the map
\begin{equation}
t\mapsto \Vert \mathfrak{B}_{t}\Vert _{2}=\Vert \Omega _{t}^{-1/2}B_{t}\Vert
_{2}^{2}
\end{equation}%
is monotonously decreasing (Lemma \ref{existence_flow_5 copy(1)} (i)), by
Corollary \ref{Corollary existence copy(1)} for $\alpha =1/2$, one deduces a
first \emph{explicit} upper bound on Hilbert--Schmidt norm $\Vert B_{t}\Vert
_{2}$:%
\begin{equation}
\Vert B_{t}\Vert _{2}^{2}\leq \Vert \mathfrak{B}_{0}\Vert _{1}^{\frac{1}{%
1+2^{-n}}}\Vert B_{0}\Vert _{2}^{\frac{2}{2^{n}+1}}\left( \frac{2^{n-2}}{%
\mathrm{e}t}\right) ^{\frac{1}{1+2^{-n}}}<\infty
\end{equation}%
for any $t>0$ and $n\in \mathbb{N}$. In particular, for any $\delta \in
(0,1) $ and as $t\rightarrow \infty $, $\Vert B_{t}\Vert _{2}^{2}=o\left(
t^{-\delta }\right) $. Unfortunately, this last estimate is not good enough
to obtain the square--integrability of the map $t\mapsto \Vert B_{t}\Vert
_{2}$ on $\left[ 0,\infty \right) $. This is nevertheless proven in the next
lemma:

\begin{lemma}[Square--Integrability of $\Vert B_{t}\Vert _{2}$]
\label{lemma integrability copy(1)}\mbox{ }\newline
Assume Conditions A1--A4. Then the map $t\mapsto \Vert B_{t}\Vert _{2}$ is
square--integrable on $\left[ 0,\infty \right) $:
\begin{equation}
\forall t\geq s\geq 0:\qquad 4\int_{s}^{t}\left\Vert B_{\tau }\right\Vert
_{2}^{2}\mathrm{d}\tau \leq \Vert \mathfrak{B}_{s}\Vert _{1}-\Vert \mathfrak{%
B}_{t}\Vert _{1}\ .  \label{inequalitybis}
\end{equation}
\end{lemma}

\noindent \textbf{Proof.} Using $\Omega _{t}\geq 4\mathfrak{B}_{t}\geq 4%
\mathfrak{B}_{t,\mu }$ and $\Vert \mathfrak{B}_{t}\Vert _{1}\leq \Vert
\mathfrak{B}_{0}\Vert _{1}<\infty $ (Lemma \ref{existence_flow_5 copy(1)}),
for all $t>0$, Equality (\ref{eq-3bis}) implies that
\begin{equation}
\partial _{t}q_{\mu }(t)\leq -4\mathrm{tr}\left\{ B_{t}\bar{B}_{t}\right\}
+4\mu \Vert \mathfrak{B}_{t}\Vert _{1}\ .  \label{inequalityinequality}
\end{equation}%
In the limit $\mu \rightarrow 0$ (cf. (\ref{eq-24})), this differential
inequality yields the square--integrability on the whole positive real line $%
\left[ 0,\infty \right) $ of the map $t\mapsto \Vert B_{t}\Vert _{2}$ as
well as (\ref{inequalitybis}), which includes $s=0$ because of (\ref{eq-7}%
).\hfill $\Box $

Much stronger decays on the Hilbert--Schmidt norm $\Vert B_{t}\Vert _{2}$
can be obtained, for instance under Condition A6, that is, under the
assumption that
\begin{equation}
\Omega _{0}\geq 4B_{0}(\Omega _{0}^{\mathrm{t}})^{-1}\bar{B}_{0}+\mu \mathbf{%
1=}4\mathfrak{B}_{0}+\mu \mathbf{1}
\end{equation}%
for some constant $\mu >0$. Indeed, in this case, the map $t\mapsto \Vert
B_{t}\Vert _{2}$ decays exponentially to zero, as $t\rightarrow \infty $,
because A6 is conserved for all times, by Lemma \ref{section extension gap}.
More precisely, one gets\ the following asymptotics on the Hilbert--Schmidt
norm $\Vert B_{t}\Vert _{2}$ :

\begin{lemma}[Asymptotics of $\Vert B_{t}\Vert _{2}$ under a gap condition]
\label{lemma integrability copy(2)}\mbox{ }\newline
Assume Conditions A1--A2 and A6, that is, $\Omega _{0}\geq 4\mathfrak{B}%
_{0}+\mu \mathbf{1}$ for some $\mu >0$. Then
\begin{equation}
\forall \alpha \in \mathbb{R},t\geq \max (0,\tfrac{\alpha }{2\mu }):\qquad
\Vert \Omega _{t}^{\alpha }B_{t}\Vert _{2}\leq \mu ^{\alpha }\mathrm{e}%
^{-2t\mu }\Vert B_{0}\Vert _{2}\ .
\end{equation}%
In particular, $\Vert B_{t}\Vert _{2}$ decays exponentially to zero, as $%
t\rightarrow \infty $.
\end{lemma}

\noindent \textbf{Proof.} If $\Omega _{0}\geq 4\mathfrak{B}_{0}+\mu \mathbf{1%
}$ for some $\mu >0$, then $\Omega _{t}\geq \mu \mathbf{1}$ for all $t\geq 0$
because of Lemma \ref{section extension gap}. Therefore, we get the
assertion by using the upper bound
\begin{equation}
\Vert \Omega _{t}^{2\alpha }\mathrm{e}^{-4t\Omega _{t}}\Vert _{\mathrm{op}%
}\leq \ \underset{\omega \geq \mu }{\sup }\left\{ \omega ^{2\alpha }\mathrm{e%
}^{-4t\omega }\right\} =\mu ^{2\alpha }\mathrm{e}^{-4t\mu }
\label{petit inequalitycool}
\end{equation}
for any $t\geq \max (0,\frac{\alpha }{2\mu })$ together with Inequalities (%
\ref{norm continuity of B}) and (\ref{petit inequality}).\hfill $\Box $

Therefore, under the gap condition A6, the map $t\mapsto \Vert B_{t}\Vert
_{2}$ is clearly integrable on $\left[ 0,\infty \right) $. We would like to
prove this property under weaker conditions than A6. A minimal requirement
is to assume A1--A4. In fact, we strengthen A4 by assuming next that $\Omega
_{0}^{-1}B_{0}\in \mathcal{L}^{2}(\mathfrak{h})$ is a Hilbert--Schmidt
operator.

As explained in Section \ref{section illustration1} via Proposition \ref%
{lemma example 2}, this property is \emph{pivotal} with respect to the
integrability of the map $t\mapsto \Vert B_{t}\Vert _{2}$ on $\left[
0,\infty \right) $. It is conserved by the flow, which equivalently means
that
\begin{equation}
\forall t\geq 0:\qquad \mathfrak{E}_{t}:=B_{t}(\Omega _{t}^{\mathrm{t}})^{-2}%
\bar{B}_{t}\in \mathcal{L}^{1}(\mathfrak{h})  \label{def Ct}
\end{equation}%
is a trace--class operator for all times. Indeed, the following assertion
holds:

\begin{lemma}[Conservation by the flow of $\Omega _{0}^{-1}B_{0}\in \mathcal{%
L}^{2}(\mathfrak{h})$]
\label{new lemma 2}\mbox{ }\newline
Assume Conditions A1--A3 and $\Omega _{0}^{-1}B_{0}\in \mathcal{L}^{2}(%
\mathfrak{h})$. Then
\begin{equation}
\forall t\geq s\geq 0:\qquad \Vert \mathfrak{E}_{t}\Vert _{1}\leq \Vert
\mathfrak{E}_{s}\Vert _{1}\exp \left\{ 32\int_{s}^{t}\Vert \mathfrak{B}%
_{\tau }\Vert _{1}\mathrm{d}\tau \right\} <\infty \ .
\end{equation}%
In particular, $(\Omega _{t}^{-1}B_{t})_{t\geq 0}\subset \mathcal{L}^{2}(%
\mathfrak{h})$ is a family of Hilbert--Schmidt operators.
\end{lemma}

\noindent \textbf{Proof.} First, observe that $\Omega _{0}^{-1}B_{0}\in
\mathcal{L}^{2}(\mathfrak{h})$ implies $(\Omega _{0}^{\mathrm{t}})^{-1}\bar{B%
}_{0}\in \mathcal{L}^{2}(\mathfrak{h})$, by Lemma \ref{lemma existence 3
copy(2)} and $B_{0}=B_{0}^{\mathrm{t}}$. In particular, $\mathfrak{E}_{0}\in
\mathcal{L}^{1}(\mathfrak{h})$ since
\begin{equation}
\Vert \mathfrak{E}_{0}\Vert _{1}:=\Vert (\Omega _{0}^{\mathrm{t}})^{-1}\bar{B%
}_{0}\Vert _{2}^{2}=\Vert \Omega _{0}^{-1}B_{0}\Vert _{2}^{2}<\infty \ .
\label{continuity0}
\end{equation}%
Moreover, $B_{0}\in \mathcal{L}^{2}(\mathfrak{h})$ and $\Omega
_{0}^{-1}B_{0}\in \mathcal{L}^{2}(\mathfrak{h})$ also yield $\Omega
_{0}^{-1/2}B_{0}\in \mathcal{L}^{2}(\mathfrak{h})$ because%
\begin{eqnarray}
\Vert \Omega _{0}^{-1/2}B_{0}\Vert _{2} &\leq &\Vert \Omega _{0}^{-1/2}%
\mathbb{P}_{\Omega _{0}\leq \mathbf{1}}B_{0}\Vert _{2}+\Vert \Omega
_{0}^{-1/2}\left( \mathbf{1}-\mathbb{P}_{\Omega _{0}\leq \mathbf{1}}\right)
B_{0}\Vert _{2}  \notag \\
&\leq &\Vert \Omega _{0}^{-1}B_{0}\Vert _{2}+\Vert B_{0}\Vert _{2}\ ,
\label{inequality idiote}
\end{eqnarray}%
where the operator $\mathbb{P}_{\Omega _{0}\leq \mathbf{1}}$ is the spectral
projection of the positive self--adjoint operator $\Omega _{0}=\Omega
_{0}^{\ast }\geq 0$ on the interval $\left[ 0,1\right] $.

Let $\mu >0$ and set%
\begin{equation}
\forall t\geq 0:\qquad \mathfrak{E}_{t,\mu }:=B_{t}\left( \Omega _{t}^{%
\mathrm{t}}+\mu \right) ^{-2}\bar{B}_{t}\geq 0\ .  \label{eq-2bis}
\end{equation}%
Similar to (\ref{eq-2}), $\mathfrak{E}_{t,\mu }$ is a bounded operator as $%
\mu >0$ and $\Omega _{t}^{\mathrm{t}}\geq 0$, by Lemma \ref{new lemma 2
copy(2)}. From Corollary \ref{Corollary existence} and Lemma \ref{lemma
existence 3 copy(4)} we also observe that the function
\begin{equation}
p_{\mu }(t):=\mathrm{tr}\left\{ \mathfrak{E}_{t,\mu }\right\} =\Vert
\mathfrak{E}_{t,\mu }\Vert _{1}
\end{equation}%
satisfies, for any strictly positive $t>0$, the differential inequality
\begin{eqnarray}
\partial _{t}p_{\mu }(t) &=&16\mathrm{tr}\left\{ B_{t}\left( \Omega _{t}^{%
\mathrm{t}}+\mu \mathbf{1}\right) ^{-2}\bar{B}_{t}B_{t}\left( \Omega _{t}^{%
\mathrm{t}}+\mu \mathbf{1}\right) ^{-1}\bar{B}_{t}\right\}
\label{eq-3bisbis} \\[0.01in]
&&+16\mathrm{tr}\left\{ B_{t}\left( \Omega _{t}^{\mathrm{t}}+\mu \mathbf{1}%
\right) ^{-1}\bar{B}_{t}B_{t}\left( \Omega _{t}^{\mathrm{t}}+\mu \mathbf{1}%
\right) ^{-2}\bar{B}_{t}\right\}  \notag \\
&&-4\mathrm{tr}\left\{ \Omega _{t}B_{t}\left( \Omega _{t}^{\mathrm{t}}+\mu
\mathbf{1}\right) ^{-2}\bar{B}_{t}+B_{t}\Omega _{t}^{\mathrm{t}}\left(
\Omega _{t}^{\mathrm{t}}+\mu \mathbf{1}\right) ^{-2}\bar{B}_{t}\right\}
\notag \\
&\leq &32\Vert \mathfrak{B}_{t}\Vert _{\mathrm{op}}\ \mathrm{tr}\{\mathfrak{E%
}_{t,\mu }\}\leq 32\Vert \mathfrak{B}_{t}\Vert _{1}\ p_{\mu }(t)\ ,
\end{eqnarray}%
by using the cyclicity of the trace (Lemma \ref{lemma existence 3 copy(3)}),
the positivity of the self--adjoint operators $\Omega _{t},\Omega _{t}^{%
\mathrm{t}}\geq 0$, and Lemma \ref{existence_flow_5 copy(1)} (i). Therefore,
thanks to Gr{ø}nwall's Lemma, we obtain that
\begin{equation}
\forall t\geq s>0:\qquad \Vert \mathfrak{E}_{t,\mu }\Vert _{1}\leq \Vert
\mathfrak{E}_{s,\mu }\Vert _{1}\exp \left\{ 32\int_{s}^{t}\Vert \mathfrak{B}%
_{\tau }\Vert _{1}\mathrm{d}\tau \right\} \ ,  \label{eq totosup1}
\end{equation}%
for all $\mu >0$. Similar to (\ref{continuity1}), note that
\begin{equation}
\left\vert p_{\mu }(t)-p_{\mu }(s)\right\vert \leq \Vert \mathfrak{E}_{t,\mu
}-\mathfrak{E}_{s,\mu }\Vert _{1}\leq 2\mu ^{-3}\Vert \Omega _{t}^{\mathrm{t}%
}-\Omega _{s}^{\mathrm{t}}\Vert _{1}\Vert B_{0}\Vert _{2}^{2}+2\mu
^{-2}\Vert B_{0}\Vert _{2}\Vert B_{t}-B_{s}\Vert _{2}  \label{continuity2}
\end{equation}%
for all $\mu >0$ and $s,t\geq 0$. Combined with Theorem \ref{lemma existence
3} (i)--(ii), it shows the continuity of the function $p_{\mu }$:
\begin{equation}
\underset{t\rightarrow s}{\lim }\ p_{\mu }(t)=p_{\mu }(s)\leq p_{0}(s)=\Vert
\mathfrak{E}_{s}\Vert _{1}  \label{continuity2bis}
\end{equation}%
for any $t,s\geq 0$. The latter implies (\ref{eq totosup1}) for $s=0$:%
\begin{equation}
\forall t\geq s\geq 0:\qquad \Vert \mathfrak{E}_{t,\mu }\Vert _{1}\leq \Vert
\mathfrak{E}_{s,\mu }\Vert _{1}\exp \left\{ 32\int_{s}^{t}\Vert \mathfrak{B}%
_{\tau }\Vert _{1}\mathrm{d}\tau \right\} <\infty \ ,  \label{eq totosup2}
\end{equation}%
see (\ref{continuity0}). In particular, $\ker (\Omega _{t}^{\mathrm{t}})\cap
\mathrm{Ran}(B_{t})=\emptyset $ as already mentioned in the proof of Lemma %
\ref{existence_flow_5 copy(1)}. Moreover, similar to (\ref{eq-23-1}) and (%
\ref{eq-24}), we infer from (\ref{eq totosup2}) together with the monotone
convergence theorem that, for all $t\geq 0$,
\begin{equation}
\lim_{\mu \rightarrow 0^{+}}\Vert \mathfrak{E}_{t,\mu }-\mathfrak{E}%
_{t,0}\Vert _{1}=0\ ,  \label{eq totosup1bis}
\end{equation}%
with $\mathfrak{E}_{t}\equiv \mathfrak{E}_{t,0}$. Using this and (\ref{eq
totosup2}) we thus deduce the upper bound of the lemma, which yields $\Omega
_{t}^{-1}B_{t}\in \mathcal{L}^{2}(\mathfrak{h})$ as
\begin{equation}
\forall t\geq 0:\qquad \Vert \mathfrak{E}_{t}\Vert _{1}:=\Vert (\Omega _{t}^{%
\mathrm{t}})^{-1}\bar{B}_{t}\Vert _{2}^{2}=\Vert \Omega _{t}^{-1}B_{t}\Vert
_{2}^{2}<\infty \ ,
\end{equation}%
similar to establishing (\ref{continuity0}).\hfill $\Box $

Lemma \ref{new lemma 2} indicates that the integrability of the map $%
t\mapsto \Vert \mathfrak{B}_{t}\Vert _{1}$ on the positive real line $\left[
0,\infty \right) $ should also be an important property for a
\textquotedblleft good\textquotedblright\ asymptotics of the flow. To obtain
this, we need, at least, the assumption
\begin{equation}
\mathbf{1}\geq 4B_{0}(\Omega _{0}^{\mathrm{t}})^{-2}\bar{B}_{0}=4\mathfrak{E}%
_{0}\ .  \label{assumption faible}
\end{equation}%
In this case, the behavior of the map $t\mapsto \Vert \mathfrak{E}_{t}\Vert
_{1}$ becomes better as it is monotonically decreasing:

\begin{lemma}[Conservation by the flow of $B_{0}(\Omega _{0}^{\mathrm{t}%
})^{-2}\bar{B}_{0}<\mathbf{1}/(4+\mathrm{r})$]
\label{new lemma 2 copy(1)}\mbox{ }\newline
Assume Conditions A1--A3, $\Omega _{0}^{-1}B_{0}\in \mathcal{L}^{2}(%
\mathfrak{h})$, and $\mathfrak{E}_{0}\leq \mathbf{1}/\left( 4+\mathrm{r}%
\right) $ for $\mathrm{r}\geq 0$. \newline
\emph{(i)} The family $(\mathfrak{E}_{t})_{t\geq 0}\subset \mathcal{L}^{1}(%
\mathfrak{h})$ defined by (\ref{def Ct}) satisfies: $\mathfrak{E}_{t}\leq
\mathbf{1}/\left( 4+\mathrm{r}\right) $.\newline
\emph{(ii)} The map $t\mapsto \Vert \mathfrak{E}_{t}\Vert _{1}$ from $%
\mathbb{R}_{0}^{+}$ to $\mathbb{R}_{0}^{+}$ is monotonically decreasing. In
particular, $\Vert \mathfrak{E}_{t}\Vert _{1}\leq \Vert \mathfrak{E}%
_{0}\Vert _{1}$ for all $t\geq 0$.
\end{lemma}

\noindent \textbf{Proof.} \underline{(i):} Recall that $\Omega
_{0}^{-1}B_{0}\in \mathcal{L}^{2}(\mathfrak{h})$ yields Condition A4, that
is, $\Omega _{0}^{-1/2}B_{0}\in \mathcal{L}^{2}(\mathfrak{h})$. By Theorems %
\ref{lemma existence 3}, \ref{section extension copy(4)} and Corollary \ref%
{Corollary existence} , the trace--class operator $\mathfrak{E}_{t,\mu }\in
\mathcal{L}^{1}(\mathfrak{h})$ defined by (\ref{eq-2bis}) for any $t\geq 0$
and $\mu >0$ has derivative (at least in the strong sense in $\mathcal{D}%
\left( \Omega _{0}\right) $) equal to
\begin{equation}
\forall t\geq s>0:\qquad \partial _{t}\mathfrak{E}_{t,\mu }=-4\mathfrak{B}%
_{t,\mu }-\mathcal{G}_{t,\mu }\mathfrak{E}_{t,\mu }-\mathfrak{E}_{t,\mu }%
\mathcal{G}_{t,\mu }\ \ ,  \label{derivative de C}
\end{equation}%
where
\begin{equation}
\mathcal{G}_{t,\mu }:=2\Omega _{t}-16\mathfrak{B}_{t,\mu }-2\mu \quad
\mathrm{and}\quad \mathfrak{B}_{t,\mu }:=B_{t}\left( \Omega _{t}^{\mathrm{t}%
}+\mu \mathbf{1}\right) ^{-1}\bar{B}_{t}\geq 0\ ,
\end{equation}%
see (\ref{eq-2}). Therefore, we introduce, as before, the evolution operator
$\mathfrak{M}_{t,s}$ defined by the non--autonomous evolution equation%
\begin{equation}
\forall t>s\geq 0:\qquad \partial _{t}\mathfrak{M}_{t,s}=-\mathcal{G}_{t,\mu
}\mathfrak{M}_{t,s}\ ,\quad \mathfrak{M}_{s,s}:=\mathbf{1}\ .
\label{eq evolution last lemma}
\end{equation}%
Indeed, the (possibly unbounded) generator%
\begin{equation}
\mathcal{G}_{t,\mu }=2\Omega _{0}-32\int_{0}^{t}B_{\tau }\bar{B}_{\tau }%
\mathrm{d}\tau -16\mathfrak{B}_{t,\mu }-2\mu
\end{equation}%
is the sum of a positive operator $2\Omega _{0}\geq 0$ and a bounded one
\begin{equation}
\mathcal{C}_{t,\mu }:=-32\int_{0}^{t}B_{\tau }\bar{B}_{\tau }\mathrm{d}\tau
-16\mathfrak{B}_{t,\mu }-2\mu
\end{equation}%
with operator norm
\begin{equation}
\left\Vert \mathcal{C}_{t,\mu }\right\Vert _{\mathrm{op}}\leq 32t\Vert
B_{0}\Vert _{2}+16\Vert \mathfrak{B}_{0}\Vert _{1}+2\mu \ ,
\label{bounded operator}
\end{equation}%
see (\ref{a priori estimate}), Lemma \ref{existence_flow_5 copy(1)} (i), and
Theorem \ref{section extension copy(4)}. As $\left\{ \mathcal{C}_{t,\mu
}\right\} _{t\geq 0}$ is also continuous with respect to the norm topology
for any $\mu >0$ (Theorem \ref{lemma existence 3} (i)--(ii)), the evolution
operator $\mathfrak{M}_{t,s}$ solving (\ref{eq evolution last lemma}) exists
and is unique, bounded uniformly in $s,t$ on compact sets, and norm
continuous for any $t>s$, see arguments given in Lemmata \ref{lemma
existence 1} and \ref{lemma existence 1 copy(1)}. Furthermore, on the domain
$\mathcal{D}\left( \Omega _{0}\right) $,
\begin{equation}
\forall t>0,\ t\geq s\geq 0:\qquad \partial _{s}\mathfrak{M}_{t,s}=\mathfrak{%
M}_{t,s}\mathcal{G}_{s,\mu }\ ,\quad \mathfrak{M}_{t,t}:=\mathbf{1}\ ,
\label{eq evolution last lemma+1}
\end{equation}%
whereas one verifies that
\begin{equation}
\forall t>s\geq 0:\qquad \partial _{s}\mathfrak{M}_{t,s}^{\ast }=\mathcal{G}%
_{s,\mu }\mathfrak{M}_{t,s}^{\ast }\ ,\quad \mathfrak{M}_{t,t}^{\ast }:=%
\mathbf{1}\ ,  \label{eq evolution last lemma+2}
\end{equation}%
in the strong sense in $\mathfrak{h}$ (as $\mathfrak{M}_{t,s}^{\ast }%
\mathfrak{h}\subseteq \mathcal{D}(\Omega _{0})$ for $t>s$). Note that one
proves (\ref{eq evolution last lemma+2}) by observing that $\mathfrak{M}%
_{t,s}$ has a representation in term of a series constructed from the
integral equation%
\begin{equation}
\forall t\geq s\geq 0:\qquad \mathfrak{M}_{t,s}=\mathrm{e}^{-2\left(
t-s\right) \Omega _{0}}-\int_{s}^{t}\mathfrak{M}_{t,\tau }\mathcal{C}_{\tau
,\mu }\mathrm{e}^{-2\left( \tau -s\right) \Omega _{0}}\mathrm{d}\tau \ ,
\end{equation}%
similar to the series (\ref{eq aup adjoint}) with $(-\mathcal{C}_{t,\mu
})_{t\geq 0}$ replacing $(2\Delta _{t})_{t\geq 0}$. See also Lemma \ref%
{lemma existence 1 copy(1)}. We omit the details.

Using $\mathfrak{E}_{t,\mu }\mathfrak{h}\subseteq \mathcal{D}(\Omega _{0})$
for $t>0$ (as $B_{t>0}\mathfrak{h}\subseteq \mathcal{D}(\Omega _{0})$), we
deduce from (\ref{derivative de C}) and (\ref{eq evolution last lemma+1})--(%
\ref{eq evolution last lemma+2}) that, for any $t>s>0$,%
\begin{equation}
\partial _{s}\{\mathfrak{M}_{t,s}\mathfrak{E}_{s,\mu }\mathfrak{M}%
_{t,s}^{\ast }\}=-4\mathfrak{M}_{t,s}\mathfrak{B}_{s,\mu }\mathfrak{M}%
_{t,s}^{\ast }\ .  \label{eq evolution last lemma+3}
\end{equation}%
This last derivative is verified by using the upper bound%
\begin{eqnarray}
&&\left\Vert \left( \epsilon ^{-1}\left( \mathfrak{M}_{t,s+\epsilon }%
\mathfrak{E}_{s+\epsilon ,\mu }\mathfrak{M}_{t,s+\epsilon }^{\ast }-%
\mathfrak{M}_{t,s}\mathfrak{E}_{s,\mu }\mathfrak{M}_{t,s}^{\ast }\right) -%
\mathfrak{M}_{t,s}\mathfrak{E}_{s,\mu }\partial _{s}\left\{ \mathfrak{M}%
_{t,s}^{\ast }\right\} \right. \right.  \notag \\
&&\left. \left. -\mathfrak{M}_{t,s}\partial _{s}\left\{ \mathfrak{E}_{s,\mu
}\right\} \mathfrak{M}_{t,s}^{\ast }-\partial _{s}\left\{ \mathfrak{M}%
_{t,s}\right\} \mathfrak{E}_{s,\mu }\mathfrak{M}_{t,s}^{\ast }\right)
\varphi \right\Vert  \notag \\
&\leq &\left\Vert \mathfrak{M}_{t,s+\epsilon }\right\Vert _{\mathrm{op}%
}\left\Vert \mathfrak{E}_{s+\epsilon ,\mu }-\mathfrak{E}_{s,\mu }\right\Vert
_{\mathrm{op}}\left\Vert \epsilon ^{-1}\left( \mathfrak{M}_{t,s+\epsilon
}^{\ast }-\mathfrak{M}_{t,s}^{\ast }\right) \varphi \right\Vert  \notag \\
&&+\left\Vert \mathfrak{M}_{t,s+\epsilon }\right\Vert _{\mathrm{op}%
}\left\Vert \mathfrak{E}_{s,\mu }\right\Vert _{\mathrm{op}}\left\Vert
\epsilon ^{-1}\left( \mathfrak{M}_{t,s+\epsilon }^{\ast }-\mathfrak{M}%
_{t,s}^{\ast }\right) -\partial _{s}\left\{ \mathfrak{M}_{t,s}^{\ast
}\right\} \varphi \right\Vert  \notag \\
&&+\left\Vert \left( \mathfrak{M}_{t,s+\epsilon }-\mathfrak{M}_{t,s}\right)
\mathfrak{E}_{s,\mu }\partial _{s}\left\{ \mathfrak{M}_{t,s}^{\ast }\right\}
\varphi \right\Vert  \notag \\
&&+\left\Vert \mathfrak{M}_{t,s+\epsilon }\right\Vert _{\mathrm{op}%
}\left\Vert \left( \epsilon ^{-1}\left( \mathfrak{E}_{s+\epsilon ,\mu }-%
\mathfrak{E}_{s,\mu }\right) -\partial _{s}\left\{ \mathfrak{E}_{s,\mu
}\right\} \right) \mathfrak{M}_{t,s}^{\ast }\varphi \right\Vert  \notag \\
&&+\left\Vert \left( \mathfrak{M}_{t,s+\epsilon }-\mathfrak{M}_{t,s}\right)
\partial _{s}\left\{ \mathfrak{E}_{s,\mu }\right\} \mathfrak{M}_{t,s}^{\ast
}\varphi \right\Vert  \notag \\
&&+\left\Vert \left( \epsilon ^{-1}\left( \mathfrak{M}_{t,s+\epsilon }-%
\mathfrak{M}_{t,s}\right) -\partial _{s}\left\{ \mathfrak{M}_{t,s}\right\}
\right) \mathfrak{E}_{s,\mu }\mathfrak{M}_{t,s}^{\ast }\varphi \right\Vert \
,  \label{upper bound chiante}
\end{eqnarray}%
for any $\varphi \in \mathfrak{h}$, $t>s>0$, and sufficiently small $%
\left\vert \epsilon \right\vert >0$. Indeed, by (\ref{continuity2}), $(%
\mathfrak{E}_{s,\mu })_{s\geq 0}$ is norm continuous, $\mathfrak{M}%
_{t,s}^{\ast }$ and its adjoint are both strongly continuous and uniformally
bounded in $t,s$ on compact sets. Therefore, it follows from (\ref%
{derivative de C}), (\ref{eq evolution last lemma+1})--(\ref{eq evolution
last lemma+2}), and (\ref{upper bound chiante}) in the limit $\epsilon
\rightarrow 0$ that (\ref{eq evolution last lemma+3}) holds and we thus
arrive at the equality%
\begin{equation}
\forall t\geq s>0:\qquad \mathfrak{E}_{t,\mu }=\mathfrak{M}_{t,s}\mathfrak{E}%
_{s,\mu }\mathfrak{M}_{t,s}^{\ast }-4\int_{s}^{t}\mathfrak{M}_{t,\tau }%
\mathfrak{B}_{\tau ,\mu }\mathfrak{M}_{t,\tau }^{\ast }\mathrm{d}\tau \ .
\label{eq toto3}
\end{equation}%
Combining (\ref{eq evolution last lemma+1})--(\ref{eq evolution last lemma+2}%
) with the operator inequalities $\Omega _{t}\geq 4\mathfrak{B}_{t}\geq 4%
\mathfrak{B}_{t,\mu }$, we obtain that, for any $t>s\geq 0$,%
\begin{eqnarray}
\partial _{s}\left\{ \mathfrak{M}_{t,s}\mathfrak{M}_{t,s}^{\ast }\right\}
&=&2\mathfrak{M}_{t,s}\mathcal{G}_{s,\mu }\mathfrak{M}_{t,s}^{\ast }\geq -16%
\mathfrak{M}_{t,s}\mathfrak{B}_{s,\mu }\mathfrak{M}_{t,s}^{\ast }-4\mu
\mathfrak{M}_{t,s}\mathfrak{M}_{t,s}^{\ast }  \notag \\
&\geq &\left( 4+\mathrm{r}\right) \left( -4\mathfrak{M}_{t,s}\mathfrak{B}%
_{s,\mu }\mathfrak{M}_{t,s}^{\ast }-\mu \mathfrak{M}_{t,s}\mathfrak{M}%
_{t,s}^{\ast }\right) \ ,
\end{eqnarray}%
with $\mathrm{r}\geq 0$. Hence, for any $t\geq s\geq 0$ and $\mathrm{r}\geq
0 $,
\begin{equation}
-4\int_{s}^{t}\mathfrak{M}_{t,\tau }\mathfrak{B}_{\tau ,\mu }\mathfrak{M}%
_{t,\tau }^{\ast }\mathrm{d}\tau \leq \frac{1}{\left( 4+\mathrm{r}\right) }%
\left( \mathbf{1}-\mathfrak{M}_{t,s}\mathfrak{M}_{t,s}^{\ast }\right) +\mu
\int_{s}^{t}\mathfrak{M}_{t,\tau }\mathfrak{M}_{t,\tau }^{\ast }\mathrm{d}%
\tau \ .  \label{eq toto4}
\end{equation}%
Inserting this inequality into (\ref{eq toto3}) we get, for any $t\geq s>0$,
that%
\begin{eqnarray}
(4+\mathrm{r})\mathfrak{E}_{t,\mu } &\leq &\mathfrak{M}_{t,s}\left( (4+%
\mathrm{r})\mathfrak{E}_{s,\mu }\right) \mathfrak{M}_{t,s}^{\ast }+\mathbf{1}%
-\mathfrak{M}_{t,s}\mathfrak{M}_{t,s}^{\ast }  \notag \\
&&+(4+\mathrm{r})\mu \int_{s}^{t}\mathfrak{M}_{t,\tau }\mathfrak{M}_{t,\tau
}^{\ast }\mathrm{d}\tau \ .  \label{eq toto5}
\end{eqnarray}%
Next, we take the limit $s\rightarrow 0$. The evolution operator $\mathfrak{M%
}_{t,s}\in \mathcal{B}(\mathfrak{h})$, i.e., the unique solution of (\ref{eq
evolution last lemma}), is jointly strongly continuous in $s$ and $t$ (see
Lemma \ref{lemma existence 1} (ii)) and is uniformly bounded. Moreover, by (%
\ref{continuity2}), the trace--class operator $\mathfrak{E}_{t,\mu }\in C[%
\mathbb{R}_{0}^{+};\mathcal{L}^{1}(\mathfrak{h})]$ is continuous in $%
\mathcal{L}^{1}(\mathfrak{h})$. It follows that Inequality (\ref{eq toto5})
also holds for $s=0$ and using the operator inequalities
\begin{equation}
(4+\mathrm{r})\mathfrak{E}_{0,\mu }\leq (4+\mathrm{r})B_{0}(\Omega _{0}^{%
\mathrm{t}})^{-2}\bar{B}_{0}\leq \mathbf{1}
\end{equation}%
we deduce that
\begin{equation}
\forall t\geq 0:\quad (4+\mathrm{r})\mathfrak{E}_{t,\mu }\leq \mathbf{1}+(4+%
\mathrm{r})\mu \int_{0}^{t}\mathfrak{M}_{t,\tau }\mathfrak{M}_{t,\tau
}^{\ast }\mathrm{d}\tau \ .  \label{inequ}
\end{equation}%
Note that (\ref{bounded operator}) implies%
\begin{equation}
\left\Vert \mathfrak{M}_{t,s}\right\Vert _{\mathrm{op}}\leq \exp \left(
32t^{2}\Vert B_{0}\Vert _{2}+16t\Vert \mathfrak{B}_{0}\Vert _{1}+\mu
t\right) \ ,  \label{uni bounded}
\end{equation}%
for any $t\geq s\geq 0$ and $\mu >0$. So, by (\ref{eq totosup1bis}) and (\ref%
{uni bounded}), we infer from the limit $\mu \rightarrow 0^{+}$ in
Inequality (\ref{inequ}) that $\mathfrak{E}_{t}\leq \mathbf{1}/\left( 4+%
\mathrm{r}\right) $.

\underline{(ii):} We proceed by rewriting (\ref{eq-3bisbis}) as
\begin{eqnarray}
\partial _{t}p_{\mu }(t) &=&4\mu p_{\mu }(t)-4\mathrm{tr}\left\{ \left(
\mathbf{1}-4B_{t}\left( \Omega _{t}^{\mathrm{t}}+\mu \mathbf{1}\right) ^{-2}%
\bar{B}_{t}\right) B_{t}\left( \Omega _{t}^{\mathrm{t}}+\mu \mathbf{1}%
\right) ^{-1}\bar{B}_{t}\right\}  \notag \\
&&-4\mathrm{tr}\left\{ \left( \Omega _{t}-4B_{t}\left( \Omega _{t}^{\mathrm{t%
}}+\mu \mathbf{1}\right) ^{-1}\bar{B}_{t}\right) B_{t}\left( \Omega _{t}^{%
\mathrm{t}}+\mu \mathbf{1}\right) ^{-2}\bar{B}_{t}\right\} \ .  \notag \\
&&  \label{flow detail}
\end{eqnarray}%
We now use Assertion (i), Lemma \ref{existence_flow_5 copy(1)} (ii), and the
cyclicity of the trace (Lemma \ref{lemma existence 3 copy(3)}) to infer from
(\ref{flow detail}) the differential inequality%
\begin{equation}
\forall t>0:\qquad \partial _{t}p_{\mu }(t)\leq 4\mu p_{\mu }(t)\ .
\label{ineq gronwall}
\end{equation}%
Thanks to Gr{ø}nwall's Lemma and the continuity of the function $p_{\mu }$
(see (\ref{continuity2bis})), we then obtain from (\ref{ineq gronwall}) that
\begin{equation}
\forall t\geq s\geq 0:\qquad p_{\mu }(t)\leq p_{\mu }(s)\mathrm{e}^{4\mu
\left( t-s\right) }\ .
\end{equation}%
The latter is combined with (\ref{eq totosup1bis}) in the limit $\mu
\rightarrow 0^{+}$ to deduce the inequality $\Vert \mathfrak{E}_{t}\Vert
_{1}\leq \Vert \mathfrak{E}_{s}\Vert _{1}$ for all $t\geq s\geq 0$. \hfill $%
\Box $

Therefore, using the upper bound of Corollary \ref{Corollary existence
copy(1)} for $\alpha =1$ together with Lemma \ref{new lemma 2 copy(1)} (ii)
one gets, for any $t\geq 1$ and $n\in \mathbb{N}$, the upper bound%
\begin{equation}
\Vert B_{t}\Vert _{2}\leq \Vert \mathfrak{E}_{0}\Vert _{1}^{\frac{1}{%
2(1+2^{-n})}}\Vert B_{0}\Vert _{2}^{\frac{1}{2^{n}+1}}\left( \frac{2^{n-1}}{%
\mathrm{e}t}\right) ^{\frac{1}{1+2^{-n}}}\ .
\end{equation}%
In particular, for any $\delta \in (0,1)$ and as $t\rightarrow \infty $, $%
\Vert B_{t}\Vert _{2}=o\left( t^{-\delta }\right) $. In fact, one can obtain
$\Vert B_{t}\Vert _{2}=o\left( t^{-1}\right) $ as well as the integrability
of the map $t\mapsto \Vert \mathfrak{B}_{t}\Vert _{1}$ under a slightly
stronger assumption than (\ref{assumption faible}):

\begin{corollary}[Integrability of $\Vert \mathfrak{B}_{t}\Vert _{1}$]
\label{lemma integrability copy(3)}\mbox{ }\newline
Assume Conditions A1--A3, $\Omega _{0}^{-1}B_{0}\in \mathcal{L}^{2}(%
\mathfrak{h})$, and $\mathbf{1}\geq (4+\mathrm{r})B_{0}(\Omega _{0}^{\mathrm{%
t}})^{-2}\bar{B}_{0}$ for some $\mathrm{r}>0$. Then the map $t\mapsto \Vert
\mathfrak{B}_{t}\Vert _{1}$ is integrable on $\left[ 0,\infty \right) $:
\begin{equation}
\forall t\geq s\geq 0:\qquad \frac{4\mathrm{r}}{4+\mathrm{r}}%
\int_{s}^{t}\Vert \mathfrak{B}_{\tau }\Vert _{1}\mathrm{d}\tau \leq \Vert
\mathfrak{E}_{s}\Vert _{1}-\Vert \mathfrak{E}_{t}\Vert _{1}\ .
\label{inequality autre}
\end{equation}%
Additionally, $\Vert B_{t}\Vert _{2}=o\left( t^{-1}\right) $, as $%
t\rightarrow \infty $.
\end{corollary}

\noindent \textbf{Proof.} By using (\ref{flow detail}), the positivity of
the self--adjoint operators $\Omega _{t},\Omega _{t}^{\mathrm{t}}\geq 0$,
the cyclicity of the trace (Lemma \ref{lemma existence 3 copy(3)}), and the
operator inequalities $\Omega _{t}\geq 4\mathfrak{B}_{t}\geq 4\mathfrak{B}%
_{t,\mu }$ (Lemma \ref{existence_flow_5 copy(1)} (ii), Theorem \ref{section
extension copy(4)}) and $\mathfrak{E}_{t,\mu }\leq \mathfrak{E}_{t}\leq
\mathbf{1}/(4+\mathrm{r})$ (Lemma \ref{new lemma 2 copy(1)} (i)), we obtain
the differential inequality
\begin{equation}
\forall t\geq s>0:\qquad \partial _{t}p_{\mu }(t)\leq 4\mu \Vert \mathfrak{E}%
_{t,\mu }\Vert _{1}-\frac{4\mathrm{r}}{4+\mathrm{r}}\Vert \mathfrak{B}%
_{t,\mu }\Vert _{1}\ .  \label{differential inequality}
\end{equation}%
By Lemma \ref{existence_flow_5 copy(1)} (i), Theorem \ref{section extension
copy(4)} and Lemma \ref{new lemma 2 copy(1)} (ii), recall that%
\begin{equation}
\forall t\geq 0:\quad \Vert \mathfrak{B}_{t}\Vert _{1}\leq \Vert \mathfrak{B}%
_{0}\Vert _{1}\quad \mathrm{and}\quad \Vert \mathfrak{E}_{t,\mu }\Vert
_{1}\leq \Vert \mathfrak{E}_{t}\Vert _{1}\leq \Vert \mathfrak{E}_{0}\Vert
_{1}\ ,  \label{differential inequalitybis}
\end{equation}%
whereas $p_{\mu }$ is continuous at zero. Therefore, (\ref{differential
inequality}) in the limit $\mu \rightarrow 0^{+}$ (cf. (\ref{eq-24}) and (%
\ref{eq totosup1bis})) implies that the map $t\mapsto \Vert \mathfrak{B}%
_{t}\Vert _{1}$ is integrable on $\left[ 0,\infty \right) $ and satisfies (%
\ref{inequality autre}). Moreover, by combining Lemma \ref{lemma
integrability copy(1)} with (\ref{inequality autre}) and (\ref{differential
inequalitybis}) we obtain, for all $T\geq 0$, that%
\begin{eqnarray}
\int_{T}^{\infty }\left( s-T\right) \left\Vert B_{s}\right\Vert _{2}^{2}%
\mathrm{d}s &=&\int_{T}^{\infty }\int_{T}^{\infty }\mathbf{1}\left[ s\geq t%
\right] \left\Vert B_{s}\right\Vert _{2}^{2}\mathrm{d}t\ \mathrm{d}s  \notag
\\
&=&\int_{T}^{\infty }\int_{t}^{\infty }\left\Vert B_{s}\right\Vert _{2}^{2}%
\mathrm{d}s\ \mathrm{d}t\leq \frac{1}{4}\int_{T}^{\infty }\Vert \mathfrak{B}%
_{t}\Vert _{1}\mathrm{d}t  \notag \\
&\leq &\frac{4+\mathrm{r}}{16\mathrm{r}}\Vert \mathfrak{E}_{T}\Vert _{1}\leq
\frac{4+\mathrm{r}}{16\mathrm{r}}\Vert \mathfrak{E}_{0}\Vert _{1}\ .
\label{contradiction 0}
\end{eqnarray}%
Since the map $s\mapsto \Vert B_{s}\Vert _{2}^{2}$ is monotonically
decreasing on $\left[ 0,\infty \right) $, we further observe that
\begin{equation}
\int_{n}^{n+1}\left( s-1\right) \left\Vert B_{s}\right\Vert _{2}^{2}\mathrm{d%
}s\geq \left( n-1\right) \left\Vert B_{n+1}\right\Vert _{2}^{2}
\end{equation}%
for all $n\in \mathbb{N}$. Therefore,
\begin{equation}
\underset{n=2}{\overset{\infty }{\sum }}\left( n-2\right) \left\Vert
B_{n}\right\Vert _{2}^{2}=\underset{n=1}{\overset{\infty }{\sum }}\left(
n-1\right) \left\Vert B_{n+1}\right\Vert _{2}^{2}\leq \int_{1}^{\infty
}\left( s-1\right) \left\Vert B_{s}\right\Vert _{2}^{2}\mathrm{d}s<\infty \ .
\label{contradiction 1}
\end{equation}%
Now, suppose that $\left\Vert B_{n}\right\Vert _{2}\geq \epsilon n^{-1}$ for
some $\epsilon >0$. Then%
\begin{equation}
\underset{n=2}{\overset{\infty }{\sum }}\left( n-2\right) \left\Vert
B_{n}\right\Vert _{2}^{2}\geq \epsilon \underset{n=2}{\overset{\infty }{\sum
}}\frac{n-2}{n^{2}}=\infty \ ,
\end{equation}%
in contradiction to (\ref{contradiction 1}), and thus $\Vert B_{t}\Vert
_{2}=o(t^{-1})$. \hfill $\Box $

In the last proof, note that we use (\ref{contradiction 0}) which is
equivalent to
\begin{equation}
\int_{0}^{t}\mathrm{d}\tau _{1}\int_{\tau _{1}}^{\infty }\mathrm{d}\tau
_{2}\left\Vert B_{\tau _{2}}\right\Vert _{2}^{2}\leq \frac{4+\mathrm{r}}{16%
\mathrm{r}}\Vert \mathfrak{E}_{0}\Vert _{1}<\infty \ ,
\label{inequality sympa-autre}
\end{equation}%
see (\ref{inequalitybis}) and (\ref{inequality autre}). However, this
inequality does not\ necessarily imply the integrability of the
Hilbert--Schmidt norm $\Vert B_{t}\Vert _{2}$. A sufficient condition is $%
\Omega _{0}^{-1-\varepsilon }B_{0}\in \mathcal{L}^{2}(\mathfrak{h})$ for
some $\varepsilon >0$. Indeed, by Corollary \ref{Corollary existence copy(1)}
for $\alpha =1+\varepsilon $ and $n\in \mathbb{N}$ such that $%
2^{-n}<\varepsilon $,%
\begin{equation}
\Vert B_{t}\Vert _{2}\leq \Vert \Omega _{t}^{-1-\varepsilon }B_{t}\Vert
_{2}^{\frac{1}{1+2^{-n}}}\Vert B_{0}\Vert _{2}^{\frac{1}{2^{n}+1}}\left(
\frac{2^{n-1}\left( 1+\varepsilon \right) }{\mathrm{e}t}\right) ^{\frac{%
1+\varepsilon }{1+2^{-n}}}\ .  \label{inequality avec eps}
\end{equation}%
The latter implies the integrability of the map $t\mapsto \Vert B_{t}\Vert
_{2}$ on $[0,\infty )$ if, for instance, $\Vert \Omega _{t}^{-1-\varepsilon
}B_{t}\Vert _{2}$ is uniformly bounded for all times. That is what we prove
below under the assumption that $\Omega _{0}^{-1-\varepsilon }B_{0}$ is a
Hilbert--Schmidt operator:

\begin{lemma}[Conservation by the flow of $\Omega _{0}^{-1-\protect%
\varepsilon }B_{0}\in \mathcal{L}^{2}(\mathfrak{h})$]
\label{lemma integrability copy(6)}\mbox{ }\newline
Assume Conditions A1--A3, $\mathbf{1}\geq 4B_{0}(\Omega _{0}^{\mathrm{t}%
})^{-2}\bar{B}_{0}$, and $\Omega _{0}^{-1-\varepsilon }B_{0}\in \mathcal{L}%
^{2}(\mathfrak{h})$ for some $\varepsilon \in (0,1/2)$. Then
\begin{equation}
\forall t\geq s\geq 0:\qquad \Vert \Omega _{t}^{-1-\varepsilon }B_{t}\Vert
_{2}\leq \Vert \Omega _{s}^{-1-\varepsilon }B_{s}\Vert _{2}\exp \left\{
8\int_{s}^{t}\Vert \mathfrak{B}_{\tau }\Vert _{1}\mathrm{d}\tau \right\} \ .
\end{equation}
\end{lemma}

\noindent \textbf{Proof.} Using an inequality like (\ref{inequality idiote})
we observe that $B_{0}\in \mathcal{L}^{2}(\mathfrak{h})$ and $\Omega
_{0}^{-1-\varepsilon }B_{0}\in \mathcal{L}^{2}(\mathfrak{h})$ with $%
\varepsilon >0$ imply that $\Omega _{0}^{-1/2}B_{0}\in \mathcal{L}^{2}(%
\mathfrak{h})$ and $\Omega _{0}^{-1}B_{0}\in \mathcal{L}^{2}(\mathfrak{h})$,
which in turn implies that $(\Omega _{0}^{\mathrm{t}})^{-1}\bar{B}_{0}\in
\mathcal{L}^{2}(\mathfrak{h})$, by Lemma \ref{lemma existence 3 copy(2)} and
$B_{0}=B_{0}^{\mathrm{t}}$. Since $\Omega _{t}=\Omega _{t}^{\ast }\geq 0$,
the function $h_{\mu }$ defined for all $\mu >0$ and $t\geq 0$ by%
\begin{equation}
h_{\mu }(t):=\Vert \left( \Omega _{t}+\mu \mathbf{1}\right) ^{-1-\varepsilon
}B_{t}\Vert _{2}^{2}=\mathrm{tr}(B_{t}(\Omega _{t}^{\mathrm{t}}+\mu \mathbf{1%
})^{-2-2\varepsilon }\bar{B}_{t})
\end{equation}%
is uniformly bounded in time by
\begin{equation}
h_{\mu }(t)\leq \mu ^{-2-2\varepsilon }\Vert B_{t}\Vert _{2}^{2}\leq \mu
^{-2-2\varepsilon }\Vert B_{0}\Vert _{2}^{2}\ .
\end{equation}%
We proceed by using functional calculus to observe that
\begin{equation}
\left( \Omega _{t}^{\mathrm{t}}+\mu \mathbf{1}\right) ^{-2\varepsilon }=%
\frac{\sin \left( 2\pi \varepsilon \right) }{\pi }\int_{0}^{\infty }\lambda
^{-2\varepsilon }\left( \Omega _{t}^{\mathrm{t}}+\left( \lambda +\mu \right)
\mathbf{1}\right) ^{-1}\mathrm{d}\lambda  \label{eq toto nouveau2bis}
\end{equation}%
for any $\mu >0$ and $\varepsilon \in (0,1/2)$, see \cite[(1.4.2) and 1.4.7
(d)]{Caps}. Note that $B_{t}(\Omega _{t}^{\mathrm{t}}+\mu \mathbf{1}%
)^{-1}\in \mathcal{L}^{2}(\mathfrak{h})$ and
\begin{equation}
\int_{0}^{\infty }\lambda ^{-2\varepsilon }\Vert \left( \Omega _{t}^{\mathrm{%
t}}+\left( \lambda +\mu \right) \mathbf{1}\right) ^{-1}\Vert _{\mathrm{op}}\
\mathrm{d}\lambda \leq \int_{0}^{\infty }\frac{1}{\lambda ^{2\varepsilon
}\left( \lambda +\mu \right) }\mathrm{d}\lambda <\infty \ ,  \label{eq sup}
\end{equation}%
because $B_{t}\in \mathcal{L}^{2}(\mathfrak{h})$, $\Omega _{t}^{\mathrm{t}%
}\geq 0$, $\varepsilon \in (0,1/2)$, and $\mu >0$. So, using Lemma \ref%
{lemma existence 3 copy(5)} we thus obtain that
\begin{equation}
h_{\mu }(t)=\frac{\sin \left( 2\pi \varepsilon \right) }{\pi }%
\int_{0}^{\infty }\lambda ^{-2\varepsilon }\xi _{\mu ,\lambda }(t)\ \mathrm{d%
}\lambda \ ,  \label{eq toto nouveau1 correct0}
\end{equation}%
where $\xi _{\mu ,\lambda }$ is the function defined, for any $\mu >0$ and $%
\lambda ,t\geq 0$, by%
\begin{equation}
\xi _{\mu ,\lambda }(t):=\mathrm{tr}\left( B_{t}\left( \Omega _{t}^{\mathrm{t%
}}+\mu \mathbf{1}\right) ^{-2}\left( \Omega _{t}^{\mathrm{t}}+\left( \lambda
+\mu \right) \mathbf{1}\right) ^{-1}\bar{B}_{t}\right) \ .
\end{equation}%
This positive function exists as
\begin{equation}
\forall \mu >0,\ \lambda ,t\geq 0:\qquad 0\leq \xi _{\mu ,\lambda }(t)\leq
\mu ^{-2}\left( \lambda +\mu \right) ^{-1}\Vert B_{0}\Vert _{2}^{2}\ .
\label{bound idiote}
\end{equation}%
For all $\mu ,t>0$ and $\lambda \geq 0$, its derivative equals%
\begin{eqnarray}
\partial _{t}\xi _{\mu ,\lambda }(t) &=&16\mathrm{tr}\left\{ B_{t}\left(
\Omega _{t}^{\mathrm{t}}+\mu \mathbf{1}\right) ^{-2}\left( \Omega _{t}^{%
\mathrm{t}}+\left( \lambda +\mu \right) \mathbf{1}\right) ^{-1}\bar{B}%
_{t}B_{t}\left( \Omega _{t}^{\mathrm{t}}+\left( \lambda +\mu \right) \mathbf{%
1}\right) ^{-1}\bar{B}_{t}\right\}  \notag \\[0.01in]
&&+16\mathrm{tr}\left\{ B_{t}\left( \Omega _{t}^{\mathrm{t}}+\mu \mathbf{1}%
\right) ^{-2}\bar{B}_{t}B_{t}\left( \Omega _{t}^{\mathrm{t}}+\mu \mathbf{1}%
\right) ^{-1}\left( \Omega _{t}^{\mathrm{t}}+\left( \lambda +\mu \right)
\mathbf{1}\right) ^{-1}\bar{B}_{t}\right\}  \notag \\
&&+16\mathrm{tr}\left\{ B_{t}\left( \Omega _{t}^{\mathrm{t}}+\mu \mathbf{1}%
\right) ^{-1}\bar{B}_{t}B_{t}\left( \Omega _{t}^{\mathrm{t}}+\mu \mathbf{1}%
\right) ^{-2}\left( \Omega _{t}^{\mathrm{t}}+\left( \lambda +\mu \right)
\mathbf{1}\right) ^{-1}\bar{B}_{t}\right\}  \notag \\
&&-4\mathrm{tr}\left\{ \Omega _{t}B_{t}\left( \Omega _{t}^{\mathrm{t}}+\mu
\mathbf{1}\right) ^{-2}\left( \Omega _{t}^{\mathrm{t}}+\left( \lambda +\mu
\right) \mathbf{1}\right) ^{-1}\bar{B}_{t}\right\}  \notag \\
&&-4\mathrm{tr}\left\{ B_{t}\Omega _{t}^{\mathrm{t}}\left( \Omega _{t}^{%
\mathrm{t}}+\mu \mathbf{1}\right) ^{-2}\left( \Omega _{t}^{\mathrm{t}%
}+\left( \lambda +\mu \right) \mathbf{1}\right) ^{-1}\bar{B}_{t}\right\} \ ,
\label{eq-3bisbis-new}
\end{eqnarray}%
because of Corollary \ref{Corollary existence}, and Lemmata \ref{lemma
existence 3 copy(3)} and \ref{lemma existence 3 copy(4)}. By using the
positivity of the self--adjoint operators $\Omega _{t},\Omega _{t}^{\mathrm{t%
}}\geq 0$ and the cyclicity of the trace (Lemma \ref{lemma existence 3
copy(3)}) together with the inequalities (\ref{V.217}), $\left\Vert
Y\right\Vert _{\mathrm{op}}\leq \left\Vert Y\right\Vert _{1}$, $\Omega
_{t}\geq 4\mathfrak{B}_{t}\geq 4\mathfrak{B}_{t,\mu }$ (Lemma \ref%
{existence_flow_5 copy(1)} (ii)), and $\mathfrak{E}_{t,\mu }\leq \mathfrak{E}%
_{t}\leq \mathbf{1}/4$ (Lemma \ref{new lemma 2 copy(1)} (i)), we can then
bound from above the derivative of $\xi _{\mu ,\lambda }$ by%
\begin{equation}
\partial _{t}\xi _{\mu ,\lambda }(t)\leq 4\left( 4\left\Vert \mathfrak{B}%
_{t}\right\Vert _{1}+\mu \right) \xi _{\mu ,\lambda }(t)\leq 4\left(
4\left\Vert \mathfrak{B}_{t}\right\Vert _{1}+\mu \right) \xi _{\mu ,\lambda
}(t)  \label{eq toto nouveau1 correct}
\end{equation}%
for all strictly positive $\mu ,t>0$ and $\lambda \geq 0$. By (\ref{bound
idiote}) and Lemma \ref{existence_flow_5 copy(1)} (i), note that, for any $%
\mu >0$ and $\varepsilon \in (0,1/2)$,
\begin{equation}
\lambda ^{-2\varepsilon }\left\vert \partial _{t}\xi _{\mu ,\lambda }\left(
t\right) \right\vert \leq 4\left( 4\left\Vert \mathfrak{B}_{0}\right\Vert
_{1}+\mu \right) \mu ^{-2}\left\Vert B_{0}\right\Vert _{2}^{2}\lambda
^{-2\varepsilon }\left( \lambda +\mu \right) ^{-1}\ ,
\end{equation}%
which is an integrable upper bound with respect to $\lambda \in (0,\infty )$%
, see (\ref{eq sup}). Therefore, we invoke Lebesgue's dominated convergence
theorem to deduce that
\begin{equation}
\partial _{t}h_{\mu }\left( t\right) =\frac{\sin \left( 2\pi \varepsilon
\right) }{\pi }\int_{0}^{\infty }\lambda ^{-2\varepsilon }\partial _{t}\xi
_{\mu ,\lambda }\left( t\right) \mathrm{d}\lambda \ ,
\end{equation}%
which, combined with (\ref{eq toto nouveau1 correct0}) and (\ref{eq toto
nouveau1 correct}), implies the differential inequality%
\begin{equation}
\forall t>0:\qquad \partial _{t}h_{\mu }\left( t\right) \leq 4\left(
4\left\Vert \mathfrak{B}_{t}\right\Vert _{1}+\mu \right) h_{\mu }\left(
t\right) \ .
\end{equation}%
Using again Gr{ø}nwall's Lemma we thus find the upper bound%
\begin{equation}
\forall t\geq s>0:\qquad h_{\mu }(t)\leq h_{\mu }(s)\exp \left\{
4\int_{s}^{t}\left( 4\left\Vert \mathfrak{B}_{\tau }\right\Vert _{1}+\mu
\right) \mathrm{d}\tau \right\} \ .  \label{eq toto nouveau2bisbisbis}
\end{equation}%
Now, since, by (\ref{eq toto nouveau2bis}),
\begin{eqnarray}
&&\left( \Omega _{t}^{\mathrm{t}}+\mu \mathbf{1}\right) ^{-2-2\varepsilon
}-\left( \Omega _{0}^{\mathrm{t}}+\mu \mathbf{1}\right) ^{-2-2\varepsilon }
\\
&=&\frac{\sin \left( 2\pi \varepsilon \right) }{\pi }\int_{0}^{\infty
}\lambda ^{-2\varepsilon }\frac{1}{\left( \Omega _{t}^{\mathrm{t}}+\left(
\lambda +\mu \right) \mathbf{1}\right) }\left( \Omega _{0}^{\mathrm{t}%
}-\Omega _{t}^{\mathrm{t}}\right) \frac{1}{\left( \Omega _{0}^{\mathrm{t}%
}+\left( \lambda +\mu \right) \mathbf{1}\right) \left( \Omega _{0}^{\mathrm{t%
}}+\mu \mathbf{1}\right) ^{2}}\mathrm{d}\lambda  \notag \\
&&+\left( \Omega _{t}^{\mathrm{t}}+\mu \mathbf{1}\right) ^{-2\varepsilon
}\left( \left( \Omega _{t}^{\mathrm{t}}+\mu \mathbf{1}\right) ^{-2}-\left(
\Omega _{0}^{\mathrm{t}}+\mu \mathbf{1}\right) ^{-2}\right) \ ,  \notag
\end{eqnarray}%
similar to (\ref{continuity1}) or (\ref{continuity2}), one gets%
\begin{equation}
\left\vert h_{\mu }(s)-h_{\mu }(0)\right\vert \leq 3\mu ^{-3-2\varepsilon
}\Vert \Omega _{s}^{\mathrm{t}}-\Omega _{0}^{\mathrm{t}}\Vert _{1}\Vert
B_{0}\Vert _{2}^{2}+2\mu ^{-2-2\varepsilon }\Vert B_{0}\Vert _{2}\Vert
B_{s}-B_{0}\Vert _{2}  \label{continuity3}
\end{equation}%
for all $\mu ,s>0$. By Theorem \ref{lemma existence 3} (i)--(ii), it
demonstrates the continuity of the function $h_{\mu }$ at zero and the
inequality (\ref{eq toto nouveau2bisbisbis}) also holds for $s=0$:
\begin{equation}
\forall t\geq s\geq 0:\qquad h_{\mu }\left( t\right) \leq h_{\mu }(s)\exp
\left\{ 4\int_{s}^{t}\left( 4\left\Vert \mathfrak{B}_{\tau }\right\Vert
_{1}+\mu \right) \mathrm{d}\tau \right\} \ ,  \label{eq toto nouveau3}
\end{equation}%
with $h_{\mu }(0)\leq h_{0}(0)<\infty $, by assumption. Similar to (\ref%
{eq-23-1}) and (\ref{eq-24}), we infer from (\ref{eq toto nouveau3}) and the
monotone convergence theorem that, for all $t\geq 0$,
\begin{equation}
\lim_{\mu \rightarrow 0^{+}}h_{\mu }(t)=\Vert \Omega _{t}^{-1-\varepsilon
}B_{t}\Vert _{2}^{2}\leq h_{0}(0)\exp \left\{ 16\int_{0}^{t}\left\Vert
\mathfrak{B}_{\tau }\right\Vert _{1}\mathrm{d}\tau \right\} <\infty \ .
\end{equation}%
Then, the assertion of this lemma follows by passing to the limit $\mu
\rightarrow 0^{+}$ in (\ref{eq toto nouveau3}). \hfill $\Box $

Since
\begin{equation}
\int_{0}^{1}\left\Vert B_{t}\right\Vert _{2}\mathrm{d}t\leq \left\Vert
B_{0}\right\Vert _{2}  \label{contradiction0}
\end{equation}%
(see (\ref{a priori estimate}) and Theorem \ref{section extension copy(4)}),
by (\ref{inequality avec eps}) for $\varepsilon \in (0,1/2)$ and Lemma \ref%
{lemma integrability copy(6)} combined with Corollary \ref{lemma
integrability copy(3)}, we directly deduce the integrability of the map $%
t\mapsto \Vert B_{t}\Vert _{2}$ on the positive real line $\left[ 0,\infty
\right) $:

\begin{corollary}[Integrability of $\Vert B_{t}\Vert _{2}$]
\label{lemma integrability}\mbox{ }\newline
Assume Conditions A1--A3, $\Omega _{0}^{-1-\varepsilon }B_{0}\in \mathcal{L}%
^{2}(\mathfrak{h})$, and $\mathbf{1}\geq \left( 4+\mathrm{r}\right)
B_{0}(\Omega _{0}^{\mathrm{t}})^{-2}\bar{B}_{0}$ for some $\mathrm{r}%
,\varepsilon >0$. Then, for any $\delta \in (0,1+\varepsilon )\cap (0,3/2)$
and as $t\rightarrow \infty $, $\Vert B_{t}\Vert _{2}=o\left( t^{-\delta
}\right) $. In particular, the map $t\mapsto \Vert B_{t}\Vert _{2}$ is
integrable on $\left[ 0,\infty \right) $.
\end{corollary}

Of course, the larger the positive parameter $\varepsilon >0$, the better
the asymptotics of the map $t\mapsto \Vert B_{t}\Vert _{2}$. As an example,
we give the study for $\varepsilon =1/2$ but one should be able to use,
recursively, all the arguments of this section to find $\Vert B_{t}\Vert
_{2}=o\left( t^{-\delta }\right) $ for any $\delta \in (0,1+\varepsilon )$,
provided $\Omega _{0}^{-1-\varepsilon }B_{0}\in \mathcal{L}^{2}(\mathfrak{h}%
) $ with $\varepsilon >0$.

\begin{lemma}[Conservation by the flow of $\Omega _{0}^{-3/2}B_{0}\in
\mathcal{L}^{2}(\mathfrak{h})$]
\label{lemma integrability copy(4)}\mbox{ }\newline
Assume Conditions A1--A3, $\Omega _{0}^{-3/2}B_{0}\in \mathcal{L}^{2}(%
\mathfrak{h})$, and $\mathbf{1}\geq \left( 4+\mathrm{r}\right) B_{0}(\Omega
_{0}^{\mathrm{t}})^{-2}\bar{B}_{0}$ for some $\mathrm{r}>0$. Then the
operator $\mathfrak{S}_{t}:=B_{t}(\Omega _{t}^{\mathrm{t}})^{-3}\bar{B}%
_{t}\in \mathcal{L}^{1}(\mathfrak{h})$ is trace--class:
\begin{equation}
\forall t\geq s\geq 0:\qquad \Vert \mathfrak{S}_{t}\Vert _{1}\leq \Vert
\mathfrak{S}_{s}\Vert _{1}\exp \left\{ 16\int_{s}^{t}\Vert \mathfrak{B}%
_{\tau }\Vert _{1}\mathrm{d}\tau \right\} \ .
\end{equation}
\end{lemma}

\noindent \textbf{Proof.} Let $\mu >0$ and set%
\begin{equation}
\forall t\geq 0:\qquad \mathfrak{S}_{t,\mu }:=B_{t}\left( \Omega _{t}^{%
\mathrm{t}}+\mu \right) ^{-3}\bar{B}_{t}\geq 0\ .
\end{equation}%
Similar to (\ref{eq-2}) or (\ref{eq-2bis}), $\mathfrak{S}_{t,\mu }$ is a
bounded operator as $\mu >0$ and $\Omega _{t}^{\mathrm{t}}\geq 0$, see Lemma %
\ref{new lemma 2 copy(2)}. Via Corollary \ref{Corollary existence} and Lemma %
\ref{lemma existence 3 copy(4)} the function $g_{\mu }$ defined by
\begin{equation}
\forall t\geq 0:\qquad g_{\mu }(t):=\mathrm{tr}\left\{ \mathfrak{S}_{t,\mu
}\right\} =\Vert \mathfrak{S}_{t,\mu }\Vert _{1}=\xi _{\mu ,0}(t)
\end{equation}%
satisfies the differential inequality%
\begin{equation}
\forall t>0:\qquad \partial _{t}g_{\mu }(t)\leq 4\left( 4\left\Vert
\mathfrak{B}_{t}\right\Vert _{1}+\mu \right) \ g_{\mu }(t)\ ,
\end{equation}%
see (\ref{eq-3bisbis-new})--(\ref{eq toto nouveau1 correct}) for $\lambda =0$%
. So, using as above Gr{ø}nwall's Lemma, it means that%
\begin{equation}
\forall t\geq s>0:\qquad g_{\mu }(t)\leq g_{\mu }(s)\exp \left\{
4\int_{s}^{t}\left( 4\Vert \mathfrak{B}_{\tau }\Vert _{1}+\mu \right)
\mathrm{d}\tau \right\} \ .  \label{eq totosup1-new}
\end{equation}%
An inequality similar to (\ref{continuity1}) or (\ref{continuity2}) or (\ref%
{continuity3}) shows the continuity of the function $g_{\mu }$ at zero and (%
\ref{eq totosup1-new}) can be extended by continuity to $s=0$. Similar to (%
\ref{eq-23-1}) and (\ref{eq-24}), we infer from (\ref{eq totosup1-new})
together with the monotone convergence theorem that, for all $t\geq s\geq 0$%
,
\begin{equation}
\lim_{\mu \rightarrow 0^{+}}\Vert \mathfrak{S}_{t,\mu }\Vert _{1}=\Vert
\mathfrak{S}_{t}\Vert _{1}<\infty \ ,  \label{eq totosup1-newbis}
\end{equation}%
with $\mathfrak{S}_{t}\equiv \mathfrak{S}_{t,0}$. The lemma thus follows
from (\ref{eq totosup1-new}) and (\ref{eq totosup1-newbis}). \hfill $\Box $

Therefore, using the upper bound of Corollary \ref{Corollary existence
copy(1)} for $\alpha =3/2$ together with Lemma \ref{lemma integrability
copy(4)} one gets, for any $t\geq 1$ and $\delta \in (0,3/2)$, that%
\begin{equation}
\Vert B_{t}\Vert _{2}=o(t^{-\delta }),\qquad \mathrm{as\ }t\rightarrow
\infty \ .
\end{equation}%
In fact, in the same way the asymptotics $\Vert B_{t}\Vert _{2}=o(t^{-1})$\
is shown in Corollary \ref{lemma integrability copy(3)}, one can verify
under the assumptions of Lemma \ref{lemma integrability copy(4)} that%
\begin{equation}
\Vert B_{t}\Vert _{2}=o(t^{-3/2}),\qquad \mathrm{as\ }t\rightarrow \infty \ .
\end{equation}

\section{Technical Proofs on the Boson Fock Space\label{Section technical
proofs copy(1)}}

\setcounter{equation}{0}%
Here we give the following proofs: the proof of Theorem \ref{theorem
important 2bis} in Section \ref{Section proof theroem important 2bis}, the
proof of Theorem \ref{theorem Ht=unitary orbite} in Section \ref{section
unitarly equi H}, the proof of Theorem \ref{theorem important 4bis} in
Section \ref{Section theorem important 4bis}, the proof of Theorem \ref%
{theorem important 3} in Section \ref{section proof thm important 3}. All
these proofs are broken up into several lemmata, which sometime yield
information beyond the contents of the above theorems.

Additionally to A1--A2, we always assume without loss of generality
Conditions A3--A4. These assumptions are however not necessary in Sections %
\ref{Section proof theroem important 2bis}--\ref{section unitarly equi H}
and only used for convenience. Indeed, A3--A4 ensure the existence of the
family $(\Omega _{t},B_{t})_{t\geq 0}$ solving (\ref{flow equation-quadratic}%
) for all times as well as the positivity of $\Omega _{t}\geq 0$ (i.e., $%
T_{+}=\infty $), which implies
\begin{equation}
\forall t\geq 0:\qquad \Vert B_{t}\Vert _{2}\leq \Vert B_{0}\Vert _{2}\ .
\label{a priori estimatebis}
\end{equation}%
See (\ref{a priori estimate}) and Theorem \ref{section extension copy(4)}.
If only A4 is not satisfied then $t\in \left[ 0,\infty \right) $ is replaced
by $t\in \lbrack 0,T_{+})$ in Sections \ref{Section proof theroem important
2bis}--\ref{section unitarly equi H}, see (\ref{temps xi})--(\ref{temps
xibis}). If A3--A4 do not hold then in arguments of Sections \ref{Section
proof theroem important 2bis}--\ref{section unitarly equi H} one replaces $%
t\in \left[ 0,\infty \right) $ and (\ref{a priori estimatebis}) by
respectively $\left[ 0,T\right] $ and%
\begin{equation}
\forall t\in \left[ 0,T\right] :\qquad \Vert B_{t}\Vert _{2}\leq \mathrm{e}%
^{8rT}\Vert B_{0}\Vert _{2}\ ,
\end{equation}%
for any $T\in (0,T_{\max })$ and some $r\equiv r_{T}>0$. See (\ref%
{B_explicit_solution}), (\ref{upper bound for W}), and (\ref{Tmax})--(\ref%
{Tmaxbis}).

\subsection{Existence and Uniqueness of the Unitary Propagator\label{Section
proof theroem important 2bis}}

To prove the existence of the unitary propagator $\mathrm{U}_{t,s}$ solution
of (\ref{flow equationbis}) with infinitesimal generator $\mathrm{G}_{t}$ (%
\ref{generator quadraticbis}), we need to verify three conditions, namely
B1, B2, and B3, see Section \ref{Section Non-autonomous evolution}.
Condition B1 is directly satisfied because, by Proposition \ref{lemma
example 1 copy(1)}, our generator $\mathrm{G}_{t}$ is self--adjoint, see\ (%
\ref{stability condition in Hilbert space}). To verify Conditions B2--B3, a
closed auxiliary operator $\Theta $ has to be fixed. The simplest choice is
to take the particle number operator $\mathrm{N}$ (\ref{particle number
operator}) plus the identity $\mathbf{1}$, i.e., $\Theta =\mathrm{N}+\mathbf{%
1}$. Then, establishing B2 and B3 amounts to prove that the relative norms $%
\Vert \mathrm{G}_{t}(\mathrm{N}+\mathbf{1})^{-1}\Vert _{\mathrm{op}}$ and $%
\Vert \lbrack \mathrm{N},\mathrm{G}_{t}](\mathrm{N}+\mathbf{1})^{-1}\Vert _{%
\mathrm{op}}$ are uniformly bounded for all times $t\geq 0$, with the
function $\Vert \mathrm{G}_{t}(\mathrm{N}+\mathbf{1})^{-1}\Vert _{\mathrm{op}%
}$ being continuous on $[0,\infty )$. We thus infer from Theorem \ref{flow
equation thm 5} the existence and uniqueness of a unitary propagator $%
\mathrm{U}_{t,s}$ solution, in the strong topology on the domain $\mathcal{D}%
\left( \mathrm{N}\right) $, of the non--autonomous evolution equation (\ref%
{flow equationbis}) with infinitesimal generator $G_{t}=\mathrm{G}_{t}$ (\ref%
{generator quadraticbis}).

We start by giving some general estimates used many times in our paper.

\begin{lemma}[Relative norms w.r.t $\mathrm{N}$--diagonal operators]
\label{lemma estimates}\mbox{ }\newline
Let $\theta =\theta ^{\ast }\geq 0$ be a positive, invertible operator on $%
\mathfrak{h}$ and $X,Y=Y^{\mathrm{t}}\in \mathcal{L}^{2}\left( \mathfrak{h}%
\right) $ with respective second quantizations%
\begin{eqnarray}
\Theta &:=&\mathbf{1+}\underset{k,\ell }{\sum }\left\{ \theta \right\}
_{k,\ell }a_{k}^{\ast }a_{\ell }\ , \\
\mathbb{X} &:=&\sum_{k,\ell }\left\{ X\right\} _{k,\ell }a_{k}^{\ast
}a_{\ell }\ , \\
\mathbb{Y} &:=&\sum_{k,\ell }\left\{ Y\right\} _{k,\ell }a_{k}a_{\ell }\ .
\end{eqnarray}%
Then the norms of $\mathbb{X}$, $\mathbb{Y}$, and $\mathbb{Y}^{\ast }$
relative to $\Theta $ are bounded by:
\begin{eqnarray}
\Vert \mathbb{X}\Theta ^{-1}\Vert _{\mathrm{op}} &\leq &\Vert \theta
^{-1/2}X^{\ast }X\theta ^{-1/2}\Vert _{2}^{1/2}+\Vert \theta ^{-1/2}X\theta
^{-1/2}\Vert _{2}\ ,  \label{norm1} \\
\Vert \mathbb{Y}\Theta ^{-1}\Vert _{\mathrm{op}} &\leq &\Vert \theta
^{-1/2}Y(\theta ^{\mathrm{t}})^{-1/2}\Vert _{2}\ ,  \label{norm2} \\
\Vert \mathbb{Y}^{\ast }\Theta ^{-1}\Vert _{\mathrm{op}} &\leq &\Vert \theta
^{-1/2}Y(\theta ^{\mathrm{t}})^{-1/2}\Vert _{2}+2\Vert \theta ^{-1/2}Y^{\ast
}Y\theta ^{-1/2}\Vert _{2}^{1/2}  \notag \\
&&+\sqrt{2}\Vert Y\Vert _{2}\ .  \label{norm3}
\end{eqnarray}
\end{lemma}

\noindent \textbf{Proof.} By the spectral theorem, there is a representation
of $\mathfrak{h}$ as $L^{2}(\mathcal{A},\mathrm{d}\mathfrak{a})$, a suitable
probability measure $\mathfrak{a}$ on a set $\mathcal{A}$ such that $\theta $
is a multiplication operator, i.e., there is a Borel--measurable function $%
\vartheta :\mathcal{A}\rightarrow \mathbb{R}_{0}^{+}$ such that
\begin{equation*}
\forall \varphi \in L^{2}(\mathcal{A},\mathrm{d}\mathfrak{a}),\ x\in
\mathcal{A}:\qquad \left( \theta \varphi \right) \left( x\right) =\vartheta
_{x}\varphi \left( x\right) \ .
\end{equation*}%
Since all norms in (\ref{norm1})--(\ref{norm3}) are unitarily invariant (up
to the appearance of the transpose $\theta ^{\mathrm{t}}$ of $\theta $ in (%
\ref{norm2}) and (\ref{norm3})), we may assume without loss of generality
that (we write $\mathrm{d}\mathfrak{a}_{x}\mathrm{d}\mathfrak{a}_{y}=:%
\mathrm{d}^{2}\mathfrak{a}_{x,y}$ and $\mathrm{d}\mathfrak{a}_{x}\mathrm{d}%
\mathfrak{a}_{y}\mathrm{d}\mathfrak{a}_{w}\mathrm{d}\mathfrak{a}_{z}=:%
\mathrm{d}^{4}\mathfrak{a}_{x,y,w,z}$)%
\begin{eqnarray}
\Theta &:=&\mathbf{1+}\int \vartheta _{x}a_{x}^{\ast }a_{x}\mathrm{d}%
\mathfrak{a}_{x}\ , \\
\mathbb{X} &:=&\int X_{x,y}a_{x}^{\ast }a_{y}\mathrm{d}^{2}\mathfrak{a}%
_{x,y}\ , \\
\mathbb{Y} &:=&\int Y_{x,y}a_{x}a_{y}\mathrm{d}^{2}\mathfrak{a}_{x,y}\ ,
\end{eqnarray}%
with $Y^{\mathrm{t}}=Y$. Now, for any $\varphi \in \mathcal{F}_{b}$, we
estimate that
\begin{eqnarray}
\left\Vert \mathbb{Y}\varphi \right\Vert &\leq &\int \left\vert
Y_{x,y}\right\vert \left\Vert a_{x}a_{y}\varphi \right\Vert \mathrm{d}^{2}%
\mathfrak{a}_{x,y}  \notag \\
&\leq &\left( \int \frac{\left\vert Y_{x,y}\right\vert ^{2}}{\vartheta
_{x}\vartheta _{y}}\mathrm{d}^{2}\mathfrak{a}_{x,y}\right) ^{1/2}\left( \int
\vartheta _{x}\vartheta _{y}\left\Vert a_{x}a_{y}\varphi \right\Vert ^{2}%
\mathrm{d}^{2}\mathfrak{a}_{x,y}\right) ^{1/2}  \notag \\
&=&\Vert \theta ^{-1/2}Y\theta ^{-1/2}\Vert _{2}\left( \int \vartheta
_{y}\Vert \left( \Theta -\mathbf{1}\right) ^{1/2}a_{y}\varphi \Vert ^{2}%
\mathrm{d}\mathfrak{a}_{y}\right) ^{1/2}  \notag \\
&\leq &\Vert \theta ^{-1/2}Y\theta ^{-1/2}\Vert _{2}\ \Vert \left( \Theta -%
\mathbf{1}\right) \varphi \Vert \ ,
\end{eqnarray}%
which proves (\ref{norm2}) for $\theta ^{\mathrm{t}}=\theta $. Next, we
observe that, by the Canonical Commutations Relations (CCR),%
\begin{eqnarray}
\Vert \mathbb{X}\varphi \Vert ^{2} &=&\int \overline{X_{x,y}}X_{w,z}\langle
a_{y}\varphi |a_{x}a_{w}^{\ast }a_{z}\varphi \rangle \mathrm{d}^{4}\mathfrak{%
a}_{x,y,w,z}  \notag \\
&=&\int \overline{X_{x,y}}X_{w,z}\langle a_{w}a_{y}\varphi
|a_{x}a_{z}\varphi \rangle \mathrm{d}^{4}\mathfrak{a}_{x,y,w,z}  \notag \\
&&+\int \left( X^{\ast }X\right) _{y,z}\langle a_{y}\varphi |a_{z}\varphi
\rangle \mathrm{d}^{2}\mathfrak{a}_{y,z}  \notag \\
&\leq &\left( \int \frac{|X_{x,y}|^{2}}{\vartheta _{x}\vartheta _{y}}\mathrm{%
d}^{2}\mathfrak{a}_{x,y}\right) \left( \int \vartheta _{x}\vartheta
_{y}\left\Vert a_{x}a_{y}\varphi \right\Vert ^{2}\mathrm{d}^{2}\mathfrak{a}%
_{x,y}\right)  \notag \\
&&+\left( \int \frac{|\left( X^{\ast }X\right) _{y,z}|^{2}}{\vartheta
_{y}\vartheta _{z}}\mathrm{d}^{2}\mathfrak{a}_{y,z}\right) ^{1/2}\left( \int
\vartheta _{y}\left\Vert a_{y}\varphi \right\Vert ^{2}\mathrm{d}\mathfrak{a}%
_{y}\right)  \notag \\
&\leq &\Vert \theta ^{-1/2}X\theta ^{-1/2}\Vert _{2}^{2}\ \Vert \left(
\Theta -\mathbf{1}\right) \varphi \Vert ^{2}  \notag \\
&&+\Vert \theta ^{-1/2}X^{\ast }X\theta ^{-1/2}\Vert _{2}\ \Vert \left(
\Theta -\mathbf{1}\right) ^{1/2}\varphi \Vert ^{2}\ ,
\end{eqnarray}%
from which (\ref{norm1}) is immediate. Finally, the CCR imply that%
\begin{eqnarray}
\left\Vert \mathbb{Y}^{\ast }\varphi \right\Vert ^{2} &=&\int Y_{x,y}%
\overline{Y_{w,z}}\langle \varphi |a_{x}a_{y}a_{w}^{\ast }a_{z}^{\ast
}\varphi \rangle \mathrm{d}^{4}\mathfrak{a}_{x,y,w,z}  \notag \\
&=&\int Y_{x,y}\overline{Y_{w,z}}\langle a_{z}a_{w}\varphi
|a_{x}a_{y}\varphi \rangle \mathrm{d}^{4}\mathfrak{a}_{x,y,w,z}  \notag \\
&&+\int \left( Y^{\ast }Y^{\mathrm{t}}+Y^{\ast }Y+\overline{Y}Y^{\mathrm{t}}+%
\overline{Y}Y\right) _{x,y}\langle a_{x}\varphi |a_{y}\varphi \rangle
\mathrm{d}^{2}\mathfrak{a}_{x,y}  \notag \\
&&+\left( \mathrm{tr}\left( \overline{Y}Y\right) +\mathrm{tr}\left( Y^{\ast
}Y\right) \right) \Vert \varphi \Vert ^{2}  \notag \\
&\leq &\Vert \theta ^{-1/2}Y\theta ^{-1/2}\Vert _{2}^{2}\ \Vert \left(
\Theta -\mathbf{1}\right) \varphi \Vert ^{2}  \notag \\
&&+4\Vert \theta ^{-1/2}Y^{\ast }Y\theta ^{-1/2}\Vert _{2}\ \Vert \left(
\Theta -\mathbf{1}\right) ^{1/2}\varphi \Vert ^{2}  \notag \\
&&+2\Vert Y\Vert _{2}^{2}\ \Vert \varphi \Vert ^{2},
\end{eqnarray}%
which yields (\ref{norm3}) for $\theta ^{\mathrm{t}}=\theta $. \hfill $\Box $

We analyze now the relative norm estimates related to the generator $\mathrm{%
G}_{t}$ introduced in (\ref{generator quadraticbis}) with respect to the
self--adjoint, invertible operator $\Theta =\mathrm{N}+\mathbf{1}$. In the
following lemmata, we assume Conditions A1--A4 defined in\ Sections \ref%
{Section bosonic quadratic operators}--\ref{Section main result}, which
ensure the existence of operators $\Omega _{t}$ and $B_{t}$ solution of the
system (\ref{flow equation-quadratic}) of differential equations for all
times, see Theorem \ref{theorem important 1 -1}. See also the introduction
of Section \ref{Section technical proofs copy(1)} about A1--A4.

\begin{lemma}[Verification of Conditions B2--B3 with $\Theta =\mathrm{N}+%
\mathbf{1}$]
\label{lemma unitary existence 1}\mbox{ }\newline
Assume Conditions A1--A4 and let
\begin{equation}
\mathcal{Y}:=\mathcal{D}\left( \mathrm{N}\right) =\left\{ \varphi \in
\mathcal{F}_{b}\
\Big |%
\ \left\Vert \varphi \right\Vert _{\mathcal{Y}}:=\left\Vert \left( \mathrm{N}%
+\mathbf{1}\right) \varphi \right\Vert <\infty \right\} \ .
\end{equation}%
Then, for any $t\geq 0$, $\mathcal{Y}\subset \mathcal{D}\left( \mathrm{G}%
_{t}\right) \ $with $\mathrm{G}_{t}\in C[\mathbb{R}_{0}^{+};\mathcal{B}(%
\mathcal{Y},\mathcal{F}_{b})]$, i.e., Condition B2 holds true. Furthermore,
the norm $\left\Vert \left[ \mathrm{N},\mathrm{G}_{t}\right] \right\Vert _{%
\mathcal{B}\left( \mathcal{Y}\right) }$ is bounded on $[0,\infty )$ by%
\begin{equation}
\Vert \left[ \mathrm{N},\mathrm{G}_{t}\right] \Vert _{\mathcal{B}\left(
\mathcal{Y}\right) }:=\Vert \left[ \mathrm{N},\mathrm{G}_{t}\right] \left(
\mathrm{N}+\mathbf{1}\right) ^{-1}\Vert _{\mathrm{op}}\leq 22\left\Vert
B_{t}\right\Vert _{2}\leq 22\left\Vert B_{0}\right\Vert _{2}\ ,
\label{condition 3bis}
\end{equation}%
i.e., Condition B3 holds true.
\end{lemma}

\noindent \textbf{Proof.} First, we analyze the behavior of the function $%
\Vert \mathrm{G}_{t}(\mathrm{N}+\mathbf{1})^{-1}\Vert _{\mathrm{op}}$. From (%
\ref{generator quadraticbis}) observe that $\mathrm{G}_{t}=2i(\mathbb{B}%
_{t}^{\ast }\mathbb{-B}_{t})$ with
\begin{equation}
\mathbb{B}_{t}:=\sum_{k,\ell }\left\{ \bar{B}_{t}\right\} _{k,\ell
}a_{k}a_{\ell }\ .  \label{definition de blackboard Bt}
\end{equation}%
Hence, the relative norm of $\mathrm{G}_{t}$ with respect to $\Theta =%
\mathrm{N}+\mathbf{1}$ is bounded by
\begin{equation}
\Vert \mathrm{G}_{t}\left( \mathrm{N}+\mathbf{1}\right) ^{-1}\Vert _{\mathrm{%
op}}\leq 2\Vert \mathbb{B}_{t}\left( \mathrm{N}+\mathbf{1}\right) ^{-1}\Vert
_{\mathrm{op}}+2\Vert \mathbb{B}_{t}^{\ast }\left( \mathrm{N}+\mathbf{1}%
\right) ^{-1}\Vert _{\mathrm{op}}\ .  \label{inegality BB 0}
\end{equation}%
By using Lemma \ref{lemma estimates} with $\theta =\mathbf{1}$ and $Y=\bar{B}%
_{t}$, the operators $\mathbb{B}_{t}\left( \mathrm{N}+\mathbf{1}\right)
^{-1} $ and $\mathbb{B}_{t}^{\ast }\left( \mathrm{N}+\mathbf{1}\right) ^{-1}$
are bounded respectively by
\begin{equation}
\Vert \mathbb{B}_{t}\left( \mathrm{N}+\mathbf{1}\right) ^{-1}\Vert _{\mathrm{%
op}}\leq \Vert B_{t}\Vert _{2}\quad \mathrm{and}\quad \Vert \mathbb{B}%
_{t}^{\ast }\left( \mathrm{N}+\mathbf{1}\right) ^{-1}\Vert _{\mathrm{op}%
}\leq (3+\sqrt{2})\Vert B_{t}\Vert _{2}\ .  \label{inegality BB 02}
\end{equation}%
Combined with (\ref{a priori estimatebis}) and the inequalities of (\ref%
{inegality BB 02}), the upper bound (\ref{inegality BB 0}) implies, for any $%
t\geq 0$, that
\begin{equation}
\Vert \mathrm{G}_{t}\left( \mathrm{N}+\mathbf{1}\right) ^{-1}\Vert _{\mathrm{%
op}}\leq 11\Vert B_{t}\Vert _{2}\leq 11\Vert B_{0}\Vert _{2}\ .
\label{inegality BB 3}
\end{equation}%
(We use $(8+2\sqrt{2})<11$.) Additionally, by substituting $(B_{t}-B_{s})$
for the operator $B_{t}$ in the last inequality (\ref{inegality BB 3}) we
also obtain, for all $t,s\geq 0$, that%
\begin{equation}
\Vert \left\{ \mathrm{G}_{t}-\mathrm{G}_{s}\right\} \left( \mathrm{N}+%
\mathbf{1}\right) ^{-1}\Vert _{\mathrm{op}}\leq 11\Vert B_{t}-B_{s}\Vert
_{2}\ .  \label{continuity generator}
\end{equation}%
By Theorem \ref{theorem important 1 -1} (ii), the operator $B_{t}$ is
continuous in the Hilbert--Schmidt topology for any $t\geq 0$. Consequently $%
\mathrm{G}_{t}\in C[\mathbb{R}_{0}^{+};\mathcal{B}(\mathcal{Y},\mathcal{F}%
_{b})]$ with $\mathcal{Y}=\mathcal{D}\left( \mathrm{N}\right) $, and
Condition B2 is therefore verified.

Finally, straightforward computations using the CCR show that the commutator
\begin{equation}
\frac{1}{i}\left[ \mathrm{N},\mathrm{G}_{t}\right] =4\underset{k,\ell }{\sum
}\left\{ B_{t}\right\} _{k,\ell }a_{k}^{\ast }a_{\ell }^{\ast }+\left\{ \bar{%
B}_{t}\right\} _{k,\ell }a_{k}a_{\ell }=4\left( \mathbb{B}_{t}^{\ast }%
\mathbb{+B}_{t}\right)  \label{comutator of G et N}
\end{equation}%
is also a well--defined self--adjoint quadratic operator (Proposition \ref%
{lemma example 1 copy(1)}). Therefore, we can substitute $2B_{t}$ for $B_{t}$
in (\ref{inegality BB 3}) to get, for all $t\geq 0$, that
\begin{equation}
\Vert \left[ \mathrm{N},\mathrm{G}_{t}\right] \left( \mathrm{N}+\mathbf{1}%
\right) ^{-1}\Vert _{\mathrm{op}}\leq 22\Vert B_{t}\Vert _{2}\leq 22\Vert
B_{0}\Vert _{2}\ ,  \label{condition 3bisbis}
\end{equation}%
i.e., Condition B3.\hfill $\Box $

The above lemma concludes the preliminary steps necessary to apply Theorem %
\ref{flow equation thm 5}.

\begin{lemma}[Existence and uniqueness of $\mathrm{U}_{t,s}$]
\label{Section exist unique U quadratic}\mbox{ }\newline
Under Conditions A1--A4, there is a unique, bounded evolution operator $(%
\mathrm{U}_{t,s})_{t\geq s\geq 0}$ satisfying on the domain $\mathcal{D}%
\left( \mathrm{N}\right) $ the non--autonomous evolutions%
\begin{equation}
\forall t\geq s\geq 0:\qquad \left\{
\begin{array}{llll}
\partial _{t}\mathrm{U}_{t,s}=-i\mathrm{G}_{t}\mathrm{U}_{t,s} & , & \mathrm{%
U}_{s,s}:=\mathbf{1} & . \\
\partial _{s}\mathrm{U}_{t,s}=i\mathrm{U}_{t,s}\mathrm{G}_{s} & , & \mathrm{U%
}_{t,t}:=\mathbf{1} & .%
\end{array}%
\right.  \label{flow equationbis-new}
\end{equation}%
Moreover, $\mathrm{U}_{t,s}$ conserves the domains $\mathcal{D}\left(
\mathrm{N}\right) $ and $\mathcal{D}\left( \mathrm{N}^{2}\right) $ for all $%
t\geq s\geq 0$ as%
\begin{eqnarray}
\Vert (\mathrm{N}+\mathbf{1})\mathrm{U}_{t,s}(\mathrm{N}+\mathbf{1}%
)^{-1}\Vert _{\mathrm{op}} &\leq &\mathrm{\exp }\left\{ 22\int_{s}^{t}\Vert
B_{\tau }\Vert _{2}\mathrm{d}\tau \right\} <\infty \ ,
\label{conservation of domain N by U} \\
\Vert (\mathrm{N}+\mathbf{1)}^{2}\mathrm{U}_{t,s}(\mathrm{N}+\mathbf{1)}%
^{-2}\Vert _{\mathrm{op}} &\leq &\mathrm{\exp }\left\{ 132\int_{s}^{t}\Vert
B_{\tau }\Vert _{2}\mathrm{d}\tau \right\} <\infty \ .
\label{conservation of domain N2 by U}
\end{eqnarray}%
The bounded operator family%
\begin{equation}
\left\{ (\mathrm{N}+\mathbf{1})\mathrm{U}_{t,s}(\mathrm{N}+\mathbf{1}%
)^{-1}\right\} _{t\geq s\geq 0}  \label{bounded operator family}
\end{equation}%
is also jointly strongly continuous in $s$ and $t$.
\end{lemma}

\noindent \textbf{Proof.} From Lemma \ref{lemma unitary existence 1},
Conditions B2--B3 are satisfied with $\mathcal{Y}=\mathcal{D}\left( \mathrm{N%
}\right) $, whereas Condition B1 with $m=1$ and $\beta _{0}\left( t\right)
=0 $ is a direct consequence of the self--adjointness of $\mathrm{G}_{t}$,
see Proposition \ref{lemma example 1 copy(1)} and (\ref{stability condition
in Hilbert space}). Therefore, we can apply Theorem \ref{flow equation thm 5}
with the closed operator $\Theta $ being the unbounded self--adjoint
operator $(\mathrm{N}+\mathbf{1})$ and the Banach space $\mathcal{X}$ being
the Hilbert space $\mathcal{F}_{b}$. In this specific case, the bounded
operator $\mathrm{U}_{\lambda ,t,s}$ has%
\begin{equation}
G_{t,\lambda }=\mathrm{G}_{t,\lambda }:=\frac{\lambda \mathrm{G}_{t}}{%
\lambda \mathbf{1}+i\mathrm{G}_{t}}\in \mathcal{B}\left( \mathcal{F}%
_{b}\right)  \label{almost unitary generator}
\end{equation}%
as infinitesimal generator, which is the Yosida approximation of the
unbounded self--adjoint operator $\mathrm{G}_{t}$, see (\ref{dyson series 0}%
)--(\ref{dyson series}). This evolution operator $\mathrm{U}_{\lambda ,t,s}$
converges, as $\lambda \rightarrow \infty $, to an evolution operator $%
\mathrm{U}_{t,s}$ in the strong sense on the boson Fock space $\mathcal{F}%
_{b}$ with $\Vert \mathrm{U}_{t,s}\Vert \leq 1$. This strong convergence on $%
\mathcal{F}_{b}$ becomes a norm convergence on the dense domain $\mathcal{D}%
\left( \mathrm{N}\right) \subset \mathcal{F}_{b}$ since, by Lemma \ref%
{existence flow with W 4},%
\begin{equation}
\underset{\lambda \rightarrow \infty }{\lim }\Vert \{\mathrm{U}_{\lambda
,t,s}-\mathrm{U}_{t,s}\}(\mathrm{N}+\mathbf{1})^{-1}\Vert _{\mathrm{op}}=0
\label{norm convegence of unitary propagators}
\end{equation}%
for all $t\geq s\geq 0$. By using Lemma \ref{existence flow with W 4bis}
with $m=1$, $\beta _{0}\left( t\right) =0$, and
\begin{equation}
\gamma _{1}\left( t\right) :=\Vert \left[ \mathrm{N},\mathrm{G}_{t}\right] (%
\mathrm{N}+\mathbf{1})^{-1}\Vert _{\mathrm{op}}\leq 22\Vert B_{t}\Vert
_{2}\leq 22\Vert B_{0}\Vert _{2}  \label{gamma t specific case}
\end{equation}%
(see (\ref{condition 3bis})), the bounded operator $\mathrm{U}_{t,s}$ also
satisfies Inequality (\ref{conservation of domain N by U}), i.e., it
conserves the domain $\mathcal{D}\left( \mathrm{N}\right) $. Furthermore,
for all $t\geq s\geq 0$, the evolution operator $\mathrm{U}_{t,s}$ is a
solution in the strong topology on $\mathcal{D}\left( \mathrm{N}\right) $ of
the non--autonomous evolution equation (\ref{flow equationbis-new}), by
Theorem \ref{flow equation thm 5}. The strong continuity of the bounded
operator family (\ref{bounded operator family}) results from\ Lemma \ref%
{existence flow with W 4bis}.

Now, we prove (\ref{conservation of domain N2 by U}), which means that $%
\mathrm{U}_{t,s}$ conserves the domain $\mathcal{D}\left( \mathrm{N}%
^{2}\right) $. The latter can be seen by using again Lemma \ref{existence
flow with W 4bis} provided we show that%
\begin{equation}
\lbrack (\mathrm{N}+\mathbf{1)}^{2},\mathrm{G}_{t}](\mathrm{N}+\mathbf{1)}%
^{-2}\in \mathcal{B}\left( \mathcal{F}_{b}\right) \ .
\end{equation}%
Note that
\begin{equation}
\lbrack (\mathrm{N}+\mathbf{1)}^{2},\mathrm{G}_{t}](\mathrm{N}+\mathbf{1)}%
^{-2}=[\mathrm{N}^{2},\mathrm{G}_{t}](\mathrm{N}+\mathbf{1)}^{-2}+2[\mathrm{N%
},\mathrm{G}_{t}](\mathrm{N}+\mathbf{1)}^{-2}\ .  \label{sup000}
\end{equation}%
In view of (\ref{gamma t specific case}), it suffices to bound the norm
\begin{equation}
\Vert \lbrack \mathrm{N}^{2},\mathrm{G}_{t}](\mathrm{N}+\mathbf{1)}%
^{-2}\Vert _{\mathrm{op}}
\end{equation}%
and we shall use the equality%
\begin{equation}
\lbrack \mathrm{N}^{2},\mathrm{G}_{t}](\mathrm{N}+\mathbf{1)}^{-2}=\mathrm{N}%
[\mathrm{N},\mathrm{G}_{t}](\mathrm{N}+\mathbf{1)}^{-2}+[\mathrm{N},\mathrm{G%
}_{t}]\mathrm{N}(\mathrm{N}+\mathbf{1)}^{-2}\ .  \label{sup00}
\end{equation}%
The commutator equality%
\begin{equation}
\left[ XY,Z\right] =X\left[ Y,Z\right] +\left[ X,Z\right] Y
\end{equation}%
for any unbounded operators $X,Y,Z$ only holds, a priori, on a common core
for the three operators in the equality. It turns out that both operators in
the r.h.s of (\ref{sup00}) are, for all $t\geq 0$, bounded respectively by
\begin{equation}
\Vert \lbrack \mathrm{N},\mathrm{G}_{t}]\mathrm{N}(\mathrm{N}+\mathbf{1)}%
^{-2}\Vert _{\mathrm{op}}\leq \Vert \lbrack \mathrm{N},\mathrm{G}_{t}](%
\mathrm{N}+\mathbf{1)}^{-1}\Vert _{\mathrm{op}}\leq 22\Vert B_{t}\Vert _{2}
\label{sup0}
\end{equation}%
and
\begin{equation}
\Vert \mathrm{N}[\mathrm{N},\mathrm{G}_{t}](\mathrm{N}+\mathbf{1)}^{-2}\Vert
_{\mathrm{op}}\leq 66\Vert B_{t}\Vert _{2}\ .  \label{sup1}
\end{equation}%
Indeed, (\ref{sup0}) corresponds to (\ref{gamma t specific case}). The proof
of the second bound uses the equality%
\begin{equation}
\mathrm{N}[\mathrm{N},\mathrm{G}_{t}](\mathrm{N}+\mathbf{1)}^{-2}=[\mathrm{N}%
,\mathrm{G}_{t}]\mathrm{N}(\mathrm{N}+\mathbf{1)}^{-2}+[\mathrm{N},[\mathrm{N%
},\mathrm{G}_{t}]]\mathrm{N}(\mathrm{N}+\mathbf{1)}^{-2}\ .  \label{sup2}
\end{equation}%
From straightforward computations using (\ref{comutator of G et N}) and the
CCR, we observe that
\begin{equation}
\frac{1}{i}\left[ \mathrm{N},\left[ \mathrm{N},\mathrm{G}_{t}\right] \right]
=8\underset{k,\ell }{\sum }\left\{ B_{t}\right\} _{k,\ell }a_{k}^{\ast
}a_{\ell }^{\ast }-\left\{ \bar{B}_{t}\right\} _{k,\ell }a_{k}a_{\ell
}=8\left( \mathbb{B}_{t}^{\ast }\mathbb{-B}_{t}\right)
\end{equation}%
is a well--defined self--adjoint quadratic operator (Proposition \ref{lemma
example 1 copy(1)}) satisfying
\begin{equation}
\Vert \left[ \mathrm{N},\left[ \mathrm{N},\mathrm{G}_{t}\right] \right]
\left( \mathrm{N}+\mathbf{1}\right) ^{-1}\Vert _{\mathrm{op}}\leq 44\Vert
B_{t}\Vert _{2}\leq 44\Vert B_{0}\Vert _{2}\ ,  \label{latter inequality01-3}
\end{equation}%
see (\ref{inegality BB 3}) when we substitute $4B_{t}$ for $B_{t}$. Using (%
\ref{sup0}), (\ref{sup2}), and (\ref{latter inequality01-3}) we arrive at (%
\ref{sup1}), which is combined with (\ref{sup00}) and (\ref{sup0}) to obtain
the upper bound
\begin{equation}
\Vert \lbrack \mathrm{N}^{2},\mathrm{G}_{t}](\mathrm{N}+\mathbf{1)}%
^{-2}\Vert _{\mathrm{op}}\leq 88\Vert B_{t}\Vert _{2}\ .
\end{equation}%
By (\ref{gamma t specific case}) and (\ref{sup000}), it follows that
\begin{equation}
\gamma _{2}(t):=\Vert \lbrack (\mathrm{N}+\mathbf{1)}^{2},\mathrm{G}_{t}](%
\mathrm{N}+\mathbf{1)}^{-2}\Vert _{\mathrm{op}}\leq 132\Vert B_{t}\Vert
_{2}\leq 132\Vert B_{0}\Vert _{2}\ ,
\label{conservation of domain N1/2 by G}
\end{equation}%
because of (\ref{a priori estimatebis}). By Lemma \ref{existence flow with W
4bis}, the latter in turn implies (\ref{conservation of domain N2 by U}%
).\hfill $\Box $

Observe that Lemma \ref{Section exist unique U quadratic} neither implies
that the adjoint $\mathrm{U}_{t,s}^{\ast }$ of the evolution operator $%
\mathrm{U}_{t,s}$ conserves the dense domain $\mathcal{D}\left( \mathrm{N}%
\right) $ of the particle number operator $\mathrm{N}$, nor that it is
jointly strongly continuous in $s$ and $t$ for all $t\geq s\geq 0$. Since
these properties are necessary below, in particular, to prove the unitarity
of $\mathrm{U}_{t,s}$, we establish them in the next lemma.

\begin{lemma}[Properties of the bounded operator $\mathrm{U}_{t,s}^{\ast }$]

\label{Section exist unique U quadratic copy(1)}\mbox{ }\newline
Under Conditions A1--A4, the adjoint $\mathrm{U}_{t,s}^{\ast }$ of the
evolution operator $\mathrm{U}_{t,s}$ is jointly strongly continuous in $s$
and $t$ for all $t\geq s\geq 0$ and a strong solution on the domain $%
\mathcal{D}\left( \mathrm{N}\right) $ of the non--autonomous evolution
equations%
\begin{equation}
\forall t\geq s\geq 0:\qquad \left\{
\begin{array}{llll}
\partial _{s}\mathrm{U}_{t,s}^{\ast }=-i\mathrm{G}_{s}\mathrm{U}_{t,s}^{\ast
} & , & \mathrm{U}_{t,t}:=\mathbf{1} & . \\
\partial _{t}\mathrm{U}_{t,s}^{\ast }=i\mathrm{U}_{t,s}^{\ast }\mathrm{G}_{t}
& , & \mathrm{U}_{s,s}^{\ast }:=\mathbf{1} & .%
\end{array}%
\right.  \label{flow equationbis-new-adjoint}
\end{equation}%
Moreover, $\mathrm{U}_{t,s}^{\ast }$ conserves the domains $\mathcal{D}%
\left( \mathrm{N}\right) $ and $\mathcal{D}\left( \mathrm{N}^{2}\right) $
for all $t\geq s\geq 0$ as
\begin{eqnarray}
\Vert (\mathrm{N}+\mathbf{1})\mathrm{U}_{t,s}^{\ast }(\mathrm{N}+\mathbf{1}%
)^{-1}\Vert _{\mathrm{op}} &\leq &\mathrm{\exp }\left\{ 22\int_{s}^{t}\Vert
B_{\tau }\Vert _{2}\mathrm{d}\tau \right\} <\infty \ ,
\label{conservation of domain N by Ubis} \\
\Vert (\mathrm{N}+\mathbf{1)}^{2}\mathrm{U}_{t,s}^{\ast }(\mathrm{N}+\mathbf{%
1)}^{-2}\Vert _{\mathrm{op}} &\leq &\mathrm{\exp }\left\{
132\int_{s}^{t}\Vert B_{\tau }\Vert _{2}\mathrm{d}\tau \right\} <\infty \ .
\label{conservation of domain N2}
\end{eqnarray}
\end{lemma}

\noindent \textbf{Proof.} We first observe that $\mathrm{U}_{\lambda
,t,s}^{\ast }$ has a norm convergent representation as a Dyson series
similar to (\ref{dyson series}):%
\begin{eqnarray}
\mathrm{U}_{\lambda ,t,s}^{\ast } &=&\mathbf{1}+\underset{n=1}{\overset{%
\infty }{\sum }}i^{n}\int_{s}^{t}\mathrm{d}\tau _{1}\cdots \int_{s}^{\tau
_{n-1}}\mathrm{d}\tau _{n}\mathrm{G}_{\tau _{n},\lambda }^{\ast }\cdots
\mathrm{G}_{\tau _{1},\lambda }^{\ast }  \notag \\
&=&\mathbf{1}+\underset{n=1}{\overset{\infty }{\sum }}i^{n}\int_{s}^{t}%
\mathrm{d}\tau _{1}\cdots \int_{\tau _{n-1}}^{t}\mathrm{d}\tau _{n}\mathrm{G}%
_{\tau _{1},\lambda }^{\ast }\cdots \mathrm{G}_{\tau _{n},\lambda }^{\ast }\
.  \label{eq2}
\end{eqnarray}%
Since the bounded generator
\begin{equation}
\mathrm{G}_{t,\lambda }^{\ast }=\frac{\lambda \mathrm{G}_{t}}{\lambda
\mathbf{1}-i\mathrm{G}_{t}}\in \mathcal{B}\left( \mathcal{F}_{b}\right)
\end{equation}%
fulfills the same estimates as $\mathrm{G}_{t,\lambda }$, there exists a
bounded operator $\mathrm{\tilde{U}}_{t,s}\in \mathcal{B}\left( \mathcal{F}%
_{b}\right) $ such that
\begin{equation}
\underset{\lambda \rightarrow \infty }{\lim }\Vert \{\mathrm{U}_{\lambda
,t,s}^{\ast }-\mathrm{\tilde{U}}_{t,s}\}(\mathrm{N}+\mathbf{1})^{-1}\Vert _{%
\mathrm{op}}=0\ ,  \label{norm convegence of unitary propagators-adjoint}
\end{equation}%
just as we did for $\mathrm{U}_{\lambda ,t,s}$ and $\mathrm{U}_{t,s}$ in
Lemma \ref{Section exist unique U quadratic} and%
\begin{eqnarray}
\Vert (\mathrm{N}+\mathbf{1})\mathrm{\tilde{U}}_{t,s}(\mathrm{N}+\mathbf{1}%
)^{-1}\Vert _{\mathrm{op}} &\leq &\exp \left\{ 22\int_{s}^{t}\gamma
_{1}\left( \tau \right) \mathrm{d}\tau \right\} \ , \\
\Vert (\mathrm{N}+\mathbf{1)}^{2}\mathrm{\tilde{U}}_{t,s}(\mathrm{N}+\mathbf{%
1)}^{-2}\Vert _{\mathrm{op}} &\leq &\exp \left\{ 132\int_{s}^{t}\gamma
_{2}\left( \tau \right) \mathrm{d}\tau \right\} \ ,
\end{eqnarray}%
see (\ref{gamma t specific case}) and (\ref{conservation of domain N1/2 by G}%
). Moreover, $(\mathrm{\tilde{U}}_{t,s})_{t\geq s\geq 0}$ is jointly
strongly continuous in $s$ and $t$ (cf. Lemma \ref{lemma unitary propagator}%
) and it is a strong solution on the domain $\mathcal{D}\left( \mathrm{N}%
\right) $ of\ (\ref{flow equationbis-new-adjoint}), similar to Lemma \ref%
{Lemma differentiability} by inverting the role of $s$ and $t$, see (\ref%
{eq2}).

It remains to prove that $\mathrm{\tilde{U}}_{t,s}=\mathrm{U}_{t,s}^{\ast }$
for all $t\geq s\geq 0$. It is in this case trivial. Indeed, for any $t\geq
s\geq 0$ and $\varphi ,\psi \in \mathcal{F}_{b}$,
\begin{equation}
\langle (\mathrm{\tilde{U}}_{t,s}-\mathrm{U}_{t,s}^{\ast })\varphi |\psi
\rangle =\langle (\mathrm{\tilde{U}}_{t,s}-\mathrm{U}_{\lambda ,t,s}^{\ast
})\varphi |\psi \rangle +\langle \varphi |(\mathrm{U}_{\lambda ,t,s}-\mathrm{%
U}_{t,s})\psi \rangle \ .  \label{eq trivial self adjoint}
\end{equation}%
As explained in the proof of Lemma \ref{Section exist unique U quadratic},\
the evolution operator $\mathrm{U}_{\lambda ,t,s}$ strongly converges, as $%
\lambda \rightarrow \infty $, to the evolution operator $\mathrm{U}_{t,s}$,
whereas $\mathrm{U}_{\lambda ,t,s}^{\ast }$ converges strongly to $\mathrm{%
\tilde{U}}_{t,s}$. Therefore, we infer from (\ref{eq trivial self adjoint})
that $\mathrm{\tilde{U}}_{t,s}=\mathrm{U}_{t,s}^{\ast }$ for any $t\geq
s\geq 0$. \hfill $\Box $

Now, we are in position to prove the unitarity of the operators $\mathrm{U}%
_{t,s}$ for $t\geq s\geq 0$.

\begin{lemma}[Unitarity of the evolution operator $\mathrm{U}_{t,s}$]
\label{lemma unitary existence 2}\mbox{ }\newline
Under Conditions A1--A4, the evolution operator $\mathrm{U}_{t,s}$ is
unitary for all $t\geq s\geq 0$, i.e., the family $(\mathrm{U}_{t,s})_{t\geq
s\geq 0}$ forms a unitary propagator.
\end{lemma}

\noindent \textbf{Proof. }We infer from Lemmata \ref{Section exist unique U
quadratic}--\ref{Section exist unique U quadratic copy(1)} that, for any $%
t\geq s\geq 0$,
\begin{equation}
(\mathbf{1}-\mathrm{U}_{t,s}\mathrm{U}_{t,s}^{\ast })(\mathrm{N}+\mathbf{1}%
)^{-1}=\int_{s}^{t}\partial _{\tau }\left\{ \mathrm{U}_{t,\tau }\mathrm{U}%
_{t,\tau }^{\ast }\right\} (\mathrm{N}+\mathbf{1})^{-1}\mathrm{d}\tau =0\ ,
\end{equation}%
and%
\begin{equation}
(\mathbf{1}-\mathrm{U}_{t,s}^{\ast }\mathrm{U}_{t,s})(\mathrm{N}+\mathbf{1}%
)^{-1}=-\int_{s}^{t}\partial _{\tau }\left\{ \mathrm{U}_{\tau ,s}^{\ast }%
\mathrm{U}_{\tau ,s}\right\} (\mathrm{N}+\mathbf{1})^{-1}\mathrm{d}\tau =0\ .
\end{equation}%
Thus, for all $t\geq s\geq 0$, $\mathrm{U}_{t,s}^{\ast }\mathrm{U}_{t,s}=%
\mathrm{U}_{t,s}\mathrm{U}_{t,s}^{\ast }=\mathbf{1}$ on the dense domain $%
\mathcal{D}\left( \mathrm{N}\right) \subset \mathcal{F}_{b}$, which
immediately implies the unitarity of $\mathrm{U}_{t,s}\in \mathcal{B}\left(
\mathcal{F}_{b}\right) $. \hfill $\Box $

The proof of Theorem \ref{theorem important 2bis} (i) is now complete by
combining Lemma \ref{Section exist unique U quadratic} with Lemma \ref{lemma
unitary existence 2}. See also the introduction of Section \ref{Section
technical proofs copy(1)} about A1--A4. In fact, by Lemmata \ref{Section
exist unique U quadratic}--\ref{lemma unitary existence 2} one deduces the
following theorem:

\begin{theorem}[Natural extension of the evolution operator $\mathrm{U}%
_{t,s} $]
\label{theorem Ht=unitary orbite copy(1)}\mbox{ }\newline
Under Conditions A1--A4, let $\mathbf{U}_{t,s}:=\mathrm{U}_{t,s}$ for $t\geq
s\geq 0$, whereas for $s\geq t\geq 0$, $\mathbf{U}_{t,s}:=\mathrm{U}%
_{s,t}^{\ast }$. Then, $(\mathbf{U}_{t,s})_{s,t\in \mathbb{R}_{0}^{+}}$
forms a unitary propagator: \newline
\emph{(i)} For any $s,t\in \mathbb{R}_{0}^{+}$, $\mathbf{U}_{t,s}=\mathbf{U}%
_{s,t}^{\ast }$ is a unitary operator. \newline
\emph{(ii)} It satisfies the cocycle property\ $\mathbf{U}_{t,x}\mathbf{U}%
_{x,s}=\mathbf{U}_{t,s}$ for any $s,x,t\in \mathbb{R}_{0}^{+}$.\newline
\emph{(iii)} It is jointly strongly continuous in $s$ and $t$ for all $%
s,t\in \mathbb{R}_{0}^{+}$.\newline
\emph{(iv)} It conserves the domains $\mathcal{D}\left( \mathrm{N}\right) $
and $\mathcal{D}\left( \mathrm{N}^{2}\right) $ for all $t\geq s\geq 0$, see (%
\ref{conservation of domain N by U})--(\ref{conservation of domain N2 by U})
and (\ref{conservation of domain N by Ubis})--(\ref{conservation of domain
N2}). \newline
\emph{(v)} It solves, in the strong sense in $\mathcal{D}\left( \mathrm{N}%
\right) $, the non--autonomous evolution equations%
\begin{equation}
\forall s,t\in \mathbb{R}_{0}^{+}:\qquad \left\{
\begin{array}{llll}
\partial _{t}\mathbf{U}_{t,s}=-i\mathrm{G}_{t}\mathbf{U}_{t,s} & , & \mathbf{%
U}_{s,s}:=\mathbf{1} & . \\
\partial _{s}\mathbf{U}_{t,s}=i\mathbf{U}_{t,s}\mathrm{G}_{s} & , & \mathbf{U%
}_{t,t}:=\mathbf{1} & .%
\end{array}%
\right.
\end{equation}
\end{theorem}

\noindent \textbf{Proof.} Combine Lemmata \ref{Section exist unique U
quadratic}--\ref{lemma unitary existence 2}. In fact, the family $(\mathbf{U}%
_{t,s})_{s,t\in \mathbb{R}_{0}^{+}}$ can directly be derived from arguments
proving Lemmata \ref{existence_flow_5}--\ref{Lemma differentiability}.
Indeed, the generator $\mathrm{G}_{t}=\mathrm{G}_{t}^{\ast }$ is
self--adjoint and so, it satisfies the Kato quasi--stability condition B1
(Section \ref{Section Non-autonomous evolution}) even with \emph{non--ordered%
} times, see (\ref{stability condition in Hilbert space}).\hfill $\Box $

Note that this theorem is only given here for the interested reader, but it
is not essential for our proofs. Actually, we only use below the families $(%
\mathrm{U}_{t,s})_{t\geq s\geq 0}$ and $(\mathrm{U}_{t,s}^{\ast })_{t\geq
s\geq 0}$ to keep explicit the unitarity of operators $\mathrm{U}_{t,s},%
\mathrm{U}_{t,s}^{\ast }$ -- instead of the cocycle property of $(\mathbf{U}%
_{t,s})_{s,t\in \mathbb{R}_{0}^{+}}$ -- because we focus on the
diagonalization of \emph{self--adjoint} quadratic boson operators.

We conclude the proof of Theorem \ref{theorem important 2bis} by analyzing
the effect of the unitary operator $\mathrm{U}_{t,s},$ on
annihilation/creation operators $a_{k},a_{k}^{\ast }$. The aim is to relate
our method to the so--called Bogoliubov $\mathbf{u}$--$\mathbf{v}$
transformation \cite{Bogoliubov1,BruZagrebnov8}, but in a more general
setting (see \cite{Berezin} and Section \ref{section bogoliubov}). The main
assertion, which yields Theorem \ref{theorem important 2bis} (ii), is the
following lemma:

\begin{lemma}[The Bogoliubov $\mathbf{u}$--$\mathbf{v}$ transformation]
\label{lemma uv-Bogoliubov transformation}\mbox{ }\newline
Under Conditions A1--A4, for all $t\geq s\geq 0$ and $k\in \mathbb{N}$, the
operators
\begin{eqnarray}
a_{t,k} &:=&\mathrm{U}_{t,s}a_{s,k}\mathrm{U}_{t,s}^{\ast }=\underset{\ell }{\sum }\left\{ \mathbf{u}_{t,s}\right\} _{k,\ell }a_{s,\ell }+\left\{ \mathbf{v}_{t,s}\right\} _{k,\ell }a_{s,\ell }^{\ast }\ ,
\label{generalized Bog transf 4} \\
a_{s,k} &:=&\mathrm{U}_{t,s}^{\ast }a_{t,k}\mathrm{U}_{t,s}=\underset{\ell }{\sum }\left\{ \mathbf{u}_{t,s}^{\ast }\right\} _{k,\ell }a_{t,\ell }-\left\{
\mathbf{v}_{t,s}^{\mathrm{t}}\right\} _{k,\ell }a_{t,\ell }^{\ast }
\label{generalized Bog transf 3bis}
\end{eqnarray}%
are well--defined on $\mathcal{D}(\mathrm{N}^{1/2})$ and satisfy the CCR.
Here, $a_{0,k}:=a_{k}$ and the operators%
\begin{equation}
\begin{array}{l}
\mathbf{u}_{t,s}:=\mathbf{1}+\underset{n=1}{\overset{\infty }{\sum }}4^{2n}%
\displaystyle\int_{s}^{t}\mathrm{d}\tau _{1}\cdots \int_{s}^{\tau _{2n-1}}%
\mathrm{d}\tau _{2n}B_{\tau _{2n}}\bar{B}_{\tau _{2n-1}}\cdots B_{\tau _{2}}%
\bar{B}_{\tau _{1}}\ , \\
\\
\mathbf{v}_{t,s}:=-\underset{n=0}{\overset{\infty }{\sum }}4^{2n+1}%
\displaystyle\int_{s}^{t}\mathrm{d}\tau _{1}\cdots \int_{s}^{\tau _{2n}}%
\mathrm{d}\tau _{2n+1}B_{\tau _{2n+1}}\bar{B}_{\tau _{2n}}\cdots \bar{B}%
_{\tau _{2}}B_{\tau _{1}}\ ,%
\end{array}
\label{generalized Bog transf 2bis}
\end{equation}%
with $\tau _{0}:=t$, satisfy for all $t\geq s\geq 0$: $\mathbf{v}_{t,s}\in
\mathcal{L}^{2}(\mathfrak{h})$ and
\begin{eqnarray}
\mathbf{u}_{t,s}\mathbf{u}_{t,s}^{\ast }-\mathbf{v}_{t,s}\mathbf{v}%
_{t,s}^{\ast } &=&\mathbf{1}\ ,\text{\qquad }\mathbf{u}_{t,s}\mathbf{v}%
_{t,s}^{\mathrm{t}}=\mathbf{v}_{t,s}\mathbf{u}_{t,s}^{\mathrm{t}}\ ,
\label{CCRu,v} \\
\mathbf{u}_{t,s}^{\ast }\mathbf{u}_{t,s}-\mathbf{v}_{t,s}^{\mathrm{t}}%
\mathbf{\bar{v}}_{t,s} &=&\mathbf{1}\ ,\qquad \mathbf{u}_{t,s}^{\ast }%
\mathbf{v}_{t,s}=\mathbf{v}_{t,s}^{\mathrm{t}}\mathbf{\bar{u}}_{t,s}\ .
\label{bog2bis}
\end{eqnarray}
\end{lemma}

\noindent \textbf{Proof.} First, we observe that $(\mathbf{u}_{t,s}-\mathbf{1%
})\in \mathcal{L}^{2}(\mathfrak{h})$ and $\mathbf{v}_{t,s}\in \mathcal{L}%
^{2}(\mathfrak{h})$, for all $t\geq s\geq 0$, because one directly checks
from (\ref{generalized Bog transf 2bis}) that
\begin{equation}
1+\Vert \mathbf{u}_{t,s}-\mathbf{1}\Vert _{2}\leq \cosh \left\{
4\int_{s}^{t}\Vert B_{\tau }\Vert _{2}\mathrm{d}\tau \right\} \ ,\text{\quad
}\Vert \mathbf{v}_{t,s}\Vert _{2}\leq \sinh \left\{ 4\int_{s}^{t}\Vert
B_{\tau }\Vert _{2}\mathrm{d}\tau \right\} \ .
\label{generalized Bog transf 2bisbis}
\end{equation}%
Furthermore, because $(B_{t})_{t\geq 0}\in C[\mathbb{R}_{0}^{+};\mathcal{L}%
^{2}(\mathfrak{h})]$ (see Theorem \ref{theorem important 1 -1} (ii), (v)),
the operator family $(\mathbf{u}_{t,s}-\mathbf{1},\mathbf{v}_{t,s})_{t\geq
s} $ is a solution in $\mathcal{L}^{2}(\mathfrak{h})\times \mathcal{L}^{2}(%
\mathfrak{h})$ of the system of differential equations
\begin{equation}
\forall t\geq s\geq 0:\qquad \left\{
\begin{array}{cccc}
\partial _{t}\mathbf{u}_{t,s}=-4\mathbf{v}_{t,s}\bar{B}_{t} & , & \quad
\mathbf{u}_{s,s}:=\mathbf{1} & . \\
\partial _{t}\mathbf{v}_{t,s}=-4\mathbf{u}_{t,s}B_{t} & , & \quad \mathbf{v}%
_{s,s}:=0 & .%
\end{array}%
\right.  \label{generalized Bog transf 2}
\end{equation}%
It follows that%
\begin{equation}
\partial _{t}\left\{ \mathbf{u}_{t,s}\mathbf{u}_{t,s}^{\ast }-\mathbf{v}%
_{t,s}\mathbf{v}_{t,s}^{\ast }\right\} =0\text{\quad \textrm{and}\quad }%
\partial _{t}\left\{ \mathbf{u}_{t,s}\mathbf{v}_{t,s}^{\mathrm{t}}-\mathbf{v}%
_{t,s}\mathbf{u}_{t,s}^{\mathrm{t}}\right\} =0\ ,
\end{equation}%
for all $t\geq s\geq 0$, which imply (\ref{CCRu,v}). On the other hand,
observe that
\begin{equation}
\begin{array}{l}
\mathbf{u}_{t,s}=\mathbf{1}+\underset{n=1}{\overset{\infty }{\sum }}4^{2n}%
\displaystyle\int_{s}^{t}\mathrm{d}\tau _{1}\cdots \int_{\tau _{2n-1}}^{t}%
\mathrm{d}\tau _{2n}B_{\tau _{1}}\bar{B}_{\tau _{2}}\cdots B_{\tau _{2n-1}}%
\bar{B}_{\tau _{2n}}\ , \\
\\
\mathbf{v}_{t,s}=-\underset{n=0}{\overset{\infty }{\sum }}4^{2n+1}%
\displaystyle\int_{s}^{t}\mathrm{d}\tau _{1}\cdots \int_{\tau _{2n}}^{t}%
\mathrm{d}\tau _{2n+1}B_{\tau _{1}}\bar{B}_{\tau _{2}}\cdots \bar{B}_{\tau
_{2n}}B_{\tau _{2n+1}}\ .%
\end{array}%
\end{equation}%
It yields on $\mathcal{L}^{2}(\mathfrak{h})$ and for all $t\geq s\geq 0$ the
system of differential equations
\begin{equation}
\forall t\geq s\geq 0:\qquad \left\{
\begin{array}{cccc}
\partial _{s}\mathbf{u}_{t,s}=4B_{s}\mathbf{\bar{v}}_{t,s} & , & \quad
\mathbf{u}_{t,t}:=\mathbf{1} & . \\
\partial _{s}\mathbf{v}_{t,s}=4B_{s}\mathbf{\bar{u}}_{t,s} & , & \quad
\mathbf{v}_{t,t}:=0 & .%
\end{array}%
\right.
\end{equation}%
Therefore, for all $t\geq s\geq 0$,
\begin{equation}
\partial _{t}\left\{ \mathbf{u}_{t,s}^{\ast }\mathbf{u}_{t,s}-\mathbf{v}%
_{t,s}^{\mathrm{t}}\mathbf{\bar{v}}_{t,s}\right\} =0\text{\quad \textrm{and}%
\quad }\partial _{t}\left\{ \mathbf{u}_{t,s}^{\ast }\mathbf{v}_{t,s}-\mathbf{%
v}_{t,s}^{\mathrm{t}}\mathbf{\bar{u}}_{t,s}\right\} =0
\end{equation}%
from which (\ref{bog2bis}) follows.

Now, for each $k\in \mathbb{N}$, we define the operator $\tilde{a}_{t,k}$
acting on the boson Fock space $\mathcal{F}_{b}$ by
\begin{equation}
\tilde{a}_{t,k}:=\underset{\ell }{\sum }\left\{ \mathbf{u}_{t}\right\}
_{k,\ell }a_{\ell }+\left\{ \mathbf{v}_{t}\right\} _{k,\ell }a_{\ell }^{\ast
}=a(\mathbf{u}_{t}^{\ast }\varphi _{k})+a^{\ast }(\mathbf{v}_{t}^{\mathrm{t}}%
\bar{\varphi}_{k})  \label{generalized Bog transf 0}
\end{equation}%
with $\mathbf{u}_{t}:=\mathbf{u}_{t,0}$ and $\mathbf{v}_{t}:=\mathbf{v}%
_{t,0} $. Recall that $\left\{ \varphi _{k}\right\} _{k=1}^{\infty }$ is
some real orthonormal basis in $\mathcal{D}\left( \Omega _{0}\right)
\subseteq \mathfrak{h}$ and $a_{k}:=a\left( \varphi _{k}\right) $ is the
standard boson annihilation operator acting on the boson Fock space $%
\mathcal{F}_{b}$. Because of (\ref{CCRu,v})--(\ref{generalized Bog transf
2bisbis}), the operator family $\{\tilde{a}_{t,k},\tilde{a}_{t,k}^{\ast
}\}_{k=1}^{\infty }$ satisfies the CCR and all operators $\tilde{a}_{t,k}$
and $\tilde{a}_{t,k}^{\ast }$ are well--defined on the domain $\mathcal{D}%
\left( \mathrm{N}^{1/2}\right) $. Indeed, a straightforward computation
shows that%
\begin{eqnarray}
\mathrm{N}_{t} &:=&\underset{k}{\sum }\tilde{a}_{t,k}^{\ast }\tilde{a}_{t,k}=\mathrm{N}+\left\Vert \mathbf{v}_{t}\right\Vert _{2}^{2}+\underset{\ell ,m}{\sum }\left\{ \mathbf{v}_{t}\mathbf{v}_{t}^{\ast }+\mathbf{u}_{t}^{\ast }\mathbf{u}_{t}-\mathbf{1}\right\} _{m,\ell }a_{m}^{\ast }a_{\ell }  \notag \\
&&+\underset{\ell ,m}{\sum }\left\{ \mathbf{u}_{t}^{\ast }\mathbf{v}_{t}\right\} _{m,\ell }a_{m}^{\ast }a_{\ell }^{\ast }+\left\{ \mathbf{u}_{t}^{\mathrm{t}}\mathbf{\bar{v}}_{t}\right\} _{m,\ell }a_{\ell }a_{m}\ ,
\end{eqnarray}%
which, by similar arguments as in Lemma \ref{lemma estimates}, implies that%
\begin{eqnarray}
&&\Vert \left( \mathrm{N}_{t}+\mathbf{1}\right) ^{1/2}\left( \mathrm{N}+%
\mathbf{1}\right) ^{-1/2}\Vert _{\mathrm{op}}
\label{conservation of particle number + constant} \\
&\leq &2\left\Vert \mathbf{v}_{t}\right\Vert _{2}+(1+\left\Vert \mathbf{u}%
_{t}-1\right\Vert _{2}^{1/2})(1+\left\Vert \left( \mathbf{u}_{t}-\mathbf{1}%
\right) \right\Vert _{2}^{1/2}+2\left\Vert \mathbf{v}_{t}\right\Vert
_{2}^{1/2})\ ,  \notag
\end{eqnarray}%
where $\left\Vert \mathbf{u}_{t}-\mathbf{1}\right\Vert _{2}$ and $\left\Vert
\mathbf{v}_{t}\right\Vert _{2}$ are bounded by (\ref{generalized Bog transf
2bisbis}), see also Lemma \ref{lemma existence 3 copy(2)}. In particular, (%
\ref{conservation of particle number + constant}) yields%
\begin{equation}
\Vert \tilde{a}_{t,k}^{\#}\left( \mathrm{N}+\mathbf{1}\right) ^{-1/2}\Vert _{%
\mathrm{op}}\leq \Vert \left( \mathrm{N}_{t}+\mathbf{1}\right) ^{1/2}\left(
\mathrm{N}+\mathbf{1}\right) ^{-1/2}\Vert _{\mathrm{op}}<\infty \ ,
\label{bogo transf new1}
\end{equation}%
where $\tilde{a}_{t,k}^{\#}$ denotes either $\tilde{a}_{t,k}$ or $\tilde{a}%
_{t,k}^{\ast }$ for any $k\in \mathbb{N}$ and $t\geq 0$.

Next, we define the strong derivative $\partial _{t}\tilde{a}_{t,k}$ for any
positive time $t\geq 0$ and all $k\in \mathbb{N}$. To this end, we set%
\begin{eqnarray}
R_{k,\ell }\left( \delta \right) &:=&\delta ^{-1}\int_{0}^{\delta }\left\{
\mathbf{v}_{t+\tau }\bar{B}_{t+\tau }-\mathbf{v}_{t}\bar{B}_{t}\right\}
_{k,\ell }\mathrm{d}\tau \ ,  \label{coefficient1} \\
\tilde{R}_{k,\ell }\left( \delta \right) &:=&\delta ^{-1}\int_{s}^{\delta
}\left\{ \mathbf{u}_{t+\tau }B_{t+\tau }-\mathbf{u}_{t}B_{t}\right\}
_{k,\ell }\mathrm{d}\tau \mathrm{\ },  \label{coefficient2}
\end{eqnarray}%
where $\delta >0$, and observe that
\begin{align}
& \delta ^{-1}\left( \tilde{a}_{t+\delta ,k}-\tilde{a}_{t,k}\right) +4%
\underset{\ell }{\sum }\left\{ \mathbf{v}_{t}\bar{B}_{t}\right\} _{k,\ell
}a_{\ell }+\left\{ \mathbf{u}_{t}B_{t}\right\} _{k,\ell }a_{\ell }^{\ast }
\notag \\
& =-4\underset{\ell }{\sum }\left( R_{k,\ell }\left( \delta \right) a_{\ell
}+\tilde{R}_{k,\ell }\left( \delta \right) a_{\ell }^{\ast }\right) \ ,
\label{bogo0}
\end{align}%
by the fundamental theorem of calculus. Introducing the definitions
\begin{equation}
F_{\delta }:=\underset{\ell }{\sum }R_{k,\ell }\left( \delta \right) a_{\ell
}\ ,\qquad \tilde{F}_{\delta }^{\ast }:=\underset{\ell }{\sum }\tilde{R}%
_{k,\ell }\left( \delta \right) a_{\ell }^{\ast }\ ,
\end{equation}%
and using the equality
\begin{equation}
\tilde{F}_{\delta }\tilde{F}_{\delta }^{\ast }=\underset{\ell }{\sum }%
\left\vert \tilde{R}_{k,\ell }\left( \delta \right) \right\vert ^{2}%
\cdot%
\mathbf{1}+\tilde{F}_{\delta }^{\ast }\tilde{F}_{\delta }\ ,
\end{equation}%
we obtain, as in Lemma \ref{lemma estimates}, that
\begin{equation}
\Vert (F_{\delta }+\tilde{F}_{\delta }^{\ast })(\mathrm{N}+\mathbf{1}%
)^{-1/2}\Vert _{\mathrm{op}}\leq 2\left( \Vert R\left( \delta \right) \Vert
_{2}+\Vert \tilde{R}\left( \delta \right) \Vert _{2}\right) \ ,
\label{bogo1}
\end{equation}%
where $R\left( \delta \right) $ and $\tilde{R}\left( \delta \right) $ are
the Hilbert--Schmidt operators defined by their coefficients (\ref%
{coefficient1}) and (\ref{coefficient2}), respectively. Since $\mathbf{u}%
_{t,s}$, $\mathbf{v}_{t,s}$, and $B_{t}$ are all continuous in $t$ with
respect to the Hilbert--Schmidt topology, (\ref{bogo1}) inserted into (\ref%
{bogo0}) implies that the limit%
\begin{equation}
\partial _{t}\tilde{a}_{t,k}:=\underset{\delta \rightarrow 0}{\lim }\left\{
\delta ^{-1}\left( \tilde{a}_{t+\delta ,k}-\tilde{a}_{t,k}\right) \right\}
=-4\underset{\ell }{\sum }\left\{ \mathbf{v}_{t}\bar{B}_{t}\right\} _{k,\ell
}a_{\ell }+\left\{ \mathbf{u}_{t}B_{t}\right\} _{k,\ell }a_{\ell }^{\ast }
\label{generalized Bog transf 2+1}
\end{equation}%
holds true on $\mathcal{D}(\mathrm{N}^{1/2})$. Moreover, a simple
computation using (\ref{bogo transf new1}) and
\begin{align}
& \Vert \left( \mathrm{N}+\mathbf{1}\right) ^{1/2}\mathrm{G}_{t}\left(
\mathrm{N}+\mathbf{1}\right) ^{-2}\Vert _{\mathrm{op}} \\
& \leq \Vert \mathrm{G}_{t}\left( \mathrm{N}+\mathbf{1}\right) ^{-1}\Vert _{%
\mathrm{op}}+\Vert \left[ \mathrm{N},\mathrm{G}_{t}\right] \left( \mathrm{N}+%
\mathbf{1}\right) ^{-1}\Vert _{\mathrm{op}}<\infty  \notag
\end{align}%
(cf. (\ref{inegality BB 3}) and (\ref{condition 3bisbis})) shows that the
evolution equations%
\begin{equation}
\forall k\in \mathbb{N},\ t\geq 0:\qquad \partial _{t}\tilde{a}_{t,k}=i\left[
\tilde{a}_{t,k},\mathrm{G}_{t}\right] \ ,\quad \tilde{a}_{0,k}=a_{k}\ ,
\label{diff eq bo}
\end{equation}%
hold true on $\mathcal{D}(\mathrm{N}^{2})\subset \mathcal{D}(\mathrm{N}%
^{1/2})$.

Observe now that, for all $k\in \mathbb{N}$ and $t\geq 0$,
\begin{equation}
\left[ \mathrm{N}\mathbf{,}\tilde{a}_{t,k}\right] =\underset{\ell }{\sum }%
\left\{ -\mathbf{u}_{t}\right\} _{k,\ell }a_{\ell }+\left\{ \mathbf{v}%
_{t}\right\} _{k,\ell }a_{\ell }^{\ast }\ ,
\end{equation}%
and it is thus straightforward to check that%
\begin{eqnarray}
\left\Vert (\mathrm{N}+\mathbf{1)}\tilde{a}_{t,k}(\mathrm{N}+\mathbf{1)}%
^{-2}\right\Vert _{\mathrm{op}} &\leq &\Vert \tilde{a}_{t,k}(\mathrm{N}+%
\mathbf{1)}^{-1}\Vert _{\mathrm{op}}+\left\Vert \left[ \mathrm{N}\mathbf{,}%
\tilde{a}_{t,k}\right] (\mathrm{N}+\mathbf{1)}^{-2}\right\Vert _{\mathrm{op}}
\notag \\
&<&\infty \ ,
\end{eqnarray}%
using (\ref{bogo transf new1}). By (\ref{diff eq bo}), for all vectors $\psi
\in \mathcal{D}(\mathrm{N}^{2})$, it implies that
\begin{equation}
\partial _{t}\left\{ \mathrm{U}_{t,s}^{\ast }\tilde{a}_{t,k}\mathrm{U}%
_{t,s}\psi \right\} =\mathrm{U}_{t,s}^{\ast }\left( \partial _{t}\tilde{a}%
_{t,k}-i\left[ \tilde{a}_{t,k},\mathrm{G}_{t}\right] \right) \mathrm{U}%
_{t,s}\psi =0  \label{deri bog}
\end{equation}%
for all $t\geq s\geq 0$, using Lemmata \ref{Section exist unique U quadratic}%
--\ref{Section exist unique U quadratic copy(1)}. Hence, we deduce from (\ref%
{deri bog}) that, for all $\psi \in \mathcal{D}(\mathrm{N}^{2})$,%
\begin{equation}
\forall k\in \mathbb{N},\ t\geq s\geq 0:\qquad \tilde{a}_{t,k}\psi =\mathrm{U%
}_{t,s}\tilde{a}_{s,k}\mathrm{U}_{t,s}^{\ast }\psi \ .
\label{extension equality}
\end{equation}%
The domain $\mathcal{D}(\mathrm{N}^{2})$ is a core for each closed operator
of the family $(\tilde{a}_{t,k})_{k\in \mathbb{N},t\geq 0}$, see (\ref%
{generalized Bog transf 0}). Therefore, since, by (\ref{bogo transf new1}),
\begin{equation}
\tilde{a}_{t,k}(\mathrm{N}+\mathbf{1)}^{-1/2}\in \mathcal{B}\left( \mathcal{F%
}_{b}\right) \ ,
\end{equation}%
for any vector $\varphi \in \mathcal{D}(\mathrm{N}^{1/2})$ there is a
sequence $\{\psi _{n}\}_{n=0}^{\infty }\subset \mathcal{D}(\mathrm{N}^{2})$
converging to $\varphi $ such that $\{\tilde{a}_{t,k}\psi
_{n}\}_{n=0}^{\infty }\subset \mathcal{F}_{b}$ converges to $\tilde{a}%
_{t,k}\varphi \in \mathcal{F}_{b}$. On the other hand, since $\mathrm{U}%
_{t,s}$ is unitary and $a_{k}$ is defined as a closed operator, $(\mathrm{U}%
_{t,s}\tilde{a}_{s,k}\mathrm{U}_{t,s}^{\ast })_{k\in \mathbb{N},t\geq s\geq
0}$ is a family of closed operators and (\ref{extension equality}) implies
the equalities
\begin{equation}
\forall n,k\in \mathbb{N},\ t\geq s\geq 0:\qquad \tilde{a}_{t,k}\psi _{n}=%
\mathrm{U}_{t,s}\tilde{a}_{s,k}\mathrm{U}_{t,s}^{\ast }\psi _{n}\ .
\end{equation}%
It follows that $\varphi \in \mathcal{D}(\mathrm{U}_{t,s}\tilde{a}_{s,k}%
\mathrm{U}_{t,s}^{\ast })$ for all $k\in \mathbb{N}$, and%
\begin{equation}
\forall k\in \mathbb{N},\ t\geq s\geq 0:\qquad \tilde{a}_{t,k}\varphi =%
\mathrm{U}_{t,s}\tilde{a}_{s,k}\mathrm{U}_{t,s}^{\ast }\varphi \ ,
\end{equation}%
for any $\varphi \in \mathcal{D}(\mathrm{N}^{1/2})$. \hfill $\Box $

In other words, the isospectral flow defined via $\mathrm{U}_{t,s}$ is a
(time--dependent) Bogoliubov $\mathbf{u}$--$\mathbf{v}$ unitary
transformation, see also Theorem \ref{flow equation thm 5 copy(1)}. This
last lemma concludes the proof of Theorem \ref{theorem important 2bis}.

In the last proof, note that we have obtained the inclusion
\begin{equation}
\mathcal{D}(\mathrm{N}^{1/2})\subset \mathcal{D}(\mathrm{U}_{t}a_{k}\mathrm{U%
}_{t}^{\ast })\ ,
\end{equation}%
by closedness of the operators $\tilde{a}_{t,k}$ and $\mathrm{U}_{t}a_{k}%
\mathrm{U}_{t}^{\ast }$ together with (\ref{bogo transf new1}) and (\ref%
{extension equality}). In fact, $\mathrm{U}_{t,s}$ and $\mathrm{U}%
_{t,s}^{\ast }$ conserve the domain $\mathcal{D}\left( \mathrm{N}%
^{1/2}\right) $ for all $t\geq s\geq 0$, as
\begin{equation}
\Vert (\mathrm{N}+\mathbf{1)}^{1/2}\mathrm{U}_{t,s}^{\#}(\mathrm{N}+\mathbf{%
1)}^{-1/2}\Vert _{\mathrm{op}}\leq \mathrm{\exp }\left\{ 88\int_{s}^{t}\Vert
B_{\tau }\Vert _{2}\mathrm{d}\tau \right\} <\infty \ ,
\label{conservation domain U N1/2}
\end{equation}%
where $\mathrm{U}_{t,s}^{\#}$ denotes either $\mathrm{U}_{t,s}$ or $\mathrm{U%
}_{t,s}^{\ast }$. It directly implies, again, that $\mathcal{D}(\mathrm{N}%
^{1/2})\subset \mathcal{D}(\mathrm{U}_{t}a_{k}\mathrm{U}_{t}^{\ast })$.
Inequality (\ref{conservation domain U N1/2}) can be seen by using Lemma \ref%
{existence_flow_5} provided we show that%
\begin{equation}
\lbrack (\mathrm{N}+\mathbf{1)}^{1/2},\mathrm{G}_{t}](\mathrm{N}+\mathbf{1)}%
^{-1/2}\in \mathcal{B}\left( \mathcal{F}_{b}\right) \ .
\end{equation}%
However, the latter demands several estimations using the equality%
\begin{equation}
\left( \mathrm{N}+\mathbf{1}\right) ^{1/2}=\frac{1}{\pi }\int_{0}^{\infty }%
\frac{\mathrm{d}\lambda }{\sqrt{\lambda }}\frac{\mathrm{N}+\mathbf{1}}{%
\mathrm{N}+\left( \lambda +1\right) \mathbf{1}}
\end{equation}%
on the domain $\mathcal{D}\left( \mathrm{N}^{1/2}\right) $, see \cite[%
(1.4.2) and 1.4.7 (e)]{Caps}.

\subsection{Brocket--Wegner Flow on Quadratic Boson Operators\label{section
unitarly equi H}}

\noindent The proof of Theorem \ref{theorem Ht=unitary orbite} is carried
out by using the fact that
\begin{equation}
\partial _{t}\left( \mathrm{U}_{t,s}^{\ast }\left( \mathrm{H}_{t}+i\lambda
\mathbf{1}\right) ^{-1}\mathrm{U}_{t,s}\right) =0  \label{derivative}
\end{equation}%
for positive times $t\geq s$ and $\lambda \in \mathbb{R}$, because formally $%
\partial _{t}\mathrm{H}_{t}=i[\mathrm{H}_{t},\mathrm{G}_{t}]$ and $\partial
_{t}\mathrm{U}_{t,s}=-i\mathrm{G}_{t}\mathrm{U}_{t,s}$. Equation (\ref%
{derivative}) yields
\begin{equation}
\left( \mathrm{H}_{t}+i\lambda \mathbf{1}\right) ^{-1}=\mathrm{U}%
_{t,s}\left( \mathrm{H}_{s}+i\lambda \mathbf{1}\right) ^{-1}\mathrm{U}%
_{t,s}^{\ast }\ ,  \label{derivative2}
\end{equation}%
which in turn implies the sought equality $\mathrm{H}_{t}=\mathrm{U}_{t,s}%
\mathrm{H}_{s}\mathrm{U}_{t,s}^{\ast }$.

Unfortunately, the proof is not as straightforward as it looks like. Indeed,
the infinitesimal generator $\mathrm{G}_{t}$ (\ref{generator quadraticbis})
of the unitary propagator $\mathrm{U}_{t,s}$ is unbounded. The derivative $%
\partial _{t}\mathrm{H}_{t}$ of the time--dependent quadratic operator $%
\mathrm{H}_{t}$ (\ref{hamiltonbis}) is also unbounded because it is a
quadratic operator formally given by a commutator $[\mathrm{H}_{t},\mathrm{G}%
_{t}]$ of two unbounded operators. Therefore, the computation of the
derivative in (\ref{derivative}) has to be performed carefully. In fact,
Equality (\ref{derivative}) is only proven for strictly positive times $%
t\geq s>0$ on the dense domain $\mathcal{D}\left( \mathrm{N}\right) $ of the
particle number operator $\mathrm{N}$ (\ref{particle number operator}) and
the statement (\ref{derivative2}) for $s=0$ is shown by strong continuity.

We break the proof of Theorem \ref{theorem Ht=unitary orbite} into several
lemmata. For the reader's convenience we start by giving two trivial
statements for unbounded operators, as they are used several times.

\begin{lemma}[Resolvent and unbounded self--adjoint operators]
\label{lemma resolvent}\mbox{ }\newline
Let $\lambda \in \mathbb{R}$, $\Theta $ be any invertible operator on a
Hilbert space $\mathcal{X}$, and $X$, $Y$ be two self--adjoint operators on $%
\mathcal{X}$ with domains $\mathcal{D}\left( X\right) ,\mathcal{D}\left(
Y\right) \subseteq \mathcal{X}$, respectively. Then one has:\newline
\emph{(i)} $\Theta (i\lambda \mathbf{1}+X)^{-1}\Theta ^{-1}=(i\lambda
\mathbf{1}+\Theta X\Theta ^{-1})^{-1}$. \newline
\emph{(ii)} If $\Theta $ is unitary and $(i\lambda \mathbf{1}+X)^{-1}=\Theta
(i\lambda \mathbf{1}+Y)^{-1}\Theta ^{\ast }$, then $X=\Theta Y\Theta ^{\ast
} $.
\end{lemma}

We give now three simple but pivotal lemmata. First, we show that the
bounded resolvent of $\mathrm{H}_{t}$ defined in (\ref{hamiltonbis})
conserves the dense domain $\mathcal{D}(\mathrm{N})$ of the particle number
operator $\mathrm{N}$ (\ref{particle number operator}). Secondly, we prove
the boundedness of the difference $(\mathrm{H}_{t}-\mathrm{H}_{s})$ of
quadratic operators with respect to the particle number operator $\mathrm{N}$%
. Thirdly, we prove the resolvent identity
\begin{equation}
\left( \mathrm{H}_{t}+i\lambda \mathbf{1}\right) ^{-1}-\left( \mathrm{H}%
_{s}+i\lambda \mathbf{1}\right) ^{-1}=\left( \mathrm{H}_{t}+i\lambda \mathbf{%
1}\right) ^{-1}\left( \mathrm{H}_{s}-\mathrm{H}_{t}\right) \left( \mathrm{H}%
_{s}+i\lambda \mathbf{1}\right) ^{-1}\ .  \label{trivial 0}
\end{equation}%
Observe that this last statement is not completely obvious. Indeed, recall
that, for all $\varphi \in \mathcal{F}_{b}$,%
\begin{equation}
\left( \mathrm{H}_{t}+i\lambda \mathbf{1}\right) \left( \mathrm{H}%
_{t}+i\lambda \mathbf{1}\right) ^{-1}\varphi =\varphi \ ,  \label{trivial 1}
\end{equation}%
but obviously, the converse equality
\begin{equation}
\left( \mathrm{H}_{t}+i\lambda \mathbf{1}\right) ^{-1}\left( \mathrm{H}%
_{t}+i\lambda \mathbf{1}\right) \varphi =\varphi  \label{trivial 2}
\end{equation}%
only holds for all $\varphi \in \mathcal{D}\left( \mathrm{H}_{t}\right) $.

\begin{lemma}[Conservation of $\mathcal{D}(\mathrm{N})$ by $\left( \mathrm{H}%
_{t}+i\mathbf{1}\right) ^{-1}$, $\mathrm{H}_{t}(\mathrm{H}_{t}+i\protect%
\lambda \mathbf{1})^{-1}$]
\label{lemma domaine conservation resolvent}\mbox{ }\newline
Under Conditions A1--A4 and for any $\lambda \in \mathbb{R}$ satisfying $%
|\lambda |>11\Vert B_{t}\Vert _{2}$ at $t\geq 0$,
\begin{equation}
\Vert (\mathrm{N}+\mathbf{1})(\mathrm{H}_{t}+i\lambda \mathbf{1})^{-1}(%
\mathrm{N}+\mathbf{1})^{-1}\Vert _{\mathrm{op}}\leq \frac{1}{|\lambda
|-11\Vert B_{t}\Vert _{2}}\ ,
\end{equation}%
whereas%
\begin{equation}
\Vert (\mathrm{N}+\mathbf{1})\mathrm{H}_{t}(\mathrm{H}_{t}+i\lambda \mathbf{1%
})^{-1}(\mathrm{N}+\mathbf{1})^{-1}\Vert _{\mathrm{op}}\leq \frac{|\lambda
|\left( 1+11\Vert B_{t}\Vert _{2}\right) }{|\lambda |-11\Vert B_{t}\Vert _{2}%
}\ .
\end{equation}
\end{lemma}

\noindent \textbf{Proof.} From (\ref{generator quadraticbis}) and Lemma \ref%
{lemma resolvent} (i), applied to $\Theta =\mathrm{N}+\mathbf{1}$ and to the
self-adjoint (Proposition \ref{lemma example 1 copy(1)}) operator $X=\mathrm{%
H}_{t}$ defined by (\ref{hamiltonbis}), we obtain
\begin{equation}
(\mathrm{N}+\mathbf{1})(\mathrm{H}_{t}+i\lambda \mathbf{1})^{-1}(\mathrm{N}+%
\mathbf{1})^{-1}=%
\Big(%
\mathrm{H}_{t}-i\mathrm{G}_{t}(\mathrm{N}+\mathbf{1})^{-1}+i\lambda \mathbf{1%
}%
\Big)%
^{-1}\ .  \label{condition lambda0}
\end{equation}%
Because of the upper bound (\ref{inegality BB 3}), the operator $\mathrm{G}%
_{t}\left( \mathrm{N}+\mathbf{1}\right) ^{-1}$ is bounded and the inequality
$|\lambda |>11\Vert B_{t}\Vert _{2}$ implies that
\begin{equation}
|\lambda |>11\Vert B_{t}\Vert _{2}\geq \Vert \mathrm{G}_{t}\left( \mathrm{N}+%
\mathbf{1}\right) ^{-1}\Vert _{\mathrm{op}}\ .  \label{condition lambda}
\end{equation}%
Since%
\begin{equation}
\underset{n=0}{\overset{\infty }{\sum }}\frac{\Vert \mathrm{G}_{t}\left(
\mathrm{N}+\mathbf{1}\right) ^{-1}\Vert _{\mathrm{op}}^{n}}{|\lambda |^{n+1}}%
=\frac{1}{|\lambda |-\Vert \mathrm{G}_{t}\left( \mathrm{N}+\mathbf{1}\right)
^{-1}\Vert _{\mathrm{op}}}\ ,
\end{equation}%
under condition (\ref{condition lambda}), we obtain the norm convergence of
the Neumann series
\begin{equation}
(\mathrm{H}_{t}-i\mathrm{G}_{t}(\mathrm{N}+\mathbf{1})^{-1}+i\lambda \mathbf{%
1})^{-1}=\underset{n=0}{\overset{\infty }{\sum }}\left( \mathrm{H}%
_{t}+i\lambda \mathbf{1}\right) ^{-1}\left\{ i\mathrm{G}_{t}\left( \mathrm{N}%
+\mathbf{1}\right) ^{-1}\left( \mathrm{H}_{t}+i\lambda \mathbf{1}\right)
^{-1}\right\} ^{n}  \label{condition lambdabis}
\end{equation}%
for any $t\geq 0$. In particular, by (\ref{condition lambda0}), we obtain
the first statement of the lemma.

Additionally, we infer from (\ref{condition lambda0}) that%
\begin{equation}
(\mathrm{N}+\mathbf{1})\mathrm{H}_{t}(\mathrm{H}_{t}+i\lambda \mathbf{1}%
)^{-1}(\mathrm{N}+\mathbf{1})^{-1}=\frac{\mathrm{H}_{t}-i\mathrm{G}_{t}(%
\mathrm{N}+\mathbf{1})^{-1}}{\mathrm{H}_{t}-i\mathrm{G}_{t}(\mathrm{N}+%
\mathbf{1})^{-1}+i\lambda \mathbf{1}}\ .
\end{equation}%
Hence, by using (\ref{condition lambdabis}) together with the obvious
inequality
\begin{equation}
\Vert \mathrm{H}_{t}(\mathrm{H}_{t}+i\lambda \mathbf{1})^{-1}\Vert _{\mathrm{%
op}}\leq 1\ ,
\end{equation}%
we obtain the second upper bound of the lemma under condition (\ref%
{condition lambda}). \hfill $\Box $

\begin{lemma}[Boundedness of the difference $(\mathrm{H}_{t}-\mathrm{H}_{s})$
in $\mathcal{D}(\mathrm{N})$]
\label{Strong limit hamiltonian 1bis}\mbox{ }\newline
Assume Conditions A1--A4. Then, for all $t,s\geq 0$,
\begin{equation}
\Vert (\mathrm{H}_{t}-\mathrm{H}_{s})(\mathrm{N}+\mathbf{1)}^{-1}\Vert _{%
\mathrm{op}}\leq 40\int_{s}^{t}\Vert B_{\tau }\Vert _{2}^{2}\mathrm{d}\tau
+(4+\sqrt{2})\Vert B_{t}-B_{s}\Vert _{2}\ .  \label{V.316a}
\end{equation}
\end{lemma}

\noindent \textbf{Proof.} By combining the definition (\ref{hamiltonbis})
with (\ref{flow equation-quadratic})--(\ref{flow
equation-quadratic-constante}) (cf. Theorem \ref{theorem important 1 -1}),
we get the equality%
\begin{eqnarray}
\mathrm{H}_{t}-\mathrm{H}_{s} &=&8\int_{s}^{t}\left\Vert B_{\tau
}\right\Vert _{2}^{2}\mathrm{d}\tau -16\underset{k,\ell }{\sum }%
\int_{s}^{t}\left\{ B_{\tau }\bar{B}_{\tau }\right\} _{k,\ell }\mathrm{d}%
\tau \ a_{k}^{\ast }a_{\ell }  \notag \\
&&+\underset{k,\ell }{\sum }\left\{ B_{t}-B_{s}\right\} _{k,\ell
}a_{k}^{\ast }a_{\ell }^{\ast }+\left\{ \bar{B}_{t}-\bar{B}_{s}\right\}
_{k,\ell }a_{k}a_{\ell }  \label{difference}
\end{eqnarray}%
for all $t\geq s\geq 0$. Therefore, applying Lemma \ref{lemma estimates}
with $\theta =\mathbf{1}$,
\begin{equation}
X=X^{\ast }=\Omega _{t}-\Omega _{s}=-16\int_{s}^{t}B_{\tau }\bar{B}_{\tau }%
\mathrm{d}\tau \ ,
\end{equation}%
and the Hilbert--Schmidt operator $Y=\bar{B}_{t}-\bar{B}_{s}\in \mathcal{L}%
^{2}\left( \mathfrak{h}\right) $, we obtain (\ref{V.316a}). \hfill $\Box $

\begin{lemma}[Analysis of the difference of resolvents in $\mathcal{D}(%
\mathrm{N})$]
\label{lemma domaine conservation resolvent copy(3)}\mbox{ }\newline
Under Conditions A1--A4 and for any $\lambda \in \mathbb{R}$ satisfying $%
|\lambda |>11\Vert B_{0}\Vert _{2}$, all times $t,s\geq 0$, and $\varphi \in
\mathcal{D}\left( \mathrm{N}\right) $,
\begin{equation}
\{\left( \mathrm{H}_{t}+i\lambda \mathbf{1}\right) ^{-1}-\left( \mathrm{H}%
_{s}+i\lambda \mathbf{1}\right) ^{-1}\}\varphi =\left( \mathrm{H}%
_{t}+i\lambda \mathbf{1}\right) ^{-1}\left( \mathrm{H}_{s}-\mathrm{H}%
_{t}\right) \left( \mathrm{H}_{s}+i\lambda \mathbf{1}\right) ^{-1}\varphi \ .
\label{statement problem}
\end{equation}
\end{lemma}

\noindent \textbf{Proof.} Clearly, by (\ref{trivial 1}),%
\begin{equation}
\left( \mathrm{H}_{t}+i\lambda \mathbf{1}\right) ^{-1}-\left( \mathrm{H}%
_{s}+i\lambda \mathbf{1}\right) ^{-1}=\{\left( \mathrm{H}_{t}+i\lambda
\mathbf{1}\right) ^{-1}\left( \mathrm{H}_{s}+i\lambda \mathbf{1}\right) -%
\mathbf{1}\}\left( \mathrm{H}_{s}+i\lambda \mathbf{1}\right) ^{-1}.
\label{equality penible 3}
\end{equation}%
By Lemma \ref{lemma domaine conservation resolvent}, the resolvent $\left(
\mathrm{H}_{s}+i\lambda \mathbf{1}\right) ^{-1}$ conserves the domain $%
\mathcal{D}\left( \mathrm{N}\right) $ under the condition that $|\lambda
|>11\Vert B_{0}\Vert _{2}\geq 11\Vert B_{t}\Vert _{2}$\ (cf. (\ref{a priori
estimatebis})). In particular, for any $s\geq 0$ and $\varphi \in \mathcal{F}%
_{b}$,%
\begin{equation}
\left( \mathrm{H}_{s}+i\lambda \mathbf{1}\right) ^{-1}\left( \mathrm{N}+%
\mathbf{1}\right) ^{-1}\varphi \in \mathcal{D}\left( \mathrm{N}\right) \cap
\mathcal{D}\left( \mathrm{H}_{s}\right) \ .  \label{domain1}
\end{equation}%
On the other hand, observe from\ Lemma \ref{Strong limit hamiltonian 1bis}
that, for all $t,s\geq 0$,%
\begin{equation}
\mathrm{H}_{t}(\mathrm{N}+\mathbf{1)}^{-1}=\mathrm{H}_{s}(\mathrm{N}+\mathbf{%
1)}^{-1}+\mathbb{D}_{t}\ ,  \label{equality penible 1}
\end{equation}%
with the \emph{bounded} operator $\mathbb{D}_{t}\in \mathcal{B}(\mathcal{F}%
_{b})$ defined by%
\begin{equation}
\mathbb{D}_{t}:=(\mathrm{H}_{t}-\mathrm{H}_{s})(\mathrm{N}+\mathbf{1)}^{-1}\
.  \label{equality penible 2}
\end{equation}%
In other words, for all $t,s\geq 0$,%
\begin{equation}
\mathcal{D}\left( \mathrm{N}\right) \cap \mathcal{D}\left( \mathrm{H}%
_{s}\right) =\mathcal{D}\left( \mathrm{N}\right) \cap \mathcal{D}\left(
\mathrm{H}_{t}\right) \ .  \label{domain2}
\end{equation}%
Therefore, for all $t,s\geq 0$ and any $\varphi \in \mathcal{F}_{b}$, (\ref%
{domain1}) and (\ref{domain2}) obviously yield%
\begin{equation}
\left( \mathrm{H}_{s}+i\lambda \mathbf{1}\right) ^{-1}\left( \mathrm{N}+%
\mathbf{1}\right) ^{-1}\varphi \in \mathcal{D}\left( \mathrm{N}\right) \cap
\mathcal{D}\left( \mathrm{H}_{t}\right) \ ,
\end{equation}%
which, by (\ref{trivial 2}), in turn implies that%
\begin{eqnarray}
&&\left( \mathrm{H}_{s}+i\lambda \mathbf{1}\right) ^{-1}\left( \mathrm{N}+%
\mathbf{1}\right) ^{-1}\varphi \\
&=&\left( \mathrm{H}_{t}+i\lambda \mathbf{1}\right) ^{-1}\left( \mathrm{H}%
_{t}+i\lambda \mathbf{1}\right) \left( \mathrm{H}_{s}+i\lambda \mathbf{1}%
\right) ^{-1}\left( \mathrm{N}+\mathbf{1}\right) ^{-1}\varphi \ .  \notag
\end{eqnarray}%
Combining this last equality with (\ref{equality penible 3}) we obtain%
\begin{align}
& \{\left( \mathrm{H}_{t}+i\lambda \mathbf{1}\right) ^{-1}-\left( \mathrm{H}%
_{s}+i\lambda \mathbf{1}\right) ^{-1}\}\left( \mathrm{N}+\mathbf{1}\right)
^{-1}\varphi \\
& =-\left( \mathrm{H}_{t}+i\lambda \mathbf{1}\right) ^{-1}\mathbb{D}_{t}(%
\mathrm{N}+\mathbf{1)}\left( \mathrm{H}_{s}+i\lambda \mathbf{1}\right)
^{-1}\left( \mathrm{N}+\mathbf{1}\right) ^{-1}\varphi  \notag
\end{align}%
for all $\varphi \in \mathcal{F}_{b}$, any $t,s\geq 0$, and under the
condition that $|\lambda |>11\Vert B_{0}\Vert _{2}$.\hfill $\Box $

\begin{remark}
\label{remark utiles}The operator on the right side of (\ref{statement
problem}) is bounded because the operator on the left side is. In
particular, Lemma \ref{lemma domaine conservation resolvent copy(3)} can be
extended to all $\varphi \in \mathcal{F}_{b}$. This is however not necessary
for our proofs.
\end{remark}

We are now in position to show the norm continuity of the time--dependent
resolvent $\left( \mathrm{H}_{t}+i\lambda \mathbf{1}\right) ^{-1}$ on the
dense domain $\mathcal{D}\left( \mathrm{N}\right) $ of the particle number
operator $\mathrm{N}$.

\begin{lemma}[Continuity of $\left( \mathrm{H}_{t}+i\mathbf{1}\right) ^{-1}$]

\label{lemma domaine conservation resolvent copy(2)}\mbox{ }\newline
Under Conditions A1--A4 and for any $\lambda \in \mathbb{R}$ satisfying $%
|\lambda |>11\Vert B_{0}\Vert _{2}$, the bounded operator%
\begin{equation}
\left( \mathrm{H}_{t}+i\lambda \mathbf{1}\right) ^{-1}\in C[\mathbb{R}%
_{0}^{+};\mathcal{B}(\mathcal{Y},\mathcal{F}_{b})]
\end{equation}%
with $\mathcal{Y}=\mathcal{D}\left( \mathrm{N}\right) $ being the dense
domain of $\mathrm{N}$. In particular, the resolvent $\left( \mathrm{H}%
_{t}+i\lambda \mathbf{1}\right) ^{-1}$ is strongly continuous for all $t\geq
0$.
\end{lemma}

\noindent \textbf{Proof.} By Theorem \ref{theorem important 1 -1} (ii), we
already know that the operator $B_{t}\in \mathcal{L}^{2}(\mathfrak{h})$ is
continuous in the Hilbert--Schmidt topology for any $t\geq 0$. Therefore,
Lemma \ref{Strong limit hamiltonian 1bis} yields in the limit $t\rightarrow
s $ ($t,s\geq 0$) that
\begin{equation}
\underset{t\rightarrow s}{\lim }\Vert (\mathrm{H}_{t}-\mathrm{H}_{s})(%
\mathrm{N}+\mathbf{1)}^{-1}\Vert _{\mathrm{op}}=0\ .
\label{continuity of H in domain N}
\end{equation}%
On the other hand, Lemma \ref{lemma domaine conservation resolvent copy(3)}
leads, for any $t\geq s\geq 0$, to the inequality
\begin{align}
& \Vert \{\left( \mathrm{H}_{t}+i\lambda \mathbf{1}\right) ^{-1}-\left(
\mathrm{H}_{s}+i\lambda \mathbf{1}\right) ^{-1}\}\left( \mathrm{N}+\mathbf{1}%
\right) ^{-1}\Vert _{\mathrm{op}}  \notag \\
& \leq \Vert \left( \mathrm{H}_{t}+i\lambda \mathbf{1}\right) ^{-1}\Vert \
\Vert \left( \mathrm{H}_{t}-\mathrm{H}_{s}\right) \left( \mathrm{N}+\mathbf{1%
}\right) ^{-1}\Vert _{\mathrm{op}}  \notag \\
& \Vert \left( \mathrm{N}+\mathbf{1}\right) \left( \mathrm{H}_{s}+i\lambda
\mathbf{1}\right) ^{-1}\left( \mathrm{N}+\mathbf{1}\right) ^{-1}\Vert _{%
\mathrm{op}}\ .  \label{continuity of H in domain N 1}
\end{align}%
Combined with the limit (\ref{continuity of H in domain N}) and Lemma \ref%
{lemma domaine conservation resolvent} for $|\lambda |>11\Vert B_{0}\Vert
_{2}\geq 11\Vert B_{t}\Vert _{2}$\ (cf. (\ref{a priori estimatebis})),
Inequality (\ref{continuity of H in domain N 1}) in turn implies in the
limit $t\rightarrow s\ $($t,s\geq 0$) that
\begin{equation}
\underset{t\rightarrow s}{\lim }\Vert \{\left( \mathrm{H}_{t}+i\lambda
\mathbf{1}\right) ^{-1}-\left( \mathrm{H}_{s}+i\lambda \mathbf{1}\right)
^{-1}\}\left( \mathrm{N}+\mathbf{1}\right) ^{-1}\Vert _{\mathrm{op}}=0\ .
\end{equation}%
Since the domain $\mathcal{D}\left( \mathrm{N}\right) $ is dense and the
resolvent $\left( \mathrm{H}_{t}+i\lambda \mathbf{1}\right) ^{-1}$ is
bounded under the condition that $\lambda \in \mathbb{R}$ is not zero, the
previous limit yields the strong continuity of the resolvent $\left( \mathrm{%
H}_{t}+i\lambda \mathbf{1}\right) ^{-1}$ for all $t\geq 0$. \hfill $\Box $

We now prove that the quadratic form associated with the resolvent $\left(
\mathrm{H}_{t}+i\lambda \mathbf{1}\right) ^{-1}$, for any real number $%
\lambda $ that fulfills $|\lambda |>11\Vert B_{0}\Vert _{2}$, satisfies a
differential equation on the set $\mathcal{D}\left( \mathrm{N}\right) \times
\mathcal{D}\left( \mathrm{N}\right) $.

\begin{lemma}[Evolution equation on resolvents]
\label{lemma domaine conservation resolvent copy(1)}\mbox{ }\newline
Under Conditions A1--A4 and for $\lambda \in \mathbb{R}$ such that $|\lambda
|>11\Vert B_{0}\Vert _{2}$, the resolvent $\left( \mathrm{H}_{t}+i\lambda
\mathbf{1}\right) ^{-1}$ satisfies, for any $\varphi ,\psi \in \mathcal{D}%
\left( \mathrm{N}\right) $, the differential equation%
\begin{equation}
\forall t\geq s>0:\qquad \langle \psi |(\partial _{t}Y_{t})\varphi \rangle
=\langle \psi |(i\left[ Y_{t},\mathrm{G}_{t}\right] )\varphi \rangle \
,\quad Y_{s}:=\left( \mathrm{H}_{s}+i\lambda \mathbf{1}\right) ^{-1}\ .
\end{equation}
\end{lemma}

\noindent \textbf{Proof.} For any $\varphi \in \mathcal{F}_{b}$ and all
strictly positive times $t>0$, we define the derivative $\partial _{t}%
\mathrm{H}_{t}$ on the domain $\mathcal{D}\left( \mathrm{N}\right) $ to be
the limit operator%
\begin{equation}
\partial _{t}(\mathrm{H}_{t}\left( \mathrm{N}+\mathbf{1}\right) ^{-1}\varphi
):=\underset{\epsilon \rightarrow 0}{\lim }\left\{ \epsilon ^{-1}\left(
\mathrm{H}_{t+\epsilon }-\mathrm{H}_{t}\right) \left( \mathrm{N}+\mathbf{1}%
\right) ^{-1}\varphi \right\} \ .  \label{derivative0}
\end{equation}%
This derivative is well--defined. Indeed, from\ Theorem \ref{lemma existence
3} (ii) the Hilbert--Schmidt operator $B_{t}$ is locally Lipschitz
continuous in $\mathcal{L}^{2}(\mathfrak{h})$ on $(0,\infty )$.
Consequently, we infer from Lemma \ref{Strong limit hamiltonian 1bis} that,
for all $t>0$,
\begin{equation}
\underset{\epsilon \rightarrow 0}{\lim }\Vert \epsilon ^{-1}(\mathrm{H}%
_{t+\epsilon }-\mathrm{H}_{t})(\mathrm{N}+\mathbf{1)}^{-1}\Vert _{\mathrm{op}%
}<\infty \ .  \label{flow equation resolvent 00}
\end{equation}%
In fact, we can apply Corollary \ref{Corollary existence} and Lemma \ref%
{lemma estimates} with $\theta =\mathbf{1}$,
\begin{equation}
X=X^{\ast }=\epsilon ^{-1}\left\{ \Omega _{t+\epsilon }-\Omega _{t}\right\}
+16B_{t}\bar{B}_{t}
\end{equation}%
and the Hilbert--Schmidt operator%
\begin{equation}
Y=\epsilon ^{-1}\left\{ \bar{B}_{t+\epsilon }-\bar{B}_{t}\right\} +2\left\{
\Omega _{t}^{\mathrm{t}}\bar{B}_{t}+\bar{B}_{t}\Omega _{t}\right\} \in
\mathcal{L}^{2}(\mathfrak{h})\ ,
\end{equation}%
for $t>0$ and $\epsilon \geq -t$, in order to show that the derivative $%
\partial _{t}\mathrm{H}_{t}$ defined by (\ref{derivative0}) is equal on the
dense domain $\mathcal{D}\left( \mathrm{N}\right) $ to%
\begin{eqnarray}
\partial _{t}\mathrm{H}_{t} &=&-2\underset{k,\ell }{\sum }\left\{ \Omega
_{t}B_{t}+B_{t}\Omega _{t}^{\mathrm{t}}\right\} _{k,\ell }a_{k}^{\ast
}a_{\ell }^{\ast }+\left\{ \Omega _{t}^{\mathrm{t}}\bar{B}_{t}+\bar{B}%
_{t}\Omega _{t}\right\} _{k,\ell }a_{k}a_{\ell }  \notag \\
&&-16\underset{k,\ell }{\sum }\left\{ B_{t}\bar{B}_{t}\right\} _{k,\ell
}a_{k}^{\ast }a_{\ell }+8\left\Vert B_{t}\right\Vert _{2}^{2}\ .
\label{derivative H_t}
\end{eqnarray}%
This operator is self--adjoint for any $t>0$, by Proposition \ref{lemma
example 1 copy(1)}, because $\Omega _{t}B_{t},B_{t}\Omega _{t}^{\mathrm{t}%
}\in \mathcal{L}^{2}(\mathfrak{h})$, see Lemma \ref{section extension
copy(1)}.

Now, from (\ref{hamiltonbis})--(\ref{generator quadraticbis}) combined with
Equation (\ref{derivative H_t}), the derivative $\partial _{t}\mathrm{H}_{t}$
is formally equal to%
\begin{equation}
\partial _{t}\mathrm{H}_{t}=i\left[ \mathrm{H}_{t},\mathrm{G}_{t}\right]
\label{flow equation resolvent 0}
\end{equation}%
and furthermore, $(\partial _{t}\mathrm{H}_{t})(\mathrm{N}+\mathbf{1)}%
^{-1}\in \mathcal{B}(\mathcal{F}_{b})$ is a bounded operator for all $t>0$.
Therefore,
\begin{align}
& \mathrm{H}_{t}\mathrm{G}_{t}\left( \mathrm{H}_{t}+i\lambda \mathbf{1}%
\right) ^{-1}\left( \mathrm{N}+\mathbf{1}\right) ^{-1}
\label{flow equation resolvent 0bis} \\
& =\mathrm{G}_{t}\mathrm{H}_{t}\left( \mathrm{H}_{t}+i\lambda \mathbf{1}%
\right) ^{-1}\left( \mathrm{N}+\mathbf{1}\right) ^{-1}+(\partial _{t}\mathrm{%
H}_{t})\left( \mathrm{H}_{t}+i\lambda \mathbf{1}\right) ^{-1}\left( \mathrm{N%
}+\mathbf{1}\right) ^{-1},  \notag
\end{align}%
where both operators in the r.h.s. of the equation are bounded for any $t>0$
and $|\lambda |>11\Vert B_{0}\Vert _{2}\geq 11\Vert B_{t}\Vert _{2}$,
because of (\ref{a priori estimatebis}), (\ref{inegality BB 3}), and Lemma %
\ref{lemma domaine conservation resolvent}. In particular, the operator
\begin{equation*}
\mathrm{H}_{t}\mathrm{G}_{t}\left( \mathrm{H}_{t}+i\lambda \mathbf{1}\right)
^{-1}\left( \mathrm{N}+\mathbf{1}\right) ^{-1}\in \mathcal{B}(\mathcal{F}%
_{b})
\end{equation*}%
is also bounded, and Equality (\ref{flow equation resolvent 0bis}) yields
\begin{equation}
\left( \mathrm{H}_{t}+i\lambda \mathbf{1}\right) ^{-1}(\partial _{t}\mathrm{H%
}_{t})\left( \mathrm{H}_{t}+i\lambda \mathbf{1}\right) ^{-1}\left( \mathrm{N}%
+\mathbf{1}\right) ^{-1}=[\left( \mathrm{H}_{t}+i\lambda \mathbf{1}\right)
^{-1},\mathrm{G}_{t}]\left( \mathrm{N}+\mathbf{1}\right) ^{-1}
\label{flow equation resolvent 01}
\end{equation}%
for all strictly positive times $t>0$.

On the other hand, for any $\varphi ,\psi \in \mathcal{F}_{b}$ and all $t>0$%
, we define the derivative%
\begin{align}
& (\langle \left( \mathrm{N}+\mathbf{1}\right) ^{-1}\psi |\partial
_{t}\{\left( \mathrm{H}_{t}+i\lambda \mathbf{1}\right) ^{-1}\}\left( \mathrm{%
N}+\mathbf{1}\right) ^{-1}\varphi \rangle )
\label{flow equation resolvent 02} \\
& :=\underset{\epsilon \rightarrow 0}{\lim }\langle \left( \mathrm{N}+%
\mathbf{1}\right) ^{-1}\psi |\epsilon ^{-1}\{\left( \mathrm{H}_{t+\epsilon
}+i\lambda \mathbf{1}\right) ^{-1}-\left( \mathrm{H}_{t}+i\lambda \mathbf{1}%
\right) ^{-1}\}\left( \mathrm{N}+\mathbf{1}\right) ^{-1}\varphi \rangle \ .
\notag
\end{align}%
For any $\varphi ,\psi \in \mathcal{F}_{b}$, Lemma \ref{lemma domaine
conservation resolvent copy(3)} yields the inequality%
\begin{align}
& |\langle \left( \mathrm{N}+\mathbf{1}\right) ^{-1}\psi |\{\left( \mathrm{H}%
_{t+\epsilon }+i\lambda \mathbf{1}\right) ^{-1}-\left( \mathrm{H}%
_{t}+i\lambda \mathbf{1}\right) ^{-1}\}\left( \mathrm{N}+\mathbf{1}\right)
^{-1}\varphi \rangle  \notag \\
& +\langle \left( \mathrm{N}+\mathbf{1}\right) ^{-1}\psi |\left( \mathrm{H}%
_{t}+i\lambda \mathbf{1}\right) ^{-1}\left( \mathrm{H}_{t+\epsilon }-\mathrm{%
H}_{t}\right) \left( \mathrm{H}_{t}+i\lambda \mathbf{1}\right) ^{-1}\left(
\mathrm{N}+\mathbf{1}\right) ^{-1}\varphi \rangle |  \notag \\
& \leq \Vert \left( \mathrm{H}_{t+\epsilon }-\mathrm{H}_{t}\right) \left(
\mathrm{N}+\mathbf{1}\right) ^{-1}\Vert _{\mathrm{op}}\Vert \left( \mathrm{N}%
+\mathbf{1}\right) \left( \mathrm{H}_{t}+i\lambda \mathbf{1}\right)
^{-1}\left( \mathrm{N}+\mathbf{1}\right) ^{-1}\Vert _{\mathrm{op}}  \notag \\
& \quad \Vert \{\left( \mathrm{H}_{t+\epsilon }-i\lambda \mathbf{1}\right)
^{-1}-\left( \mathrm{H}_{t}-i\lambda \mathbf{1}\right) ^{-1}\}\left( \mathrm{%
N}+\mathbf{1}\right) ^{-1}\Vert _{\mathrm{op}}\ \Vert \psi \Vert \ \Vert
\varphi \Vert \ .  \label{flow equation resolvent 04}
\end{align}%
Therefore, since $|\lambda |>11\Vert B_{0}\Vert _{2}\geq 11\Vert B_{t}\Vert
_{2}$\ for any $t\geq 0$, we use the upper bound (\ref{flow equation
resolvent 04}) in the limit $\epsilon \rightarrow 0$ together with Lemmata %
\ref{lemma domaine conservation resolvent}, \ref{lemma domaine conservation
resolvent copy(2)}, and (\ref{derivative0})--(\ref{flow equation resolvent
00}), in order to prove that the derivative (\ref{flow equation resolvent 02}%
) at any strictly positive time $t>0$ is equal to%
\begin{align}
& \langle \left( \mathrm{N}+\mathbf{1}\right) ^{-1}\psi |\partial
_{t}\{\left( \mathrm{H}_{t}+i\lambda \mathbf{1}\right) ^{-1}\}\left( \mathrm{%
N}+\mathbf{1}\right) ^{-1}\varphi \rangle \\
& =-\langle \left( \mathrm{N}+\mathbf{1}\right) ^{-1}\psi |\left( \mathrm{H}%
_{t}+i\lambda \mathbf{1}\right) ^{-1}\left( \partial _{t}\mathrm{H}%
_{t}\right) \left( \mathrm{H}_{t}+i\lambda \mathbf{1}\right) ^{-1}\left(
\mathrm{N}+\mathbf{1}\right) ^{-1}\varphi \rangle \ ,  \notag
\end{align}%
which in turn implies the lemma because of Equation (\ref{flow equation
resolvent 01}).\hfill $\Box $

Now, we prove Theorem \ref{theorem Ht=unitary orbite} by using the
resolvents of $\mathrm{H}_{t}$ and $\mathrm{H}_{s}$ and Lemma \ref{lemma
resolvent} (ii).

\begin{lemma}[Unitary transformation of $\mathrm{H}_{s}$ via $\mathrm{U}%
_{t,s}$]
\label{lemma resolvent 3}\mbox{ }\newline
Under Conditions A1--A4, $\mathrm{H}_{t}=\mathrm{U}_{t,s}\mathrm{H}_{s}%
\mathrm{U}_{t,s}^{\ast }$, where $\mathrm{U}_{t,s}$ is the unitary
propagator defined in Theorem \ref{theorem important 2bis} for any $t\geq
s\geq 0$.
\end{lemma}

\noindent \textbf{Proof.} Let $|\lambda |>11\Vert B_{0}\Vert _{2}\geq
11\Vert B_{t}\Vert _{2}$\ (cf. (\ref{a priori estimatebis})) and $t\geq s>0$%
. For any vectors $\varphi ,\psi \in \mathcal{F}_{b}$, observe that%
\begin{eqnarray}
&&\left\langle \left( \mathrm{N}+\mathbf{1}\right) ^{-1}\psi \right\vert
\left. \partial _{t}\left( \mathrm{U}_{t,s}^{\ast }\left( \mathrm{H}%
_{t}+i\lambda \mathbf{1}\right) ^{-1}\mathrm{U}_{t,s}\left( \mathrm{N}+%
\mathbf{1}\right) ^{-1}\right) \varphi \right\rangle  \notag \\
&=&\left\langle \left( \mathrm{N}+\mathbf{1}\right) ^{-1}\psi \right\vert
\partial _{t}(\mathrm{U}_{t,s}^{\ast }\left( \mathrm{N}+\mathbf{1}\right)
^{-1})  \notag \\
&&\left. \left\{ \left( \mathrm{N}+\mathbf{1}\right) \left( \mathrm{H}%
_{t}+i\lambda \mathbf{1}\right) ^{-1}\left( \mathrm{N}+\mathbf{1}\right)
^{-1}\right\} \left\{ \left( \mathrm{N}+\mathbf{1}\right) \mathrm{U}%
_{t,s}\left( \mathrm{N}+\mathbf{1}\right) ^{-1}\right\} \varphi \right\rangle
\notag \\
&&+\left\langle \left( \mathrm{N}+\mathbf{1}\right) ^{-1}\psi \right\vert
\left. \left\{ \mathrm{U}_{t,s}^{\ast }\left( \mathrm{H}_{t}+i\lambda
\mathbf{1}\right) ^{-1}\right\} \partial _{t}(\mathrm{U}_{t,s}\left( \mathrm{%
N}+\mathbf{1}\right) ^{-1})\varphi \right\rangle  \notag \\
&&+\left\langle \left( \mathrm{N}+\mathbf{1}\right) ^{-1}\left\{ \left(
\mathrm{N}+\mathbf{1}\right) \mathrm{U}_{t,s}\left( \mathrm{N}+\mathbf{1}%
\right) ^{-1}\right\} \psi \right\vert \partial _{t}(\left( \mathrm{H}%
_{t}+i\lambda \mathbf{1}\right) ^{-1})  \notag \\
&&\left. \left( \mathrm{N}+\mathbf{1}\right) ^{-1}\left\{ \left( \mathrm{N}+%
\mathbf{1}\right) \mathrm{U}_{t,s}\left( \mathrm{N}+\mathbf{1}\right)
^{-1}\right\} \varphi \right\rangle \ .  \label{eq sup encore new 0}
\end{eqnarray}%
Each of the five operators in braces $\{\cdot \}$ within this equation is
\emph{bounded} because of (\ref{conservation of domain N by U}), (\ref%
{conservation of domain N by Ubis}), and Lemma \ref{lemma domaine
conservation resolvent}. Moreover, Lemmata \ref{Section exist unique U
quadratic}--\ref{Section exist unique U quadratic copy(1)} tells us that%
\begin{equation}
\forall t\geq s\geq 0:\qquad \left\{
\begin{array}{ll}
\partial _{t}\{\mathrm{U}_{t,s}\left( \mathrm{N}+\mathbf{1}\right) ^{-1}\}=-i%
\mathrm{G}_{t}\mathrm{U}_{t,s}\left( \mathrm{N}+\mathbf{1}\right) ^{-1} & ,
\\
\partial _{t}\{\mathrm{U}_{t,s}^{\ast }\left( \mathrm{N}+\mathbf{1}\right)
^{-1}\}=i\mathrm{U}_{t,s}^{\ast }\mathrm{G}_{t}\left( \mathrm{N}+\mathbf{1}%
\right) ^{-1} & ,%
\end{array}%
\right.  \label{eq sup encore new 1bis0}
\end{equation}%
in the strong sense in $\mathcal{F}_{b}$, whereas, by Lemma \ref{lemma
domaine conservation resolvent copy(1)},
\begin{align}
& \langle \left( \mathrm{N}+\mathbf{1}\right) ^{-1}\psi |(\partial
_{t}\{\left( \mathrm{H}_{t}+i\lambda \mathbf{1}\right) ^{-1}\})\left(
\mathrm{N}+\mathbf{1}\right) ^{-1}\varphi \rangle
\label{eq sup encore new 1bis} \\
& =\langle \left( \mathrm{N}+\mathbf{1}\right) ^{-1}\psi |(i[\left( \mathrm{H%
}_{t}+i\lambda \mathbf{1}\right) ^{-1},\mathrm{G}_{t}])\left( \mathrm{N}+%
\mathbf{1}\right) ^{-1}\varphi \rangle ,  \notag
\end{align}%
for all vectors $\varphi ,\psi \in \mathcal{F}_{b}$ and all strictly
positive times $t>0$. Therefore, it follows from Equation (\ref{eq sup
encore new 0}) combined with the derivatives (\ref{eq sup encore new 1bis0}%
)--(\ref{eq sup encore new 1bis}) that%
\begin{equation}
\langle \left( \mathrm{N}+\mathbf{1}\right) ^{-1}\psi |\partial _{t}\{%
\mathrm{U}_{t,s}^{\ast }\left( \mathrm{H}_{t}+i\lambda \mathbf{1}\right)
^{-1}\mathrm{U}_{t,s}\}\left( \mathrm{N}+\mathbf{1}\right) ^{-1}\varphi
\rangle =0  \label{eq sup encore new 1}
\end{equation}%
for any $\varphi ,\psi \in \mathcal{F}_{b}$ and all $t\geq s>0$, i.e.,
\begin{equation}
\langle \left( \mathrm{N}+\mathbf{1}\right) ^{-1}\psi |(\mathrm{U}%
_{t,s}^{\ast }\left( \mathrm{H}_{t}+i\lambda \mathbf{1}\right) ^{-1}\mathrm{U%
}_{t,s}-\left( \mathrm{H}_{s}+i\lambda \mathbf{1}\right) ^{-1})\left(
\mathrm{N}+\mathbf{1}\right) ^{-1}\varphi \rangle =0\ .
\label{eq sup encore new 2}
\end{equation}%
The vector
\begin{equation}
\left( \mathrm{N}+\mathbf{1}\right) \left\{ \mathrm{U}_{t,s}^{\ast }\left(
\mathrm{H}_{t}+i\lambda \mathbf{1}\right) ^{-1}\mathrm{U}_{t,s}-\left(
\mathrm{H}_{s}+i\lambda \mathbf{1}\right) ^{-1}\right\} \left( \mathrm{N}+%
\mathbf{1}\right) ^{-1}\varphi  \label{eq sup encore new 3}
\end{equation}%
is well--defined because of Inequalities (\ref{conservation of domain N by U}%
), (\ref{conservation of domain N by Ubis}), and Lemma \ref{lemma domaine
conservation resolvent}. Consequently, by taking $\psi \in \mathcal{F}_{b}$
equal to (\ref{eq sup encore new 3}) in (\ref{eq sup encore new 2}) we
obtain
\begin{equation}
\mathrm{U}_{t,s}^{\ast }\left( \mathrm{H}_{t}+i\lambda \mathbf{1}\right)
^{-1}\mathrm{U}_{t,s}\left( \mathrm{N}+\mathbf{1}\right) ^{-1}=\left(
\mathrm{H}_{s}+i\lambda \mathbf{1}\right) ^{-1}\left( \mathrm{N}+\mathbf{1}%
\right) ^{-1}\ .  \label{eq sup encore new 4}
\end{equation}%
Since the domain $\mathcal{D}\left( \mathrm{N}\right) $ is dense and the
resolvent $\left( \mathrm{H}_{t}+i\lambda \mathbf{1}\right) ^{-1}$ as well
as the unitary operator $\mathrm{U}_{t,s}$ are bounded, we infer from (\ref%
{eq sup encore new 4}) that
\begin{equation}
\mathrm{U}_{t,s}^{\ast }\left( \mathrm{H}_{t}+i\lambda \mathbf{1}\right)
^{-1}\mathrm{U}_{t,s}=\left( \mathrm{H}_{s}+i\lambda \mathbf{1}\right) ^{-1}
\label{eq sup encore new 5}
\end{equation}%
for all strictly positive times $t\geq s>0$. It thus remains to prove this
equality for all $t\geq s=0$.

The unitary operator $\mathrm{U}_{t,s}$ satisfies the cocycle property\ $%
\mathrm{U}_{t,x}\mathrm{U}_{x,s}=\mathrm{U}_{t,s}$ for any $t\geq x\geq
s\geq 0$ and so, Equation (\ref{eq sup encore new 5}) is equivalent to the
equality%
\begin{equation}
\mathrm{U}_{t}^{\ast }\left( \mathrm{H}_{t}+i\lambda \mathbf{1}\right) ^{-1}%
\mathrm{U}_{t}=\mathrm{U}_{s}^{\ast }\left( \mathrm{H}_{s}+i\lambda \mathbf{1%
}\right) ^{-1}\mathrm{U}_{s}  \label{eq sup encore new 6}
\end{equation}%
for all $t\geq s>0$, where we recall that $\mathrm{U}_{t}:=\mathrm{U}_{t,0}$%
. To perform the limit $s\rightarrow 0$ in the previous equation, note that,
for any $\varphi \in \mathcal{F}_{b}$,
\begin{eqnarray}
&&\Vert \{\mathrm{U}_{s}^{\ast }\left( \mathrm{H}_{s}+i\lambda \mathbf{1}%
\right) ^{-1}\mathrm{U}_{s}-\left( \mathrm{H}_{0}+i\lambda \mathbf{1}\right)
^{-1}\}\varphi \Vert  \notag \\
&\leq &\Vert \mathrm{U}_{s}^{\ast }\Vert _{\mathrm{op}}\ \Vert \left(
\mathrm{H}_{s}+i\lambda \mathbf{1}\right) ^{-1}\Vert _{\mathrm{op}}\Vert
\left( \mathrm{U}_{s}-\mathbf{1}\right) \varphi \Vert  \notag \\
&&+\Vert \mathrm{U}_{s}^{\ast }\Vert _{\mathrm{op}}\ \Vert \{\left( \mathrm{H%
}_{s}+i\lambda \mathbf{1}\right) ^{-1}-\left( \mathrm{H}_{0}+i\lambda
\mathbf{1}\right) ^{-1}\}\varphi \Vert  \notag \\
&&+\Vert \left( \mathrm{U}_{s}^{\ast }-\mathbf{1}\right) \left( \mathrm{H}%
_{0}+i\lambda \mathbf{1}\right) ^{-1}\varphi \Vert \ .
\label{eq sup encore new 7}
\end{eqnarray}%
The unitary operators $\mathrm{U}_{t,s}$ and $\mathrm{U}_{t,s}^{\ast }$ are
jointly strongly continuous in $s$ and $t$ with $\mathrm{U}_{s,s}^{\ast }=%
\mathrm{U}_{s,s}:=\mathbf{1}$ for all $t\geq s\geq 0$, see Lemmata \ref%
{Section exist unique U quadratic}--\ref{lemma unitary existence 2}.
Furthermore, the resolvent $\left( \mathrm{H}_{t}+i\lambda \mathbf{1}\right)
^{-1}$ is strongly continuous for any $t\geq 0$, by Lemma \ref{lemma domaine
conservation resolvent copy(2)}. Consequently, we combine Equality (\ref{eq
sup encore new 6}) in the limit $s\rightarrow 0$ with Inequality (\ref{eq
sup encore new 7}) to verify (\ref{eq sup encore new 5}) for all $t\geq s=0$%
. In other words, we obtain in this way the desired statement:%
\begin{equation}
\forall t\geq s\geq 0:\qquad \mathrm{U}_{t,s}^{\ast }\left( \mathrm{H}%
_{t}+i\lambda \mathbf{1}\right) ^{-1}\mathrm{U}_{t,s}=\left( \mathrm{H}%
_{s}+i\lambda \mathbf{1}\right) ^{-1}\ .  \label{eq sup encore new 8}
\end{equation}%
The lemma then follows from Lemma \ref{lemma resolvent} (ii) because, as
already explained, $\mathrm{U}_{t,s}$ is unitary, by Lemma \ref{lemma
unitary existence 2}.\hfill $\Box $

\subsection{Quasi $\mathrm{N}$--Diagonalization of Quadratic Boson Operators
\label{Section theorem important 4bis}}

The limits of $\Omega _{t}$, $C_{t}$, and $\mathrm{H}_{t}$, as $t\rightarrow
\infty $, can all be obtained by using the square integrability of $\Vert
B_{t}\Vert _{2}$ on $[0,\infty )$. Sufficient conditions to get this
property are given by Theorem \ref{theorem important 1 -3} (i). Therefore,
the proof of Theorem \ref{theorem important 4bis} is broken in three lemmata
by always using the assumption that $\Vert B_{t}\Vert _{2}$ is
square--integrable on $[0,\infty )$.

\begin{lemma}[Limits of $\Omega _{t}$ and $C_{t}$ for infinite times]
\label{lemma omega t infinity}\mbox{ }\newline
Assume Conditions A1--A4 and the square--integrability of $t\mapsto \Vert
B_{t}\Vert _{2}$. Then $(\Omega _{0}-\Omega _{t})_{t\geq 0}\in \mathcal{L}%
^{1}(\mathfrak{h})$ converges in $\mathcal{L}^{1}(\mathfrak{h})$ to $(\Omega
_{0}-\Omega _{\infty })$, where%
\begin{equation}
\Omega _{\infty }:=\Omega _{0}-16\int_{0}^{\infty }B_{\tau }\bar{B}_{\tau }%
\mathrm{d}\tau =\Omega _{\infty }^{\ast }\geq 0
\end{equation}%
on the domain $\mathcal{D}\left( \Omega _{0}\right) $, and the limit $%
t\rightarrow \infty $ of real numbers $(C_{t})_{t\geq 0}$ (cf. (\ref{flow
equation-quadratic-constante})) equals%
\begin{equation}
C_{\infty }:=8\int_{0}^{\infty }\left\Vert B_{\tau }\right\Vert _{2}^{2}%
\mathrm{d}\tau +C_{0}<\infty \ .
\end{equation}
\end{lemma}

\noindent \textbf{Proof.} The operators $\Delta _{t}:=\Omega _{0}-\Omega
_{t}\in \mathcal{L}^{1}(\mathfrak{h})$ for all $t\geq 0$ form a Cauchy
sequence on the Banach space $\mathcal{L}^{1}(\mathfrak{h})$, as $%
t\rightarrow \infty $, because the map $t\mapsto \Vert B_{t}\Vert _{2}$ is
square--integrable on $[0,\infty )$. Indeed, integrating (\ref{flow
equation-quadratic}) observe that%
\begin{equation}
\forall t,s\in \mathbb{R}_{0}^{+}:\qquad \Omega _{t}=\Omega
_{s}-16\int_{s}^{t}B_{\tau }\bar{B}_{\tau }\mathrm{d}\tau \geq 0\ ,
\label{important equality}
\end{equation}%
which implies that
\begin{equation}
\forall t\geq s\geq 0:\qquad \left\Vert \Delta _{t}-\Delta _{s}\right\Vert
_{1}\leq 16\int_{s}^{t}\left\Vert B_{\tau }\right\Vert _{2}^{2}\mathrm{d}%
\tau \ .  \label{limit delta eq 1}
\end{equation}%
In other words, by completeness of $\mathcal{L}^{1}(\mathfrak{h})$, there is
a trace--class operator
\begin{equation}
\Delta _{\infty }:=16\int_{0}^{\infty }B_{\tau }\bar{B}_{\tau }\mathrm{d}%
\tau \in \mathcal{L}^{1}(\mathfrak{h})
\end{equation}%
to which $\Delta _{t}$ converges in trace--norm. In particular, the domain
of $\Omega _{\infty }:=\Omega _{0}-\Delta _{\infty }$ equals $\mathcal{D}%
\left( \Omega _{0}\right) $. The positivity of $\Omega _{\infty }$ is also a
direct consequence of $\Omega _{t}\geq 0$. The existence of $C_{\infty }$ as
well as its explicit form are obvious because of (\ref{flow
equation-quadratic-constante}). \hfill $\Box $

The convergence of the operator $\Omega _{t}$ to $\Omega _{\infty }$ given
in Lemma \ref{lemma omega t infinity} means via (\ref{flow
equation-quadratic}) that one can extend (\ref{important equality}) to
infinite times:%
\begin{equation}
\forall t,s\in \mathbb{R}_{0}^{+}\cup \{\infty \}:\qquad \Omega _{t}=\Omega
_{s}-16\int_{s}^{t}B_{\tau }\bar{B}_{\tau }\mathrm{d}\tau \ .
\label{limit delta eq 3}
\end{equation}

Next, we analyze Theorem \ref{lemma constante of motion} (i), (iii), and
Corollary \ref{lemma constante of motion copy(3)} in the limit $t\rightarrow
\infty $, when A4 holds (i.e., $T_{+}=\infty $). In particular, the limit
operator $\Omega _{\infty }$ given in Lemma \ref{lemma omega t infinity} can
explicitly be computed in the special case $\Omega _{0}B_{0}=B_{0}\Omega
_{0}^{\mathrm{t}}$.

\begin{lemma}[Limit operator $\Omega _{\infty }$ and constant of motion]
\label{lemma omega t infinity2}\mbox{ }\newline
Assume Conditions A1--A4 and the square--integrability of $t\mapsto \Vert
B_{t}\Vert _{2}$. Let $s\in \mathbb{R}_{0}^{+}\cup \{\infty \}$ and $%
B_{\infty }:=0$.\newline
\emph{(i)}
\begin{equation}
\Omega _{\infty }^{2}=\Omega _{s}^{2}-8B_{s}\bar{B}_{s}+32\int_{s}^{\infty
}B_{\tau }\Omega _{\tau }^{\mathrm{t}}\bar{B}_{\tau }\mathrm{d}\tau \ .
\end{equation}%
\emph{(ii)} If $\Omega _{0}B_{0}=B_{0}\Omega _{0}^{\mathrm{t}}$ then
\begin{equation}
\Omega _{s}B_{s}=B_{s}\Omega _{s}^{\mathrm{t}},\quad \Omega _{\infty
}=\{\Omega _{s}^{2}-4B_{s}\bar{B}_{s}\}^{1/2}\ .
\end{equation}%
\emph{(iii)} The constant of motion of the flow in the limit $t\rightarrow
\infty $ equals%
\begin{equation}
\mathrm{tr}\left( \Omega _{\infty }^{2}-\Omega _{s}^{2}+4B_{s}\bar{B}%
_{s}\right) =0\ .  \label{constant of motion infinity}
\end{equation}
\end{lemma}

\noindent \textbf{Proof.} \underline{(i):} Using Theorem \ref{lemma
constante of motion} (i) we observe that
\begin{equation}
\left\Vert \Omega _{t}^{2}-\Omega _{s}^{2}\right\Vert _{1}\leq 8\left\Vert
B_{s}\right\Vert _{2}^{2}+8\left\Vert B_{t}\right\Vert
_{2}^{2}+32\int_{s}^{t}\left\Vert B_{\tau }\Omega _{\tau }^{\mathrm{t}}\bar{B%
}_{\tau }\right\Vert _{1}\mathrm{d}\tau  \label{eq sympa}
\end{equation}%
for any $t\geq s\geq 0$. Since, for all $t>0$,
\begin{equation}
\left\Vert B_{t}\Omega _{t}^{\mathrm{t}}\bar{B}_{t}\right\Vert
_{1}=\left\Vert \bar{B}_{t}\Omega _{t}B_{t}\right\Vert _{1}=\Vert \Omega
_{t}^{1/2}B_{t}\Vert _{2}^{2}
\end{equation}%
(Lemma \ref{lemma existence 3 copy(2)}), we apply the upper bound of Lemma %
\ref{section extension copy(2)} for $\alpha =1/2$ and $n=1$ to get the
inequality%
\begin{equation}
\forall t>0:\qquad \left\Vert B_{t}\Omega _{t}^{\mathrm{t}}\bar{B}%
_{t}\right\Vert _{1}\leq \frac{1}{2\mathrm{e}t}\Vert B_{0}\Vert _{2}\Vert
B_{t}\Vert _{2}\ .
\end{equation}%
Therefore, since $t\mapsto \Vert B_{t}\Vert _{2}$ is square--integrable on $%
[0,\infty )$, the map $t\mapsto \Vert B_{t}\Omega _{t}^{\mathrm{t}}\bar{B}%
_{t}\Vert _{2}$ is also integrable as $t\rightarrow \infty $ and, by (\ref%
{eq sympa}), the operator family $(\Omega _{t}^{2}-\Omega _{s}^{2})_{t\geq
s}\in \mathcal{L}^{1}(\mathfrak{h})$ for fixed $s>0$ is a Cauchy sequence on
the Banach space $\mathcal{L}^{1}(\mathfrak{h})$, as $t\rightarrow \infty $.
By completeness of $\mathcal{L}^{1}(\mathfrak{h})$ together with Theorem \ref%
{lemma constante of motion} (i) (here $T_{+}=\infty $) and the resolvent
identity (\ref{V.197}), the positive operator family $(\Omega
_{t}^{2})_{t\geq 0}$ converges in the norm resolvent sense to the positive
operator, defined for all $s\geq 0$, by
\begin{equation}
\Xi _{\infty }:=\Omega _{s}^{2}-8B_{s}\bar{B}_{s}+32\int_{s}^{\infty
}B_{\tau }\Omega _{\tau }^{\mathrm{t}}\bar{B}_{\tau }\mathrm{d}\tau \geq 0\ ,
\label{assertion plus 0}
\end{equation}%
with domain $\mathcal{D}\left( \Xi _{\infty }\right) =\mathcal{D}\left(
\Omega _{0}^{2}\right) $.

On the other hand, we observe that, for any $t\geq 0$,
\begin{equation}
\left( 16\int_{t}^{\infty }B_{\tau }\bar{B}_{\tau }\mathrm{d}\tau \right)
\mathfrak{h}\subset \mathcal{D}\left( \Omega _{0}\right) \ .
\label{assertion plus}
\end{equation}%
Indeed, using (\ref{limit delta eq 3}) and the closedness of the bounded
resolvent $\left( \Omega _{\infty }+\mathbf{1}\right) ^{-1}$ together with
Lemmata \ref{lemma existence 3 copy(3)} (i) and \ref{lemma existence 3
copy(2)} we arrive at the upper bound%
\begin{eqnarray}
\left\Vert \Omega _{\infty }\int_{t}^{\infty }B_{\tau }\bar{B}_{\tau }%
\mathrm{d}\tau \right\Vert _{\mathrm{op}} &\leq &\left\Vert \Omega _{\infty
}\left( \Omega _{\infty }+\mathbf{1}\right) ^{-1}\right\Vert _{\mathrm{op}%
}\int_{t}^{\infty }\left\Vert B_{\tau }\Omega _{\tau }^{\mathrm{t}}\bar{B}%
_{\tau }\right\Vert _{1}\mathrm{d}\tau  \notag \\
&&+16\left\Vert \Omega _{\infty }\left( \Omega _{\infty }+\mathbf{1}\right)
^{-1}\right\Vert _{\mathrm{op}}\left( \int_{t}^{\infty }\left\Vert B_{\tau
}\right\Vert _{2}^{2}\mathrm{d}\tau \right) ^{2}  \notag \\
&&+\left\Vert \Omega _{\infty }\left( \Omega _{\infty }+\mathbf{1}\right)
^{-1}\right\Vert _{\mathrm{op}}\int_{t}^{\infty }\left\Vert B_{\tau
}\right\Vert _{2}^{2}\mathrm{d}\tau \ .
\end{eqnarray}%
Since the maps $t\mapsto \Vert B_{t}\Vert _{2}^{2}$ and $t\mapsto \left\Vert
B_{t}\Omega _{t}^{\mathrm{t}}\bar{B}_{t}\right\Vert _{1}$ are both
integrable on $[0,\infty )$, the last upper bound yields Assertion (\ref%
{assertion plus}). Therefore, we can use (\ref{V.197}), (\ref{equality utile}%
)--(\ref{equality utilebis}), and (\ref{limit delta eq 3}) together with (%
\ref{assertion plus}) and obtain a similar upper bound as (\ref{chiante1}),
that is,%
\begin{equation}
\left\Vert \left( \Omega _{\infty }^{2}+\mathbf{1}\right) ^{-1}-\left(
\Omega _{t}^{2}+\mathbf{1}\right) ^{-1}\right\Vert _{\mathrm{op}}\leq
128\int_{t}^{\infty }\left\Vert B_{\tau }\right\Vert _{2}^{2}\mathrm{d}\tau
\left( 1+8\int_{t}^{\infty }\left\Vert B_{\tau }\right\Vert _{2}^{2}\mathrm{d%
}\tau \right) \ .
\end{equation}%
By square--integrability of the map $t\mapsto \Vert B_{t}\Vert _{2}$ on $%
[0,\infty )$, it follows that the positive operator family $(\Omega
_{t}^{2})_{t\geq 0}$ converges in the norm resolvent sense to $\Omega
_{\infty }^{2}$, which is also defined on the domain $\mathcal{D}\left(
\Omega _{0}^{2}\right) $, by (\ref{limit delta eq 3}) and (\ref{assertion
plus}). As a consequence, $\Omega _{\infty }^{2}$ equals the operator $\Xi
_{\infty }$ defined by (\ref{assertion plus 0}).\smallskip

\underline{(ii):} If $\Omega _{0}B_{0}=B_{0}\Omega _{0}^{\mathrm{t}}$ then,
for all $t,s\geq 0$, $\Omega _{t}B_{t}=B_{t}\Omega _{t}^{\mathrm{t}}$ and $%
\Omega _{t}^{2}-4B_{t}\bar{B}_{t}=\Omega _{s}^{2}-4B_{s}\bar{B}_{s}$, see
Corollary \ref{lemma constante of motion copy(3)} with $T_{+}=\infty $.
Since $\Vert B_{t}\Vert _{2}$ converges to zero in the limit $t\rightarrow
\infty $ and the sequence $(\Omega _{\infty }^{2}-\Omega _{t}^{2})_{t\geq
0}\in \mathcal{L}^{1}(\mathfrak{h})$ converges in $\mathcal{L}^{1}(\mathfrak{%
h})$ to the zero operator (cf. (i)), we deduce the second assertion using
the trivial equality
\begin{equation}
\Omega _{\infty }^{2}-\Omega _{s}^{2}+4B_{s}\bar{B}_{s}=\{\Omega _{\infty
}^{2}-\Omega _{t}^{2}\}+\{\Omega _{t}^{2}-\Omega _{s}^{2}+4B_{s}\bar{B}_{s}\}
\label{eq trivial}
\end{equation}%
for all $t,s\geq 0$.\smallskip

\underline{(iii):} Similarly, (\ref{eq trivial}) together with Theorem \ref%
{lemma constante of motion} (iii) and the trace--class convergence of $%
(\Omega _{\infty }^{2}-\Omega _{t}^{2})_{t\geq 0}$ (cf. (i)) also implies (%
\ref{constant of motion infinity}). \hfill $\Box $

Therefore, Theorem \ref{theorem important 4bis} (i) is a direct consequence
of Lemmata \ref{lemma omega t infinity}--\ref{lemma omega t infinity2}
combined with Theorem \ref{theorem important 1 -3} (i). It remains to prove
its second statement (ii). This is easily performed in the next two lemmata,
which conclude the proof of Theorem \ref{theorem important 4bis}.

\begin{lemma}[Limit of $\mathrm{H}_{t}\mathbf{\ }$as $t\rightarrow \infty $
on the domain $\mathcal{D}(\mathrm{N})$]
\label{lemma H t infinity}\mbox{ }\newline
Assume Conditions A1--A4 and the square--integrability of $t\mapsto \Vert
B_{t}\Vert _{2}$. Then
\begin{equation}
\underset{t\rightarrow \infty }{\lim }\left\Vert (\mathrm{H}_{\infty }-%
\mathrm{H}_{t})(\mathrm{N}+\mathbf{1)}^{-1}\right\Vert _{\mathrm{op}}=0\ ,
\end{equation}%
where the self--adjoint quadratic boson operator $\mathrm{H}_{\infty }$ is
defined by
\begin{equation}
\mathrm{H}_{\infty }:=\underset{k,\ell }{\sum }\{\Omega _{\infty }\}_{k,\ell
}a_{k}^{\ast }a_{\ell }+C_{\infty }\ .
\end{equation}
\end{lemma}

\noindent \textbf{Proof.} Because of (\ref{limit delta eq 3}) and the
square--integrability of the map $t\mapsto \Vert B_{t}\Vert _{2}$ on $%
[0,\infty )$, we can extend Lemma \ref{Strong limit hamiltonian 1bis} to all
$t,s\in \mathbb{R}_{0}^{+}\cup \{\infty \}$ with $B_{\infty }:=0$ and
obviously deduce the assertion. \hfill $\Box $

\begin{lemma}[$\mathrm{H}_{t}\rightarrow \mathrm{H}_{\infty }$ in the strong
resolvent sense]
\label{lemma convergence in strong resolvent sence of H}\mbox{ }\newline
Assume Conditions A1--A4 and the square--integrability of $t\mapsto \Vert
B_{t}\Vert _{2}$. Then, for any non--zero real $\lambda \neq 0$ and any
vector $\varphi \in \mathcal{F}_{b}$,
\begin{equation}
\underset{t\rightarrow \infty }{\lim }\left\Vert \{(\mathrm{H}_{\infty
}+i\lambda \mathbf{1})^{-1}-(\mathrm{H}_{t}+i\lambda \mathbf{1}%
)^{-1}\}\varphi \right\Vert =0\ .
\end{equation}
\end{lemma}

\noindent \textbf{Proof.} For any non--zero real $\lambda \neq 0$, there is $%
T_{\lambda }\geq 0$ such that $|\lambda |>11\Vert B_{t}\Vert _{2}$\ for all $%
t>T_{\lambda }$ because, by assumption, the Hilbert--Schmidt norm $\Vert
B_{t}\Vert _{2}$ vanishes when $t\rightarrow \infty $. If the map $t\mapsto
\Vert B_{t}\Vert _{2}$ is square--integrable on $[0,\infty )$ then Lemmata %
\ref{lemma domaine conservation resolvent}--\ref{lemma domaine conservation
resolvent copy(3)} are satisfied for $t=\infty $ and all $s\geq 0$ because
of Lemma \ref{lemma omega t infinity}, see also (\ref{limit delta eq 3}).
Therefore, similar to (\ref{continuity of H in domain N 1}), we obtain for
any $t>T_{\lambda }$ the inequality%
\begin{eqnarray}
&&\left\Vert \{(\mathrm{H}_{\infty }+i\lambda \mathbf{1})^{-1}-(\mathrm{H}%
_{t}+i\lambda \mathbf{1})^{-1}\}(\mathrm{N}+\mathbf{1})^{-1}\right\Vert _{%
\mathrm{op}}  \notag \\
&\leq &\Vert \left( \mathrm{H}_{\infty }+i\lambda \mathbf{1}\right)
^{-1}\Vert _{\mathrm{op}}\Vert \left( \mathrm{H}_{\infty }-\mathrm{H}%
_{t}\right) \left( \mathrm{N}+\mathbf{1}\right) ^{-1}\Vert _{\mathrm{op}}
\notag \\
&&\Vert \left( \mathrm{N}+\mathbf{1}\right) \left( \mathrm{H}_{t}+i\lambda
\mathbf{1}\right) ^{-1}\left( \mathrm{N}+\mathbf{1}\right) ^{-1}\Vert _{%
\mathrm{op}}\ ,  \label{eq converge resolvent 1}
\end{eqnarray}%
which in turn yields the limit
\begin{equation}
\underset{t\rightarrow \infty }{\lim }\left\Vert \{(\mathrm{H}_{\infty
}+i\lambda \mathbf{1})^{-1}-(\mathrm{H}_{t}+i\lambda \mathbf{1})^{-1}\}(%
\mathrm{N}+\mathbf{1})^{-1}\right\Vert _{\mathrm{op}}=0
\label{eq converge resolvent 4}
\end{equation}%
for any non--zero real $\lambda \neq 0$, see Lemma \ref{lemma H t infinity}.
The domain $\mathcal{D}(\mathrm{N})$ is dense in $\mathcal{F}_{b}$ and both
resolvent are bounded. It is then straightforward to verify the strong
convergence of resolvents from (\ref{eq converge resolvent 4}). \hfill $\Box
$

\subsection{$\mathrm{N}$--Diagonalization of Quadratic Boson Operators\label%
{section proof thm important 3}}

To obtain the limits $t\rightarrow \infty $ of the bounded operators $%
\mathbf{u}_{t,s}$, $\mathbf{v}_{t,s}$ (see (\ref{generalized Bog transf 2bis}%
)) and $\mathrm{U}_{t,s}$ (Theorem \ref{theorem important 2bis}), as well as
the equality $\mathrm{H}_{\infty }=\mathrm{U}_{\infty ,s}\mathrm{H}_{s}%
\mathrm{U}_{\infty ,s}^{\ast }$, one needs the integrability of the map $%
t\mapsto \Vert B_{t}\Vert _{2}$ on $[0,\infty )$. Sufficient conditions to
ensure this are given by Theorem \ref{theorem important 1 -3}. So, we break
up the proof of Theorem \ref{theorem important 3} into several lemmata,
always using the integrability of the map $t\mapsto \Vert B_{t}\Vert _{2}$
on $[0,\infty )$.

\begin{lemma}[Limits of $\mathbf{u}_{t,s}$ and $\mathbf{v}_{t,s}$ as $%
t\rightarrow \infty $]
\label{lemma unitary asymptotics 1}\mbox{ }\newline
Assume Conditions A1--A4 and the integrability of $t\mapsto \Vert B_{t}\Vert
_{2}$. Then $\mathbf{u}_{t,s}$ and $\mathbf{v}_{t,s}$ converge in $\mathcal{L%
}^{2}(\mathfrak{h})$ respectively to%
\begin{equation}
\mathbf{u}_{\infty ,s}:=\mathbf{1}+\underset{n=1}{\overset{\infty }{\sum }}%
4^{2n}\int_{s}^{\infty }\mathrm{d}\tau _{1}\cdots \int_{s}^{\tau _{2n-1}}%
\mathrm{d}\tau _{2n}B_{\tau _{2n}}\bar{B}_{\tau _{2n-1}}\cdots B_{\tau _{2}}%
\bar{B}_{\tau _{1}}  \label{def 1}
\end{equation}%
and%
\begin{equation}
\mathbf{v}_{\infty ,s}:=-\underset{n=0}{\overset{\infty }{\sum }}%
4^{2n+1}\int_{s}^{\infty }\mathrm{d}\tau _{1}\cdots \int_{s}^{\tau _{2n}}%
\mathrm{d}\tau _{2n+1}B_{\tau _{2n+1}}\bar{B}_{\tau _{2n}}\cdots \bar{B}%
_{\tau _{2}}B_{\tau _{1}}\ ,  \label{def 2}
\end{equation}%
with $\mathbf{u}_{\infty ,\infty }=\mathbf{1}$, $\mathbf{v}_{\infty ,\infty
}=0$,
\begin{eqnarray}
\mathbf{u}_{\infty ,s}\mathbf{u}_{\infty ,s}^{\ast }-\mathbf{v}_{\infty ,s}%
\mathbf{v}_{\infty ,s}^{\ast } &=&\mathbf{1}\ ,\text{\qquad }\mathbf{u}%
_{\infty ,s}\mathbf{v}_{\infty ,s}^{\mathrm{t}}=\mathbf{v}_{\infty ,s}%
\mathbf{u}_{\infty ,s}^{\mathrm{t}}\ , \\
\mathbf{u}_{\infty ,s}^{\ast }\mathbf{u}_{\infty ,s}-\mathbf{v}_{\infty ,s}^{%
\mathrm{t}}\mathbf{\bar{v}}_{\infty ,s} &=&\mathbf{1}\ ,\qquad \mathbf{u}%
_{\infty ,s}^{\ast }\mathbf{v}_{\infty ,s}=\mathbf{v}_{\infty ,s}^{\mathrm{t}%
}\mathbf{\bar{u}}_{\infty ,s}\ ,
\end{eqnarray}%
and
\begin{align}
\Vert \mathbf{u}_{\infty ,s}-\mathbf{1}\Vert _{2}& \leq \cosh \left\{
4\int_{s}^{\infty }\Vert B_{\tau }\Vert _{2}\ \mathrm{d}\tau \right\} -1\ ,
\\
\Vert \mathbf{v}_{\infty ,s}\Vert _{2}& \leq \sinh \left\{ 4\int_{s}^{\infty
}\Vert B_{\tau }\Vert _{2}\ \mathrm{d}\tau \right\} \ .
\end{align}
\end{lemma}

\noindent \textbf{Proof.} By using (\ref{generalized Bog transf 2}), written
as two integral equations, and (\ref{generalized Bog transf 2bisbis}) we get
the upper bound
\begin{eqnarray}
&&\left\Vert \mathbf{u}_{t_{1},s}-\mathbf{u}_{t_{2},s}\right\Vert
_{2}+\left\Vert \mathbf{v}_{t_{1},s}-\mathbf{v}_{t_{2},s}\right\Vert _{2}
\notag \\
&\leq &\left( \exp \left\{ 4\int_{t_{1}}^{t_{2}}\left\Vert B_{\tau
}\right\Vert _{2}\ \mathrm{d}\tau \right\} -1\right) \exp \left\{
4\int_{s}^{t_{1}}\left\Vert B_{\tau }\right\Vert _{2}\ \mathrm{d}\tau
\right\}
\end{eqnarray}%
for any $t_{2}\geq t_{1}\geq s\geq 0$. In fact, by (\ref{generalized Bog
transf 2bis}) and (\ref{def 1})--(\ref{def 2}), this inequality also holds
for $t_{2}=\infty $, provided the map $t\mapsto \Vert B_{t}\Vert _{2}$ is
integrable on $[0,\infty )$. Consequently, the Hilbert--Schmidt operators $(%
\mathbf{u}_{t,s}-\mathbf{1})$ and $\mathbf{v}_{t,s}$ defined by (\ref%
{generalized Bog transf 2bis}) converge in $\mathcal{L}^{2}(\mathfrak{h})$,
respectively to the operators $(\mathbf{u}_{\infty ,s}-\mathbf{1})$ and $%
\mathbf{v}_{\infty ,s}$ defined above. The properties of $\mathbf{u}_{\infty
,s}$ and $\mathbf{v}_{\infty ,s}$ given in the lemma are straightforward to
verify and we omit the details. Also, the operators $\mathbf{u}_{\infty
,\infty }$ and $\mathbf{v}_{\infty ,\infty }$ are defined by taking the
limit $s\rightarrow \infty $. It is easy to check that $\mathbf{u}_{\infty
,\infty }=\mathbf{1}$ and $\mathbf{v}_{\infty ,\infty }=0$. \hfill $\Box $

This lemma corresponds to the first statement (i) of Theorem \ref{theorem
important 3}, using Theorem \ref{theorem important 1 -3}. Now, we prove its
second statement (ii), again assuming the integrability of the map $t\mapsto
\Vert B_{t}\Vert _{2}$ on $[0,\infty )$.

\begin{lemma}[Limits of $\mathrm{U}_{t,s}$ and $\mathrm{U}_{t,s}^{\ast }$ as
$t\rightarrow \infty $]
\label{lemma unitary asymptotics 1 copy(1)}\mbox{ }\newline
Assume Conditions A1--A4 and the integrability of $t\mapsto \Vert B_{t}\Vert
_{2}$. Then, as $t\rightarrow \infty $, the unitary operator $\mathrm{U}%
_{t,s}$ strongly converges to a strongly continuous in $s\in \mathbb{R}%
_{0}^{+}\cup \left\{ \infty \right\} $ unitary operator $\mathrm{U}_{\infty
,s}$ satisfying the non--autonomous evolution equation
\begin{equation}
\forall s\geq 0:\qquad \partial _{s}\mathrm{U}_{\infty ,s}=i\mathrm{U}%
_{\infty ,s}\mathrm{G}_{s}\ ,\quad \mathrm{U}_{\infty ,\infty }=\mathbf{1}\ ,
\end{equation}%
on the domain $\mathcal{D}\left( \mathrm{N}\right) $ and the cocycle
property $\mathrm{U}_{\infty ,s}=\mathrm{U}_{\infty ,x}\mathrm{U}_{x,s}$ for
$x\geq s\geq 0$. Moreover, its adjoint $\mathrm{U}_{t,s}^{\ast }$ also
converges in the strong topology to $\mathrm{U}_{\infty ,s}^{\ast }$ when $%
t\rightarrow \infty $.
\end{lemma}

\noindent \textbf{Proof.} From straightforward estimations using Lemmata \ref%
{Section exist unique U quadratic}, \ref{lemma unitary existence 2}, and (%
\ref{inegality BB 3}) we obtain that, for any $t_{2}\geq s_{2}\geq s_{1}$
and $t_{2}\geq t_{1}\geq s_{1}$,%
\begin{eqnarray}
\Vert (\mathrm{U}_{t_{2},s_{2}}-\mathrm{U}_{t_{1},s_{1}})(\mathrm{N}+\mathbf{%
1})^{-1}\Vert _{\mathrm{op}} &\leq &11\int_{s_{1}}^{s_{2}}\Vert B_{\tau
}\Vert _{2}\ \mathrm{d}\tau  \label{flow equation convergence unitary 1} \\
&&+\frac{1}{2}\exp \left\{ 22\int_{s_{1}}^{t_{1}}\Vert B_{\tau }\Vert _{2}\
\mathrm{d}\tau \right\}  \notag \\
&&\times \left( \exp \left\{ 22\int_{t_{1}}^{t_{2}}\Vert B_{\tau }\Vert
_{2}\ \mathrm{d}\tau \right\} -1\right) \ .  \notag
\end{eqnarray}%
In particular, if the map $t\mapsto \Vert B_{t}\Vert _{2}$ is integrable on $%
[0,\infty )$ then by taking the limit $\min \left\{ t_{1},t_{2}\right\}
\rightarrow \infty $\ and choosing $s_{1}:=s_{2}:=s$ in (\ref{flow equation
convergence unitary 1}) we show that the operator
\begin{equation}
\{\mathrm{U}_{t_{2},s}-\mathrm{U}_{t_{1},s}\}(\mathrm{N}+\mathbf{1})^{-1}
\end{equation}%
converges to zero in norm. Hence, by standard arguments (cf. (\ref{Cauchy
sequence 1})--(\ref{Cauchy sequence 3})), there exists a bounded operator $%
\mathrm{U}_{\infty ,s}$ defined, for any $\varphi \in \mathcal{F}_{b}$ and $%
s\geq 0$, by the strong limit
\begin{equation}
\mathrm{U}_{\infty ,s}\varphi :=\underset{t\rightarrow \infty }{\lim }%
\left\{ \mathrm{U}_{t,s}\varphi \right\} \ .
\label{flow equation convergence unitary 1bisbis}
\end{equation}%
In particular, by the unitarity of the operator $\mathrm{U}_{t,s}$ for all $%
t\geq s\geq 0$,
\begin{equation}
\forall s\geq 0:\qquad \left\Vert \mathrm{U}_{\infty ,s}\right\Vert _{%
\mathrm{op}}\leq 1\ .  \label{bound for limit unitarity}
\end{equation}%
From (\ref{flow equation convergence unitary 1}) we also obtain
\begin{equation}
\mathrm{U}_{\infty ,\infty }\varphi :=\underset{s\rightarrow \infty }{\lim }%
\mathrm{U}_{\infty ,s}\varphi =\underset{s\rightarrow \infty }{\lim }\left\{
\underset{t\rightarrow \infty }{\lim }\mathrm{U}_{t,s}\varphi \right\} =%
\underset{t\rightarrow \infty }{\lim }\mathrm{U}_{t,t}\varphi =\mathbf{1}\ .
\label{flow equation convergence unitary 1bis}
\end{equation}

By using similar arguments as in the proof of Lemma \ref{lemma unitary
propagator}, one additionally verifies that $\mathrm{U}_{\infty ,s}$ is
strongly continuous in $s$ and $\mathrm{U}_{\infty ,s}=\mathrm{U}_{\infty ,x}%
\mathrm{U}_{x,s}$ for any $x\geq s\geq 0$. Moreover, for any sufficiently
small parameter $\left\vert \epsilon \right\vert >0$, $t>s+\epsilon $, and $%
\varphi \in \mathcal{D}\left( \mathrm{N}^{2}\right) \subseteq \mathcal{D}%
\left( \mathrm{G}_{s}\right) $ (cf. (\ref{inegality BB 3})),%
\begin{eqnarray}
&&\Vert \{\epsilon ^{-1}(\mathrm{U}_{\infty ,s+\epsilon }-\mathrm{U}_{\infty
,s})\varphi -i\mathrm{U}_{\infty ,s}\mathrm{G}_{s}\}\varphi \Vert  \notag \\
&\leq &\Vert \{\mathrm{U}_{\infty ,s}-\mathrm{U}_{t,s}\}\mathrm{G}%
_{s}\varphi \Vert +\epsilon ^{-1}\int_{s}^{s+\epsilon }\Vert \{\mathrm{G}%
_{\tau }-\mathrm{G}_{s}\}\varphi \Vert \mathrm{d}\tau  \notag \\
&&+\epsilon ^{-1}\int_{s}^{s+\epsilon }\Vert \{\mathrm{U}_{t,\tau }-\mathrm{U%
}_{t,s}\}\left( \mathrm{N}+\mathbf{1}\right) ^{-1}\Vert _{\mathrm{op}}\Vert
\mathrm{G}_{s}\left( \mathrm{N}+\mathbf{1}\right) ^{-1}\Vert _{\mathrm{op}}%
\mathrm{d}\tau  \notag \\
&&+\epsilon ^{-1}\int_{s}^{s+\epsilon }\Vert \{\mathrm{U}_{t,\tau }-\mathrm{U%
}_{t,s}\}\left( \mathrm{N}+\mathbf{1}\right) ^{-1}\Vert _{\mathrm{op}}\Vert %
\left[ \mathrm{N},\mathrm{G}_{s}\right] \left( \mathrm{N}+\mathbf{1}\right)
^{-1}\Vert _{\mathrm{op}}\mathrm{d}\tau  \notag \\
&&+\epsilon ^{-1}\Vert \{\mathrm{U}_{\infty ,s}-\mathrm{U}_{t,s}\}\varphi
\Vert +\epsilon ^{-1}\Vert \{\mathrm{U}_{\infty ,s+\epsilon }-\mathrm{U}%
_{t,s+\epsilon }\}\varphi \Vert \ .
\label{flow equation convergence unitary 2}
\end{eqnarray}%
Since $\mathrm{G}_{t}\in C[\mathbb{R}_{0}^{+};\mathcal{B}(\mathcal{Y},%
\mathcal{F}_{b})]$ and
\begin{equation}
\left[ \mathrm{N},\mathrm{G}_{s}\right] \left( \mathrm{N}+\mathbf{1}\right)
^{-1}\in \mathcal{B}\left( \mathcal{F}_{b}\right)
\end{equation}%
(cf. Lemma \ref{lemma unitary existence 1}), we take $t=t(\epsilon
)\rightarrow \infty $ as $\epsilon \rightarrow 0$ in (\ref{flow equation
convergence unitary 2}) and use (\ref{flow equation convergence unitary 1})
together with standard arguments to get that $\partial _{s}\mathrm{U}%
_{\infty ,s}=i\mathrm{U}_{\infty ,s}\mathrm{G}_{s}$ for any $\varphi \in
\mathcal{D}\left( \mathrm{N}^{2}\right) \subseteq \mathcal{D}\left( \mathrm{G%
}_{s}\right) $. Remark that $\mathcal{D}\left( \mathrm{N}^{2}\right) $ is
clearly a core for the particle number operator $\mathrm{N}$. As a
consequence, for any vector $\varphi \in \mathcal{D}(\mathrm{N})$ there is a
sequence $\{\varphi _{n}\}_{n=0}^{\infty }\subset \mathcal{D}(\mathrm{N}%
^{2}) $ converging to $\varphi $ such that $\{\mathrm{N}\varphi
_{n}\}_{n=0}^{\infty }\subset \mathcal{F}_{b}$ converges to $\mathrm{N}%
\varphi \in \mathcal{F}_{b}$. Since, for any $n\in \mathbb{N}$, $s\geq 0$,
and sufficiently small $\left\vert \epsilon \right\vert >0$,%
\begin{eqnarray}
\Vert (\mathrm{U}_{\infty ,s+\epsilon }-\mathrm{U}_{\infty ,s})\left(
\varphi -\varphi _{n}\right) \Vert &\leq &\Vert (\mathrm{U}_{\infty
,s+\epsilon }-\mathrm{U}_{\infty ,s})\left( \mathrm{N}+\mathbf{1}\right)
^{-1}\Vert _{\mathrm{op}}  \notag \\
&&\times \left( \Vert \mathrm{N}\left( \varphi -\varphi _{n}\right) \Vert
+\Vert \varphi -\varphi _{n}\Vert \right)  \notag \\
&&
\end{eqnarray}%
and%
\begin{equation}
\Vert \mathrm{U}_{\infty ,s}\mathrm{G}_{s}\left( \varphi -\varphi
_{n}\right) \Vert \leq \Vert \mathrm{G}_{s}\left( \mathrm{N}+\mathbf{1}%
\right) ^{-1}\Vert _{\mathrm{op}}+\left( \Vert \mathrm{N}\left( \varphi
-\varphi _{n}\right) \Vert +\Vert \varphi -\varphi _{n}\Vert \right) \ ,
\end{equation}%
we infer from\ $\partial _{s}\mathrm{U}_{\infty ,s}\varphi _{n}=i\mathrm{U}%
_{\infty ,s}\mathrm{G}_{s}\varphi _{n}$, Lemma \ref{lemma unitary existence
1}, and (\ref{flow equation convergence unitary 1}) that the derivative $%
\partial _{s}\mathrm{U}_{\infty ,s}=i\mathrm{U}_{\infty ,s}\mathrm{G}_{s}$
holds on the domain $\mathcal{D}\left( \mathrm{N}\right) $, for any $s\geq 0$%
.

Now, the fact that $\mathrm{U}_{t,s}^{\ast }$ strongly converges to a
bounded operator $\mathrm{U}_{\infty ,s}^{\ast }$ results from standard
arguments using the inequality
\begin{equation}
\Vert (\mathrm{U}_{t_{2},s}^{\ast }-\mathrm{U}_{t_{1},s}^{\ast })(\mathrm{N}+%
\mathbf{1})^{-1}\Vert _{\mathrm{op}}\leq 11\int_{t_{1}}^{t_{2}}\Vert B_{\tau
}\Vert _{2}\ \mathrm{d}\tau \ ,  \label{unitarity +3}
\end{equation}%
which is deduced, for any $t_{2}\geq t_{1}\geq s$, from (\ref{inegality BB 3}%
) and Lemma \ref{Section exist unique U quadratic copy(1)}. Moreover,
similar to (\ref{flow equation convergence unitary 1bis}),
\begin{equation}
\mathrm{U}_{\infty ,\infty }^{\ast }\varphi :=\underset{s\rightarrow \infty }%
{\lim }\mathrm{U}_{\infty ,s}^{\ast }\varphi =\mathbf{1}\ .
\label{flow equation convergence unitary 1bis+1}
\end{equation}%
See also similar arguments to establishing Lemma \ref{Section exist unique U
quadratic copy(1)}, in particular (\ref{eq trivial self adjoint}).

Finally, the unitarity of $\mathrm{U}_{\infty ,s}$ is straightforward to
verify. Indeed, using the unitary of the operator $\mathrm{U}_{t,s}$ for all
$t\geq s\geq 0$,
\begin{eqnarray}
\left\Vert \left( \mathrm{U}_{\infty ,s}^{\ast }\mathrm{U}_{\infty ,s}-%
\mathbf{1}\right) \varphi \right\Vert &=&\left\Vert \left( \mathrm{U}%
_{\infty ,s}^{\ast }\mathrm{U}_{\infty ,s}-\mathrm{U}_{t,s}^{\ast }\mathrm{U}%
_{t,s}\right) \varphi \right\Vert  \notag \\
&\leq &\left\Vert \left( \mathrm{U}_{\infty ,s}-\mathrm{U}_{t,s}\right)
\varphi \right\Vert +\left\Vert \left( \mathrm{U}_{\infty ,s}^{\ast }-%
\mathrm{U}_{t,s}^{\ast }\right) \mathrm{U}_{\infty ,s}\varphi \right\Vert \ ,
\notag \\
&&  \label{unitarity +1}
\end{eqnarray}%
for any $\varphi \in \mathcal{F}_{b}$, and%
\begin{eqnarray}
\left\Vert \left( \mathrm{U}_{\infty ,s}\mathrm{U}_{\infty ,s}^{\ast }-%
\mathbf{1}\right) \varphi \right\Vert &=&\left\Vert \left( \mathrm{U}%
_{\infty ,s}\mathrm{U}_{\infty ,s}^{\ast }-\mathrm{U}_{t,s}\mathrm{U}%
_{t,s}^{\ast }\right) \varphi \right\Vert  \notag \\
&\leq &\left\Vert \mathrm{U}_{\infty ,s}^{\ast }-\mathrm{U}_{t,s}^{\ast
}\varphi \right\Vert +\left\Vert \left( \mathrm{U}_{\infty ,s}-\mathrm{U}%
_{t,s}\right) \mathrm{U}_{\infty ,s}^{\ast }\varphi \right\Vert \ .  \notag
\\
&&  \label{unitarity +2}
\end{eqnarray}%
Therefore, since, as $t\rightarrow \infty $, $\mathrm{U}_{t,s}$ (resp. $%
\mathrm{U}_{t,s}^{\ast }$) strongly converges to $\mathrm{U}_{\infty ,s}$
(resp. $\mathrm{U}_{t,s}^{\ast }$), Equations (\ref{flow equation
convergence unitary 1bis}), (\ref{flow equation convergence unitary 1bis+1}%
), (\ref{unitarity +1}) and (\ref{unitarity +2}) yield%
\begin{equation}
\forall s\in \mathbb{R}_{0}^{+}\cup \left\{ \infty \right\} :\qquad \mathrm{U%
}_{\infty ,s}^{\ast }\mathrm{U}_{\infty ,s}=\mathrm{U}_{\infty ,s}\mathrm{U}%
_{\infty ,s}^{\ast }=\mathbf{1}\ .
\end{equation}%
\hfill $\Box $

We continue the proof of Theorem \ref{theorem important 3} by showing that $%
\mathrm{U}_{\infty ,s}$ realizes a Bogoliubov $\mathbf{u}$--$\mathbf{v}$
transformation for any $s\in \mathbb{R}_{0}^{+}\cup \left\{ \infty \right\} $
(Theorem \ref{theorem important 3} (iii)).

\begin{lemma}[The Bogoliubov $\mathbf{u}$--$\mathbf{v}$ transformation]
\label{lemma unitary asymptotics 1 copy(2)}\mbox{ }\newline
Assume Conditions A1--A4 and the integrability of $t\mapsto \Vert B_{t}\Vert
_{2}$. Then, for all $s\in \mathbb{R}_{0}^{+}\cup \left\{ \infty \right\} $,
the unitary operator $\mathrm{U}_{\infty ,s}$ satisfies on $\mathcal{D}(%
\mathrm{N}^{1/2})$:%
\begin{equation}
\forall k\in \mathbb{N}:\quad \mathrm{U}_{\infty ,s}a_{s,k}\mathrm{U}%
_{\infty ,s}^{\ast }=\underset{\ell }{\sum }\left\{ \mathbf{u}_{\infty
,s}\right\} _{k,\ell }a_{\ell }+\left\{ \mathbf{v}_{\infty ,s}\right\}
_{k,\ell }a_{\ell }^{\ast }\ ,
\end{equation}%
where the annihilation operator $a_{s,k}$ is defined in\ Lemma \ref{lemma
uv-Bogoliubov transformation}.
\end{lemma}

\noindent \textbf{Proof.} Inequality (\ref{condition 3bis}) tells us that%
\begin{equation}
\Vert \lbrack \mathrm{N},\mathrm{G}_{t}](\mathrm{N}+\mathbf{1})^{-1}\Vert _{%
\mathrm{op}}\leq 22\Vert B_{t}\Vert _{2}  \label{inegality BB 3-bisbis}
\end{equation}%
and the upper bound (\ref{conservation of domain N by Ubis}) yields
\begin{equation}
\Vert (\mathrm{N}+\mathbf{1})\mathrm{U}_{t,s}^{\ast }(\mathrm{N}+\mathbf{1}%
)^{-1}\Vert _{\mathrm{op}}\leq \exp \left\{ 22\int_{s}^{\infty }\Vert
B_{\tau }\Vert _{2}\ \mathrm{d}\tau \right\} \ ,
\label{inegality BB 3-bisbisbis}
\end{equation}%
provided the map $t\mapsto \Vert B_{t}\Vert _{2}$ is integrable on $%
[0,\infty )$. So, we can follow similar arguments as the ones given in the
proof of Lemma \ref{existence flow with W 4bis}. In particular, we can
express the uniformly bounded operator
\begin{equation}
(\mathrm{N}+\mathbf{1})\mathrm{U}_{t,s}^{\ast }(\mathrm{N}+\mathbf{1})^{-1}
\end{equation}%
in terms of $\mathrm{U}_{t,s}^{\ast }$ and the bounded operator
\begin{equation}
\lbrack \mathrm{N},\mathrm{G}_{t}](\mathrm{N}+\mathbf{1})^{-1}\in \mathcal{B}%
\left( \mathcal{F}_{b}\right) \ .
\end{equation}%
Indeed, using (\ref{inegality BB 3-bisbis})--(\ref{inegality BB 3-bisbisbis}%
) and the non--autonomous evolution equations (\ref{flow
equationbis-new-adjoint}), we observe that the bounded operators
\begin{equation}
\mathrm{U}_{t,s}^{\ast }\qquad \mathrm{and}\qquad \left( \mathrm{N}+\mathbf{1%
}\right) \mathrm{U}_{t,s}^{\ast }\left( \mathrm{N}+\mathbf{1}\right) ^{-1}
\end{equation}%
satisfy the equality%
\begin{eqnarray}
&&\left( \left( \mathrm{N}+\mathbf{1}\right) \mathrm{U}_{t,s}^{\ast }\left(
\mathrm{N}+\mathbf{1}\right) ^{-1}-\mathrm{U}_{t,s}^{\ast }\right) \left(
\mathrm{N}+\mathbf{1}\right) ^{-1}  \notag \\
&=&-\int_{s}^{t}\partial _{\tau }\left\{ \mathrm{U}_{\tau ,s}^{\ast }\left(
\mathrm{N}+\mathbf{1}\right) ^{-1}\left( \mathrm{N}+\mathbf{1}\right) ^{2}%
\mathrm{U}_{t,\tau }^{\ast }\left( \mathrm{N}+\mathbf{1}\right)
^{-2}\right\} \mathrm{d}\tau  \notag \\
&=&i\int_{s}^{t}\mathrm{U}_{\tau ,s}^{\ast }\left[ \mathrm{N},\mathrm{G}%
_{\tau }\right] \left( \mathrm{N}+\mathbf{1}\right) ^{-1}\left( \mathrm{N}+%
\mathbf{1}\right) \mathrm{U}_{t,\tau }^{\ast }\left( \mathrm{N}+\mathbf{1}%
\right) ^{-2}\mathrm{d}\tau \ .  \label{GammaUgamma as a serie0}
\end{eqnarray}%
In other words, as $\mathcal{D}\left( \mathrm{N}\right) $ is a dense subset
of $\mathcal{F}_{b}$, one obtains
\begin{eqnarray}
(\mathrm{N}+\mathbf{1})\mathrm{U}_{t,s}^{\ast }\left( \mathrm{N}+\mathbf{1}%
\right) ^{-1} &=&\mathrm{U}_{t,s}^{\ast }+\underset{n=1}{\overset{\infty }{%
\sum }}i^{n}\int_{s}^{t}\mathrm{d}\tau _{1}\cdots \int_{s}^{\tau _{n-1}}%
\mathrm{d}\tau _{n}  \label{GammaUgamma as a serie} \\
&&\underset{j=1}{\overset{n}{\prod }}\mathrm{U}_{\tau _{j},s}^{\ast }\left\{ %
\left[ \mathrm{N},\mathrm{G}_{\tau _{j}}\right] \left( \mathrm{N}+\mathbf{1}%
\right) ^{-1}\right\} \mathrm{U}_{\tau _{j-1},\tau _{j}}^{\ast }  \notag
\end{eqnarray}%
on the whole Hilbert space $\mathcal{F}_{b}$, where $\tau _{0}:=t$. By (\ref%
{inegality BB 3-bisbis}) and the unitarity of $\mathrm{U}_{t,s}^{\ast }$ for
all $t\geq s\geq 0$, note that%
\begin{eqnarray}
&&\left\Vert \underset{n=N}{\overset{\infty }{\sum }}i^{n}\int_{s}^{t}%
\mathrm{d}\tau _{1}\cdots \int_{s}^{\tau _{n-1}}\mathrm{d}\tau _{n}\underset{%
j=1}{\overset{n}{\prod }}\mathrm{U}_{\tau _{j},s}^{\ast }\left\{ \left[
\mathrm{N},\mathrm{G}_{\tau _{j}}\right] \left( \mathrm{N}+\mathbf{1}\right)
^{-1}\right\} \mathrm{U}_{\tau _{j-1},\tau _{j}}^{\ast }\right\Vert _{%
\mathrm{op}}  \notag \\
&\leq &\frac{\left( 22\int_{s}^{\infty }\Vert B_{\tau }\Vert _{2}\mathrm{d}%
\tau \right) ^{N}}{N!}\exp \left\{ 22\int_{s}^{\infty }\Vert B_{\tau }\Vert
_{2}\mathrm{d}\tau \right\}  \label{GammaUgamma as a seriebis}
\end{eqnarray}%
for any $N\geq 1$. In other words, the series (\ref{GammaUgamma as a serie})
is norm convergent, uniformly in $t>s$. Observe that the particle number
operator $\mathrm{N}$ is a closed operator. Consequently, using the limit (%
\ref{flow equation convergence unitary 1bisbis}) and the upper bound (\ref%
{inegality BB 3-bisbis}) together with the unitarity of $\mathrm{U}%
_{t,s}^{\ast }$ and (\ref{GammaUgamma as a serie}) we get
\begin{equation}
(\mathrm{N}+\mathbf{1})\mathrm{U}_{\infty ,s}^{\ast }\left( \mathrm{N}+%
\mathbf{1}\right) ^{-1}\varphi =\underset{t\rightarrow \infty }{\lim }%
\left\{ \left( \mathrm{N}+\mathbf{1}\right) \mathrm{U}_{t,s}^{\ast }\left(
\mathrm{N}+\mathbf{1}\right) ^{-1}\varphi \right\}
\label{commutateur limit 0}
\end{equation}%
for all $s\geq 0$ and any vector $\varphi \in \mathcal{F}_{b}$, where
\begin{equation}
\Vert \left( \mathrm{N}+\mathbf{1}\right) \mathrm{U}_{\infty ,s}^{\ast
}\left( \mathrm{N}+\mathbf{1}\right) ^{-1}\Vert _{\mathrm{op}}\leq \exp
\left\{ 22\int_{s}^{\infty }\Vert B_{\tau }\Vert _{2}\ \mathrm{d}\tau
\right\} <\infty \ .  \label{commutateur limit 0bis}
\end{equation}%
Now, for any $k\in \mathbb{N}$, we define on $\mathcal{D}\left( \mathrm{N}%
^{1/2}\right) $ the asymptotic annihilation operator $a_{\infty ,k}$ to be%
\begin{equation}
a_{\infty ,k}:=\sum_{\ell }\left\{ \mathbf{u}_{\infty ,s}\right\} _{k,\ell
}a_{\ell }+\left\{ \mathbf{v}_{\infty ,s}\right\} _{k,\ell }a_{\ell }^{\ast
}=a(\mathbf{u}_{\infty }^{\ast }\varphi _{k})+a^{\ast }(\mathbf{v}_{\infty
}^{\mathrm{t}}\bar{\varphi}_{k})\ ,  \label{generalized Bog transf 0bis}
\end{equation}%
where the bounded operators $\mathbf{u}_{\infty ,s}$ and $\mathbf{v}_{\infty
,s}$ are defined in Lemma \ref{lemma unitary asymptotics 1} and $\mathbf{u}%
_{\infty }:=\mathbf{u}_{\infty ,0}$, $\mathbf{v}_{\infty }:=\mathbf{v}%
_{\infty ,0}$. Recall that $\left\{ \varphi _{k}\right\} _{k=1}^{\infty }$
is some real orthonormal basis in $\mathcal{D}\left( \Omega _{0}\right)
\subseteq \mathfrak{h}$ and $a_{k}:=a\left( \varphi _{k}\right) $ is the
standard boson annihilation operator acting on the boson Fock space $%
\mathcal{F}_{b}$. We observe next that, for all $\varphi \in \mathcal{F}_{b}$
and $k\in \mathbb{N}$,%
\begin{eqnarray}
&&\Vert (a_{\infty ,k}-\mathrm{U}_{\infty ,s}a_{s,k}\mathrm{U}_{\infty
,s}^{\ast })(\mathrm{N}+\mathbf{1})^{-1}\varphi \Vert  \notag \\
&\leq &\Vert (\mathrm{U}_{t,s}-\mathrm{U}_{\infty ,s})a_{s,k}(\mathrm{N}+%
\mathbf{1})^{-1}(\mathrm{N}+\mathbf{1})\mathrm{U}_{\infty ,s}^{\ast }(%
\mathrm{N}+\mathbf{1})^{-1}\varphi \Vert  \notag \\
&&+\Vert \mathrm{U}_{t,s}\Vert _{\mathrm{op}}\Vert a_{s,k}(\mathrm{N}+%
\mathbf{1})^{-1}\Vert _{\mathrm{op}}\Vert (\mathrm{N}+\mathbf{1})(\mathrm{U}%
_{t,s}^{\ast }-\mathrm{U}_{\infty ,s}^{\ast })(\mathrm{N}+\mathbf{1}%
)^{-1}\varphi \Vert  \notag \\
&&+\Vert (a_{\infty ,k}-\mathrm{U}_{t,s}a_{s,k}\mathrm{U}_{t,s}^{\ast })(%
\mathrm{N}+\mathbf{1})^{-1}\Vert _{\mathrm{op}}\ .
\label{commutator limit 0000}
\end{eqnarray}%
Because of (\ref{bogo transf new1}), for all $s\geq 0$ and $k\in \mathbb{N}$%
,
\begin{equation}
\Vert a_{s,k}(\mathrm{N}+\mathbf{1})^{-1}\Vert _{\mathrm{op}}<\infty \ .
\label{commutator limit 000obis}
\end{equation}%
Furthermore, by (\ref{generalized Bog transf 4}), that is,
\begin{equation}
a_{t,k}:=a(\mathbf{u}_{t}^{\ast }\varphi _{k})+a^{\ast }(\mathbf{v}_{t}^{%
\mathrm{t}}\bar{\varphi}_{k})=\mathrm{U}_{t,s}a_{s,k}\mathrm{U}_{t,s}^{\ast }
\end{equation}%
on $\mathcal{D}(\mathrm{N}^{1/2})$, straightforward estimations as in Lemma %
\ref{lemma estimates} imply that, for all $t\geq s\geq 0$ and $k\in \mathbb{N%
}$,
\begin{equation}
\Vert (a_{\infty ,k}-\mathrm{U}_{t,s}a_{s,k}\mathrm{U}_{t,s}^{\ast })(%
\mathrm{N}+\mathbf{1})^{-1}\Vert _{\mathrm{op}}\leq \Vert \mathbf{u}_{\infty
,s}-\mathbf{u}_{t,s}\Vert _{2}+2\Vert \mathbf{v}_{\infty ,s}-\mathbf{v}%
_{t,s}\Vert _{2}\ .  \label{commutator limit 0001}
\end{equation}%
Hence, for all $s\geq 0$ and $k\in \mathbb{N}$, we combine Lemma \ref{lemma
unitary asymptotics 1} with Inequality (\ref{commutator limit 0001}) and get
the limit
\begin{equation}
\underset{t\rightarrow \infty }{\lim }\Vert (a_{\infty ,k}-\mathrm{U}%
_{t,s}a_{s,k}\mathrm{U}_{t,s}^{\ast })(\mathrm{N}+\mathbf{1})^{-1}\Vert _{%
\mathrm{op}}=0\ ,  \label{commutator limit 000}
\end{equation}%
which, by the upper bounds (\ref{commutateur limit 0bis}), (\ref{commutator
limit 0000}), and (\ref{commutator limit 000obis}), together with Lemma \ref%
{lemma unitary asymptotics 1 copy(1)} and (\ref{commutateur limit 0}),
implies that
\begin{equation}
\forall \varphi \in \mathcal{D}\left( \mathrm{N}\right) ,\ s\geq 0,\ k\in
\mathbb{N}:\qquad a_{\infty ,k}\varphi =\mathrm{U}_{\infty ,s}a_{s,k}\mathrm{%
U}_{\infty ,s}^{\ast }\varphi \ ,  \label{equality on N3/2}
\end{equation}%
provided the map $t\mapsto \Vert B_{t}\Vert _{2}$ is integrable on $%
[0,\infty )$. The domain $\mathcal{D}(\mathrm{N})$ is a core for each
element of the family $(a_{\infty ,k})_{k\in \mathbb{N},t\geq 0}$ of closed
operators, which satisfy%
\begin{equation}
a_{\infty ,k}(\mathrm{N}+\mathbf{1)}^{-1/2}\in \mathcal{B}\left( \mathcal{F}%
_{b}\right) \ ,
\end{equation}%
see (\ref{generalized Bog transf 0bis}), Lemma \ref{lemma unitary
asymptotics 1}, and the arguments used to prove (\ref{bogo transf new1}).
Furthermore, $\mathrm{U}_{\infty ,s}$ is unitary and $a_{s,k}$ is a closed
operator, see Lemmata \ref{lemma uv-Bogoliubov transformation} and \ref%
{lemma unitary asymptotics 1 copy(1)}. Therefore, (\ref{equality on N3/2})
can be extended by continuity to $\mathcal{D}(\mathrm{N}^{1/2})$, just as we
did for (\ref{extension equality}). \hfill $\Box $

Note that the previous lemma shows that
\begin{equation}
\mathcal{D}(\mathrm{N}^{1/2})\subset \mathcal{D}(\mathrm{U}_{\infty
,s}a_{s,k}\mathrm{U}_{\infty ,s}^{\ast })\ .
\end{equation}%
This fact can directly be seen from the inequality
\begin{equation}
\Vert (\mathrm{N}+\mathbf{1)}^{1/2}\mathrm{U}_{\infty ,s}^{\ast }(\mathrm{N}+%
\mathbf{1)}^{-1/2}\Vert _{\mathrm{op}}\leq \mathrm{\exp }\left\{
88\int_{s}^{\infty }\Vert B_{\tau }\Vert _{2}\mathrm{d}\tau \right\}
\end{equation}%
for all $s\geq 0$. Assuming (\ref{conservation domain U N1/2}) one gets this
upper bound in the same way we have proven (\ref{commutateur limit 0bis}).

We are in position to conclude that $\mathrm{H}_{s}$ and $\mathrm{H}_{\infty
}$ are unitarily equivalent, i.e., to prove the fourth statement (iv) of
Theorem \ref{theorem important 3}.

\begin{lemma}[Unitary equivalence of $\mathrm{H}_{s}$ and $\mathrm{H}%
_{\infty }$]
\label{lemma convergence in strong resolvent sence of H copy(1)}\mbox{ }%
\newline
Assume Conditions A1--A4 and the integrability of $t\mapsto \Vert B_{t}\Vert
_{2}$. Then
\begin{equation*}
\forall s\geq 0:\qquad \mathrm{H}_{\infty }=\mathrm{U}_{\infty ,s}\mathrm{H}%
_{s}\mathrm{U}_{\infty ,s}^{\ast }\ ,
\end{equation*}%
where $\mathrm{U}_{\infty ,s}$ is the unitary operator defined in Lemma \ref%
{lemma unitary asymptotics 1 copy(1)}.
\end{lemma}

\noindent \textbf{Proof.} Let $\lambda >11\Vert B_{0}\Vert _{2}$. Because $%
\mathrm{U}_{t,s}$ is unitary and
\begin{equation}
\forall t\geq s\geq 0:\qquad \left( \mathrm{H}_{t}+i\lambda \mathbf{1}%
\right) ^{-1}=\mathrm{U}_{t,s}\left( \mathrm{H}_{s}+i\lambda \mathbf{1}%
\right) ^{-1}\mathrm{U}_{t,s}^{\ast }\ ,  \label{resolvant eq 1}
\end{equation}%
(see (\ref{eq sup encore new 8})), one gets, for any $\varphi \in \mathcal{F}%
_{b}$ and $t\geq s\geq 0$,%
\begin{eqnarray}
&&\Vert \{(\mathrm{H}_{\infty }+i\lambda \mathbf{1})^{-1}-\mathrm{U}_{\infty
,s}(\mathrm{H}_{s}+i\lambda \mathbf{1})^{-1}\mathrm{U}_{\infty ,s}^{\ast
}\}\varphi \Vert  \notag \\
&\leq &\Vert \{(\mathrm{H}_{\infty }+i\lambda \mathbf{1})^{-1}-(\mathrm{H}%
_{t}+i\lambda \mathbf{1})^{-1}\}\varphi \Vert  \notag \\
&&+\Vert \left( \mathrm{H}_{s}+i\lambda \mathbf{1}\right) ^{-1}\Vert _{%
\mathrm{op}}\ \Vert (\mathrm{U}_{t,s}^{\ast }-\mathrm{U}_{\infty ,s}^{\ast
})\varphi \Vert  \notag \\
&&+\Vert (\mathrm{U}_{t,s}-\mathrm{U}_{\infty ,s})(\mathrm{H}_{s}+i\lambda
\mathbf{1})^{-1}\mathrm{U}_{\infty ,s}^{\ast }\varphi \Vert \ .
\label{eq converge U 3}
\end{eqnarray}%
Consequently, by Lemmata \ref{lemma convergence in strong resolvent sence of
H} and \ref{lemma unitary asymptotics 1 copy(1)}, we obtain that, for any $%
s\geq 0$,%
\begin{equation}
(\mathrm{H}_{\infty }+i\lambda \mathbf{1})^{-1}=\mathrm{U}_{\infty ,s}(%
\mathrm{H}_{s}+i\lambda \mathbf{1})^{-1}\mathrm{U}_{\infty ,s}^{\ast }\ ,
\label{resolvant eq 3}
\end{equation}%
provided $\lambda >11\Vert B_{0}\Vert _{2}$. We finally use Lemma \ref{lemma
resolvent} (ii) to conclude the proof. \hfill $\Box $

Theorem \ref{theorem important 3} (iv) then follows from Lemma \ref{lemma
convergence in strong resolvent sence of H copy(1)} combined with Theorem %
\ref{theorem important 1 -3}.

\section{Appendix\label{Appendix}}

\setcounter{equation}{0}%
For the reader's convenience and because similar arguments are used above,
we first give in Section \ref{Section Non-autonomous evolution} a detailed
analysis of non--autonomous evolution equations for unbounded operators of
hyperbolic type on Banach spaces. In particular, we clarify Ishii's approach
\cite{Ishii1,Ishii2} to non--autonomous hyperbolic evolution equations.
Then, Section \ref{section bogoliubov} is devoted to generators of
Bogoliubov $\mathbf{u}$--$\mathbf{v}$ (unitary) transformations, whereas
Section \ref{section hilbert schmidt} should be seen as a toolbox where
useful, simple results related to Hilbert--Schmidt operators and the trace
are proven.

\subsection{Non--Autonomous Evolution Equations on Banach Spaces \label%
{Section Non-autonomous evolution}}

This section is patterned after \cite{bachcourse}.

In quantum mechanics, a well--known example of a non--autonomous evolution
equation is the time-dependent Schrödinger equation. The self--adjoint
generator $G_{t}=G_{t}^{\ast }$ acting on a Hilbert space is interpreted in
this context as a time-dependent Hamiltonian. More generally, the notion of
non--autonomous evolution equations on Banach spaces is well--known in the
general context of abstract quasi--linear evolution equations, see, e.g.,
\cite{Katobis,Caps,Schnaubelt1,Pazy} for reviews of this topic. By using two
Banach spaces $\mathcal{X}$ and $\mathcal{Y}$ with $\mathcal{Y}\subset
\mathcal{X}$ being a dense set in $\mathcal{X}$, standard sufficient
conditions for the well--posedness of non--autonomous evolution equations (%
\ref{flow equationbis}), that is,
\begin{equation}
\forall t\geq s\geq 0:\qquad \partial _{t}U_{t,s}=-iG_{t}U_{t,s}\ ,\quad
U_{s,s}:=\mathbf{1}\ ,  \label{flow equationbisbisbis}
\end{equation}%
are the following:

\begin{itemize}
\item[B1] \textit{(Kato quasi--stability).} There exist a constant $m\geq 1$
and a locally bounded, measurable map $\beta _{0}:\mathbb{R}%
_{0}^{+}\rightarrow \mathbb{R}$ such that
\begin{equation}
\left\Vert \underset{j=1}{\overset{n}{\prod }}\left( \lambda _{j}\mathbf{1}%
+iG_{t_{j}}\right) ^{-1}\right\Vert _{\mathrm{op}}\leq m\underset{j=1}{%
\overset{n}{\prod }}\frac{1}{\lambda _{j}-\beta _{0}\left( t_{j}\right) }
\label{Kato-stability}
\end{equation}%
for any family of real numbers $\left\{ t_{j},\lambda _{j}\right\} $ such
that $0\leq t_{1}\leq \ldots \leq t_{n}$ and $\lambda _{1}>\beta _{0}\left(
t_{1}\right) ,\ldots ,\lambda _{n}>\beta _{0}\left( t_{n}\right) $.

\item[B2] \textit{(Domains and continuity).} One has $\mathcal{Y}\subset
\mathcal{D}\left( G_{t}\right) \ $for any $t\geq 0$ with $G_{t}\in C[\mathbb{%
R}_{0}^{+};\mathcal{B}(\mathcal{Y},\mathcal{X})]$.

\item[B3] \textit{(Intertwining condition).} There exists a closed (linear)
operator $\Theta $ from its dense domain $\mathcal{D}\left( \Theta \right) =%
\mathcal{Y}$ to $\mathcal{X}$ such that the norm
\begin{equation}
\beta _{1}\left( t\right) :=\left\Vert \left[ \Theta ,G_{t}\right]
\right\Vert _{\mathcal{B}(\mathcal{Y},\mathcal{X})}
\end{equation}%
is bounded for any $t\in \mathbb{R}_{0}^{+}$.
\end{itemize}

\noindent If $G_{t}$ is a self--adjoint operator on a Hilbert space then the
assumption B1 is directly satisfied with $\beta _{0}\left( t\right) =0$ and $%
m=1$ since, for all $\lambda \in \mathbb{R}$,
\begin{equation}
\left\Vert \left( \lambda \mathbf{1}+iG_{t}\right) ^{-1}\right\Vert _{%
\mathrm{op}}=\frac{1}{\underset{r\in \sigma \left( G_{t}\right) }{\inf }%
\left\vert \lambda +ir\right\vert }\leq \left\vert \lambda \right\vert
^{-1}\ .  \label{stability condition in Hilbert space}
\end{equation}

Within this appendix, we present in this context a detailed analysis of the
well--posedness of non--autonomous evolution equations. A first proof was
performed by Kato with analogue assumptions \cite{Kato,Kato1973}. His idea
was to discretize the differential equation (\ref{flow equationbisbisbis})
in order to use the Hille--Yosida generation theorems. Then, by taking the
continuous limit he obtained a well--defined solution $U_{t,s}$ of (\ref%
{flow equationbisbisbis}). The strategy presented here is different since we
use the Yosida approximation, i.e., the sequence $\left\{ G_{t,\lambda
}\right\} _{\lambda \geq 0}$ of strongly continuous maps defined by
\begin{equation}
t\rightarrow G_{t,\lambda }:=\frac{\lambda G_{t}}{\lambda \mathbf{1}+iG_{t}}%
\in \mathcal{B}\left( \mathcal{X}\right) \ .  \label{dyson series 0}
\end{equation}%
Indeed, for all $t>s\geq 0$, we already know the solution in the strong
topology of the initial value problem (\ref{flow equationbisbisbis}) with $%
G_{t,\lambda }$ replacing $G_{t}.$ It is equal to the evolution operator%
\footnote{$U_{\lambda ,t,s}$ is jointly strongly continuous in $s$ and $t$
with $U_{\lambda ,t,s}=U_{\lambda ,t,x}U_{\lambda ,x,s}$ for $t\geq x\geq
s\geq 0\ $and $U_{\lambda ,s,s}=\mathbf{1}$.} $U_{\lambda ,t,s}$
well--defined by the Dyson series
\begin{equation}
U_{\lambda ,t,s}:=\mathbf{1}+\underset{n=1}{\overset{\infty }{\sum }}\left(
-i\right) ^{n}\int_{s}^{t}\mathrm{d}\tau _{1}\cdots \int_{s}^{\tau _{n-1}}%
\mathrm{d}\tau _{n}\underset{j=1}{\overset{n}{\prod }}G_{\tau _{j},\lambda }
\label{dyson series}
\end{equation}%
for all $t\geq s\geq 0$ and $\lambda \in \mathbb{R}^{+}$. Then, since $%
G_{t,\lambda }$ approximates in the strong topology the generator $G_{t}$
for large $\lambda $, we analyze the strong limit of operators $U_{\lambda
,t,s}$, as $\lambda \rightarrow \infty $, to obtain the existence of a limit
operator $U_{t,s}$ with appropriate properties.

This idea was already used in an analogue context for the hyperbolic case by
Ishii \cite{Ishii1}, see also \cite{Ishii2}. Nevertheless, we give again
this proof in detail since it is technically\textit{\ }different from
Ishii's ones \cite{Ishii1,Ishii2} (compare with Lemma \ref{existence flow
with W 4}) and also, because we use without details several times similar
arguments in our proofs above. In comparison with other methods, it uses a
simple controlled approximation of $U_{t,s}$ in terms of the Dyson series $%
U_{\lambda ,t,s}$ (\ref{dyson series}) and Conditions B1--B3 appear in a
natural way. Now, we express the main theorem of this subsection.

\begin{theorem}[$U_{t,s}$ as operator limit of Dyson series]
\label{flow equation thm 5}\mbox{ }\newline
Under Conditions B1--B3, there is a unique strong solution $(U_{t,s})_{t\geq
s\geq 0}\subset \mathcal{B}(\mathcal{X})$ on $\mathcal{Y}\subset \mathcal{D}%
(G_{t})$ of the non--autonomous evolution equations\footnote{%
The derivatives $\partial _{t}$ and $\partial _{s}$ on the borderline $t=s$
or $s=0$ have to be understood as either right or left derivatives. See
Lemma \ref{Lemma differentiability}.}%
\begin{equation}
\forall t\geq s\geq 0:\qquad \left\{
\begin{array}{llll}
\partial _{t}U_{t,s}=-iG_{t}U_{t,s} & , & U_{s,s}:=\mathbf{1} & . \\
\partial _{s}U_{t,s}=iU_{t,s}G_{s} & , & U_{t,t}:=\mathbf{1} & .%
\end{array}%
\right.
\end{equation}%
In particular, for all $t\geq s\geq 0$, $U_{t,s}\mathcal{Y}\subset \mathcal{Y%
}$ and $U_{t,s}$ is an evolution operator. Furthermore, $U_{t,s}$ is
approximated by a Dyson series since, as $\lambda \rightarrow \infty $, $%
U_{\lambda ,t,s}$ converges to $U_{t,s}$ strongly in $\mathcal{B}(\mathcal{X}%
)$ and in norm in $\mathcal{B}(\mathcal{Y},\mathcal{X})$.
\end{theorem}

The detailed proof of this theorem corresponds to Lemmata \ref%
{existence_flow_5}--\ref{Lemma differentiability}. Remark that \emph{%
additional} properties on $U_{t,s}$ are also proven within these lemmas. To
simplify our notations (for example a definition of a norm in $\mathcal{Y}$%
), we assume that $\Theta \ $is invertible but our arguments are in fact
\emph{independent} of this assumption. Also, to avoid triviality, we assume $%
G_{t}\ $is unbounded for any $t\geq 0$. Because $G_{t}\Theta ^{-1}\in
\mathcal{B}(\mathcal{X})$ (cf. B2), the operator $\Theta $ shares with $%
G_{t} $ this property, i.e., $\Theta \notin \mathcal{B}(\mathcal{X})$.
Therefore, an important ingredient of our proof is to get a uniform upper
bound of the operator norms $\Vert U_{\lambda ,t,s}\Vert _{\mathrm{op}}$ and
$\Vert \Theta U_{\lambda ,t,s}\Theta ^{-1}\Vert _{\mathrm{op}}$, see Lemma %
\ref{existence_flow_5}. For this standard question, B1 and the function $%
\gamma $ defined, for all $t\geq 0$, by%
\begin{equation}
\gamma \left( t\right) :=\beta _{0}\left( t\right) +m\beta _{1}\left(
t\right) =\beta _{0}\left( t\right) +m\left\Vert \left[ \Theta ,G_{t}\right]
\Theta ^{-1}\right\Vert _{\mathrm{op}}  \label{important function}
\end{equation}%
(cf. B3) enter into the game. This preliminary work is necessary to prove
the convergence of $U_{\lambda ,t,s}$, as $\lambda \rightarrow \infty $, to
a bounded operator $U_{t,s}$ on the Banach space $\mathcal{X}$, see Lemma %
\ref{existence flow with W 4}. We proceed by proving that $U_{t,s}$ is an
evolution operator, see Lemma \ref{lemma unitary propagator}. Then, a
crucial result before obtaining the differentiability of $U_{t,s}$ is the
study of the convergence of the bounded operator family $(\Theta U_{\lambda
,t,s}\Theta ^{-1})_{\lambda \geq 0}$, cf. Lemma \ref{existence flow with W
4bis}. Finally, we study the differentiability of $U_{t,s}$, see Lemma \ref%
{Lemma differentiability}. This concludes the proof of Theorem \ref{flow
equation thm 5}. Now, we give the promised series of lemmata with their
proofs.

\begin{lemma}[Preservation by $U_{\protect\lambda ,t,s}$\ of the Banach
space $\mathcal{Y}$]
\label{existence_flow_5}\mbox{ }\newline
Under Conditions B1--B3, for any $t\geq s\geq 0$ and $\lambda >\gamma \left(
t\right) \geq \beta _{0}\left( t\right) $, one has:%
\begin{eqnarray}
&&\left\Vert U_{\lambda ,t,s}\right\Vert _{\mathrm{op}} \leq m\exp \left\{
\int_{s}^{t}\frac{\lambda \beta _{0}\left( \tau \right) }{\lambda -\beta
_{0}\left( \tau \right) }\mathrm{d}\tau \right\} \ ,  \label{eq sup1} \\
&&\left\Vert \Theta U_{\lambda ,t,s}\Theta ^{-1}\right\Vert _{\mathrm{op}}
\leq m\exp \left\{ \int_{s}^{t}\frac{\lambda \gamma \left( \tau \right) }{\lambda -\gamma \left( \tau \right) }\mathrm{d}\tau \right\} \ .
\label{eq sup2}
\end{eqnarray}%
\end{lemma}

\noindent \textbf{Proof.} First, note that
\begin{equation}
iG_{t,\lambda }:=\frac{i\lambda G_{t}}{\lambda \mathbf{1}+iG_{t}}=\lambda
\mathbf{1}-\frac{\lambda ^{2}}{\lambda \mathbf{1}+iG_{t}}=:\lambda \mathbf{1}%
+\tilde{G}_{t,\lambda }\ .  \label{definition of G}
\end{equation}%
Also, if $\tilde{U}_{\lambda ,t,s}:=\mathrm{e}^{\lambda \left( t-s\right)
}U_{\lambda ,t,s}$ then $\partial _{t}\{\tilde{U}_{\lambda ,t,s}\}=-\tilde{G}%
_{t,\lambda }\tilde{U}_{\lambda ,t,s}$. Therefore, B1 implies the following
estimate for the Dyson expansion (\ref{dyson series}):
\begin{eqnarray}
\Vert U_{\lambda ,t,s}\Vert _{\mathrm{op}} &=&\mathrm{e}^{-\lambda \left(
t-s\right) }\Vert \tilde{U}_{\lambda ,t,s}\Vert _{\mathrm{op}}  \notag \\
&\leq &\mathrm{e}^{-\lambda \left( t-s\right) }\left( 1+\underset{n=1}{%
\overset{\infty }{\sum }}\int_{s}^{t}\mathrm{d}\tau _{1}\cdots
\int_{s}^{\tau _{n-1}}\mathrm{d}\tau _{n}%
\Big \|%
\underset{j=1}{\overset{n}{\prod }}\tilde{G}_{\tau _{j},\lambda }%
\Big \|%
_{\mathrm{op}}\right)  \notag \\
&\leq &\mathrm{e}^{-\lambda \left( t-s\right) }\left( 1+\underset{n=1}{%
\overset{\infty }{\sum }}\int_{s}^{t}\mathrm{d}\tau _{1}\cdots
\int_{s}^{\tau _{n-1}}\mathrm{d}\tau _{n}\underset{j=1}{\overset{n}{\prod }}%
\frac{\lambda ^{2}}{\lambda -\beta _{0}\left( \tau _{j}\right) }\right) \ ,
\notag \\
&&
\end{eqnarray}%
i.e., we get (\ref{eq sup1}). Moreover, $\left[ \Theta ,G_{t}\right] \Theta
^{-1}\in \mathcal{B}\left( \mathcal{X}\right) $ and, similar to Lemma \ref%
{lemma resolvent} (i) and Equation (\ref{condition lambdabis}),
\begin{eqnarray}
\Theta \left( \lambda \mathbf{1}+iG_{t}\right) ^{-1}\Theta ^{-1} &=&\frac{1}{%
\lambda \mathbf{1}+iG_{t}+i\left[ \Theta ,G_{t}\right] \Theta ^{-1}} \\
&=&\underset{n=0}{\overset{\infty }{\sum }}\left( \lambda \mathbf{1}%
+iG_{t}\right) ^{-1}\left\{ i\left[ G_{t},\Theta \right] \Theta ^{-1}\left(
\lambda \mathbf{1}+iG_{t}\right) ^{-1}\right\} ^{n}  \notag \\
&&
\end{eqnarray}%
for any family of real numbers $\left\{ t_{j},\lambda _{j}\right\} $ such
that $0\leq t_{1}\leq \ldots \leq t_{n}$ and $\lambda _{1}>\gamma \left(
t_{1}\right) ,\ldots ,\lambda _{n}>\gamma \left( t_{n}\right) $. Therefore,
it is standard to verify that%
\begin{eqnarray}
\left\Vert \underset{j=1}{\overset{n}{\prod }}\Theta \left( \lambda _{j}%
\mathbf{1}+iG_{t_{j}}\right) ^{-1}\Theta ^{-1}\right\Vert _{\mathrm{op}}
&\leq &m\underset{j=1}{\overset{n}{\prod }}\frac{1}{\lambda _{j}-\beta
_{0}\left( t_{j}\right) -m\beta _{1}\left( t_{j}\right) }  \notag \\
&=&m\underset{j=1}{\overset{n}{\prod }}\frac{1}{\lambda _{j}-\gamma \left(
t_{j}\right) }\ ,  \label{kato stability bis}
\end{eqnarray}%
from which we prove (\ref{eq sup2}) just as we did for (\ref{eq sup1}%
).\hfill $\Box $

\begin{lemma}[Convergence of $U_{\protect\lambda ,t,s}$\ when $\protect%
\lambda \rightarrow \infty $]
\label{existence flow with W 4}\mbox{ }\newline
Under Conditions B1--B3 and for any $t\geq s\geq 0$, the strong limit $%
U_{t,s}$ of $U_{\lambda ,t,s}$ when $\lambda \rightarrow \infty $ exists and
satisfies
\begin{equation}
\underset{\lambda \rightarrow \infty }{\lim }\left\Vert \left\{ U_{\lambda
,t,s}-U_{t,s}\right\} \Theta ^{-1}\right\Vert _{\mathrm{op}}=0\mathrm{\quad }%
\text{and}\mathrm{\quad }\left\Vert U_{t,s}\right\Vert _{\mathrm{op}}\leq
m\exp \left\{ \int_{s}^{t}\beta _{0}\left( \tau \right) \mathrm{d}\tau
\right\} .
\end{equation}
\end{lemma}

\noindent \textbf{Proof.} Fix an arbitrary large parameter $T$ taken such
that $T\geq t$ and let
\begin{equation}
\left\{
\begin{array}{ll}
M:=\underset{}{\underset{0\leq t\leq T}{\max }}\left\{ \left\Vert
G_{t}\Theta ^{-1}\right\Vert _{\mathrm{op}}\right\} & , \\
\gamma _{\sup }:=\underset{}{\underset{0\leq t\leq T}{\max }}\left\{ \gamma
\left( t\right) \right\} =\underset{0\leq t\leq T}{\max }\left\{ \beta
_{0}\left( t\right) +m\beta _{1}\left( t\right) \right\} & , \\
Z_{\epsilon }:=\underset{}{\max }\left\{ \left\Vert \left(
G_{t}-G_{t^{\prime }}\right) \Theta ^{-1}\right\Vert _{\mathrm{op}}\
\big|%
\ t,t^{\prime }\in \left[ 0,T+1\right] ,\left\vert t-t^{\prime }\right\vert
\leq 2\epsilon \right\} & .%
\end{array}%
\right.  \label{upper bound Jbis}
\end{equation}%
Also, by taking $G_{t<0}:=G_{0}$ we define the derivable operator $%
J_{\epsilon ,\alpha }\left( t\right) $, for any $t\geq 0$, by
\begin{equation}
J_{\epsilon ,\alpha }\left( t\right) :=\frac{1}{2\epsilon }\int_{t-\epsilon
}^{t+\epsilon }\frac{\alpha }{\alpha \mathbf{1}+iG_{\tau }}\mathrm{d}\tau \ .
\end{equation}%
By Condition B1 combined with (\ref{kato stability bis}) and (\ref{upper
bound Jbis}) for any $\alpha \geq 2\gamma _{\sup }$, observe that
\begin{equation}
\left\Vert \left\{ \mathbf{1}-J_{\epsilon ,\alpha }\left( t\right) \right\}
\Theta ^{-1}\right\Vert _{\mathrm{op}}\leq 2mM\alpha ^{-1}\ .
\label{upper bound J}
\end{equation}%
Then, via Lemma \ref{existence_flow_5} for $\alpha ,\eta ,\lambda \geq
2\gamma _{\sup }$, we directly get
\begin{eqnarray}
\left\Vert \left\{ U_{\lambda ,t,s}-U_{\eta ,t,s}\right\} \Theta
^{-1}\right\Vert _{\mathrm{op}} &\leq &\left\Vert \left\{ J_{\epsilon
,\alpha }\left( t\right) U_{\lambda ,t,s}-U_{\eta ,t,s}J_{\epsilon ,\alpha
}\left( s\right) \right\} \Theta ^{-1}\right\Vert _{\mathrm{op}}  \notag \\
&&+\frac{4m^{2}M}{\alpha }\exp \left\{ \int_{s}^{t}2\gamma \left( \tau
\right) \mathrm{d}\tau \right\} \ .  \label{full upper bound 1}
\end{eqnarray}%
Now, by using (\ref{flow equationbisbisbis}) (with $G_{t,\lambda }$
replacing $G_{t}$) one obtains
\begin{equation}
J_{\epsilon ,\alpha }\left( t\right) U_{\lambda ,t,s}-U_{\eta
,t,s}J_{\epsilon ,\alpha }\left( s\right) =\int_{s}^{t}\partial _{\tau
}\left\{ U_{\eta ,t,\tau }J_{\epsilon ,\alpha }\left( \tau \right)
U_{\lambda ,\tau ,s}\right\} \mathrm{d}\tau =\mathrm{A}+\mathrm{B}+\mathrm{C}
\label{full upper bound 2}
\end{equation}%
with
\begin{eqnarray}
\mathrm{A}:= &&i\int_{s}^{t}U_{\eta ,t,\tau }\left\{ G_{\tau ,\eta }-G_{\tau
,\lambda }\right\} J_{\epsilon ,\alpha }\left( \tau \right) U_{\lambda ,\tau
,s}\mathrm{d}\tau \ , \\
\mathrm{B}:= &&\frac{i}{2\epsilon }\int_{s}^{t}\mathrm{d}\tau \int_{\tau
-\epsilon }^{\tau +\epsilon }\mathrm{d}r\ U_{\eta ,t,\tau }\left[ \frac{%
\lambda ^{2}}{\lambda \mathbf{1}+iG_{\tau }},\frac{\alpha }{\alpha \mathbf{1}%
+iG_{r}}\right] U_{\lambda ,\tau ,s}\ , \\
\mathrm{C}:= &&\frac{i\alpha }{2\epsilon }\int_{s}^{t}U_{\eta ,t,\tau }\frac{%
1}{\alpha \mathbf{1}+iG_{\tau +\epsilon }}\left\{ G_{\tau -\epsilon
}-G_{\tau +\epsilon }\right\} \frac{1}{\alpha \mathbf{1}+iG_{\tau -\epsilon }%
}U_{\lambda ,\tau ,s}\mathrm{d}\tau \ .  \notag \\
&&
\end{eqnarray}%
So, we bound below the terms $\Vert \mathrm{A}\Theta ^{-1}\Vert _{\mathrm{op}%
}$, $\Vert \mathrm{B}\Theta ^{-1}\Vert _{\mathrm{op}}$, and $\Vert \mathrm{C}%
\Theta ^{-1}\Vert _{\mathrm{op}}$. First, by Lemma \ref{existence_flow_5}
for any $\eta ,\lambda \geq 2\gamma _{\sup }$, we get that
\begin{equation}
\left\Vert \mathrm{A}\Theta ^{-1}\right\Vert _{\mathrm{op}}\leq m^{2}\exp
\left\{ \int_{s}^{t}2\gamma \left( \tau \right) \mathrm{d}\tau \right\}
\int_{s}^{t}\left\Vert \left( G_{\tau ,\eta }-G_{\tau ,\lambda }\right)
J_{\epsilon ,\alpha }\left( \tau \right) \Theta ^{-1}\right\Vert _{\mathrm{op%
}}\mathrm{d}\tau \ .  \label{A upperbound 1}
\end{equation}%
Since we have the equality
\begin{eqnarray}
\left\{ G_{\tau ,\eta }-G_{\tau ,\lambda }\right\} J_{\epsilon ,\alpha
}\left( \tau \right) &=&\frac{\alpha \left( \eta -\lambda \right) iG_{\tau
}^{2}}{\left( \eta \mathbf{1}+iG_{\tau }\right) \left( \lambda \mathbf{1}%
+iG_{\tau }\right) \left( \alpha \mathbf{1}+iG_{\tau }\right) }  \notag \\
&&\left\{ 1+\frac{i}{2\epsilon }\int_{\tau -\epsilon }^{\tau +\epsilon
}\left( G_{r}-G_{\tau }\right) \left( \frac{1}{\alpha \mathbf{1}+iG_{r}}%
\right) \mathrm{d}r\right\}  \notag \\
&&
\end{eqnarray}%
and the estimation
\begin{equation}
\left\Vert \frac{iG_{\tau }}{\alpha \mathbf{1}+iG_{\tau }}\right\Vert _{%
\mathrm{op}}\leq 1+\alpha \left\Vert \left( \alpha \mathbf{1}+iG_{\tau
}\right) ^{-1}\right\Vert _{\mathrm{op}}\ ,  \label{petit equality}
\end{equation}%
by Condition B1 combined with (\ref{kato stability bis}) and (\ref{upper
bound Jbis}) we obtain
\begin{equation}
\left\Vert \left\{ G_{\tau ,\eta }-G_{\tau ,\lambda }\right\} J_{\epsilon
,\alpha }\left( \tau \right) \Theta ^{-1}\right\Vert _{\mathrm{op}}\leq
12m^{3}\left\{ \frac{\alpha M}{\min \left\{ \lambda ,\eta \right\} }%
+3mZ_{\epsilon }\right\}
\end{equation}%
for $\alpha ,\eta ,\lambda \geq 2\gamma _{\sup }\ $and $|\eta -\lambda |\leq
\max \{\lambda ,\eta \}$. Therefore, we infer from (\ref{A upperbound 1})
that
\begin{equation}
\left\Vert \mathrm{A}\Theta ^{-1}\right\Vert _{\mathrm{op}}\leq
12m^{5}\left\{ \frac{\alpha M}{\min \left\{ \lambda ,\eta \right\} }%
+3mZ_{\epsilon }\right\} \left( t-s\right) \exp \left\{ \int_{s}^{t}2\gamma
\left( \tau \right) \mathrm{d}\tau \right\} \ .  \label{A upperbound 2}
\end{equation}%
On the other hand, since
\begin{equation}
\left[ \frac{1}{\lambda +iG_{\tau }},\frac{1}{\alpha +iG_{r}}\right] =\frac{1%
}{\lambda +iG_{\tau }}\frac{1}{\alpha +iG_{r}}\left[ G_{r},G_{\tau }-G_{r}%
\right] \frac{1}{\alpha +iG_{r}}\frac{1}{\lambda +iG_{\tau }}\ ,
\end{equation}%
then, by using again B1, Lemma \ref{existence_flow_5}, (\ref{kato stability
bis}), (\ref{upper bound Jbis}), and (\ref{petit equality}), one gets, for
any $\alpha ,\eta ,\lambda \geq 2\gamma _{\sup }$, the upper bounds
\begin{eqnarray}
\left\Vert \mathrm{B}\Theta ^{-1}\right\Vert _{\mathrm{op}} &\leq
&48m^{4}Z_{\epsilon }\left( t-s\right) \exp \left\{ \int_{s}^{t}2\gamma
\left( \tau \right) \mathrm{d}\tau \right\} \ ,  \label{B upperbound 1} \\
\left\Vert \mathrm{C}\Theta ^{-1}\right\Vert _{\mathrm{op}} &\leq &\frac{%
2m^{4}}{\epsilon \alpha }Z_{\epsilon }\left( t-s\right) \exp \left\{
\int_{s}^{t}2\gamma \left( \tau \right) \mathrm{d}\tau \right\} \ .
\label{C upperbound 1}
\end{eqnarray}%
Therefore, we combine (\ref{A upperbound 1}), (\ref{B upperbound 1}), and (%
\ref{C upperbound 1}) with $m\geq 1$, (\ref{full upper bound 1}), and (\ref%
{full upper bound 2}) to deduce the following upper bound:
\begin{eqnarray}
\left\Vert \left\{ U_{\lambda ,t,s}-U_{\eta ,t,s}\right\} \Theta
^{-1}\right\Vert _{\mathrm{op}} &\leq &84\left( M+1\right) m^{6}\left\{
\dfrac{1}{\alpha }+\frac{\alpha }{\min \left\{ \lambda ,\eta \right\} }%
+\left( 1+\frac{1}{\epsilon \alpha }\right) Z_{\epsilon }\right\}  \notag \\
&&\times \left( t-s+1\right) \exp \left\{ \int_{s}^{t}2\gamma \left( \tau
\right) \mathrm{d}\tau \right\} \ .
\end{eqnarray}%
By taking first $\min \{\lambda ,\eta \}\rightarrow \infty $, then $\alpha
\rightarrow \infty $ and $\epsilon \downarrow 0^{+}$, we conclude that%
\begin{equation}
\underset{\min \left\{ \lambda ,\eta \right\} \rightarrow \infty }{\lim }%
\left\Vert \left\{ U_{\lambda ,t,s}-U_{\eta ,t,s}\right\} \Theta
^{-1}\right\Vert _{\mathrm{op}}=0\ ,  \label{Cauchy sequence 1}
\end{equation}%
where we have used Condition B2 when $\epsilon \downarrow 0^{+}$. The fact
that the bounded operator family $(U_{\lambda ,t,s})_{\lambda \geq 0}$ is a
Cauchy sequence in the strong topology on $\mathcal{X}$ is now standard to
verify: Let $\delta >0$ and $\varphi \in \mathcal{X}$. By density of $%
\mathcal{Y}=\mathcal{D}\left( \Theta \right) $, there exists $\psi \in
\mathcal{Y}$ such that $\Vert \varphi -\psi \Vert \leq \delta $. Then, by
Lemma \ref{existence_flow_5}, one deduces, for $\eta ,\lambda \geq 2\gamma
_{\sup }$, that
\begin{eqnarray}
\left\Vert \left\{ U_{\lambda ,t,s}-U_{\eta ,t,s}\right\} \varphi
\right\Vert &\leq &\left\Vert \left\{ U_{\lambda ,t,s}-U_{\eta ,t,s}\right\}
\Theta ^{-1}\right\Vert _{\mathrm{op}}\left\Vert \Theta \psi \right\Vert
\notag \\
&&+2\delta m\exp \left\{ \int_{s}^{t}2\gamma \left( \tau \right) \mathrm{d}%
\tau \right\} \ .  \label{Cauchy sequence 2}
\end{eqnarray}%
Hence, we combine (\ref{Cauchy sequence 2}) with (\ref{Cauchy sequence 1})
to observe that the family $(U_{\lambda ,t,s}\varphi )_{\lambda \geq 2\gamma
_{\sup }}$ forms a Cauchy sequence in the Banach space $\mathcal{X}$.
Therefore, there exists an operator $U_{t,s}$ defined by
\begin{equation}
\forall \varphi \in \mathcal{X}:\qquad U_{t,s}\varphi :=\underset{\lambda
\rightarrow \infty }{\lim }\left\{ U_{\lambda ,t,s}\varphi \right\} \in
\mathcal{X}
\end{equation}%
and satisfying
\begin{equation}
\underset{\lambda \rightarrow \infty }{\lim }\left\Vert \left\{ U_{\lambda
,t,s}-U_{t,s}\right\} \Theta ^{-1}\right\Vert _{\mathrm{op}}=0\ .
\end{equation}%
Moreover, since
\begin{equation}
\left\Vert U_{t,s}\varphi \right\Vert =\underset{\lambda \rightarrow \infty }%
{\lim }\left\Vert U_{\lambda ,t,s}\varphi \right\Vert \leq \left\Vert
\varphi \right\Vert \underset{\lambda \rightarrow \infty }{\lim }\left\{
\left\Vert U_{\lambda ,t,s}\right\Vert _{\mathrm{op}}\right\} \ ,
\label{Cauchy sequence 3}
\end{equation}%
we directly obtain the corresponding upper bound on the operator norm $\Vert
U_{t,s}\Vert _{\mathrm{op}}$ from Lemma \ref{existence_flow_5}.\hfill $\Box $

\begin{lemma}[$U_{t,s}$ as an evolution operator]
\label{lemma unitary propagator}\mbox{ }\newline
Under Conditions B1--B3, the bounded operator family $(U_{t,s})_{t\geq s\geq
0}$ is jointly strongly continuous in $s$ and $t$, and satisfies the cocycle
property $U_{t,s}=U_{t,x}U_{x,s}$ for $t\geq x\geq s\geq 0\ $with $U_{s,s}=%
\mathbf{1}$.
\end{lemma}

\noindent \textbf{Proof.} First, for any $\varphi \in \mathcal{X}\ $and $%
t\geq x\geq s\geq 0$, straightforward estimations using Lemmata \ref%
{existence_flow_5}--\ref{existence flow with W 4} show that
\begin{equation}
U_{s,s}\varphi =\underset{\lambda \rightarrow \infty }{\lim }U_{\lambda
,s,s}\varphi =\varphi
\end{equation}%
and
\begin{equation}
U_{t,s}\varphi -U_{t,x}U_{x,s}\varphi =\underset{\lambda \rightarrow \infty }%
{\lim }\left\{ U_{\lambda ,t,s}\varphi -U_{\lambda ,t,x}U_{\lambda
,x,s}\varphi \right\} =0\ .
\end{equation}%
Now, it remains to prove the strong continuity of $U_{t,s}\Theta ^{-1}.$ If $%
s\leq t$, $s^{\prime }\leq t^{\prime }$, and $\max \left\{ s^{\prime
},s\right\} <\min \left\{ t,t^{\prime }\right\} ,$ i.e., $\left( s,t\right)
\cap \left( s^{\prime },t^{\prime }\right) \neq \varnothing ,$ then, by
rewriting (\ref{flow equationbisbisbis}) (with $G_{t,\lambda }$ replacing $%
G_{t}$) as an integral equation, one obtains the inequality
\begin{eqnarray}
&&\left\Vert \left\{ U_{\lambda ,t^{\prime },s^{\prime }}-U_{\lambda
,t,s}\right\} \Theta ^{-1}\right\Vert _{\mathrm{op}}  \notag \\
&\leq &mM\left\{ \left\vert s-s^{\prime }\right\vert +\left\vert t-t^{\prime
}\right\vert \right\} \exp \left\{ \underset{\min \left\{ s,s^{\prime
}\right\} }{\overset{\max \left\{ t,t^{\prime }\right\} }{%
\mathop{\displaystyle \int}}}2\gamma \left( \tau \right) \mathrm{d}\tau
\right\}  \label{continuity of U}
\end{eqnarray}%
for $\lambda \geq 2\gamma _{\sup }\ $(\ref{upper bound Jbis}), see also
Lemma \ref{existence_flow_5}. In other words, $U_{\lambda ,t,s}\Theta ^{-1}$
is uniformly norm continuous. Therefore, by density of $\mathcal{Y}=\mathcal{%
D}\left( \Theta \right) $ combined with Lemma \ref{existence flow with W 4},
$U_{t,s}$ is jointly strongly continuous in $s$ and $t$.\hfill $\Box $

\begin{lemma}[Convergence of $\Theta U_{\protect\lambda ,t,s}\Theta ^{-1}$\
when $\protect\lambda \rightarrow \infty $]
\label{existence flow with W 4bis}\mbox{ }\newline
Under Conditions B1--B3 and for any $t\geq s\geq 0$, the strong limit $%
V_{t,s}$ of $\Theta U_{\lambda ,t,s}\Theta ^{-1}$ when $\lambda \rightarrow
\infty $ exists, is jointly strongly continuous in $s$ and $t$, and
satisfies $V_{t,s}=\Theta U_{t,s}\Theta ^{-1}$ with%
\begin{equation}
\left\Vert \Theta U_{t,s}\Theta ^{-1}\right\Vert _{\mathrm{op}}\leq m\exp
\left\{ \int_{s}^{t}\gamma \left( \tau \right) \mathrm{d}\tau \right\} \ .
\end{equation}
\end{lemma}

\noindent \textbf{Proof.} For any $t\geq s\geq 0$ and $\lambda >\gamma
\left( t\right) \geq \beta _{0}\left( t\right) $, define the bounded
operators
\begin{equation}
V_{\lambda ,t,s}:=\Theta U_{\lambda ,t,s}\Theta ^{-1}
\label{upper bound Jbisbis}
\end{equation}%
and
\begin{equation}
V_{\lambda ,t,s;\tau _{1},\ldots ,\tau _{n}}^{\left( n\right) }:=\underset{%
j=1}{\overset{n}{\prod }}U_{\lambda ,t,\tau _{j}}\left\{ \left[ \Theta
,G_{\tau _{j},\lambda }\right] \Theta ^{-1}\right\} U_{\lambda ,\tau
_{j},\tau _{j+1}}\ ,  \label{upper bound Jbisbisbis}
\end{equation}%
with $\tau _{n+1}:=s$. See Lemma \ref{existence_flow_5}. By (\ref{flow
equationbisbisbis}), observe that
\begin{eqnarray}
V_{\lambda ,t,s}-U_{\lambda ,t,s} &=&\int_{s}^{t}\partial _{\tau }\left\{
U_{\lambda ,t,\tau }V_{\lambda ,\tau ,s}\right\} \mathrm{d}\tau  \notag \\
&=&-i\int_{s}^{t}U_{\lambda ,t,\tau }\left\{ \left[ \Theta ,G_{\tau ,\lambda
}\right] \Theta ^{-1}\right\} V_{\lambda ,\tau ,s}\mathrm{d}\tau \ .
\label{GammaUgamma as a serie00}
\end{eqnarray}%
Therefore, $V_{\lambda ,t,s}$ (\ref{upper bound Jbisbis}) is directly
expressed in terms of $U_{\lambda ,t,s}$ and the bounded operator $\left[
\Theta ,G_{t,\lambda }\right] \Theta ^{-1}:$%
\begin{equation}
V_{\lambda ,t,s}=U_{\lambda ,t,s}+\underset{n=1}{\overset{\infty }{\sum }}%
\left( -i\right) ^{n}\int_{s}^{t}\mathrm{d}\tau _{1}\cdots \int_{s}^{\tau
_{n-1}}\mathrm{d}\tau _{n}V_{\lambda ,t,s;\tau _{1},\ldots ,\tau
_{n}}^{\left( n\right) }\ .  \label{V as a serie}
\end{equation}%
Since Condition B1 together with (\ref{kato stability bis}) implies that
\begin{equation}
\left\Vert \left[ \Theta ,G_{t,\lambda }\right] \Theta ^{-1}\right\Vert _{%
\mathrm{op}}\leq 4m^{2}\left\Vert \left[ \Theta ,G_{t}\right] \Theta
^{-1}\right\Vert _{\mathrm{op}}=4m^{2}\beta _{1}\left( t\right)
\label{petite upper bound}
\end{equation}%
for any $\lambda \geq 2\gamma _{\sup }$, we infer from Lemma \ref%
{existence_flow_5} that
\begin{equation}
\left\Vert V_{\lambda ,t,s;\tau _{1},\ldots ,\tau _{n}}^{\left( n\right)
}\right\Vert _{\mathrm{op}}\leq m\exp \left\{ \int_{s}^{t}2\gamma \left(
\tau \right) \mathrm{d}\tau \right\} \left( 4m^{3}\right) ^{n}\underset{j=1}{%
\overset{n}{\prod }}\beta _{1}\left( \tau _{j}\right) ,
\end{equation}%
cf. (\ref{upper bound Jbis}) and (\ref{upper bound Jbisbisbis}). Thus, for
any $\lambda \geq 2\gamma _{\sup }$ and every $N>1$,
\begin{eqnarray}
&&\left\Vert \underset{n=N}{\overset{\infty }{\sum }}\left( -i\right)
^{n}\int_{s}^{t}\mathrm{d}\tau _{1}\cdots \int_{s}^{\tau _{n-1}}\mathrm{d}%
\tau _{n}V_{\lambda ,t,s;\tau _{1},\ldots ,\tau _{n}}^{\left( n\right)
}\right\Vert _{\mathrm{op}}  \notag \\
&\leq &\frac{m\varsigma ^{N}}{N!}\exp \left\{ \int_{s}^{t}\left\{ 2\gamma
\left( \tau \right) +4m^{3}\beta _{1}\left( \tau \right) \right\} \mathrm{d}%
\tau \right\}
\end{eqnarray}%
with $\varsigma :=4m^{3}\max \left\{ \beta _{1}\left( \tau \right) :\tau \in %
\left[ s,t\right] \right\} $. In other words, the series (\ref{V as a serie}%
) is norm convergent, uniformly in $\lambda \geq 2\gamma _{\sup }$. Also,
for all $\lambda \geq 2\gamma _{\sup }$ and $\varphi \in \mathcal{X}$,
\begin{eqnarray}
\left\Vert \left[ \Theta ,G_{\lambda ,t}-G_{t}\right] \Theta ^{-1}\varphi
\right\Vert &\leq &2m\beta _{1}\left( t\right) \left\Vert \Theta \frac{G_{t}%
}{\lambda \mathbf{1}+iG_{t}}\Theta ^{-1}\varphi \right\Vert  \notag \\
&&+\left\Vert \frac{G_{t}}{\lambda \mathbf{1}+iG_{t}}\left\{ \left[ \Theta
,G_{t}\right] \Theta ^{-1}\right\} \varphi \right\Vert
\label{petite upper bound limit}
\end{eqnarray}%
and
\begin{eqnarray}
\left\Vert \frac{G_{t}}{\lambda \mathbf{1}+iG_{t}}\Theta ^{-1}\right\Vert _{%
\mathrm{op}} &\leq &2mM\lambda ^{-1}\ , \\
\left\Vert \Theta \frac{G_{t}}{\lambda \mathbf{1}+iG_{t}}\Theta
^{-2}\right\Vert _{\mathrm{op}} &\leq &2mM\lambda ^{-1}\left\{ 1+\beta
_{1}\left( t\right) \right\} \ ,
\end{eqnarray}%
see\ (\ref{upper bound Jbis}). So, by density of the subset $\mathcal{D}%
\left( \Theta \right) =\mathcal{Y}$ of $\mathcal{X}$, one obtains, for all $%
\varphi \in \mathcal{X}$, that
\begin{equation}
\underset{\lambda \rightarrow \infty }{\lim }\left\Vert \frac{G_{t}}{\lambda
\mathbf{1}+iG_{t}}\varphi \right\Vert =0
\end{equation}%
and%
\begin{equation}
\underset{\lambda \rightarrow \infty }{\lim }\left\Vert \Theta \frac{G_{t}}{%
\lambda \mathbf{1}+iG_{t}}\Theta ^{-1}\varphi \right\Vert =0\ .
\end{equation}%
Consequently, Inequality (\ref{petite upper bound limit}) implies that, for
all $\varphi \in \mathcal{X}$,
\begin{equation}
\underset{\lambda \rightarrow \infty }{\lim }\left\{ \left[ \Theta
,G_{t,\lambda }\right] \Theta ^{-1}\varphi \right\} =\left[ \Theta ,G_{t}%
\right] \Theta ^{-1}\varphi \ .  \label{petite upper boundbis}
\end{equation}%
And then, by Lemma \ref{existence flow with W 4}, we can define the operator
$V_{t,s;\tau _{1},\ldots ,\tau _{n}}^{\left( n\right) }$ by
\begin{equation}
V_{t,s;\tau _{1},\ldots ,\tau _{n}}^{\left( n\right) }\varphi :=\underset{j=1%
}{\overset{n}{\prod }}U_{t,\tau _{j}}\left\{ \left[ \Theta ,G_{\tau _{j}}%
\right] \Theta ^{-1}\right\} U_{\tau _{j},\tau _{j+1}}=\underset{\lambda
\rightarrow \infty }{\lim }V_{\lambda ,t,s;\tau _{1},\ldots ,\tau
_{n}}^{\left( n\right) }\varphi
\end{equation}%
for any $\varphi \in \mathcal{X}$, $n\in \mathbb{N}$, and $t\geq \tau
_{1}\geq \ldots \geq \tau _{n}\geq \tau _{n+1}:=s\geq 0$. Moreover, by
Lebesgue's dominated convergence theorem, there also exists an operator $%
V_{t,s}$ defined, for any $\varphi \in \mathcal{X}$ and $t\geq s\geq 0$, by%
\begin{equation}
V_{t,s}\varphi :=\underset{\lambda \rightarrow \infty }{\lim }\left\{ \Theta
U_{\lambda ,t,s}\Theta ^{-1}\varphi \right\} =U_{t,s}\varphi +\underset{n=1}{%
\overset{\infty }{\sum }}\int_{s}^{t}\mathrm{d}\tau _{1}\cdots
\int_{s}^{\tau _{n-1}}\mathrm{d}\tau _{n}V_{t,s;\tau _{1},\ldots ,\tau
_{n}}^{\left( n\right) }\varphi \ .  \label{limite de V}
\end{equation}%
Observe that the operator $V_{t,s}$ is bounded since, for any $\varphi \in
\mathcal{X}$ and $t\geq s\geq 0$,%
\begin{eqnarray}
\left\Vert V_{t,s}\varphi \right\Vert &=&\underset{\lambda \rightarrow
\infty }{\lim }\left\Vert \Theta U_{\lambda ,t,s}\Theta ^{-1}\varphi
\right\Vert  \notag \\
&\leq &\left\Vert \varphi \right\Vert \underset{\lambda \rightarrow \infty }{%
\lim }\left\{ \left\Vert \Theta U_{\lambda ,t,s}\Theta ^{-1}\right\Vert _{%
\mathrm{op}}\right\}  \notag \\
&\leq &\left\Vert \varphi \right\Vert \ m\exp \left\{ \int_{s}^{t}\gamma
\left( \tau \right) \mathrm{d}\tau \right\} \ ,  \label{upper bound V}
\end{eqnarray}%
see Lemma \ref{existence_flow_5}. Now, let $\varphi \in \mathcal{X}$ and $%
\psi _{\lambda }:=U_{\lambda ,t,s}\Theta ^{-1}\varphi \in \mathcal{Y}$. By
Lemma \ref{existence flow with W 4},
\begin{equation}
\underset{\lambda \rightarrow \infty }{\lim }\left\{ \psi _{\lambda
}\right\} =U_{t,s}\Theta ^{-1}\varphi \ ,
\end{equation}%
whereas we infer from\ (\ref{limite de V}) that
\begin{equation}
\underset{\lambda \rightarrow \infty }{\lim }\left\{ \Theta \psi _{\lambda
}\right\} =V_{t,s}\varphi \ .
\end{equation}%
Since $\Theta $ is a closed operator, the last equalities then imply that
\begin{equation}
V_{t,s}\varphi =\Theta U_{t,s}\Theta ^{-1}\varphi
\end{equation}%
for\ any\ $\varphi \in \mathcal{X}$. Finally, from (\ref{limite de V})
combined with Lemma \ref{existence flow with W 4}, (\ref{upper bound Jbis}),
and (\ref{upper bound V}), observe that, for\ any\ $\varphi \in \mathcal{X}$%
,
\begin{eqnarray}
\left\Vert \left( V_{t^{\prime },s^{\prime }}-V_{t,s}\right) \varphi
\right\Vert &\leq &\left\Vert \left( U_{t^{\prime },s^{\prime
}}-U_{t,s}\right) \varphi \right\Vert +m^{2}\gamma _{\sup }\left\{
\left\vert s-s^{\prime }\right\vert +\left\vert t-t^{\prime }\right\vert
\right\}  \notag \\
&&\exp \left\{ \underset{\min \left\{ s,s^{\prime }\right\} }{\overset{\max
\left\{ t,t^{\prime }\right\} }{\mathop{\displaystyle \int}}}\left( \beta
_{0}\left( \tau \right) +2\gamma \left( \tau \right) \right) \mathrm{d}\tau
\right\} \ .
\end{eqnarray}%
In other words, the bounded operator family $(V_{t,s})_{t\geq s\geq 0}$ is
jointly strongly continuous in $s$ and $t$.\hfill $\Box $

\begin{lemma}[Differentiability of the evolution operator $U_{t,s}$]
\label{Lemma differentiability}\mbox{ }\newline
Under Conditions B1--B3, the evolution operator $(U_{t,s})_{t\geq s\geq 0}$
is the unique strong solution on $\mathcal{Y}$ of the initial value problem
\begin{equation}
\forall t>s\geq 0:\qquad \partial _{t}U_{t,s}=-iG_{t}U_{t,s}\ ,\quad
U_{s,s}:=\mathbf{1}\ .
\end{equation}%
Furthermore, $(U_{t,s})_{t\geq s\geq 0}$ also satisfies the non--autonomous
evolution equation
\begin{equation}
\forall t>s>0:\qquad \partial _{s}U_{t,s}=iU_{t,s}G_{s}\ ,\quad U_{t,t}:=%
\mathbf{1}\ .  \label{reversed autonomous eq}
\end{equation}%
At $t=s\geq 0$, its right derivative also equals $\partial
_{t}^{+}U_{t,s}|_{t=s}=-iG_{s}$, whereas $\partial
_{s}^{-}U_{t,s}|_{s=t}=iG_{t}$ and $\partial
_{s}^{+}U_{t,s}|_{s=0}=iU_{t,0}G_{0}$ for $t>0$, all in the strong sense in $%
\mathcal{Y}$.
\end{lemma}

\noindent \textbf{Proof.} By using Lemma \ref{existence_flow_5} and (\ref%
{upper bound Jbis}), direct manipulations show that, for any $\psi \in
\mathcal{Y}$, $\epsilon >0$ and $\lambda \geq 2\gamma _{\sup }$,
\begin{eqnarray}
&&\left\Vert \left\{ \epsilon ^{-1}\left( U_{t+\epsilon ,s}-U_{t,s}\right)
+iG_{t}U_{t,s}\right\} \psi \right\Vert  \notag \\
&\leq &2mM\epsilon ^{-1}\int_{t}^{t+\epsilon }\left( \left\Vert \Theta
\left\{ U_{\tau ,s}-U_{t,s}\right\} \psi \right\Vert +\left\Vert \Theta
\left\{ U_{\lambda ,\tau ,s}-U_{\tau ,s}\right\} \psi \right\Vert \right)
\mathrm{d}\tau  \notag \\
&&+Z_{\epsilon }\left\Vert \Theta \psi \right\Vert m\exp \left\{
\int_{s}^{t}2\gamma \left( \tau \right) \mathrm{d}\tau \right\} +\epsilon
^{-1}\int_{t}^{t+\epsilon }\left\Vert \left\{ G_{\tau ,\lambda }-G_{\tau
}\right\} U_{t,s}\psi \right\Vert \mathrm{d}\tau  \notag \\
&&+\epsilon ^{-1}\left\Vert \left\{ U_{t+\epsilon ,s}-U_{\lambda ,t+\epsilon
,s}\right\} \psi \right\Vert +\epsilon ^{-1}\left\Vert \left\{ U_{\lambda
,t,s}-U_{t,s}\right\} \psi \right\Vert \ .
\label{upper bound differential equation}
\end{eqnarray}%
Since, by Condition B1 and (\ref{upper bound Jbis}), one has
\begin{equation}
\left\Vert \frac{G_{t}}{\lambda +iG_{t}}\Theta ^{-1}\right\Vert _{\mathrm{op}%
}\leq \frac{mM}{\lambda -\beta _{0}\left( t\right) }\ ,
\end{equation}%
we use again the density of $\mathcal{Y}$ for any $\varphi \in \mathcal{X}$
and $\delta >0$ by taking $\tilde{\psi}\in \mathcal{Y}$ such that $\Vert
\varphi -\tilde{\psi}\Vert \leq \delta \ $in order to get the inequality
\begin{equation}
\left\Vert \left\{ G_{t,\lambda }-G_{t}\right\} \Theta ^{-1}\varphi
\right\Vert \leq M\left\Vert \frac{G_{t}}{\lambda +iG_{t}}\varphi
\right\Vert \leq 3mM\delta +\frac{mM^{2}}{\lambda -\beta _{0}\left( t\right)
}\left\Vert \Theta \tilde{\psi}\right\Vert \ .
\label{strong convergence of G lambda}
\end{equation}%
Through Lemma \ref{existence flow with W 4bis} we observe that
\begin{equation}
\left\Vert \Theta U_{t,s}\psi \right\Vert \leq \left\Vert \Theta
U_{t,s}\Theta ^{-1}\right\Vert _{\mathrm{op}}\left\Vert \Theta \psi
\right\Vert \leq \left\Vert \Theta \psi \right\Vert m\exp \left\{
\int_{s}^{t}\gamma \left( \tau \right) \mathrm{d}\tau \right\} <\infty \ .
\end{equation}%
Consequently, by taking the limit $\lambda \rightarrow \infty \ $in (\ref%
{upper bound differential equation}) with the use of Lebesgue's dominated
convergence theorem combined with Lemmata \ref{existence_flow_5}--\ref%
{existence flow with W 4bis} for any $\psi \in \mathcal{Y}$, we get the
upper bound
\begin{eqnarray}
&&\left\Vert \left\{ \epsilon ^{-1}\left( U_{t+\epsilon ,s}-U_{t,s}\right)
+iG_{t}U_{t,s}\right\} \psi \right\Vert  \notag \\
&\leq &2mM\epsilon ^{-1}\int_{t}^{t+\epsilon }\left\Vert \Theta \left\{
U_{\tau ,s}-U_{t,s}\right\} \psi \right\Vert \mathrm{d}\tau +3mM\delta
\notag \\
&&+Z_{\epsilon }\left\Vert \Theta \psi \right\Vert m\exp \left\{
\int_{s}^{t}2\gamma \left( \tau \right) \mathrm{d}\tau \right\} \ .
\end{eqnarray}%
When the parameters $\delta $ and $\epsilon $ go to zero, this last
inequality associated with Lemma \ref{existence flow with W 4bis} shows that
\begin{equation}
\forall t\geq s\geq 0:\qquad \partial _{t}^{+}U_{t,s}=-iG_{t}U_{t,s}\ ,\quad
U_{s,s}:=\mathbf{1}\ ,
\end{equation}%
where the right derivative is in the strong sense in $\mathcal{Y}$. The left
derivative is obtained in the same way, provided $t>s$. In other words, $%
U_{t,s}$ is a strong solution on $\mathcal{Y}$ of the initial value problem $%
\partial _{t}U_{t,s}=-iG_{t}U_{t,s}$ for any $t\geq s\geq 0$.

Equation (\ref{reversed autonomous eq}) is also proven in a similar way as
follows: For any $t>s>0$, $\psi \in \mathcal{Y}$, sufficiently small $%
\left\vert \epsilon \right\vert >0$ and $\lambda \geq 2\gamma _{\sup }$, we
use straightforward estimates together with Lemmata \ref{existence_flow_5}--%
\ref{existence flow with W 4} and (\ref{upper bound Jbis}) to obtain that%
\begin{eqnarray}
&&\left\Vert \left\{ \epsilon ^{-1}\left( U_{t,s+\epsilon }-U_{t,s}\right)
-iU_{t,s}G_{s}\right\} \psi \right\Vert  \notag \\
&\leq &\epsilon ^{-1}\int_{s}^{s+\epsilon }\left( \left\Vert \left(
U_{\lambda ,t,\tau }-U_{t,\tau }\right) G_{\tau }\psi \right\Vert
+\left\Vert \left( U_{t,\tau }-U_{t,s}\right) G_{\tau }\psi \right\Vert
\right) \mathrm{d}\tau  \notag \\
&&+m\exp \left\{ \int_{s}^{t}\beta _{0}\left( \tau \right) \mathrm{d}\tau
\right\} Z_{\epsilon }\left\Vert \Theta \psi \right\Vert
\label{derivative s0} \\
&&+m\exp \left\{ \int_{s}^{t}2\gamma \left( \tau \right) \mathrm{d}\tau
\right\} \epsilon ^{-1}\int_{s}^{s+\epsilon }\left\Vert \left( G_{\tau
,\lambda }-G_{\tau }\right) \psi \right\Vert \mathrm{d}\tau  \notag \\
&&+\left\Vert \epsilon ^{-1}\left\{ U_{t,s+\epsilon }-U_{\lambda
,t,s+\epsilon }\right\} \psi \right\Vert +\left\Vert \epsilon ^{-1}\left\{
U_{\lambda ,t,s}-U_{t,s}\right\} \psi \right\Vert \ .  \notag
\end{eqnarray}%
Now, by using Lemma \ref{lemma unitary propagator}, (\ref{strong convergence
of G lambda}), and Lebesgue's dominated convergence theorem, we take the
limit $\lambda \rightarrow 0^{+}$ in (\ref{derivative s0}) to deduce that
\begin{eqnarray}
&&\left\Vert \left\{ \epsilon ^{-1}\left( U_{t,s+\epsilon }-U_{t,s}\right)
-iU_{t,s}G_{s}\right\} \psi \right\Vert  \notag \\
&\leq &\epsilon ^{-1}\int_{s}^{s+\epsilon }\left\Vert \left( U_{t,\tau
}-U_{t,s}\right) G_{\tau }\psi \right\Vert \mathrm{d}\tau \\
&&+m\exp \left\{ \int_{s}^{t}\beta _{0}\left( \tau \right) \mathrm{d}\tau
\right\} Z_{\epsilon }\left\Vert \Theta \psi \right\Vert +3m^{2}M\delta \exp
\left\{ \int_{s}^{t}2\gamma \left( \tau \right) \mathrm{d}\tau \right\} \ .
\notag
\end{eqnarray}%
Passing to the limits $\delta ,\epsilon \rightarrow 0$ and using Lemma \ref%
{lemma unitary propagator} we thus arrive at
\begin{equation}
\forall t>s>0:\qquad \partial _{s}U_{t,s}=iU_{t,s}G_{s}\ ,\quad U_{t,t}:=%
\mathbf{1}\ ,
\end{equation}%
where the derivative is in the strong sense in $\mathcal{Y}$. Using exactly
the same arguments,
\begin{equation}
\partial _{s}^{+}U_{t,s}|_{s=0}=iU_{t,0}G_{0}\qquad \mathrm{and}\qquad
\partial _{s}^{-}U_{t,s}|_{s=t}=iG_{t}\ ,
\end{equation}%
provided $t>0$.

Note that the uniqueness of the solution of the Cauchy problem (\ref{flow
equationbisbisbis}) results from the fact that any other (bounded) solution $%
\tilde{U}_{t,s}$ satisfy the equality
\begin{equation}
(\tilde{U}_{t,s}-U_{t,s})\varphi =\int_{s}^{t}\partial _{\tau }\{U_{t,\tau }%
\tilde{U}_{\tau ,s}\}\varphi \ \mathrm{d}\tau =0  \label{uniqueness eq}
\end{equation}%
for any $\varphi \in \mathcal{Y}$. The set $\mathcal{Y}$ is dense and both
operators are bounded, so the previous equality means that $U_{t,s}=\tilde{U}%
_{t,s}$.\hfill $\Box $

\subsection{Autonomous Generators of Bogoliubov Transformations\label%
{section bogoliubov}}

Generators of Bogoliubov transformations have been studied in detail in \cite%
{Bruneau-derezinski2007}. Here, we indicate the rough idea of the proof for
\emph{finite} \emph{dimension}, i.e., when $\mathfrak{h}$ is a finite
dimensional Hilbert space. More precisely, the goal of this subsection is to
give a simple proof of the fact that all Bogoliubov $\mathbf{u}$--$\mathbf{v}
$ transformations, i.e., all unitary transformations $U$ on the boson Fock
space $\mathcal{F}_{b}$ of the form
\begin{equation}
\forall \varphi \in \mathfrak{h}:\qquad Ua(\varphi )U^{\ast }=a(\mathbf{u}%
^{\ast }\varphi )+a^{\ast }(\mathbf{v}^{\mathrm{t}}\bar{\varphi})+\langle
\bar{\psi}|\bar{\varphi}\rangle  \label{bog transf}
\end{equation}%
with $\mathbf{u,v\in }\mathcal{B}(\mathfrak{h})$, $\psi \in \mathfrak{h}$ ($%
\bar{\psi}(x)=\overline{\psi (x)}$, $x\in \mathbb{R}$) can be represented as
$U=\exp (i\mathbb{Q})$, where
\begin{eqnarray}
\mathbb{Q} &=&\underset{k,\ell }{\sum }\left\{ X\right\} _{k,\ell
}a_{k}^{\ast }a_{\ell }+\left\{ Y\right\} _{k,\ell }a_{k}^{\ast }a_{\ell
}^{\ast }+\left\{ \bar{Y}\right\} _{k,\ell }a_{k}a_{\ell }  \notag \\
&&+\underset{k}{\sum }\left\{ c_{k}a_{k}^{\ast }+\bar{c}_{k}a_{k}\right\} +%
\mathrm{const}  \label{bog transf2}
\end{eqnarray}%
is a self--adjoint quadratic boson operator. In fact, below we show that
\begin{equation}
U=U_{1}\left( \mathbf{u,v}\right) \exp \left[ a(\psi )-a^{\ast }(\psi )%
\right] \ ,
\end{equation}%
where $U_{1}\left( \mathbf{u,v}\right) =\exp (i\mathbb{Q}_{1})$, with $%
\mathbb{Q}_{1}$ being a self--adjoint quadratic boson operator of the form (%
\ref{bog transf2}) for $c_{k}=0$. See Lemma \ref{lemma 2} and (\ref{sympa4}).

The unitarity of the Bogoliubov transformation $U\in \mathcal{B}(\mathcal{F}%
_{b})$ is directly related to properties of operators $\mathbf{u,v\in }%
\mathcal{B}(\mathfrak{h})$, as expressed in the following proposition:

\begin{proposition}[Properties of operators $\mathbf{u}$ and $\mathbf{v}$]
\label{lemma 4}\mbox{ }\newline
Assume that the Bogoliubov transformation $U\in \mathcal{B}(\mathcal{F}_{b})$
defined by (\ref{bog transf}) with $\mathbf{u,v}\in \mathcal{B}(\mathfrak{h}%
) $ is unitary. Then $\mathbf{v}\in \mathcal{L}^{2}(\mathfrak{h})$ is
Hilbert--Schmidt and%
\begin{eqnarray}
\mathbf{uu}^{\ast }-\mathbf{vv}^{\ast } &=&\mathbf{1}\ ,\qquad \mathbf{uv}^{%
\mathrm{t}}=\mathbf{vu}^{\mathrm{t}}\ ,  \label{eq unitary1} \\
\mathbf{u}^{\ast }\mathbf{u}-\mathbf{v}^{\mathrm{t}}\mathbf{\bar{v}} &=&%
\mathbf{1}\ ,\qquad \mathbf{u}^{\ast }\mathbf{v}=\mathbf{v}^{\mathrm{t}}%
\mathbf{\bar{u}}\ .  \label{eq unitary2}
\end{eqnarray}%
Conversely, if $\mathbf{v}\in \mathcal{L}^{2}(\mathfrak{h})$ and (\ref{eq
unitary1})--(\ref{eq unitary2}) hold then the Bogoliubov transformation $%
U\in \mathcal{B}(\mathcal{F}_{b})$ defined by (\ref{bog transf}) is unitary.
\end{proposition}

\noindent \textbf{Proof.} A Bogoliubov transformation $U\in \mathcal{B}(%
\mathcal{F}_{b})$ is unitary iff the operator family $(Ua_{k}U^{\ast
},U^{\ast }a_{k}^{\ast }U)_{k=1}^{\infty }$ satisfy the Canonical
Commutations Relations (CCR), like the operator family $(a_{k},a_{k}^{\ast
})_{k=1}^{\infty }$ defined below by (\ref{def ak}), and the vacuum $%
\left\vert 0\right\rangle \in \mathcal{D}\left( U\right) $. By (\ref{bog
transf}), this property yields the proposition through some computations. It
refers to the Shale criterion.\hfill $\Box $

Note that the homogeneous Bogoliubov transformation corresponds to $\psi =0$
in (\ref{bog transf}). In this case, by taking any \emph{real} orthonormal
basis $\{\varphi _{k}\}_{k=1}^{\infty }\subseteq \mathfrak{h}$ and using the
definition
\begin{equation}
\forall k\in \mathbb{N}:\qquad a_{k}:=a(\varphi _{k})\ ,  \label{def ak}
\end{equation}%
we observe that
\begin{equation}
\tilde{a}_{k}:=Ua(\varphi _{k})U^{\ast }=a(\mathbf{u}^{\ast }\varphi
_{k})+a^{\ast }(\mathbf{v}^{\mathrm{t}}\varphi _{k})=\underset{\ell }{\sum }%
\left\{ \mathbf{u}\right\} _{k,\ell }a_{\ell }+\left\{ \mathbf{v}\right\}
_{k,\ell }a_{\ell }^{\ast }\ ,
\end{equation}%
as for the Bogoliubov $\mathbf{u}$--$\mathbf{v}$ (unitary) transformation $%
\mathrm{U}_{t,s}$, see Theorem \ref{theorem important 2bis} (ii).

It is then natural to ask for the generator of Bogoliubov $\mathbf{u}$--$%
\mathbf{v}$ transformations when it defines a unitary operator $U\in
\mathcal{B}(\mathcal{F}_{b})$, i.e., when $\mathbf{v}\in \mathcal{L}^{2}(%
\mathfrak{h})$ and (\ref{eq unitary1})--(\ref{eq unitary2}) hold. This is
the main result of this subsection, expressed in the following theorem:

\begin{theorem}[Generator of Bogoliubov $\mathbf{u}$--$\mathbf{v}$
transformations]
\label{flow equation thm 5 copy(1)}\mbox{ }\newline
The generator $\mathbb{Q}$ of a Bogoliubov transformation $U=\exp (i\mathbb{Q%
})$ defined by (\ref{bog transf}) with $\psi \in \mathfrak{h}$ and operators
$\mathbf{u,v}$ satisfying $\mathbf{v}\in \mathcal{L}^{2}(\mathfrak{h})$ and (%
\ref{eq unitary1})--(\ref{eq unitary2}) is a self--adjoint quadratic boson
operator.
\end{theorem}

To prove this theorem, we make use of the fact that if $\mathbb{Q}_{1}=%
\mathbb{Q}_{1}^{\ast }$ and $\mathbb{Q}_{2}=\mathbb{Q}_{2}^{\ast }$ are two
self--adjoint quadratic boson operators then
\begin{equation}
\mathrm{e}^{i\mathbb{Q}_{1}}\mathrm{e}^{i\mathbb{Q}_{2}}=\mathrm{e}^{i%
\mathbb{Q}_{3}}
\end{equation}%
for some self--adjoint quadratic boson operator $\mathbb{Q}_{3}=\mathbb{Q}%
_{3}^{\ast }$. In the case of infinite dimension, note that there may, in
general, be a problem of defining $\mathbb{Q}_{3}$ because the operators $%
X_{1},X_{2}$ (cf. (\ref{bog transf2})) of respectively $\mathbb{Q}_{1}$ and $%
\mathbb{Q}_{2}$ may be unbounded with no dense common domain of definition.
Again we refer to \cite{Bruneau-derezinski2007} for the general case.

We also need three elementary lemmata. The first one concerns an explicit
computation about an elementary homogeneous Bogoliubov transformation.

\begin{lemma}[Elementary homogeneous Bogoliubov transf.]
\label{lemma 1}\mbox{ }\newline
Let $\alpha >0$ and $f\in \mathfrak{h}$ such that $\Vert f\Vert =1$. Then,
for all $\varphi \in \mathfrak{h}$,
\begin{eqnarray}
&&\exp \left[ \frac{\alpha }{2}\left( a(f)^{2}-a^{\ast }(f)^{2}\right) %
\right] a(\varphi )\exp \left[ -\frac{\alpha }{2}\left( a(f)^{2}-a^{\ast
}(f)^{2}\right) \right]  \notag \\
&=&a\left( \cosh \left( \alpha \right) \langle f|\varphi \rangle f\right)
+a^{\ast }\left( \sinh \left( \alpha \right) \langle \bar{f}|\bar{\varphi}%
\rangle f\right) +a\left( \varphi -\langle f|\varphi \rangle f\right) \ .
\notag \\
&&
\end{eqnarray}
\end{lemma}

\noindent \textbf{Proof.} We refrain from proving the self--adjointness of $%
q:=ia(f)^{2}-ia^{\ast }(f)^{2}$. For any $\varphi \in \mathfrak{h}$ and $%
t\in \lbrack 0,\alpha /2]$ with $\alpha >0$, we introduce the operator%
\begin{equation}
a_{t}\left( \varphi \right) :=\exp \left[ -itq\right] a(\varphi )\exp \left[
itq\right] \ .
\end{equation}%
Then its derivative for all $t\in \lbrack 0,\alpha /2]$ equals%
\begin{equation}
\partial _{t}a_{t}\left( \varphi \right) =i\left[ a_{t}\left( \varphi
\right) ,q\right] =\mathrm{e}^{-itq}\left[ a(\varphi ),a^{\ast }(f)^{2}%
\right] \mathrm{e}^{itq}=2\langle \varphi |f\rangle a_{t}^{\ast }\left(
f\right) \ .
\end{equation}%
So, clearly,
\begin{equation}
\forall \varphi \bot f\ :\qquad \mathrm{e}^{-i\alpha q/2}a(\varphi )\mathrm{e%
}^{i\alpha q/2}=a(\varphi )\ .  \label{VI.621}
\end{equation}%
Conversely, if $\varphi =f$ and $\left\Vert f\right\Vert =1$ then
\begin{equation}
\partial _{t}\left(
\begin{array}{c}
a_{t}\left( f\right) \\
a_{t}^{\ast }\left( f\right)%
\end{array}%
\right) =2\left(
\begin{array}{cc}
0 & 1 \\
1 & 0%
\end{array}%
\right) \left(
\begin{array}{c}
a_{t}\left( f\right) \\
a_{t}^{\ast }\left( f\right)%
\end{array}%
\right) \ ,
\end{equation}%
which in turn implies that
\begin{equation}
\left(
\begin{array}{c}
a_{\alpha /2}\left( f\right) \\
a_{\alpha /2}^{\ast }\left( f\right)%
\end{array}%
\right) =\exp \left[ \alpha \left(
\begin{array}{cc}
0 & 1 \\
1 & 0%
\end{array}%
\right) \right] \left(
\begin{array}{c}
a\left( f\right) \\
a^{\ast }\left( f\right)%
\end{array}%
\right) \ .  \label{eq. 10}
\end{equation}%
Since%
\begin{eqnarray}
\exp \left[ \alpha \left(
\begin{array}{cc}
0 & 1 \\
1 & 0%
\end{array}%
\right) \right] &=&\underset{n=0}{\overset{\infty }{\sum }}\left\{ \frac{%
\alpha ^{2n}}{\left( 2n\right) !}\left(
\begin{array}{cc}
1 & 0 \\
0 & 1%
\end{array}%
\right) +\frac{\alpha ^{2n+1}}{\left( 2n+1\right) !}\left(
\begin{array}{cc}
0 & 1 \\
1 & 0%
\end{array}%
\right) \right\}  \notag \\
&=&\left(
\begin{array}{cc}
\cosh \left( \alpha \right) & \sinh \left( \alpha \right) \\
\sinh \left( \alpha \right) & \cosh \left( \alpha \right)%
\end{array}%
\right) \ ,
\end{eqnarray}%
we infer from (\ref{eq. 10}) that
\begin{equation}
a_{\alpha /2}\left( f\right) =\cosh \left( \alpha \right) a\left( f\right)
+\sinh \left( \alpha \right) a^{\ast }\left( f\right) \ ,  \label{VI.622}
\end{equation}%
for any $\alpha >0$ and $f\in \mathfrak{h}$ such that $\Vert f\Vert =1$. The
lemma then results from (\ref{VI.621}) and (\ref{VI.622}) combined with the
antilinearity of $a\left( \varphi \right) $\ and the linearity of $a^{\ast
}\left( \varphi \right) $ with respect to $\varphi \in \mathfrak{h}$.\hfill $%
\Box $

The second step is to analyze an elementary inhomogeneous Bogoliubov
transformation where $\mathbf{u}=\mathbf{v}=0$.

\begin{lemma}[Elementary inhomogeneous Bogoliubov transf.]
\label{lemma 2}\mbox{ }\newline
Let $\psi \in \mathfrak{h}$ and $\alpha >0$. Then, for all $\varphi \in
\mathfrak{h}$,%
\begin{equation}
\exp \left[ \alpha \left( a(\psi )-a^{\ast }(\psi )\right) \right] a(\varphi
)\exp \left[ -\alpha \left( a(\psi )-a^{\ast }(\psi )\right) \right]
=a(\varphi )+\alpha \langle \bar{\psi}|\bar{\varphi}\rangle \ .
\end{equation}
\end{lemma}

\noindent \textbf{Proof.} For any $t\in \lbrack 0,\alpha ]$ with $\alpha >0$%
, define the operators%
\begin{equation}
Z:=i\left( a(\psi )-a^{\ast }(\psi )\right) \qquad \mathrm{and}\qquad
a_{t}\left( \phi \right) :=\mathrm{e}^{-itZ}a\left( \varphi \right) \mathrm{e%
}^{itZ}\ .
\end{equation}%
Then its derivative for all $t\in \lbrack 0,\alpha ]$ equals
\begin{equation}
\partial _{t}a_{t}\left( \phi \right) =\mathrm{e}^{-itZ}\left[ a\left(
\varphi \right) ,a^{\ast }(\psi )\right] \mathrm{e}^{itZ}=\langle \varphi
|\psi \rangle =\langle \bar{\psi}|\bar{\varphi}\rangle \ ,
\end{equation}%
which clearly implies the assertion.\hfill $\Box $

The proof of Theorem \ref{flow equation thm 5 copy(1)} uses the fact that
any unitary operator is the exponential of some self--adjoint operator. This
is completely standard and we shortly prove it here for completeness.

\begin{lemma}[Generator of unitary operators]
\label{lemma 3}\mbox{ }\newline
Let $u\in \mathcal{B}(\mathfrak{h})$ be unitary. Then there exists a
self--adjoint operator $\left( \mathrm{h},\mathcal{D}\left( \mathrm{h}%
\right) \right) $ acting on $\mathfrak{h}$ such that $u=\exp (i\mathrm{h}).$
\end{lemma}

\noindent \textbf{Proof.} Since $u$ is unitary, it is normal and the
spectral theorem implies that there is a realization of $u$ on some $L^{2}(%
\mathcal{A},\mathrm{d}\mathfrak{a})$ as a multiplication operator, i.e.,
there exists a measurable function $\upsilon :\mathcal{A}\rightarrow \mathbb{%
R}$ such that
\begin{equation}
\forall f\in L^{2}(\mathcal{A},\mathrm{d}\mathfrak{a}),\ x\in \mathcal{A}%
:\qquad \left( uf\right) \left( x\right) =\mathrm{e}^{i\upsilon \left(
x\right) }f\left( x\right) \ .
\end{equation}%
Defining $\mathrm{h}$ on $\mathfrak{h}$ by
\begin{equation}
\forall f\in L^{2}(\mathcal{A},\mathrm{d}\mathfrak{a}),\ x\in \mathcal{A}%
:\qquad \left( \mathrm{h}f\right) \left( x\right) :=\upsilon \left( x\right)
f\left( x\right) \ ,
\end{equation}%
one gets the self--adjoint operator sought for. \hfill $\Box $

Now, we are in position to prove Theorem \ref{flow equation thm 5 copy(1)}%
:\bigskip

\noindent \textbf{Proof of Theorem \ref{flow equation thm 5 copy(1)}.} Since
$\mathbf{v}\in \mathcal{L}^{2}(\mathfrak{h})$, this operator is compact and
its singular value decomposition is
\begin{equation}
\mathbf{v}=\underset{k}{\sum }\lambda _{k}|g_{k}\rangle \langle \bar{f}%
_{k}|\ ,
\end{equation}%
where $\{f_{k}\}_{k=1}^{N},\{g_{k}\}_{k=1}^{N}\subseteq \mathfrak{h}$ are
orthonormal bases and $\{\lambda _{k}\}_{k=1}^{N}\subset \mathbb{R}$ is a
set of real numbers satisfying%
\begin{equation}
\left\Vert \mathbf{v}\right\Vert _{2}^{2}=\mathrm{tr}\left( \mathbf{vv}%
^{\ast }\right) \mathbf{=}\underset{k}{\sum }\lambda _{k}^{2}<\infty \ .
\end{equation}%
Indeed, straightforward computations show that%
\begin{equation}
\mathbf{vv}^{\ast }=\underset{k}{\sum }\lambda _{k}^{2}|g_{k}\rangle \langle
g_{k}|
\end{equation}%
and since $\mathbf{uu}^{\ast }-\mathbf{vv}^{\ast }=\mathbf{1}$, we also
deduce that
\begin{equation}
\mathbf{uu}^{\ast }=\underset{k}{\sum }\left( \lambda _{k}^{2}+1\right)
|g_{k}\rangle \langle g_{k}|\ .  \label{eq 26}
\end{equation}%
Writing
\begin{eqnarray}
s_{k} &:=&\sinh \left( \alpha _{k}\right) :=\lambda _{k}\ , \\
c_{k} &:=&\cosh \left( \alpha _{k}\right) :=\sqrt{1+\lambda _{k}^{2}}\ ,
\end{eqnarray}%
and using some suitable orthonormal basis $\{h_{k}\}_{k=1}^{N}\subseteq
\mathfrak{h}$ we infer from Equation (\ref{eq 26}) that
\begin{equation}
\mathbf{v}=\underset{k}{\sum }s_{k}|g_{k}\rangle \langle \bar{f}_{k}|\qquad
\mathrm{and}\qquad \mathbf{u}=\underset{k}{\sum }c_{k}|g_{k}\rangle \langle
h_{k}|\ .
\end{equation}%
Next, observe that%
\begin{eqnarray}
\mathbf{u}^{\ast }\mathbf{v} &\mathbf{=}&\underset{k,\ell }{\sum }%
c_{k}s_{\ell }|h_{k}\rangle \langle g_{k}|g_{\ell }\rangle \langle \bar{f}%
_{\ell }|=\underset{k}{\sum }c_{k}s_{k}|h_{k}\rangle \langle \bar{f}_{k}|\ ,
\\
\mathbf{v}^{\mathrm{t}}\mathbf{\bar{u}} &=&\underset{k,\ell }{\sum }%
c_{k}s_{\ell }|f_{\ell }\rangle \langle \bar{g}_{\ell }|\bar{g}_{k}\rangle
\langle \bar{h}_{k}|=\underset{k}{\sum }c_{k}s_{k}|f_{k}\rangle \langle \bar{%
h}_{k}|\ .
\end{eqnarray}%
Therefore, $\mathbf{u}^{\ast }\mathbf{v}=\mathbf{v}^{\mathrm{t}}\mathbf{\bar{%
u}}$ implies the equality $f_{k}=h_{k}$ for all $k\in \mathbb{N}$ such that $%
s_{k}\neq 0$ and we conclude that%
\begin{equation}
\mathbf{v}=\underset{k}{\sum }s_{k}|g_{k}\rangle \langle \bar{h}_{k}|\qquad
\mathrm{and}\qquad \mathbf{u}=\underset{k}{\sum }c_{k}|g_{k}\rangle \langle
h_{k}|\ .  \label{sympa}
\end{equation}%
Now, we define the unitary operator $U_{1}$ to be
\begin{equation}
U_{1}:=\exp (-i\mathbb{Q}_{1})\ ,
\end{equation}%
where
\begin{equation}
\mathbb{Q}_{1}:=\underset{k}{\sum }\frac{i}{2}\alpha _{k}\left(
a(g_{k})^{2}-a^{\ast }(g_{k})^{2}\right) =\mathbb{Q}_{1}^{\ast }\ .
\end{equation}%
By Lemma \ref{lemma 1}, we have, for all $\varphi \in \mathfrak{h}$, the
equality
\begin{equation}
U_{1}a(\varphi )U_{1}^{\ast }=\underset{k}{\sum }\left\{ a(c_{k}\langle
g_{k}|\varphi \rangle g_{k})+a^{\ast }(s_{k}\langle \bar{g}_{k}|\bar{\varphi}%
\rangle g_{k})\right\} \ .  \label{sympa2}
\end{equation}%
We also define a unitary operator $\hat{u}$ by $\hat{u}\left( g_{k}\right)
=h_{k}$ for all $k\in \mathbb{N}$. By Lemma \ref{lemma 3}, there is a
self--adjoint operator $\mathrm{h}$ such that
\begin{equation}
\hat{u}:=\underset{k}{\sum }|h_{k}\rangle \langle g_{k}|=\exp (i\mathrm{h})\
.
\end{equation}%
Using the second quantization $\mathrm{d}\Gamma \left( \mathrm{h}\right) $
of $\mathrm{h}$, which is in this case a self--adjoint quadratic boson
operator, we thus note that
\begin{equation}
\forall \varphi \in \mathfrak{h}:\qquad \exp \left( i\mathrm{d}\Gamma \left(
\mathrm{h}\right) \right) a^{\ast }(\varphi )\exp \left( -i\mathrm{d}\Gamma
\left( \mathrm{h}\right) \right) =a^{\ast }(\hat{u}\varphi )\ .
\label{sympa3}
\end{equation}%
Finally, we set the unitary operator
\begin{equation}
U:=\exp \left( i\mathrm{d}\Gamma \left( \mathrm{h}\right) \right) U_{1}\exp
\left( a(\psi )-a^{\ast }(\psi )\right) \ .  \label{sympa4}
\end{equation}%
Then $U=\exp (i\mathbb{Q})$ for some self--adjoint quadratic boson operator $%
\mathbb{Q}=\mathbb{Q}^{\ast }$ and using Lemma \ref{lemma 2}, (\ref{sympa}),
(\ref{sympa2}), and (\ref{sympa3}), we arrive, for all $\varphi \in
\mathfrak{h}$, at the equalities
\begin{eqnarray}
Ua(\varphi )U^{\ast } &=&\mathrm{e}^{i\mathrm{d}\Gamma \left( \mathrm{h}%
\right) }U_{1}\left( a(\varphi )+\langle \bar{\psi}|\bar{\varphi}\rangle
\right) U_{1}^{\ast }\mathrm{e}^{-i\mathrm{d}\Gamma \left( \mathrm{h}\right)
}  \notag \\
&=&\mathrm{e}^{i\mathrm{d}\Gamma \left( \mathrm{h}\right) }\left( \underset{k%
}{\sum }\left\{ a(c_{k}\langle g_{k}|\varphi \rangle g_{k})+a^{\ast
}(s_{k}\langle \bar{g}_{k}|\bar{\varphi}\rangle g_{k})\right\} +\langle \bar{%
\psi}|\bar{\varphi}\rangle \right) \mathrm{e}^{-i\mathrm{d}\Gamma \left(
\mathrm{h}\right) }  \notag \\
&=&\underset{k}{\sum }\left\{ a(c_{k}\langle g_{k}|\varphi \rangle
h_{k})+a^{\ast }(s_{k}\langle \bar{g}_{k}|\bar{\varphi}\rangle
h_{k})\right\} +\langle \bar{\psi}|\bar{\varphi}\rangle  \notag \\
&=&a(\mathbf{u}^{\ast }\varphi )+a^{\ast }(\mathbf{v}^{\mathrm{t}}\bar{%
\varphi})+\langle \bar{\psi}|\bar{\varphi}\rangle \ ,
\end{eqnarray}%
which prove Theorem \ref{flow equation thm 5 copy(1)} because the last
equality uniquely defines the unitary operator $U$ up to an overall phase
factor. \hfill $\Box $

\subsection{Trace and Representation of Hilbert--Schmidt Operators\label%
{section hilbert schmidt}}

It is well--known that the Hilbert space $\mathcal{L}^{2}\left( \mathfrak{h}%
\right) $ of Hilbert--Schmidt operators can be identified\ with the tensor
product $\mathfrak{h}\otimes \mathfrak{h}^{\ast }$ by the natural unitary
isomorphism
\begin{equation}
\mathcal{J}%
\Big[%
\left\vert \varphi \right\rangle \left\langle \psi \right\vert
\Big]%
:=\varphi \otimes \psi \ .  \label{isomorphism}
\end{equation}%
The Hilbert space $\mathfrak{h}\otimes \mathfrak{h}^{\ast }$ has norm $%
\left\Vert \cdot \right\Vert _{2}$ defined from the usual scalar product
\begin{equation}
\left( \varphi \otimes \psi ,\varphi ^{\prime }\otimes \psi ^{\prime
}\right) _{2}:=\langle \varphi |\varphi ^{\prime }\rangle \langle \psi
^{\prime }|\psi \rangle \ ,\mathrm{\quad for\ }\varphi \otimes \psi ,\varphi
^{\prime }\otimes \psi ^{\prime }\in \mathfrak{h}\otimes \mathfrak{h}^{\ast
}\ .
\end{equation}%
The norm on $\mathcal{L}^{2}\left( \mathfrak{h}\right) $ is also denoted by $%
\left\Vert \cdot \right\Vert _{2}$ and results from the scalar product%
\begin{equation}
\left( X,Y\right) _{2}:=\mathrm{tr}(X^{\ast }Y)\ ,\mathrm{\qquad for\ }%
X,Y\in \mathcal{L}^{2}(\mathfrak{h})\ .
\end{equation}

The unitary isomorphism $\mathcal{J}$ is useful in this paper, albeit not
essential, because of the unboundedness of the self--adjoint operator $%
\Omega _{0}=\Omega _{0}^{\ast }\geq 0$. Indeed, the unitary isomorphism $%
\mathcal{J}$ allows us to interpret the differential equation
\begin{equation}
\forall t\geq 0:\qquad \partial _{t}Y_{t}=-\alpha \left(
Z_{t}Y_{t}+Y_{t}Z_{t}^{\mathrm{t}}\right) \ ,\quad Y_{0}\in \mathcal{L}%
^{2}\left( \mathfrak{h}\right) \ ,  \label{diff eq}
\end{equation}%
on $\mathcal{L}^{2}\left( \mathfrak{h}\right) $ as a \emph{non--autonomous
evolution equation}, even if $Z_{t}$ is an unbounded operator acting on $%
\mathfrak{h}$. See as an example Equations (\ref{flow equation-quadratic})
and (\ref{eq iso}). In particular, the uniqueness of a solution in $\mathcal{%
L}^{2}\left( \mathfrak{h}\right) $ of (\ref{diff eq}) is an immediate
consequence of this fact. It is illustrated for a special parabolic case in
the next lemma:

\begin{lemma}[Uniqueness of a solution of $\partial _{t}Y_{t}=-\protect%
\alpha \left( Z_{t}Y_{t}+Y_{t}Z_{t}^{\mathrm{t}}\right) $]
\label{lemma existence 3 copy(1)}\mbox{ }\newline
Let%
\begin{equation}
\forall t\in \lbrack 0,T):\qquad Z_{t}:=Z_{0}+Q_{t}\ ,\quad Q_{0}:=0\ ,
\end{equation}%
where $T\in (0,\infty ]$, $Z_{0}=Z_{0}^{\ast }\geq 0$ is a positive
(possibly unbounded) operator on $\mathfrak{h}$ and $(Q_{t})_{t\in \lbrack
0,T)}$ is a strongly continuous family of bounded, self--adjoint operators.
Then the zero operator family $(0)_{t\in \lbrack 0,T)}\in C^{1}[[0,T);%
\mathcal{L}^{2}(\mathfrak{h})]$ is the unique solution in the
Hilbert--Schmidt topology of the integral equation $X=\mathfrak{F}(X)$, where%
\begin{equation}
\forall t\in \lbrack 0,T):\qquad \left[ \mathfrak{F}(X)\right] _{t}:=-\alpha
\int_{0}^{t}\left( Z_{\tau }X_{\tau }+X_{\tau }Z_{\tau }^{\mathrm{t}}\right)
\mathrm{d}\tau  \label{problem1}
\end{equation}%
for any $\alpha >0$. The same assertion holds if $Z_{\tau }$ is replaced by $%
Z_{\tau }^{\mathrm{t}}$ in (\ref{problem1}).
\end{lemma}

\noindent \textbf{Proof.} Assume the family $(X_{t})_{t\in \lbrack 0,T)}$ of
Hilbert--Schmidt operators solves the corresponding integral equation. Then,
using the unitary isomorphism $\mathcal{J}$ we note that $\hat{X}_{t}:=%
\mathcal{J}\left( X_{t}\right) $ obeys
\begin{equation}
\forall t\in \lbrack 0,T):\qquad \hat{X}_{t}=-\alpha \int_{0}^{t}\hat{Z}%
_{\tau }\hat{X}_{\tau }\mathrm{d}\tau \ ,  \label{problem1bis}
\end{equation}%
where
\begin{equation}
\hat{Z}_{t}:=\mathcal{J}\left[ Z_{t}\left( \cdot \right) +\left( \cdot
\right) Z_{t}^{\mathrm{t}}\right] \mathcal{J}^{\ast }=Z_{t}\otimes \mathbf{1}%
+\mathbf{1}\otimes Z_{t}^{\mathrm{t}}\ .  \label{problem2}
\end{equation}%
The existence and uniqueness of the solution of (\ref{problem1bis}) now
follows from standard arguments, since $\hat{Z}_{t}=\hat{Z}_{0}+\hat{Q}_{t}$
with
\begin{equation}
\hat{Z}_{0}:=Z_{0}\otimes \mathbf{1}+\mathbf{1}\otimes Z_{0}^{\mathrm{t}%
}\geq 0\quad \mathrm{and}\quad \hat{Q}_{t}:=Q_{t}\otimes \mathbf{1}+\mathbf{1%
}\otimes Q_{t}^{\mathrm{t}}\in \mathcal{B}\left( \mathfrak{h}\right) \ .
\end{equation}%
It implies that $X_{t}=\mathcal{J}^{\ast }(\hat{X}_{t})=0$ is the unique
solution of (\ref{problem1}) in $\mathcal{L}^{2}\left( \mathfrak{h}\right) $%
. If $Z_{\tau }$ replaces $Z_{\tau }^{\mathrm{t}}$ in (\ref{problem1}), then
we only need to change $Z_{t}^{\mathrm{t}}$ by $Z_{t}^{\ast }=Z_{t}$ in (\ref%
{problem2}) and the same arguments as above yield the assertion. \hfill $%
\Box $

\noindent It clearly follows under the assumptions of Lemma \ref{lemma
existence 3 copy(1)} that the solution in $\mathcal{L}^{2}\left( \mathfrak{h}%
\right) $ of the (parabolic) non--autonomous evolution equation (\ref{diff
eq}) is unique. (Existence of a solution of (\ref{diff eq}) is in fact
standard under these assumptions.)

Now, we observe that the cyclicity of the trace
\begin{equation}
\mathrm{tr}(XY)=\mathrm{tr}(YX)<\infty  \label{cyclicity}
\end{equation}%
holds when $X,Y\in \mathcal{B}\left( \mathfrak{h}\right) $ and both $%
XY,YX\in \mathcal{L}^{1}(\mathfrak{h})$, see \cite[Cor. 3.8]{Simon}, but one
must be careful in invoking cyclicity of the trace with unbounded operators
like $\Omega _{0}=\Omega _{0}^{\ast }\geq 0$. In the next lemma, we get
around this problem by extending \cite[Cor. 3.8]{Simon}:

\begin{lemma}[Cyclicity of the trace]
\label{lemma existence 3 copy(3)}\mbox{ }\newline
\emph{(i)} For any $X,Y\in \mathcal{B}(\mathfrak{h})$ such that $XY,YX\in
\mathcal{L}^{1}(\mathfrak{h})$,%
\begin{equation}
\mathrm{tr}(XY)=\mathrm{tr}(YX)\ .  \label{trace}
\end{equation}%
In particular, if $X\in \mathcal{L}^{1}(\mathfrak{h})$ and $Y\in \mathcal{B}(%
\mathfrak{h})$ then $XY,YX\in \mathcal{L}^{1}(\mathfrak{h})$ and (\ref{trace}%
) holds. \newline
\emph{(ii)} For any $X\in \mathcal{B}(\mathfrak{h})$ and (possibly
unbounded) self--adjoint operators $Y=Y^{\ast }$ on $\mathfrak{h}$ such that
$XY,YX\in \mathcal{L}^{1}(\mathfrak{h})$,
\begin{equation}
\mathrm{tr}(XY)=\mathrm{tr}(YX)\ .
\end{equation}%
\emph{(iii)} For any (possibly unbounded) operators $X$ and $Y$ on $%
\mathfrak{h}$,
\begin{equation}
\left\Vert XY\right\Vert _{2}^{2}:=\mathrm{tr}(Y^{\ast }X^{\ast }XY)=\mathrm{%
tr}(XYY^{\ast }X^{\ast })=:\left\Vert Y^{\ast }X^{\ast }\right\Vert
_{2}^{2}\ .
\end{equation}
\end{lemma}

\noindent \textbf{Proof.} \underline{(i):} The first assertion is \cite[Cor.
3.8]{Simon} and results from the fact that the spectrums $\sigma (XY)$ and $%
\sigma (YX)$, respectively of the trace--class operators $XY$ and $YX$,
satisfy
\begin{equation}
\sigma (XY)\cup \{0\}=\sigma (YX)\cup \{0\}
\end{equation}%
with the same multiplicity for each non--zero element of the spectrums. The
fact that $X\in \mathcal{L}^{1}(\mathfrak{h})$ and $Y\in \mathcal{B}(%
\mathfrak{h})$ yield $XY,YX\in \mathcal{L}^{1}(\mathfrak{h})$ is deduced
from \cite[Thm VI.19 (b)]{ReedSimon}.\smallskip

\underline{(ii):} If\ $Y$\ is bounded then the assertion follows directly
from (i), so we may assume that $Y$ is unbounded. For $m>0$, observe that
\begin{equation}
\mathrm{tr}\left( XY-XY\mathbf{1}\left[ \left\vert Y\right\vert \leq m\right]
\right) =\underset{k=1}{\overset{\infty }{\sum }}\lambda _{k}\langle \varphi
_{k}|\mathbf{1}\left[ \left\vert Y\right\vert >m\right] \psi _{k}\rangle \ ,
\end{equation}%
using the singular value decomposition%
\begin{equation}
XY=\underset{k=1}{\overset{\infty }{\sum }}\lambda _{k}|\psi _{k}\rangle
\langle \varphi _{k}|
\end{equation}%
of the trace--class operator $XY$, where the singular values $\{\lambda
_{k}\}_{k=1}^{\infty }$ are absolutely summable and $\{\varphi
_{k}\}_{k=1}^{\infty },\{\psi _{k}\}_{k=1}^{\infty }\subset \mathfrak{h}$
are orthonormal bases. Since
\begin{equation}
\underset{m\rightarrow \infty }{\lim }\langle \varphi _{k}|\mathbf{1}\left[
\left\vert Y\right\vert >m\right] \psi _{k}\rangle =0\ ,
\end{equation}%
for all $k\in \mathbb{N}$, Lebesgue's dominated convergence theorem implies
that
\begin{equation}
\mathrm{tr}\left( XY\right) =\underset{m\rightarrow \infty }{\lim }\mathrm{tr%
}\left( XY\mathbf{1}\left[ \left\vert Y\right\vert \leq m\right] \right) \ .
\end{equation}%
Similarly,
\begin{equation}
\mathrm{tr}\left( YX\right) =\underset{m\rightarrow \infty }{\lim }\mathrm{tr%
}\left( Y\mathbf{1}\left[ \left\vert Y\right\vert \leq m\right] X\right) \ ,
\end{equation}%
and thus (i) implies that
\begin{eqnarray}
\mathrm{tr}\left( XY\right) &=&\underset{m\rightarrow \infty }{\lim }\mathrm{%
tr}\left( XY\mathbf{1}\left[ \left\vert Y\right\vert \leq m\right] \right) \\
&=&\underset{m\rightarrow \infty }{\lim }\mathrm{tr}\left( Y\mathbf{1}\left[
\left\vert Y\right\vert \leq m\right] X\right) =\mathrm{tr}\left( YX\right)
\ .  \notag
\end{eqnarray}%
\smallskip

\underline{(iii):} Suppose that $\left\Vert XY\right\Vert _{2}^{2}<\infty $.
Then, $XY\in \mathcal{L}^{2}\left( \mathfrak{h}\right) $ and also $Y^{\ast
}X^{\ast }=(XY)^{\ast }\in \mathcal{L}^{2}\left( \mathfrak{h}\right) $.
Hence, both $(Y^{\ast }X^{\ast })(XY),(XY)(Y^{\ast }X^{\ast })\in \mathcal{L}%
^{1}\left( \mathfrak{h}\right) $, and (i) implies that
\begin{equation}
\left\Vert XY\right\Vert _{2}^{2}:=\mathrm{tr}((Y^{\ast }X^{\ast })(XY))=%
\mathrm{tr}((XY)(Y^{\ast }X^{\ast }))=:\left\Vert Y^{\ast }X^{\ast
}\right\Vert _{2}^{2}\ .
\end{equation}%
\hfill $\Box $

We conclude the paper by three other elementary lemmata which are
extensively used in our proofs and are related to properties of the trace.

\begin{lemma}[Trace, transpose and complex conjugate]
\label{lemma existence 3 copy(2)}\mbox{ }\newline
For any operator $X$ on $\mathfrak{h}$ and $n\in \mathbb{N}$, $\left( X^{%
\mathrm{t}}\right) ^{n}=\left( X^{n}\right) ^{\mathrm{t}}$, $\bar{X}^{n}=%
\overline{X^{n}}$. This property can be extended to any $n\in \mathbb{Z}$
whenever $X$ is invertible. Moreover, $X\in \mathcal{L}^{1}(\mathfrak{h})$
iff $X^{\mathrm{t}}\in \mathcal{L}^{1}(\mathfrak{h})$; $X\in \mathcal{L}^{1}(%
\mathfrak{h})$ iff $\bar{X}\in \mathcal{L}^{1}(\mathfrak{h})$. In
particular, if $X\in \mathcal{L}^{1}(\mathfrak{h})$ then%
\begin{equation}
\mathrm{tr}(X^{\mathrm{t}})=\mathrm{tr}(X),\quad \mathrm{tr}(\bar{X})=%
\overline{\mathrm{tr}(X)}\ .
\end{equation}
\end{lemma}

\noindent \textbf{Proof.} Recall that the scalar product on $\mathfrak{h}$
is given by
\begin{equation}
\langle f|g\rangle :=\int_{\mathcal{M}}\overline{f\left( x\right) }g\left(
x\right) \mathrm{d}\mathfrak{a}\left( x\right) \ ,
\end{equation}%
where, for every $f\in \mathfrak{h}$, we define its complex conjugate $\bar{f%
}\in \mathfrak{h}$ by $\bar{f}\left( x\right) :=\overline{f\left( x\right) }$%
, for all $x\in \mathcal{M}$. See (\ref{scalar product}). For any operator $%
X $ on $\mathfrak{h}$, we define its transpose $X^{\mathrm{t}}$\ and its
complex conjugate $\bar{X}$ by $\langle f|X^{\mathrm{t}}g\rangle :=\langle
\bar{g}|X\bar{f}\rangle $ and $\langle f|\bar{X}g\rangle :=\overline{\langle
\bar{f}|X\bar{g}\rangle }$ for all $f,g\in \mathfrak{h}$, respectively. So,
for any $f,g\in \mathfrak{h}$ and every $n\in \mathbb{N}$,%
\begin{equation}
\langle f|\left( X^{\mathrm{t}}\right) ^{2}g\rangle =\langle \overline{X^{%
\mathrm{t}}g}|X\bar{f}\rangle =\langle \overline{X\bar{f}}|X^{\mathrm{t}%
}g\rangle =\langle \bar{g}|X^{2}\bar{f}\rangle =\langle f|\left(
X^{2}\right) ^{\mathrm{t}}g\rangle \ .
\end{equation}%
Then, the assertion $\left( X^{\mathrm{t}}\right) ^{n}=\left( X^{n}\right) ^{%
\mathrm{t}}$ for all $n\in \mathbb{N}$ follows by induction. If $X$ is
invertible then, for any $f,g\in \mathfrak{h}$,
\begin{equation}
\langle f|X^{\mathrm{t}}\left( X^{-1}\right) ^{\mathrm{t}}g\rangle =\langle
\overline{\left( X^{-1}\right) ^{\mathrm{t}}g}|X\bar{f}\rangle =\langle
\overline{X\bar{f}}|\left( X^{-1}\right) ^{\mathrm{t}}g\rangle =\langle \bar{%
g}|X^{-1}X\bar{f}\rangle =\langle f|g\rangle
\end{equation}%
and, similar to this,
\begin{equation}
\langle f|\left( X^{-1}\right) ^{\mathrm{t}}X^{\mathrm{t}}g\rangle =\langle
f|g\rangle \ .
\end{equation}%
Therefore, $\left( X^{-1}\right) ^{\mathrm{t}}=\left( X^{\mathrm{t}}\right)
^{-1}$. As a consequence, $\left( X^{\mathrm{t}}\right) ^{n}=\left(
X^{n}\right) ^{\mathrm{t}}$ for all $n\in \mathbb{Z}$. The proof of $\bar{X}%
^{n}=\overline{X^{n}}$ is performed in the same way. We omit the details.
Furthermore, by taking any orthonormal bases $\{\eta _{k}\}_{k=1}^{\infty
},\{\psi _{k}\}_{k=1}^{\infty }\subseteq \mathfrak{h}$ and $m\in \mathbb{N}$
one finds%
\begin{equation}
\underset{k=1}{\overset{m}{\sum }}\langle \eta _{k}|X^{\mathrm{t}}\psi
_{k}\rangle =\underset{k=1}{\overset{m}{\sum }}\langle \bar{\psi}_{k}|X\bar{%
\eta}_{k}\rangle \ .
\end{equation}%
It follows that $X\in \mathcal{L}^{1}(\mathfrak{h})$ iff $X^{\mathrm{t}}\in
\mathcal{L}^{1}(\mathfrak{h})$. Furthermore, if $X\in \mathcal{L}^{1}(%
\mathfrak{h})$ then we choose a \emph{real} orthonormal basis $%
\{g_{k}\}_{k=1}^{\infty }\subseteq \mathfrak{h}$ to deduce that
\begin{equation}
\mathrm{tr}(X^{\mathrm{t}})=\underset{k=1}{\overset{\infty }{\sum }}\langle
g_{k}|X^{\mathrm{t}}g_{k}\rangle =\underset{k=1}{\overset{\infty }{\sum }}%
\langle g_{k}|Xg_{k}\rangle =\mathrm{tr}(X)\ .
\end{equation}%
Similar to this, $X\in \mathcal{L}^{1}(\mathfrak{h})$ iff $\bar{X}\in
\mathcal{L}^{1}(\mathfrak{h})$, and if $X\in \mathcal{L}^{1}(\mathfrak{h})$
then $\mathrm{tr}(\bar{X})=\overline{\mathrm{tr}(X)}$.\hfill $\Box $

In the next lemma, we use a strongly differentiable family $%
(Y_{t})_{t>0}\subset \mathcal{B}\left( \mathfrak{h}\right) $\ such that $%
(\partial _{t}Y_{t})_{t>0}\subset \mathcal{B}\left( \mathfrak{h}\right) $ is
locally uniformly bounded. This means that $\left\Vert \partial
_{t}Y_{t}\right\Vert _{\mathrm{op}}$ is uniformly bounded on any compact set
of $\left( 0,\infty \right) $.

\begin{lemma}[Trace and time derivatives]
\label{lemma existence 3 copy(4)}\mbox{ }\newline
Let $(X_{t})_{t>0}\in C[\mathbb{R}^{+};\mathcal{L}^{2}(\mathfrak{h})]$ be a
family of Hilbert--Schmidt operators which is differentiable in $\mathcal{L}%
^{2}\left( \mathfrak{h}\right) $ and $(Y_{t})_{t>0}\subset \mathcal{B}\left(
\mathfrak{h}\right) $ be a strongly differentiable family\ such that $%
(\partial _{t}Y_{t})_{t>0}\subset \mathcal{B}\left( \mathfrak{h}\right) $ is
locally uniformly bounded. Then, for any $t>0$,
\begin{equation}
\partial _{t}\left\{ \mathrm{tr}(X_{t}^{\ast }Y_{t}X_{t})\right\} =\mathrm{tr%
}(\partial _{t}\left\{ X_{t}^{\ast }\right\} Y_{t}X_{t})+\mathrm{tr}%
(X_{t}^{\ast }Y_{t}\partial _{t}\left\{ X_{t}\right\} )+\mathrm{tr}%
(X_{t}^{\ast }\partial _{t}\left\{ Y_{t}\right\} X_{t})\ .
\end{equation}
\end{lemma}

\noindent \textbf{Proof.} For any $\epsilon >-t$, the equality%
\begin{eqnarray}
&&\epsilon ^{-1}\left( X_{t+\epsilon }^{\ast }Y_{t+\epsilon }X_{t+\epsilon
}-X_{t}^{\ast }Y_{t}X_{t}\right)  \notag \\
&&-\partial _{t}\left\{ X_{t}^{\ast }\right\} Y_{t}X_{t}-X_{t}^{\ast
}\partial _{t}\left\{ Y_{t}\right\} X_{t}-X_{t}^{\ast }Y_{t}\partial
_{t}\left\{ X_{t}\right\}  \notag \\
&=&X_{t}^{\ast }\left( \epsilon ^{-1}\left( Y_{t+\epsilon }-Y_{t}\right)
-\partial _{t}Y_{t}\right) X_{t}  \notag \\
&&+\left( X_{t+\epsilon }^{\ast }-X_{t}^{\ast }\right) \left\{ \epsilon
^{-1}\left( Y_{t+\epsilon }-Y_{t}\right) \right\} X_{t+\epsilon }  \notag \\
&&+X_{t}^{\ast }\left\{ \epsilon ^{-1}\left( Y_{t+\epsilon }-Y_{t}\right)
\right\} \left( X_{t+\epsilon }-X_{t}\right)  \notag \\
&&+\left( \epsilon ^{-1}\left( X_{t+\epsilon }^{\ast }-X_{t}^{\ast }\right)
-\partial _{t}X_{t}^{\ast }\right) Y_{t}X_{t+\epsilon }  \notag \\
&&+\partial _{t}\left\{ X_{t}^{\ast }\right\} Y_{t}\left( X_{t+\epsilon
}-X_{t}\right)  \notag \\
&&+X_{t}^{\ast }Y_{t}\left( \epsilon ^{-1}\left( X_{t+\epsilon
}-X_{t}\right) -\partial _{t}X_{t}\right) \ ,
\end{eqnarray}%
combined with the cyclicity of the trace (Lemma \ref{lemma existence 3
copy(3)} (i)), the Cauchy--Schwarz inequality, and the continuity of the map
$Z\mapsto Z^{\ast }$ in the Hilbert--Schmidt topology, implies the upper
bound
\begin{eqnarray}
&&\left\vert \epsilon ^{-1}\left( \mathrm{tr}(X_{t+\epsilon }^{\ast
}Y_{t+\epsilon }X_{t+\epsilon })-\mathrm{tr}(X_{t}^{\ast }Y_{t}X_{t})\right)
\right.  \notag \\
&&\left. -\mathrm{tr}(\partial _{t}\left\{ X_{t}^{\ast }\right\} Y_{t}X_{t})-%
\mathrm{tr}(X_{t}^{\ast }Y_{t}\partial _{t}\left\{ X_{t}\right\} )-\mathrm{tr%
}(X_{t}^{\ast }\partial _{t}\left\{ Y_{t}\right\} X_{t})\right\vert  \notag
\\
&\leq &\left( \left\Vert \epsilon ^{-1}\left( Y_{t+\epsilon }-Y_{t}\right)
\right\Vert _{\mathrm{op}}\left( \left\Vert X_{t+\epsilon }\right\Vert
_{2}+\left\Vert X_{t}\right\Vert _{2}\right) +\left\Vert \partial
_{t}X_{t}\right\Vert _{2}\left\Vert Y_{t}\right\Vert _{\mathrm{op}}\right)
\left\Vert X_{t+\epsilon }-X_{t}\right\Vert _{2}  \notag \\
&&+\left\Vert Y_{t}\right\Vert _{\mathrm{op}}\left( \left\Vert X_{t+\epsilon
}\right\Vert _{2}+\left\Vert X_{t}\right\Vert _{2}\right) \left\Vert
\epsilon ^{-1}\left( X_{t+\epsilon }-X_{t}\right) -\partial
_{t}X_{t}\right\Vert _{2}  \notag \\
&&+\left\vert \mathrm{tr}\left( X_{t}^{\ast }\left( \epsilon ^{-1}\left(
Y_{t+\epsilon }-Y_{t}\right) -\partial _{t}Y_{t}\right) X_{t}\right)
\right\vert \ .  \label{ine conbis}
\end{eqnarray}%
We observe that there is a constant $\mathrm{K}>0$ such that
\begin{equation}
\left\Vert \epsilon ^{-1}\left( Y_{t+\epsilon }-Y_{t}\right) \right\Vert _{%
\mathrm{op}}\leq \epsilon ^{-1}\int_{t}^{t+\epsilon }\left\Vert \partial
_{\tau }Y_{\tau }\right\Vert _{\mathrm{op}}\mathrm{d}\tau <\mathrm{K}
\label{ine conbisbis}
\end{equation}%
for sufficiently small $\left\vert \epsilon \right\vert $, because $%
(\partial _{t}Y_{t})_{t>0}\subset \mathcal{B}\left( \mathfrak{h}\right) $ is
locally uniformly bounded. Then, by computing the trace with some
orthonormal basis and using Lebesgue's dominated convergence theorem, we
obtain that
\begin{equation}
\underset{\epsilon \rightarrow 0}{\lim }\ \mathrm{tr}\left( X_{t}^{\ast
}\left\{ \epsilon ^{-1}\left( Y_{t+\epsilon }-Y_{t}\right) \right\}
X_{t}\right) =\mathrm{tr}(X_{t}^{\ast }\partial _{t}\left\{ Y_{t}\right\}
X_{t})\ .  \label{ine conbisbisbis}
\end{equation}%
Moreover, by assumption,%
\begin{equation}
\underset{\epsilon \rightarrow 0}{\lim }\left\Vert X_{t+\epsilon
}-X_{t}\right\Vert _{2}=\underset{\epsilon \rightarrow 0}{\lim }\left\Vert
\epsilon ^{-1}\left( X_{t+\epsilon }-X_{t}\right) -\partial
_{t}X_{t}\right\Vert _{2}=0\ .  \label{ine conbisbisbisbisbis}
\end{equation}%
Therefore, we arrive at the assertion by using (\ref{ine conbis})--(\ref{ine
conbisbisbisbisbis}) and the boundedness of the families $%
(X_{t})_{t>0}\subset \mathcal{L}^{2}\left( \mathfrak{h}\right) $ and $%
(Y_{t})_{t>0}\subset \mathcal{B}\left( \mathfrak{h}\right) $, respectively
in the Hilbert--Schmidt and norm topologies. \hfill $\Box $

\begin{lemma}[Trace and Fubini's theorem]
\label{lemma existence 3 copy(5)}\mbox{ }\newline
Let $\mathcal{I}\subseteq \lbrack 0,\infty )$ and $(X_{\lambda })_{\lambda
\in \mathcal{I}}\subset \mathcal{B}\left( \mathfrak{h}\right) $ be any
family of bounded operators satisfying
\begin{equation}
\int_{\mathcal{I}}\left\Vert X_{\lambda }\right\Vert _{\mathrm{op}}\ \mathrm{%
d}\lambda <\infty \ .  \label{assumption}
\end{equation}%
Then the operator%
\begin{equation}
\left( Y:=\int_{\mathcal{I}}X_{\lambda }\ \mathrm{d}\lambda \right) \in
\mathcal{B}\left( \mathfrak{h}\right)
\end{equation}%
is bounded and furthermore,%
\begin{equation}
\forall Z\in \mathcal{L}^{2}\left( \mathfrak{h}\right) :\qquad \mathrm{tr}%
\left( Z^{\ast }YZ\right) =\int_{\mathcal{I}}\mathrm{tr}\left( Z^{\ast
}X_{\lambda }Z\right) \mathrm{d}\lambda \ .
\end{equation}
\end{lemma}

\noindent \textbf{Proof.} Note first that the boundedness of $Y$ is an
immediate consequence of (\ref{assumption}) and the triangle inequality
satisfied by the operator norm. Now, by using the Cauchy--Schwarz inequality
and (\ref{assumption}) we observe that, for any $f,g\in \mathfrak{h}$,
\begin{eqnarray}
\int_{\mathcal{I}}\left( \int_{\mathcal{M}}\left\vert \overline{f\left(
x\right) }\left( X_{\lambda }g\right) \left( x\right) \right\vert \mathrm{d}%
\mathfrak{a}\left( x\right) \right) \mathrm{d}\lambda &=&\int_{\mathcal{I}}%
\big{\langle}%
\left\vert f\right\vert
\big{|}%
\left\vert X_{\lambda }g\right\vert
\big{\rangle}%
\mathrm{d}\lambda \leq \left\Vert f\right\Vert \int_{\mathcal{I}}\left\Vert
X_{\lambda }g\right\Vert \mathrm{d}\lambda  \notag \\
&\leq &\left\Vert f\right\Vert \left\Vert g\right\Vert \int_{\mathcal{I}%
}\left\Vert X_{\lambda }\right\Vert _{\mathrm{op}}\mathrm{d}\lambda <\infty
\ .
\end{eqnarray}%
Therefore, Fubini's theorem implies that%
\begin{equation}
\forall f,g\in \mathfrak{h}:\qquad \langle f|Yg\rangle =\int_{\mathcal{I}%
}\langle f|X_{\lambda }g\rangle \mathrm{d}\lambda \ .  \label{fubini1}
\end{equation}%
Taking any orthonormal basis $\{g_{k}\}_{k=1}^{\infty }\subseteq \mathfrak{h}
$ we also infer from (\ref{assumption}) that%
\begin{equation}
\forall Z\in \mathcal{L}^{2}\left( \mathfrak{h}\right) :\quad \int_{\mathcal{%
I}}\ \underset{k=1}{\overset{\infty }{\sum }}\left\vert \langle
Zg_{k}|X_{\lambda }Zg_{k}\rangle \right\vert \ \mathrm{d}\lambda \leq
\left\Vert Z\right\Vert _{2}^{2}\int_{\mathcal{I}}\left\Vert X_{\lambda
}\right\Vert _{\mathrm{op}}\mathrm{d}\lambda <\infty \ .  \label{fubini2}
\end{equation}%
Consequently, we combine (\ref{fubini1})--(\ref{fubini2}) with Fubini's
theorem to arrive at the equalities
\begin{equation}
\mathrm{tr}\left( Z^{\ast }YZ\right) =\int_{\mathcal{I}}\ \underset{k=1}{%
\overset{\infty }{\sum }}\langle Zg_{k}|X_{\lambda }Zg_{k}\rangle \mathrm{d}%
\lambda =\int_{\mathcal{I}}\mathrm{tr}\left( Z^{\ast }X_{\lambda }Z\right)
\mathrm{d}\lambda \ ,
\end{equation}%
for any $Z\in \mathcal{L}^{2}\left( \mathfrak{h}\right) $.\hfill $\Box $

\bigskip

\noindent \textbf{Acknowledgement\smallskip }

\noindent We thank W. de Siqueira Pedra, L. Fanelli, H. Grosse, J. Hoppe, G.
Lechner, A. Miclke, Phan Thành Nam and M. Walser for useful discussions and
comments and for providing important references. We also thank the referees
for their work and constructive criticisms. Support by IHP network
HPRN-CT-2002-00277 of the European Union, the grants MTM2010-16843,
MTM2014-53850, the BCAM Severo Ochoa accreditation SEV-2013-0323 of the
Spanish Ministry of Economy and Competitiveness MINECO, the grant IT641-13
and the BERC 2014-2017 program financed by the Basque government is
gratefully acknowledged. $\bigskip $

\noindent \textbf{Volker Bach} \\
Institut für Analysis und Algebra, TU Braunschweig, Pockelsstraße 11,
38106 Braunschweig
\vspace{0.5cm}

\noindent \textbf{Jean-Bernard Bru} \\ Departamento de Matem\'{a}ticas, Facultad de Ciencia y Tecnolog\'{\i}a, Universidad del Pa\'{\i}s Vasco, Apartado 644, 48080 Bilbao \\ BCAM - Basque Center for Applied Mathematics, Mazarredo, 14. 48009 Bilbao\\ IKERBASQUE, Basque Foundation for Science, 48011, Bilbao

\end{document}